\documentclass[dissertation]{uathesis}

\usepackage{graphicx}
\usepackage{epsfig}
\usepackage{natbib}			

\usepackage{aastex_hack}		
\usepackage{ifpdf}
\ifpdf
\DeclareGraphicsExtensions{.pdf,.png,.jpg,.mps}
\else
\DeclareGraphicsExtensions{.eps}
\fi

\usepackage[breaklinks = true]{hyperref}
\usepackage{color}

\completetitle{Effects of Turbulent Magnetic Fields on the
Transport and Acceleration of Energetic Charged Particles:
Numerical Simulations with Application to Heliospheric Physics}
\fullname{Fan Guo}			
\degreename{Doctor of Philosophy}	

\begin{document}

\maketitlepage
{DEPARTMENT OF PLANETARY SCIENCES}	
{2012}							


\approval
{22 August 2012}		
{Joe Giacalone}	
{Joe Giacalone}
{J. R. Jokipii}	
{Roger Yelle}		
{Jozsef Kota}		
{Ke Chiang Hsieh}		

\statementbyauthor


\incacknowledgements{acknowledgements}

\incdedication{dedication}


\tableofcontents

\listoffigures

\listoftables

\incabbreviations{abbreviations}

\incabstract{abstract}

\chapter{Introduction and Background\label{chap1}}

\section{Overview \label{overview1}}
\begin{figure}
\begin{center}
\includegraphics[width=0.6\textwidth]{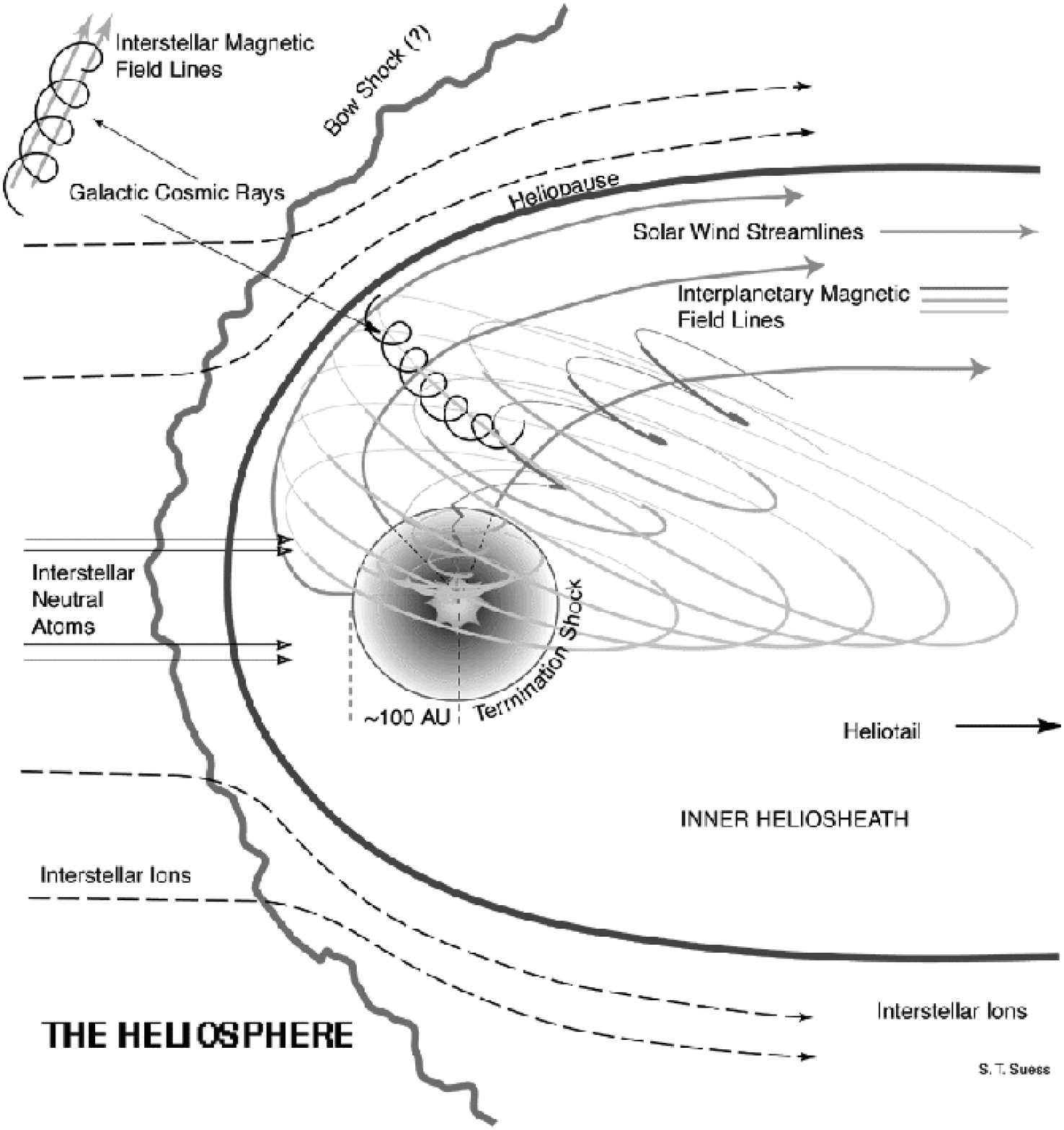}
\caption[The interaction between the interstellar medium and the heliosphere.]{The interaction between interstellar medium and the heliosphere. Figure provided courtesy of Steve T. Suess. \label{fig2}}
\end{center}
\end{figure}

 The heliosphere (Figure \ref{fig2}) is structured by plasma flows and magnetic fields. As the supersonic ``solar wind" \citep{Parker1958a} expands from the solar atmosphere, it forces plasmas and magnetic fields outward, forming the cavity that interacts with the surrounding interstellar medium. At around $100$ AU, the solar wind suddenly slows down and forms the termination shock. The heliosphere is a natural laboratory for physical processes involving charged particles and changing magnetic field. The related physical quantities are continuously measured by spacecraft and ground-based instruments. In particular, the Voyager spacecraft are currently exploring the heliopause, which is the boundary between the heliosphere and the local interstellar medium. It marks the frontier of the solar system and the farthest distance ($\sim 121$ AU at the time of this writing) that man-made instruments have ever reached.

Energetic charged particles are a minor component of space plasmas, but have important and profound effects. They are serious concerns to space weather because of their hazardous effects to astronauts and space satellites. Energetic charged particles propagate at high speeds and carry important information about their source regions and the media they propagate through. Since the early measurements by Victor Hess in $1912$, scientists have been measuring energetic particles with energies up to about $10^{21}$ eV and the electromagnetic radiation they produce.

The acceleration and transport of energetic charged particles are fundamental problems in space physics and astrophysics, in which electric and magnetic fields often play an essential role. From the first principle, the motions of charged particles in electromagnetic fields are governed by the Lorentz force. The large-scale transport and acceleration of energetic charged particles are often described by the Parker's transport equation \citep[][see Equation \ref{parker_equation}]{Parker1965}, and the effect of turbulent magnetic fields is approximated by a large-scale spatial diffusion tensor \citep{Jokipii1966,Jokipii1971}. However, recent studies suggest that the simple spatial diffusion approximation may be oversimplified and the observed behavior of energetic particles is often more complicated. For example, during some solar energetic particle events, the intensity-time profiles often show small-scale sharp variations. In addition, when the velocity distribution of charged particles is highly anisotropic, it cannot be properly described by the transport equation that assumes a quasi-isotropic distribution. It is important to adequately describe this for problems such as the evolution of the velocity distribution of energetic particles during their propagation and the acceleration of low-energy particles at shocks.
Understanding this is a challenge to theoretical studies due to the complex nature of particle trajectories in turbulent magnetic fields. Thanks to recent advances in computing capabilities, the acceleration and transport of charged particles can be studied by numerical simulations. This dissertation focuses on understanding the transport and acceleration of energetic charged particles in the existence of turbulent magnetic fields.

\section{Charged Particles in the Heliosphere  \label{chapter1-particle2}}

\begin{figure}
\begin{center}
\includegraphics[width=0.8\textwidth]{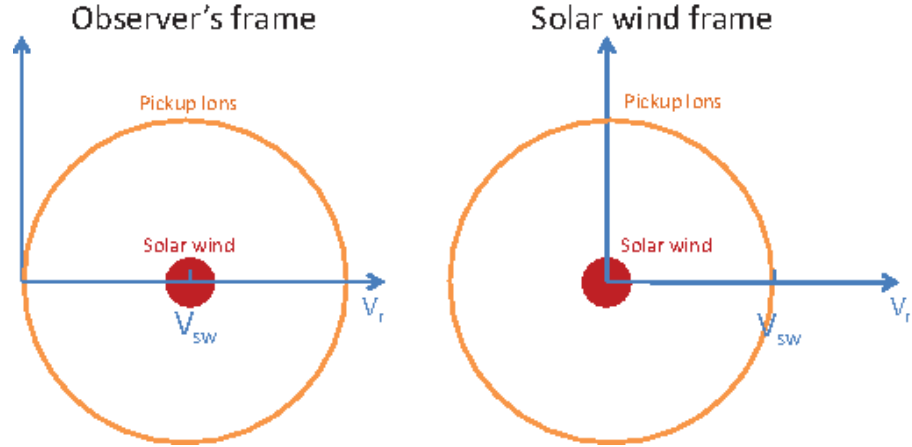}
\caption[The velocity distribution of the solar wind ions and pickup ions]{The velocity distribution of the solar wind ions and pickup ions in observer's frame and the solar wind frame. \label{pickup}}
\end{center}
\end{figure}

The heliosphere is permeated by charged particles of different origins. There are two important components of charged particles that contribute to the global dynamics inside the heliosphere, i.e., the solar wind and pickup ions. The solar wind is a continuous plasma flow coming from the solar atmosphere \citep{Parker1958a}. It is accelerated to supersonic speeds close to the Sun and propagates outward. The solar wind is often divided into two distinct components, termed as the slow solar wind and fast solar wind. The fast solar wind represents a plasma flow with a temperature about $8 \times 10^5$ K and a speed of about $800$ km/s, whereas the slow solar wind represents a hotter ($T \sim 1.5 \times 10^6$ K) and denser plasma flow with a slower speed of about $400$ km/s \citep{Meyer-Vernet2007}. The observations of the solar wind at different latitudes by the Ulysses spacecraft have shown that the slow solar wind is confined to ``streamer belt" that is about 20 degrees around the heliospheric current sheet at solar minimum. At the same time the fast solar wind entirely occupies higher latitudes. At solar maximum, the solar wind becomes more mixed and complicated \citep{McComas2003}. Pickup ions are mainly originated from interstellar neutral particles \citep{Axford1972,Vasyliunas1976,Gloeckler1995}, with minor contributions from inner source pickup ions \citep{Geiss1995,Gloeckler1998,Schwadron2000} and pickup ions from comets \citep{Ipavich1986,Gloeckler1986}. Interstellar neutral particles can freely penetrate into the heliosphere before they are ionized by charge exchange and/or solar ultraviolet radiation. Once the neutral particles are ionized, they are influenced by electric and magnetic fields (see Section \ref{HelioMagneticField}). As illustrated in Figure \ref{pickup}, when the magnetic field vector has a component that is perpendicular to the solar wind velocity vector, the electric and magnetic fields embedded in the solar wind force them to accelerate and make gyro-motions in the frame co-moving with the solar wind. This is the so called ``pickup" process. In observer's frame, the pickup ions have velocities from zero to two times of solar wind speed. After the pickup process, the gyroradii of pickup ions are several times larger than that of the solar wind particles. The gyro-motion forms a ring velocity distribution perpendicular to the ambient magnetic field. The distribution is unstable and expected to generate electromagnetic waves. The ``pickup ions" will be scattered into an isotropic distribution by the electromagnetic waves \citep{Wu1972,Wu1986} and/or background turbulence. Starting from a heliocentric distance of about $30$ AU, the pickup ions play a dominant role in the physics of the outer heliosphere by dominating the pressure of the plasma flow \citep{Richardson2009}. It should be noted that neutral particles could also influence the dynamics in the heliosphere, which has been discussed extensively \citep[][and references therein]{Zank1999}.

\begin{figure}
\begin{center}
\begin{tabular}{c}
\hspace{1mm} \epsfig{file=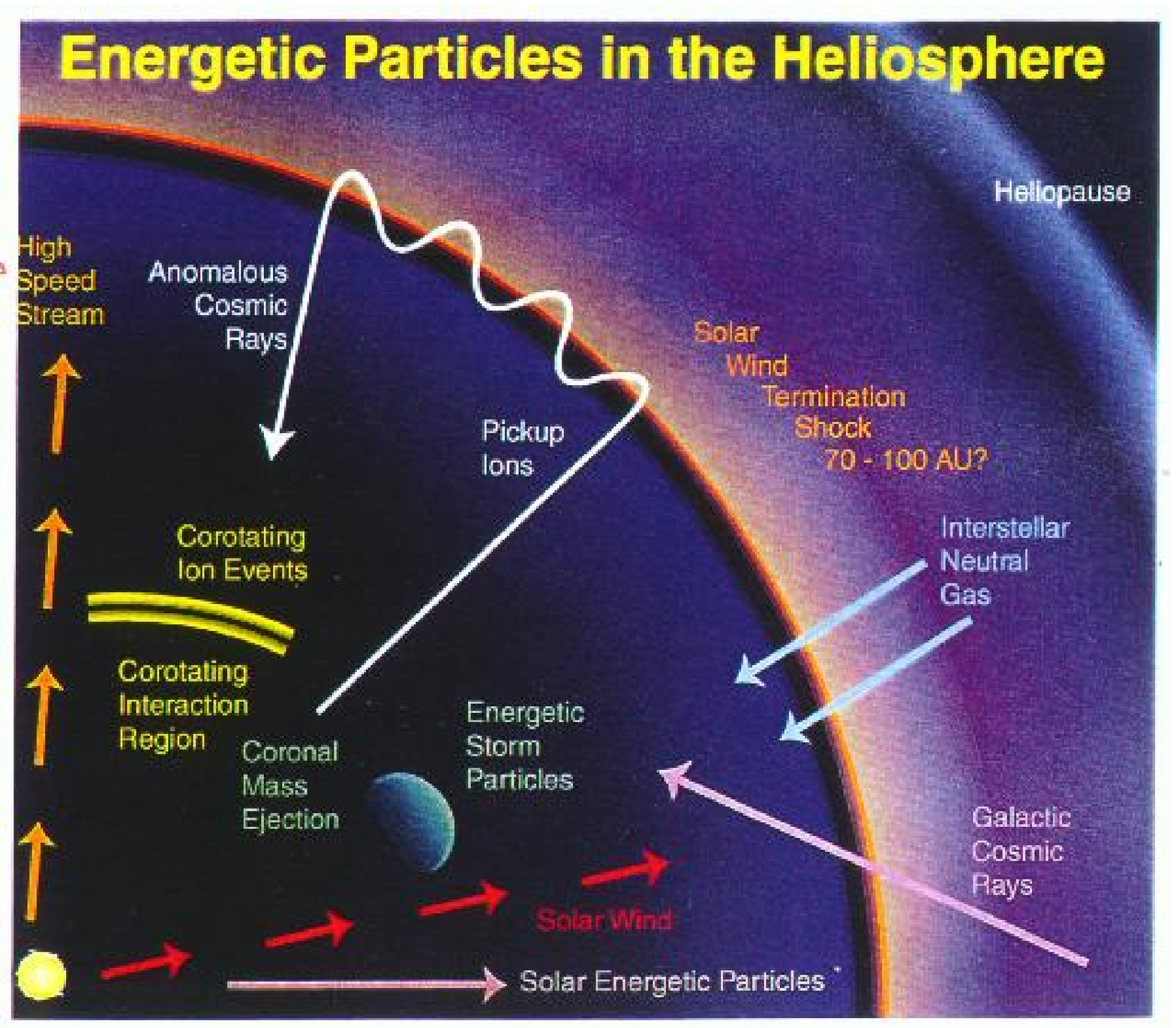,width=0.65\textwidth,clip=} \\ \vspace{4mm} 
\epsfig{file=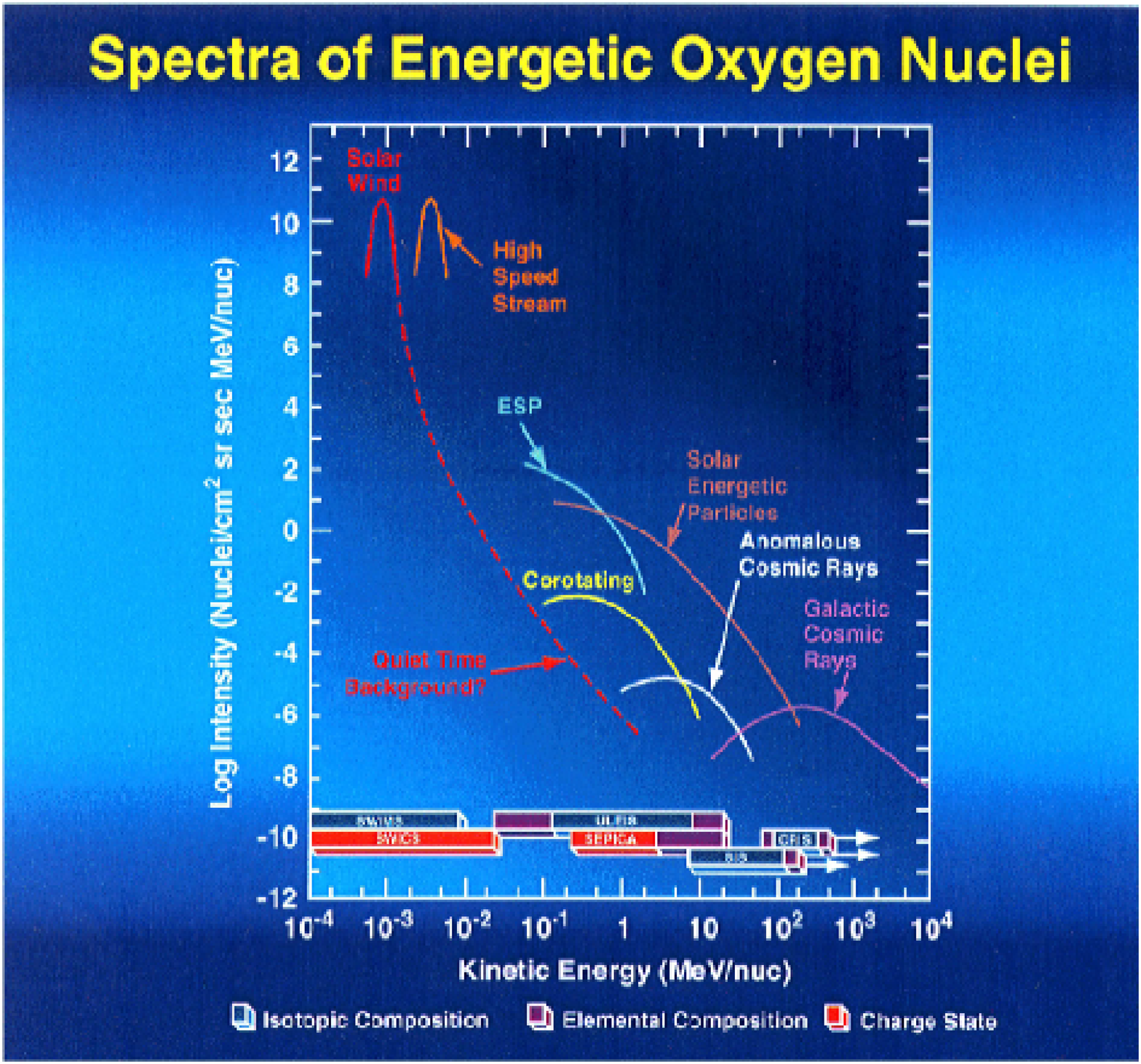,width=0.64\textwidth,clip=}
\end{tabular}
\caption[Various types of energetic charged particles in the heliosphere]{Various types of energetic charged particles in the heliosphere. The upper panel indicates related physical processes in different regions of the heliosphere. The bottom panel indicates typical energy spectra of energetic charged particles observed at $1$ AU. See text for more details. The pictures are adapted from \citet{Stone1998} with permission by Springer Science + Business Media. \label{fig3}}
\end{center}
\end{figure}

Energetic charged particles are a high-energy, non-thermal component of the plasmas in the heliosphere. They carry large kinetic energies, so their speeds are much faster than background fluid. They have significant effects on space weather and their physics is important to consider. When the energy density of energetic charged particles is large enough, they may even mediate the dynamics of plasma flows \citep[e.g.,][]{Florinski2009}.
Energetic charged particles in the heliosphere have different components depending on their acceleration sites, energy ranges, charge states, and elemental compositions, etc. Figure \ref{fig3} illustrates various types of charged particles in the heliosphere, their acceleration regions, and their typical energy spectra. In this figure, the solar wind with energies in keV range represents the kinetic energy of the background flow in the heliosphere. Solar energetic particles (SEPs) are usually observed to be from several hundred keV/nucleon to tens of MeV/nucleon for typical events, and even more than 1 GeV/nucleon in extreme events. The SEP events are usually divided into two classes based on the progress made in the last three decades, termed as ``impulsive events'' and ``gradual events'' \citep[see][and references therein]{Reames1999}. The SEPs associated with impulsive events are thought to be accelerated in solar flares. Impulsive events are characterized by the impulsive peaks in their intensity-time profiles, confined source regions in longitude, high ionization states, and overabundance in isotope ratios such as $^3$He/$^4$He and Fe/O compared with the values in the solar wind. Energetic particles related to gradual events are thought to be accelerated by collisionless shocks driven by coronal mass ejections (CMEs). The gradual events have extended intensity-time profiles because of the continuous acceleration at shock fronts. They are also more widely distributed in longitude, indicating a broadened source region \citep{Reames1999}.  Recently, this bi-model picture has been challenged. It has been found that for most events, SEPs appear to have a mixed property of the two classes of events. For example, some large gradual SEP events show a substantially high ionization charge state \citep{Mazur1999} and enhanced isotope ratios in $^3$He/$^4$He and Fe/O \citep{Cohen1999,Mason1999}. In large SEP events, flares and CMEs usually occur together, therefore one may expect that both of the processes play a role. Identifying their relations and contributions to large SEP events is still under hot debate. Energetic storm particles (ESPs) are usually associated with the passage of travelling interplanetary shocks, where the ions can often be accelerated to from several tens of keV/nucleon to MeV/nucleon, and occasionally more than $10$ MeV/nucleon. Energetic charged particles associated with corotating interaction regions (CIRs) formed by the interaction between the fast and slow solar wind streams are often observed. These particles appear to have a higher energy range compared to that of ESPs. Anomalous cosmic rays (ACRs) are thought to be originated from pickup ions \citep{Fisk1974}, and are accelerated to energies between several MeV/nucleon to $100$ MeV/nucleon at the termination shock \citep{Pesses1981}. Galactic cosmic rays (GCRs) that have energies typically larger than $10$ MeV/nucleon are coming from the outside of the heliosphere. 

One remarkable feature in Figure \ref{fig3} is that the energy spectra of the accelerated ions are often close to a power law, which indicates a common acceleration process. The acceleration of electrons is also observed, and sometimes accompanies the acceleration of ions. We will discuss the acceleration of electrons in Chapter $4$. It is worthwhile to note that many, if not all, of the energetic charged particles are associated with shock waves. For example, it is now established that many SEP events, especially gradual events are associated with CME-driven shocks \citep{Reames1999}. Energetic particles in impulsive events are accelerated in solar flares. The mechanism for particle acceleration in solar flares is not clear so far. In Chapter $4$ we show that collisionless shocks are a possible candidate for the energization of charged particles in solar flare regions. Energetic particles can be accelerated at interplanetary shocks driven by interplanetary CMEs and CIRs. ACRs are thought to be pickup ions accelerated at the solar wind termination shock \citep{Pesses1981}. GCRs are usually thought to be accelerated by astrophysical shocks such as supernova blast waves. Their spectrum is observed as the remarkable power law spectrum of cosmic rays.

After the acceleration, the energetic particles travel in and interact with the heliospheric magnetic field. Understanding the propagation of energetic particles is difficult since, in general, the motion of charged particles in a turbulent magnetic field is very complicated. The transport of energetic particles in heliospheric magnetic field is usually considered to be a diffusion process \citep{Jokipii1966,Jokipii1971}. In principle, if the motion of energetic particles is well understood, they can be used as a tracer of magnetic field structure in the heliosphere.

\section{The Heliospheric Magnetic Field and its Fluctuations  \label{HelioMagneticField}}

Since plasma flows in the heliosphere are highly electrically conductive, the heliospheric magnetic field is frozen in the background fluid to a high degree. It can be inferred from the generalized Ohm's law that the only macroscopic electric field in this situation is due to the motional electric field $\textbf{E} = - \textbf{U}_{flow} \times \textbf{B}/c$ \citep[see, e.g.,][]{Krall1973book}, where $\textbf{U}_{flow}$ is the background flow speed, $\textbf{B}$ is magnetic field vector, and $c$ is the speed of light in \textit{vacuum}. The supersonic solar wind carries the magnetic field to many AU. Because of solar rotation, the magnetic field lines of force have a spiral shape termed as the Parker spiral \citep{Parker1958a,Hundausen1972}. In spherical coordinates (radial heliocentric distance $\hat{r}$, polar angle $\hat{\theta}$, and azimuthal angle $\hat{\phi}$), the average magnetic field is given by

\begin{equation}
\textbf{B} = B_r \hat{r} + B_\phi \hat{\phi} = B_s \frac{R_s^2}{r^2} [\hat{r} - \frac{r \Omega_s \sin\theta}{V_{SW}}(1-\frac{R_s^2}{r^2}) \hat{\phi}],
\end{equation}

\noindent where $B_r$ and $B_\phi$ are the magnetic fields in the $\hat{r}$ and $\hat{\phi}$ direction respectively, $B_s$ is the radial magnetic field close to the Sun at heliocentric distance $R_s$, $V_{SW}$ is the speed of the solar wind, and $\Omega_s$ is the angular frequency of the solar rotation. In this steady-state model, $B_s$, $V_{SW}$, and $\Omega_s$ are assumed to be constants.

At a low heliocentric latitude and a distance of $1$ AU, the average angle between the direction of magnetic fields and the orientation of the solar wind flow at low latitude is about $45$ degrees. In the outer heliosphere, the plasma almost flows transverse to the mean magnetic field. At high latitude regions ($\sin\theta \sim 0$), the azimuthal component of the Parker spiral magnetic field is small. However, it is expected that at a large heliocentric distance, the transverse perturbation of magnetic field lines of force dominates since it decays as $B_{tp} \propto 1/r$. The distant magnetic field almost always transverses to the radial direction while the average magnetic field is still the Parker spiral magnetic field \citep{Jokipii1989}. The inferred large scale fluctuations have been observed by the Ulysses spacecraft \citep{Jokipii1995,Balogh1995}.

\begin{figure}
\begin{tabular}{cc}
\epsfig{file=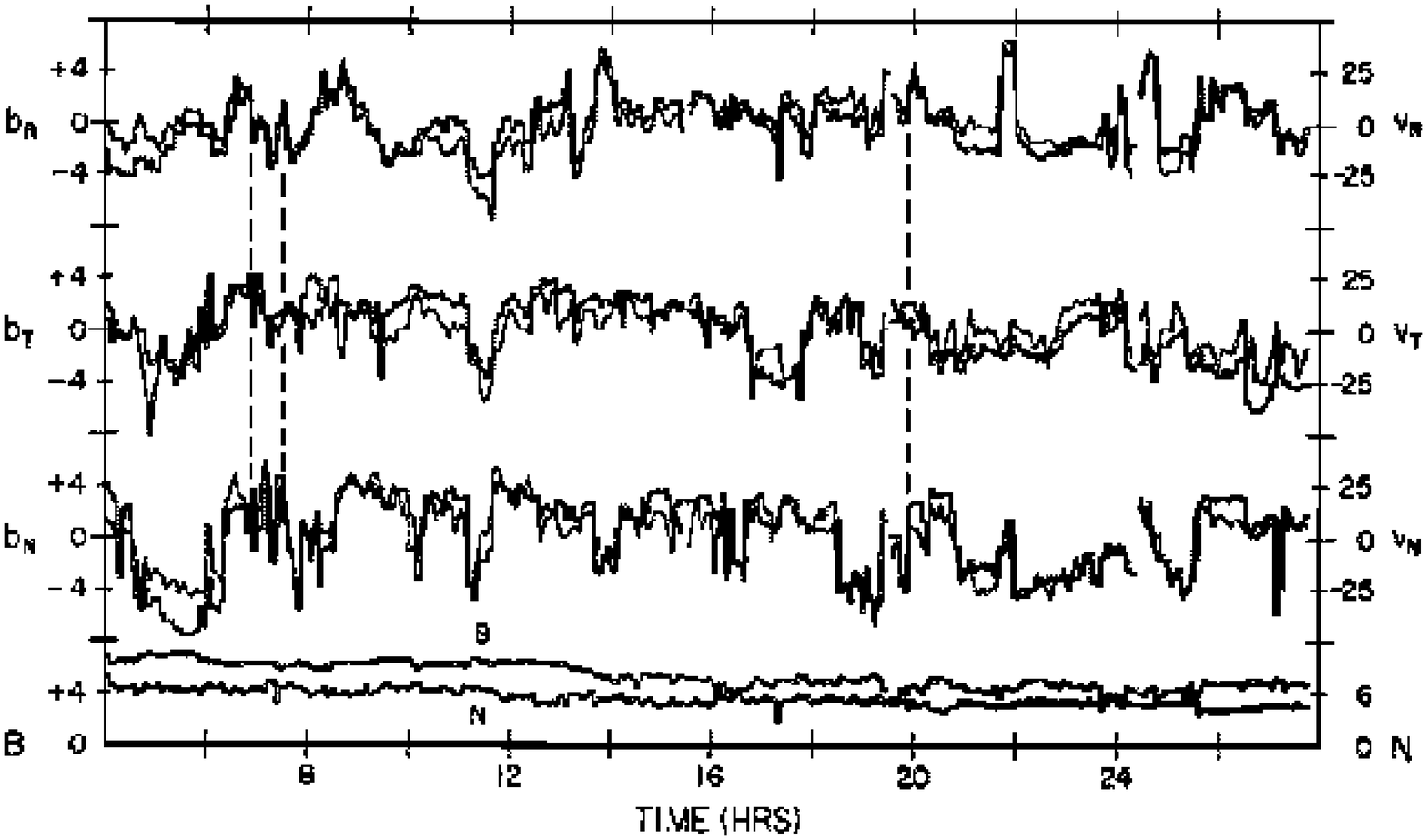,width=0.63\textwidth,clip=} &
\epsfig{file=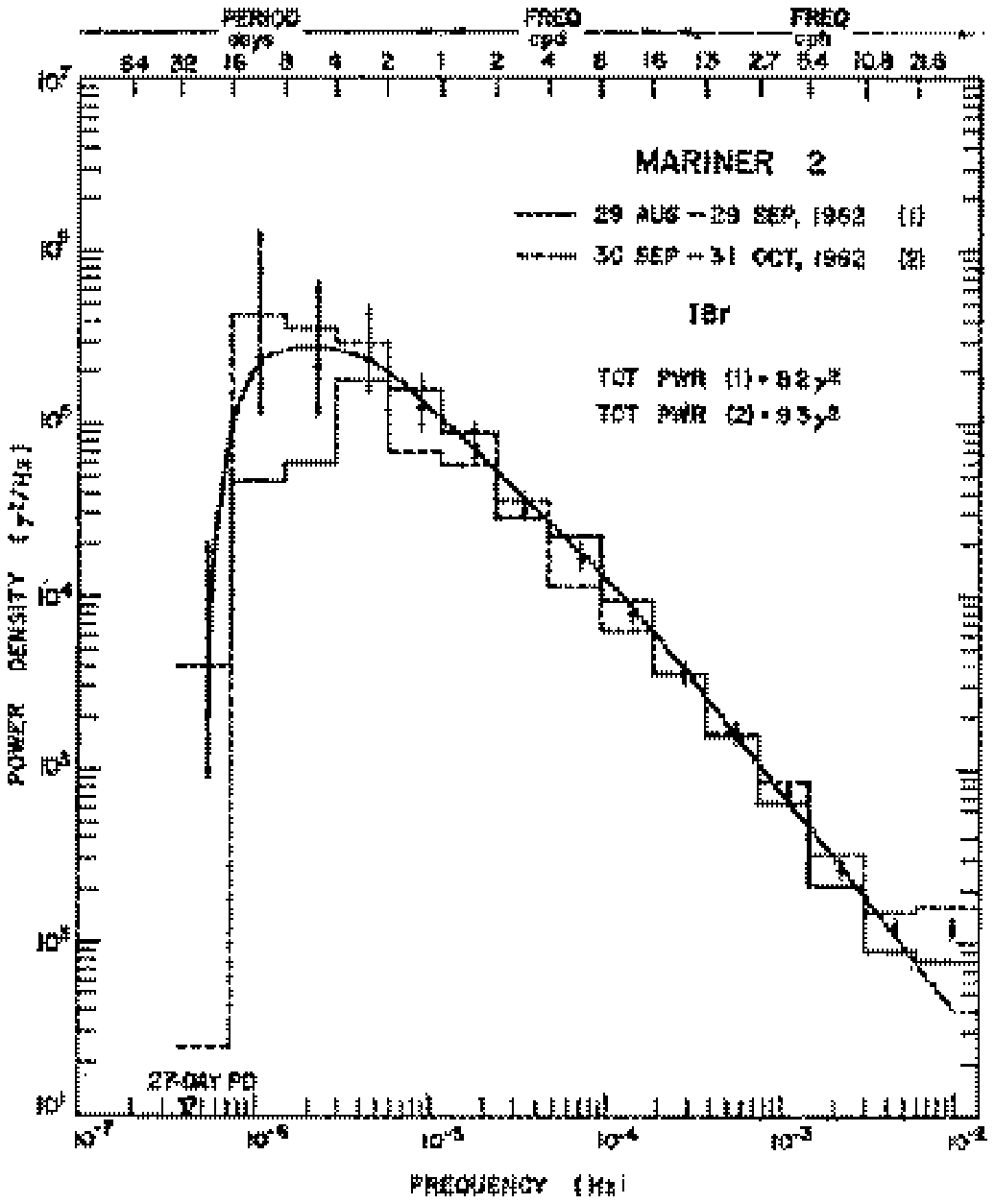,width=0.37\textwidth,clip=}
\end{tabular}
\caption[The spacecraft measurement of the Solar wind turbulence.]{The spacecraft measurement of the Solar wind turbulence. Left: An example of magnetic field and velocity measurements in the solar wind \citep[adapted from][]{Belcher1971}. Right: An example of power spectrum of magnetic field component measured in the solar wind fluctuations \citep[Figure 6 in][]{Coleman1968}. The figure is reproduced by permission of the AAS. \label{magneticturbulence}}
\end{figure}

Magnetic fields in the heliosphere are turbulent \citep{Tu1995,Goldstein1995}. It is often convenient to write the magnetic field as a summation of an average component and a turbulent component $\textbf{B} = \textbf{B}_0 + \delta\textbf{B}$. In the solar wind, the magnetic fluctuations are observed to be highly Alfvenic \citep[][Figure \ref{magneticturbulence}]{Belcher1971}, i.e., the magnetic fluctuation vector is highly correlated with the velocity fluctuation vector. Observations show that the power of turbulent magnetic field $P_B$ versus spatial wave number $k$ is close to a Kolmogorov law $P_B \propto k^{-5/3}$ \citep[][Figure \ref{magneticturbulence}]{Coleman1968}. The power spectrum suggests that most power of the fluctuations is in large spatial scales, and cascades into smaller and smaller spatial scales until dissipation effects are important. The correlation length is observed to be on the order of $10^6$ km at $1$ AU and increases in the outer heliosphere \citep{Matthaeus1982}. Recent theories, numerical simulations, and observations have revealed that the solar wind magnetic turbulence is anisotropic. In other words, most of the fluctuations have wave vectors transverse to the mean magnetic field \citep[e.g.,][]{Matthaeus1990,Goldreich1995,Matthaeus1996}.

Electric and magnetic fluctuations can also be produced by plasma instabilities. For example, freshly created pickup ions can have a ring distribution and excite ion-cyclotron waves \citep{Wu1972,Wu1986}. In the upstream medium of a collisionless shock, the streaming of energetic particles may excite electromagnetic fluctuations \citep{Lee1982,Lee1983}. The fluctuation may be important for particle acceleration at quasi-parallel shocks.

\section{Collisionless Shocks}
Collisionless shocks are believed to be important accelerators for charged particles in the heliosphere. In this section we introduce the basic concepts of collisionless shocks. The acceleration of charged particles will be discussed in Chapter 3 and Chapter 4. 

Shocks are characterized by sharp transitions in the physical quantities of medium such as flow speed, density, magnetic field, and temperature, etc. Since Coulomb collisions are too infrequent in space plasmas, the kinetic energy of plasma flow is dissipated at shocks through other mechanisms such as the interaction between particles and plasma waves. The shocks are termed as collisionless shocks. Dependent on the angle between upstream magnetic field vector and shock normal $\theta_{Bn}$, collisionless shocks can be divided into quasi-perpendicular shocks ($\theta_{Bn} > 45 ^\circ$) and quasi-parallel shocks ($\theta_{Bn} < 45 ^\circ$). In the limit of ideal magnetohydrodynamics (MHD), shocks are a kind of discontinuities. One can relate upstream and downstream medium using MHD conservation laws and Maxwell equations. The result gives the well-known jump conditions for MHD discontinuities \citep[e.g.,][]{Burgess1995}:

\begin{eqnarray}
B_{1n} &=& B_{2n} \\
U_{1n} B_{1t} &=& U_{2n} B_{2t} \\
\rho_1 U_{1n} &=& \rho_2 U_{2n} \\
\rho_1 U_{1n}^2 + P_1 + \frac{B^2_1}{8 \pi} &=& \rho_2 U_{2n}^2 + P_2 + \frac{B^2_2}{8 \pi} \\
\rho_1 U_{1n} U_{1t} - \frac{B_{1n} B_{1t}}{4 \pi} &=& \rho_2 U_{2n} U_{2t} - \frac{B_{2n} B_{2t}}{4 \pi} \\
\rho_1 U_{1n} (\frac{1}{2} U_1^2 + \frac{\gamma}{\gamma - 1}\frac{P_1}{\rho_1}) + U_{1n}\frac{B_1^2}{4\pi} - (\textbf{U}_1 \cdot \textbf{B}_1) \frac{B_{1n}}{4 \pi} &=& \rho_2 U_{2n} (\frac{1}{2} U_2^2 + \\ \nonumber \frac{\gamma}{\gamma - 1}\frac{P_2}{\rho_2}) +  U_{2n}\frac{B_2^2}{4\pi} - (\textbf{U}_2 \cdot \textbf{B}_2) \frac{B_{2n}}{4 \pi}
\end{eqnarray}

\noindent where $B$, $U$, $\rho$, and $P$ represent magnetic field, flow speed, density, and pressure, the subscript ``1'' and ``2'' specify physical quantities in upstream and downstream media, and the subscripts ``n'' and ``t'' mean the normal components and transverse components, respectively. 

The shock solutions of the jump conditions have three possibilities: slow shocks, intermediate shocks, and fast shocks, which correspond to three modes of waves in MHD. In this thesis we only discuss fast shocks. At a fast-mode shock, the flow is decelerated and compressed. The transverse component of magnetic field increases as the magnetic field get compressed at the shock. Fast shocks are most frequently observed shocks in the heliosphere, including planetary bow shocks, CME-driven shocks, most of interplanetary shocks, and the solar wind termination shock.  

\begin{figure}
\begin{center}
\includegraphics[width=0.8\textwidth]{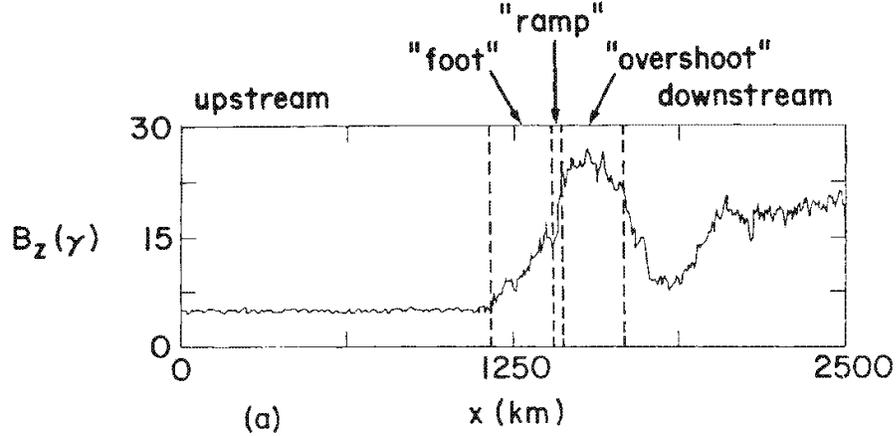}
\caption[The structure of quasi-perpendicular supercritical shocks]{The structure of quasi-perpendicular supercritical shocks. The picture is adapted from \citep{Wu1984b} with permission by Springer Science + Business Media. \label{qperp}}
\end{center}
\end{figure}

\begin{figure}
\begin{center}
\includegraphics[width=0.8\textwidth]{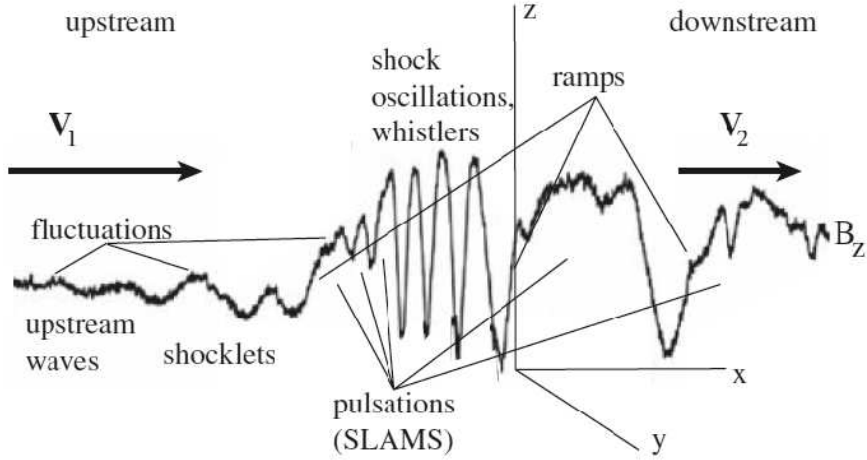}
\caption[The structure of quasi-parallel supercritical shocks]{The structure of quasi-parallel supercritical shocks. The picture is adapted from \citep{Treumann2009} with permission by Springer Science + Business Media. \label{qpara}}
\end{center}
\end{figure}

Strong shocks in the heliosphere are usually supercritical shocks where the shock Mach numbers $M_A > M_{critical}$ and $M_{critical} \leq 2.76$ depends on the shock normal angle and the plasma beta, etc \citep{Stone1985}. In these shocks a fraction of ions get reflected at the shock front, which provide an additional dissipation mechanism as resistivity cannot provide enough dissipation \citep{Tidman1971}. At small spatial scales, shocks have microscopic structures and the structures are often distinct dependent on the shock normal angle $\theta_{Bn}$. For quasi-perpendicular supercritical shocks, the shock structure is relatively well-defined. Since magnetic field vector is mainly perpendicular to the shock normal vector, after ions first encounter the shock, their gyro-motions are confined to the vicinity of the shock front. The quasi-perpendicular shocks are featured by the ``foot-ramp-overshoot" structure, as shown in Figure \ref{qperp} \citep{Leroy1982}. For quasi-parallel shocks, their micro-structures are less clear than that of perpendicular shocks. The micro-structure for quasi-parallel supercritical shocks is shown in Figure \ref{qpara}. Since the background magnetic field is mainly along the shock normal, the reflected ions can travel far upstream. It forms ion beams that excite ion-scale low-frequency waves. These waves grow in amplitude and is shortened in wavelength as they approach the shock. The structures are so called SLAMS (stands for Short Large Amplitude Magnetic Structures).

\section{The Motions of Charged Particles}
The heliosphere provides a natural laboratory to study the physics of charged particles. In this section we discuss the motions of charged particles influenced by a variety of effects. For energetic charged particles in the heliosphere, their motions are completely dominated by the Lorentz force.

When the kinetic energy density of charged particles for the problem of interest is much less than the background field, the motion of charged particles has virtually no feedback to the background field and can be approximated as test particles.
The major forces acting on a charged object include electromagnetic force and gravity.
In spherical coordinates with radial heliocentric distance $\hat{r}$, according to Newton's second law, we have

\begin{equation}
F_{total} = \frac{d \textbf{p}}{d t} =
q(\textbf{E} + \frac{1}{c} \textbf{v} \times \textbf{B}) - \frac{GM_sm}{r^2} \hat{r} + \textsl{other forces},
\end{equation}

\noindent where $\textbf{E}$ and $\textbf{B}$ are electric and magnetic field vectors, $\textbf{p}$, $\textbf{v}$, $q$, and $m$ are the momentum vector, velocity vector, electric charge, and mass of a particle, respectively, $c$ is the speed of light in \textit{vacuum}, $G$ is the gravitation constant, and $M_s$ is the mass of the Sun.

It is easy to show that for energetic charged particles such as ions and electrons in the heliosphere, their motion is dominated by the Lorentz force. For larger particles, such as charged dust grains, other effects like radiation pressure and Poynting-Robertson drag, etc. may also be important dependent on their charge and mass. In this dissertation we focus on energetic charged particles. Their motions in electromagnetic field can be described by

\begin{equation}
F_{EM} = \frac{d \textbf{p}}{d t} = q \textbf{E} + \frac{q}{c} \textbf{v} \times \textbf{B}. \label{lorentz_equation}
\end{equation}

In the simplest case with constant magnetic field $\textbf{B} = B_0 \hat{z}$ and no electric field, the equation has the solution that describes a gyro-motion around a magnetic line of force:
\begin{eqnarray}
v_x &=& v_\perp \cos(\Omega t + \phi_0) \nonumber\\
v_y &=& -v_\perp \sin(\Omega t + \phi_0) \\
v_z &=& v_\parallel \nonumber\\
x    &=& \frac{v_\perp}{\Omega} \sin(\Omega t + \phi_0) + x_0 \nonumber\\
y    &=& \frac{v_\perp}{\Omega} \cos(\Omega t + \phi_0) + y_0\\
z    &=& v_\parallel t + z_0, \nonumber
\end{eqnarray}

\noindent where $v_\perp$ and $v_\parallel$ are velocity components perpendicular and parallel to the magnetic field, respectively. The gyrofrequency $\Omega = qB/mc$. $\phi_0$, $x_0$, $y_0$ and $z_0$ are constant.

When charged particles are moving in electric and magnetic fields that are spatially and temporally dependent, the motions of energetic charged particles are very complicated in general. An important approximation is when the particle moves in a slowly varying electric and magnetic field on spatial and temporal scales much larger than the scale of gyro-motions. In this case the motion of a charged particle can be expressed as a summation of a gyro-motion and a drift motion of the guiding center. In a static electric and magnetic field, the guiding-center motion of charged particles can be expressed as \citep{Boyd2003}:

\begin{equation}
\textbf{v}_{gc} = \left[v_\parallel + \frac{cW_\perp}{qB^3} \textbf{B}\cdot(\nabla\times \textbf{B})\right] \frac{\textbf{B}}{B} + \textbf{U}_{flow\perp} + \frac{cW_\perp}{qB^3}\textbf{B}\times \nabla B + \frac{2cW_\parallel}{q B^4} \textbf{B}\times (\textbf{B}\cdot \nabla)\textbf{B},
\label{drift-motion}
\end{equation}

\noindent where $W_\perp = mv_\perp^2/2$ and $W_\parallel = mv_\parallel^2/2$ are the components of particle kinetic energy perpendicular and parallel to the magnetic field, and $\textbf{U}_{flow\perp}$ is the background flow speed component that is perpendicular to the magnetic field. The first two terms describe the particle motion parallel to the magnetic field including the original velocity parallel to the magnetic field and a small modification caused by parallel drift. The remaining terms are drift motion transverse to a magnetic field line caused by drift in the motional electric field, gradient $B$ drift, and curvature $B$ drift.

Magnetic fluctuations have profound influences on the behavior of energetic charged particles. One important effect is pitch-angle scattering by resonant interactions between charged particles and magnetic fluctuations. When the resonant condition $k_\parallel v_\parallel - \omega - \Omega = 0$ is satisfied, the particle can strongly interact with magnetic fluctuations and the pitch-angles of charged particles may be changed. Here $k_\parallel$ and $\omega$ are the wave number parallel to the magnetic field and angular frequency of the wave, and $v_\parallel$ is the parallel velocity of the particle. The resonant wave-particle interaction results in the evolution of distribution function that can be described by a pitch-angle diffusion.


\section{The Transport and Acceleration of Charged Particles}
Particle transport and acceleration are fundamental problems in space physics and astrophysics. This section gives an overview of this subject.  The large-scale behavior of energetic charged particles is usually approximated by a diffusion process, given the fact that the scattering time is short compared to the time scale of interest. For particles moving in a compression/expansion flow, energetic particles experience an increase/decrease in energy. A complete equation that describes the evolution of a nearly isotropic distribution of energetic charged particles is the well-known Parker transport equation \citep{Parker1965}. The equation describes the large-scale evolution of the distribution function $f(x_i, p, t)$ of the energetic particles with momentum $p$ dependent on the position $x_i$ and time $t$ including effects of diffusion, convection, drift, acceleration and source particles:

\begin{equation}
\frac{\partial f}{\partial t} = \frac{\partial}{\partial
  x_i}\left[\kappa_{ij}\frac{\partial f}{\partial x_j}\right] -
  U_{i}\frac{\partial f}{\partial x_i}-V_{d,i}\frac{\partial f}{\partial x_i}+
  \frac{1}{3}\frac{\partial U_{i}}{\partial x_i} \left[\frac{\partial f}{\partial \ln p} \right] +
  Q, \label{parker_equation}
\end{equation}

\noindent where $\kappa_{ij}$ is the symmetric part of the diffusion coefficient tensor, $U_i$ is the 
velocity of plasma fluid, and $Q$ is a local source. 
The drift velocity can be formally derived from the drift motion for a single particle in guiding center approximation (Equation \ref{drift-motion}) by assuming a quasi-isotropic distribution function \citep{Isenberg1979}. It is given by
$\textbf{V}_{d,i} = (p_icw_i/3q_i)\nabla\times(\textbf{B}/B^2)$, where $w_i$ is the
velocity of the particles, $c$ is the speed of light in \textit{vacuum}, and $q_i$ is the electric
charge of the particles. The drift can be included in the diffusion tensor as an anti-symmetric part 

\begin{equation}
\kappa_A = p_i c w_i /3q_i B. 
\end{equation}

The motions of energetic charged particles parallel and perpendicular to magnetic field directions are generally quite distinct. The spatial transport coefficient along the mean magnetic field can be related to the pitch-angle diffusion coefficient $D_{\mu \mu}$ \citep{Jokipii1966}. The diffusion of energetic particles transverse to the mean magnetic field is less understood. Recent test-particle simulation has shown that the perpendicular diffusion coefficient $\kappa_\perp$ can be a few percent of the parallel diffusion coefficient $\kappa_\parallel$ \citep{Giacalone1999}. The symmetric part of the spatial diffusion tensor can be related by the magnetic field vector $B_i$ and diffusion coefficient parallel and perpendicular to the magnetic field:

\begin{equation}
\kappa_{ij} = \kappa_\perp \delta_{ij} - \frac{(\kappa_\perp-\kappa_\parallel)B_iB_j}{B^2}. \label{equation-diffusion-tensor}
\end{equation}

The Parker's transport equation also describes the acceleration (deceleration) of charged particles. This equation states that the first-order energy change is due to the compression (expansion) of plasma flows. It has been shown by a series papers \citep{Krymsky1977,Axford1977,Bell1978,Blandford1978} that collisionless shocks are the acceleration sites of charged particles because of the sharp compressions. It should also be noted that particle acceleration may also happen in MHD turbulence \citep[e.g.,][]{Petrosian2004}, reconnection regions \citep[e.g.,][]{Drake2010}, and parallel electric fields due to non-ideal MHD effects \citep{Damiano2005}. These processes will not be discussed in this dissertation.

\section{Summary of the Following Chapters}

In this dissertation, we study the acceleration and transport of charged particles, focusing on the effect of fluctuating magnetic fields. The importance of turbulent magnetic fields has been recognized in many previous works. However, the previous studies often treat the effect of magnetic field as a rather simplified process. For example, the propagation of energetic particles in space is often assumed to be a simple spatial diffusion, and the particle acceleration at collisionless shock has often been considered as a process in a $1$-D planar shock, etc. While these simplified approximations were successful in describing the physical processes, these theories have been facing some difficulties in understanding the physical processes and explaining observations. 

Recently, there have been a few observations that begin to challenge this picture. For example, some spacecraft have observed SEP events in great details \citep{Mazur2000}. The observations of impulsive SEP events show fine structures in intensity-time plots on small temporal scales (hours) that are not described by a large-scale spatial diffusion \citep{Mazur2000,Chollet2007,Chollet2008b}. The Voyager spacecraft crossed the solar wind termination shock but did not observe a saturated energy spectrum of ACRs, which is predicted by the 1-D, time-steady shock theory. As we will show in this dissertation,  many features in the observations of energetic particles can be explained by considering the turbulent nature of magnetic fields. Moreover, we also discuss situations that the transport of charged particles cannot be described by the Parker's equation. An example is the process related to low-energy particles or other charged particles with high anisotropies. Since the Parker's equation only concerns a quasi-isotropic distribution of charged particles, their physical processes can not be appropriately described by the equation. Using numerical simulations, we study the transport of energetic particles with high anisotropies and the acceleration of low-energy particles.

In Chapter $2$ we study the propagation of charged particles in a turbulent magnetic field, which is similar to the propagation of impulsive SEPs in the inner heliosphere. The trajectories of energetic charged particles in the turbulent magnetic field are numerically integrated. The charged particles reached $1$ AU are collected to mimic spacecraft observations. We show that small-scale variations in the observed particle intensity (the so-called ``dropouts") and velocity dispersion observed by spacecraft can be well reproduced using this method. Our study also gives a new constraint on the error of ``onset analysis", which is a technique commonly used to study the propagation of energetic particle and infer the information of the initial injection of energetic particles. We also find that the dropouts are rarely produced in the simulations using the so-called ``two-component'' magnetic turbulence model \citep{Matthaeus1990}. The result questions the validity of this model in studying particle transport. In Chapter $3$ we study the acceleration of ions in the existence of turbulent magnetic fields. We use $3$-D hybrid simulations (kinetic ions and fluid electron) to study the acceleration of low-energy particles at parallel shocks. This gives new results for the initial acceleration of particles at shocks in fully three-dimensional electric and magnetic fields. We also use a stochastic integration method to study diffusive shock acceleration in the existence of large-scale magnetic variation. The results show that the observations by Voyager spacecraft can be explained by a $2$-D shock that includes the large-scale magnetic field variation. In Chapter $4$ we study the electron acceleration at a shock passing into a turbulent magnetic field by using a combination of hybrid simulations and test-particle electron simulations. We found that the acceleration of electrons is enhanced by including this effect. We discuss the application of this process in interplanetary shocks and flare termination shocks. We also discuss the implication of this process for SEP events. The correlation of electrons and ions in SEP events indicates that perpendicular or quasi-perpendicular shocks play an important role in accelerating charged particles. In Chapter $5$ we summarize the results and discuss the future work.

\chapter{The Effect of Turbulent Magnetic Fields on the Propagation of Solar Energetic Particles in the Inner Heliosphere\label{chapter2}}

\section{Overview of the Transport of Solar Energetic Particles \label{chapter2-intro}}
One of the most important tasks in the study of solar energetic particles (SEPs) is to understand their propagation in the heliospheric magnetic field. The large-scale transport of SEPs in the solar wind is usually studied by solving the transport equation (Equation \ref{parker_equation}) first derived by \citet{Parker1965}. The spatial diffusion tensor (Equation \ref{equation-diffusion-tensor}) can be studied by considering the statistical properties of magnetic turbulence \citep{Jokipii1966,Jokipii1971,Giacalone1999,Mattaeus2003}. The transport equation has been routinely used to fit the intensity-time profiles of SEP events for several decades \citep[e.g.,][]{Burlaga1967}. For gradual SEP events (see Section \ref{chapter1-particle2}), it has been realized that the profiles of the SEPs cannot be described by a simple spatial diffusion process since the energetic particles are continuously accelerated at the shock front, meaning that there must be at least energy changes \citep{Kahler1984}. For impulsive SEP events (see Chapter \ref{chapter1-particle2}), energetic particles are usually released from a confined region in a short duration, which provides a relatively simple case to study the transport of energetic particles in space. A long standing problem related to the transport of energetic particles is that the mean-free paths inferred from SEP events are usually much longer than those derived from the quasi-linear theory \citep{Palmer1982,Bieber1994}, which assumes that the trajectories of charged particles are weakly perturbed by magnetic fluctuations. The discrepancy between the observations and the theories is still not well resolved. Another problem that requires further investigation is the large-scale transport of charged particles normal to the magnetic field. Some analyses give a rather small cross-field diffusion so the ratio of the perpendicular diffusion coefficient to the parallel diffusion coefficient is $\kappa_\perp/\kappa_\parallel \sim 10^{-4}$ or smaller \citep{Roelof1983}. Recent numerical simulations and analytical studies give a larger value of $\kappa_\perp/\kappa_\parallel \sim 0.02 $ or larger for energetic particles moving in the heliospheric magnetic field at $1$ AU \citep{Giacalone1999,Qin2002,Mattaeus2003}.

\begin{figure}
\begin{center}
\includegraphics[width=1\textwidth]{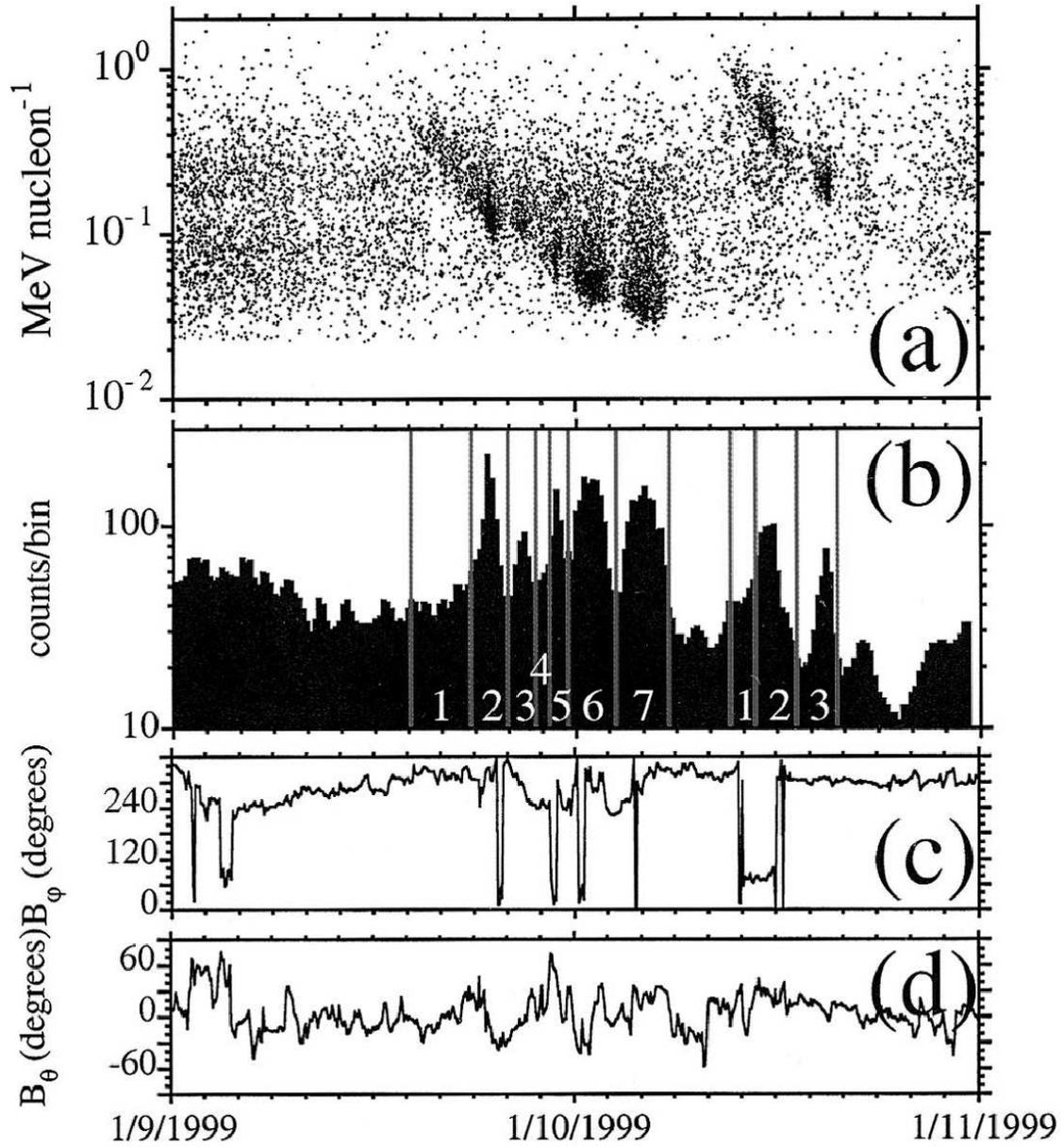}
\caption[An example of impulsive SEP events that show ``dropouts" of energetic particle intensity.]{An example of impulsive SEP events that show ``dropouts" of energetic particle intensity \citep[Figure 1 in][]{Mazur2000}. The figure is reproduced by permission of the AAS.}\label{dropoutACE}
 \end{center}
 \end{figure}

Recently, there have been several observations of SEP events that reveal some new characteristics of particle transport. For example, \citet{Mazur2000} reported that the intensity of impulsive SEP events often shows small-scale sharp variations, alternatively called the ``dropouts" of SEPs (see Figure \ref{dropoutACE}). These dropouts are commonly seen in impulsive SEP events and the typical convected distance between the dropouts is about $0.03$ AU, similar to the spatial scale of the correlation length of the solar wind turbulence. The occurrence of the dropouts does not seem to be associated with the rapid magnetic field changes as one can see from Figure \ref{dropoutACE}, meaning that it is more associated with some large-scale transport effects. The phenomena indicate that the diffusion of energetic particles transverse to the local magnetic field is very small \citep[see also,][]{Chollet2011}, and the transport of energetic particles in the solar wind is not a simple diffusion process as described by the Parker's transport equation (\ref{parker_equation}). However, Some spacecraft measurements indicate that the ratio between perpendicular to parallel diffusion coefficient $\kappa_\perp/\kappa_\parallel$ can reach values of $0.2$ or even larger \citep{Dwyer1997,Zhang2003}, which is unexpectedly large compared to those obtained from numerical simulations \citep{Giacalone1999}. Newly available data shows that impulsive SEP events are occasionally seen by all three spacecraft (STEREO A/B and ACE) with a more than $100$-degree longitudinal separation \citep{Wiedenbeck2010}. \citet{Giacalone2012}'s numerical simulations suggest that the perpendicular diffusion has to be as large as a few percent to explain these multi-spacecraft observations.  It should be noted that the motions of energetic charged particles transverse to magnetic field can be considered to consist of two parts: 1. the actual particle motion across the local magnetic field due to drift or scattering; and 2. the particle motions along meandering magnetic field lines but normal to the mean magnetic field. The observed SEP ``dropouts'' may be interpreted as that the motions of particles across the local magnetic field is small, but a large part of the perpendicular diffusion can be contributed by field-line random walk. These new observations have provided an excellent opportunity to examine and constrain the relative contributions from these two effects to the large-scale perpendicular transport.

\begin{figure}
\begin{center}
\includegraphics[width=1\textwidth]{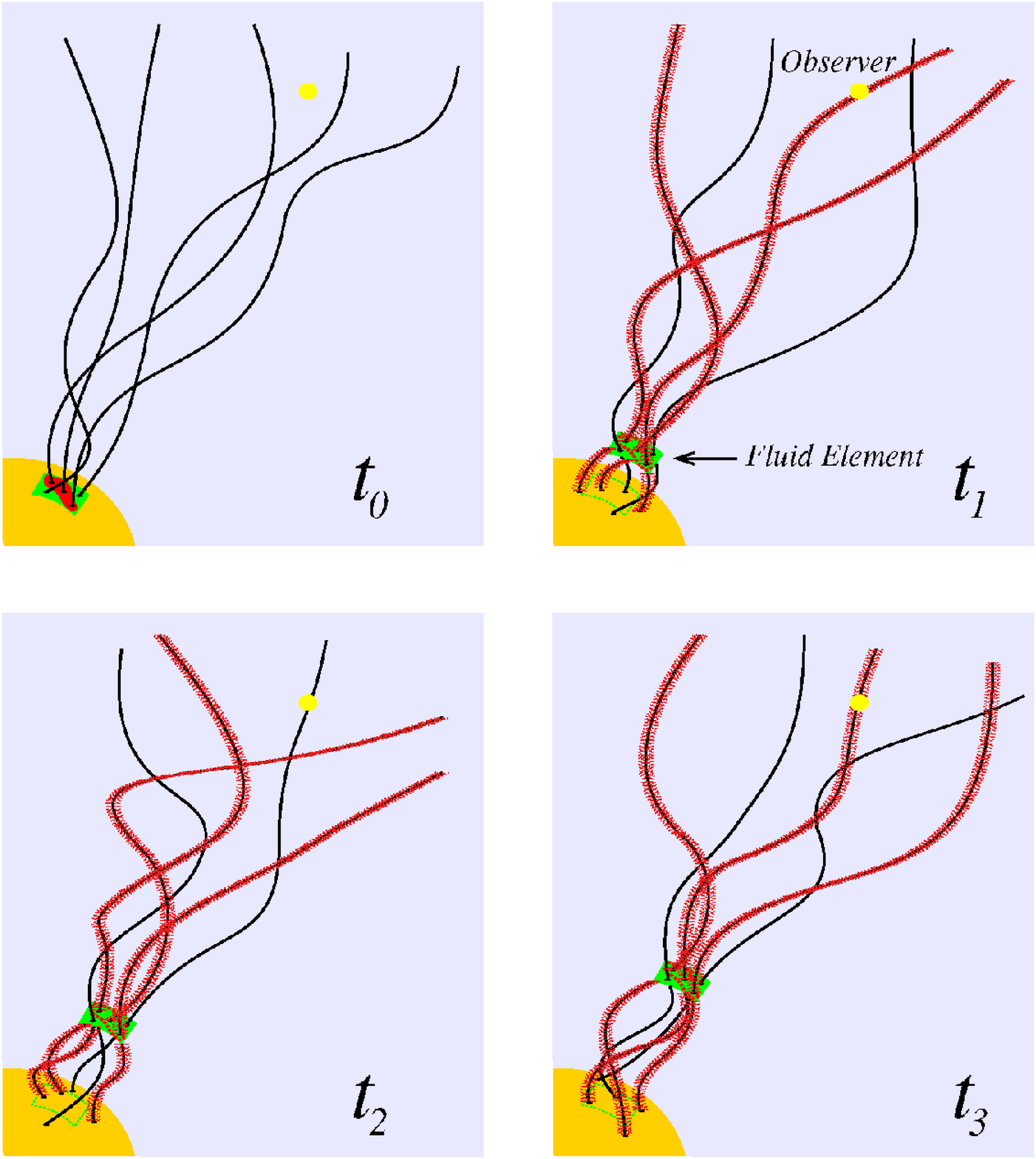}
\caption[A cartoon that illustrates the effect of meandering magnetic field lines on spacecraft observations of energetic particles.]{A cartoon that illustrates the effect of meandering magnetic field lines on the observations of energetic particles. The SEP dropouts can be seen when different field lines pass by the observer. Figure provided courtesy Joe Giacalone, University of Arizona.}\label{Fluline}.
 \end{center}
 \end{figure}

Using numerical simulations that contain large-scale turbulent magnetic fields, \citet{Giacalone2000} have demonstrated that the dropouts can be reproduced when energetic particles are released in a small source region near the Sun. The idea of the model can be illustrated by Figure \ref{Fluline}. When the source region of a SEP event is small, it just releases energetic particles into a small volume in space so that only some magnetic flux tubes are filled by energetic particles. When the field lines of force are meandering, the magnetic flux tubes will convect through spacecraft at 1 AU with a mixture of those filled by energetic particles and those that are not. The spacecraft that observe the passage of these flux tubes will see ``dropouts" of the SEP intensity. It should be noted that this model is consistent with magnetic turbulence models that allow a large perpendicular diffusion with a value of $\kappa_\perp/\kappa_\parallel \sim 0.02$ or larger due to field line random walk \citep{Giacalone1999}. The time duration between the numerically produced dropouts is several hours, which is similar to that observed in the impulsive SEP events.
It also naturally reproduces the feature that the typical spatial scale for the convected distance between the dropouts is the same as the correlation scale in the solar wind turbulence \citep{Mazur2000}. Based on the so-called ``two-component'' model (see Section \ref{chapter2-model}), \citet{Ruffolo2003} and \citet{Chuychai2007} proposed a somewhat different idea. They argued that some magnetic field lines in the solar wind can have very restricted random walk. The corresponding magnetic flux tubes connecting to the source regions are concentrated by energetic particles; For magnetic field lines that are meandering in space, the energetic particles in the associated flux tubes will diffuse away. However, this effect depends on the ``two-component" magnetic field model they use (a composition of a two-dimensional fluctuation and a one-dimensional fluctuation) and the motions of charged particles during the trapping are not explored by numerical simulations. Although previous numerical simulations that contain large-scale field-line random walk has successfully produced SEP ``dropouts'' \citep{Giacalone2000}, this model assumed an \textit{ad hoc} pitch-angle scattering that is not realistic. Physically, the pitch-angle scattering should be caused by small-scale magnetic fluctuations, which was not present in the \citet{Giacalone2000} model. One main purpose of this study is to include the effect of small-scale magnetic turbulence and examine the propagation of SEPs in a turbulent magnetic field that has a power spectrum similar to that derived from observations. The current numerical simulations directly solve the equations of motion for charged particles in turbulent magnetic fields generated by magnetic turbulence models. The results (Section \ref{chapter2-dropout}) give some new insight for the transport of energetic particles in the heliospheric magnetic fields.

\begin{figure}
\begin{center}
\includegraphics[width=0.8\textwidth]{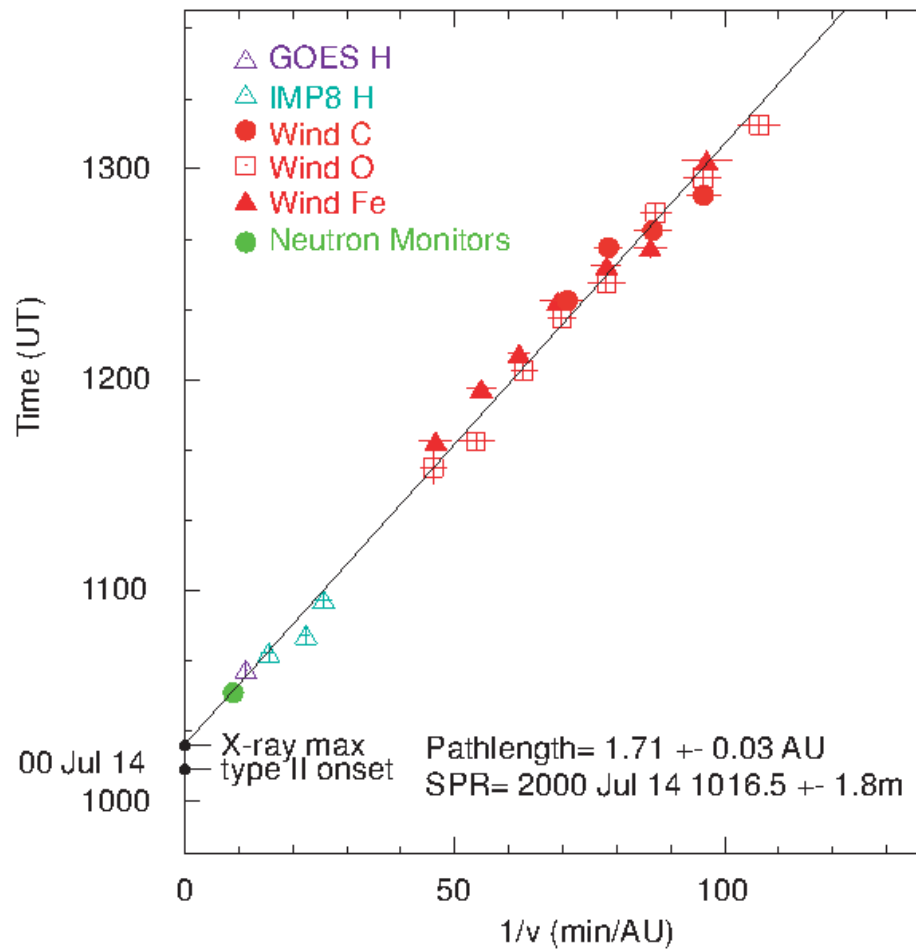}
\caption[An example of onset analyses for SEP events.]{An example of onset analyses for SEP events. The plot is adapted from \citep[Figure 8 in][]{Reames2009}. The figure is reproduced by permission of the AAS.}\label{linear}
 \end{center}
 \end{figure}

Another important issue on the transport of SEPs in interplanetary space is whether we can relate the \textit{in-situ} spacecraft observation of a particular SEP event at $1$ AU to the initial release of energetic particles in the solar atmosphere (time and/or location). Since the energetic particles suffer from spatial diffusion, they gradually lose the information about the source regions and injection times after they are released. A popular way to get the information about the location of source regions and release time is to analyse the onsets of SEP events, namely, the earliest arriving particles at a given energy \citep{Krucker2000,Tylka2003,Mewaldt2003,Kahler2006,Chollet2008a,Hill2009,
Reames2009}. Those particles have experienced the least scattering during propagation. One can obtain the apparent propagation path length $L$ and the apparent release time $t_{release}$ of SEP events by linearly fitting the onsets of the first arriving particles based on the formula

\begin{eqnarray} t - t_{release} = L / v,
\end{eqnarray}\label{equation-onset}

\noindent where $t$ is the arriving time for first-arriving particles at a given energy and $v$ is the velocity corresponding to the energy. The assumptions implicitly made in these studies are that the first-arriving particles are released impulsively and have experienced no scattering or energy change and that they have travelled exactly along the magnetic field lines with pitch-angle cosines $\mu = 1$. A main criticism of this method is that the assumption is inconsistent with the fact that the mean-free paths of energetic particles in the inner heliosphere are usually less than $1$ AU. Moreover, the mean-free path is usually energy dependent. Nevertheless, some onset analyses do show a good linear relation. In Figure \ref{linear} we show an example for the onset analysis to a SEP event, which is adapted from \citep[Figure 8 in][]{Reames2009}. One can see that the linear relation in the ``onset time" versus $c/v$ plot is quite good for the energy range they use ($>1$ MeV/nucleon). However, the apparent path length is larger than typical length of Parker spiral (1.1-1.2 AU). The feature that the fitted path length is different from the Parker spiral magnetic field lines has been found by a few authors \citep{Krucker2000,Tylka2003,Mewaldt2003,Kahler2006,Chollet2008a}. \citet{Pei2006} have demonstrated that the effect of large-scale field line meandering can significantly change the arrival times for energetic particles, and when some of the field lines are straightened radially, the energetic particles can arrive at $1$ AU faster than particles travel along the Parker spiral. This is the issue that will be further explored in this chapter. \citet{Lintunen2004}, \citet{Saiz2005} and \citet{Diaz2011} have used more sophisticated particle transport models to examine the validity of the velocity dispersion analysis and they estimate the errors contained in these analyses could be on the order of several minutes or even an hour for typical parameters at $1$ AU. However, none of these works considers the propagation of energetic particles in a turbulent magnetic field that has a power spectrum that extends to small resonant scales similar to \citet{Giacalone1999}.

In this study, we use two different types of $3$-D magnetic field turbulence models often used in studying the transport of energetic particles in space. The generated fluctuating magnetic field has a Kolmogorov-like power spectrum with wavelengths from just larger than the correlation scale, leading to large-scale field-line random walk, down through very small scales that lead to resonant pitch-angle scattering of the particles. In Section \ref{chapter2-model} we describe the magnetic turbulence models and numerical methods we use to study the propagation of energetic particles. In Section \ref{chapter2-dispersion} we use the magnetic turbulence models to study the first-arriving particles and test the validity of the velocity dispersion analysis. We estimate the errors in this technique in a variety of cases with different magnetic fluctuation amplitudes and thresholds for the onset. In Section \ref{chapter2-dropout}, we use the magnetic turbulence models to study the propagation of SEPs.  We show that the ``dropouts" of impulsive SEPs can be produced using the foot-point random motion model when the source region is small. However, for the two-component model, we find that the ``dropouts" are rarely seen for the parameters we use. The results of this chapter will be summarized in Section \ref{chapter2-summary}.

\section{Numerical Model\label{chapter2-model}}
In this study we consider the propagation of energetic particles from a spatially compact and instantaneous source in turbulent magnetic fields. We primarily use two magnetic field turbulence models that capture the main observations of magnetic field fluctuation in the solar wind: the so-called ``two-component" model \citep{Matthaeus1990} and the foot-point random motion model \citep[e.g.,][]{Jokipii1969,Jokipii1989,Giacalone2006}. This section gives a mathematical description of the turbulent magnetic field models and the numerical method for integrating the trajectories of energetic charged particles.

\subsection{Turbulent Magnetic Fluctuations} \label{magnetic-model}
In a three-dimensional Cartesian geometry ($x, y, z$), the turbulent magnetic field can be expressed as

\begin{eqnarray} \textbf{B} &=& \textbf{B}_0 + \delta\textbf{B}\nonumber\\
                            &=& B_0 \hat{z}+ \delta B_x(x,y,z,t)\hat{x}+\delta B_y(x,y,z,t)\hat{y} + \delta B_z(x,y,z,t)\hat{z}. \label{equation-mag}
\end{eqnarray}

\noindent This expression assumes a globally uniform background magnetic field $\textbf{B}_0$ and a fluctuating magnetic field component $\delta \textbf{B}$.

The two-component model is a quasi-static model for the wave-vector spectrum of magnetic fluctuation based on observations of the solar wind turbulence \citep{Matthaeus1990}. In this model, the fluctuating magnetic field is expressed as the sum of two parts: a slab component $\delta \textbf{B}^{s} = (B_x^{s}(z), B_y^{s}(z), 0)$ and a two-dimensional component $\delta \textbf{B}^{2D} = (B_x^{2D}(x, y), B_y^{2D}(x, y), 0)$. The slab component is a one-dimensional fluctuating magnetic field with all wave vectors along the direction of the uniform magnetic field $\hat{z}$, and the two-dimensional component only consists of magnetic fluctuation with wave vectors along the transverse direction $\hat{x}$ and $\hat{y}$. It has been observed that the magnetic field fluctuation has components with wave vectors nearly parallel or perpendicular to the magnetic field, with more wave power concentrated in the perpendicular directions (usually about 80\% in the solar wind). This model captures the anisotropic characteristic of the solar wind turbulence but neglects any turbulence component that propagates obliquely to the magnetic field $\textbf{B}_0$. 

Another often used model for magnetic turbulence is based on the idea that magnetic fluctuations can be generated by foot-point random motions \citep{Jokipii1969,Jokipii1989,Giacalone2006}. One can consider a Cartesian geometry with the uniform magnetic field $\textbf{B}_0$ along the $z$ direction and the source surface lying in the $x$-$y$ plane at $z=0$. Since the magnetic field lines are frozen in the surface velocity field, the magnetic field fluctuation can be produced by foot-point motions in the form of Equation \ref{equation-mag}. We assume that the surface foot-point motion is described by $\textbf{v}_{fp}(x, y, t) = \nabla \times \Psi (x, y, t)$, where $\Psi$ is an arbitrary stream function. The fluctuating component of the magnetic field anywhere is given by

\begin{eqnarray}
\delta\textbf{B}^{fp} &=& \frac{B_0}{U}\textbf{v}_{fp}(x, y, t-z/U).
\end{eqnarray}

The magnetic field at $z > 0$ is assumed to have no dynamical variation but continuously dragged outward by a background fluid (the solar wind) with a convective speed $U$. When the magnetic field is evaluated at a certain time, the magnetic field is fully three-dimensional with dependences on $x, y,$ and $z$.

\begin{figure}
\begin{tabular}{c}
\epsfig{file=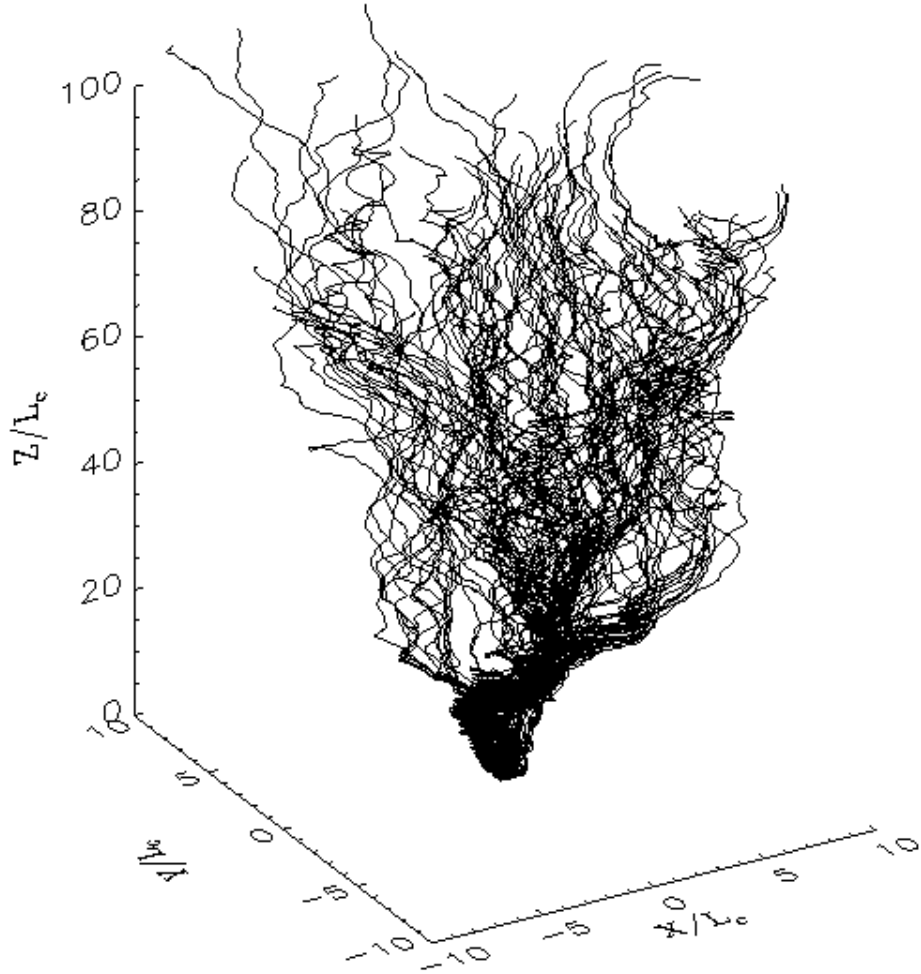,width=\textwidth,clip=}
\end{tabular}
\caption[The turbulent magnetic field lines produced by the foot-point random motion model.]{The turbulent magnetic field lines produced by the foot-point random motion model originated from $-L_c<x<L_c$ and $-L_c<y<L_c$ at $t = 0$. See the text in Section \ref{magnetic-model} for description and parameters. \label{fig2-1}}
\end{figure}

In both of these two magnetic fluctuation models the magnetic field are variable in three spatial dimensions. As demonstrated by \citet{Jokipii1993} and \citet{Jones1998}, it is important to consider particle transport in a fully three-dimensional magnetic field since particles tie on their original field lines in one-dimensional or two-dimensional magnetic field due to the presence of at least one ignorable spatial coordinate. The magnetic fluctuations can be constructed by the random phase approximation \citep[e.g.,][]{Giacalone1999} and assuming a power spectrum of magnetic field. This power spectrum can be determined from spacecraft observations. The slab component $\delta\textbf{B}^{s}$, two-dimensional component $\delta\textbf{B}^{2D}$, and fluctuating magnetic field produced by the foot-point random motion $\delta\textbf{B}^{fp}$ can be expressed as 

\begin{eqnarray}
\delta\textbf{B}^{s} = \sum^{N_{m}}_{n=1} A_n\left[\cos \alpha_n (\cos \phi_n \hat{x} + \sin \phi_n \hat{y}) + i\sin \alpha_n (-\sin\phi_n \hat{x} + \cos\phi_n \hat{y})\right] \nonumber\\
\times \exp(ik_nz + i\beta_n), \label{Bs}
\end{eqnarray}

\begin{eqnarray}
\delta\textbf{B}^{2D} = \sum^{N_{m}}_{n=1} A_n i(-\sin\phi_n \hat{x}+\cos\phi_n \hat{y}) \nonumber \\
\times \exp[i k_n (\cos\phi_n x+\sin \phi_n y) + i\beta_n], \label{B2d}
\end{eqnarray}

\begin{eqnarray}
\delta\textbf{B}^{fp} &=& (\hat{x}\frac{\partial}{\partial y} - \hat{y}\frac{\partial}{\partial x})
\times \left[\sum^{N_{m}}_{n=1} \left(-\frac{1}{k_n}\right) A_n e^{ik_n(\cos \phi_n x + \sin \phi_n y)+i\omega_n(t-z/U)+i\beta_n}\right], \label{Bfp}
\end{eqnarray}

\noindent where $\beta_n$ is the phase of each wave mode, $A_n$ is its amplitude, $\omega_n$ is its frequency, $\alpha_n$ is polarization angle, and $\phi_n$ determines spatial direction of the $k$-vector in the $x$-$y$ plane. $\beta_n$, $\alpha_n$, and $\phi_n$ are random numbers between $0$ and $2\pi$. The frequency is taken to be $\omega_n = 0.1Uk_n$. All the forms of fluctuating magnetic field satisfy the condition $\nabla \cdot \delta \textbf{B} = 0$.

The amplitude of magnetic fluctuation at wave number $k_n$ is assumed to follow a Kolomogorov-like power law:

\begin{eqnarray}
A_n^2 = \sigma^2 \frac{\Delta V_n}{1 + (k_n L_c)^\gamma} \left[\sum^{N_m}_{n=1}\frac{\Delta V_n}{1+(k_n L_c)^\gamma}\right]^{-1},
\end{eqnarray}

\noindent where $\sigma^2 = \langle \delta B^2 \rangle / B_0^2$ is the total magnetic variance and $\Delta V$ is a normalization factor. In one-dimensional, two-dimensional, and three-dimensional omnidirectional spectra,
$\Delta V_n = \Delta k_n$, $2\pi k_n \Delta k_n$, and, $4\pi k_n^2 \Delta k_n$ and $\gamma = 5/3$, $8/3$, and $11/3$, respectively.

It has been pointed out by \citet{Giacalone2006} that these two models are closely related and the two-component model can be reproduced using the foot-point random motion model by choosing a particular set of fluctuating velocity field. It should be noted that both of these two simplified models assume a quasi-static field that may not be appropriate for describing magnetic turbulence. Nonlinear structures of the magnetic turbulence, such as current sheets that could have important effects are not included. Nevertheless, these two models are very useful in studying the transport of energetic charged particles in magnetic turbulence and explaining the observations of SEP events. We also note that since these models assumes a uniform solar wind speed in Cartesian coordinates, they do not include the effects of an expanding solar wind in spherical coordinates such as adiabatic cooling and adiabatic focusing.  

In the simulations we generally use parameters similar to what is observed in the solar wind at $1$ AU. The minimum and maximum wavelengths $\lambda_{min}$ and $\lambda_{max}$ are taken to be $10^{-4}$-$10^{-5}$ AU and $1$ AU. The mean magnetic field $B_0$ is typically taken to be $5$ nT. The convection velocity of the solar wind $U$ is set to be $400$ km/s. The correlation length is assumed to be $L_c = 0.01$ AU. In figure \ref{fig2-1} we illustrate $100$ turbulent magnetic field lines origin from a surface region within $-L_c<x<L_c$ and $-L_c<y<L_c$ at $z = 0$ at time $t = 0$ produced by foot-point random motion. It is clear that the magnetic field lines are meandering in large scales. The meandering field lines originated from the compact region can have large displacements in the $x$ and $y$ directions. 

\subsection{Test Particle Simulations} \label{testparticle-model}

In order to study the propagation of energetic particles in the heliospheric magnetic field, we numerically integrate the trajectories of energetic particles in magnetic fields generated from the magnetic turbulence models described previously. In each time step, the magnetic field is calculated at the position of a charged particle. The numerical technique used to integrate the trajectories of energetic particles is the so-called Bulirsh-Stoer method, which is described in detail by \citet{Press1986}. It is highly accurate and conserves energy well. The algorithm uses an adjustable time-step method that is based on the evaluation of the local truncation error. The time step is increased if the local truncation error is smaller than $10^{-6}$ for several consecutive time steps. In the case of no electric field, the energy of a single particle in the fluctuating magnetic field is conserved to a high degree with total changes smaller than $0.01\%$ during the simulation.

\section{A Numerical Study on the Velocity Dispersion of Solar Energetic Particles \label{chapter2-dispersion}}

In this section we use both of the magnetic turbulence models described in Section \ref{magnetic-model} to study the velocity dispersions of energetic particles in the heliospheric magnetic field. The velocity dispersion is due to the fact that faster particles are detected earlier than slow particles if they are released at the same time and location. The parameters are chosen to match the magnetic field observed at $1$ AU. Protons are released impulsively at $z = 0$ with random pitch angles $\mu = v_z/v$ between 0 and 1. This injected pitch-angle distribution is different than previous studies \citep{Saiz2005}, which assume that the initial pitch angles for all the particles are $\mu = 1$. The trajectories of the charged particles are integrated until they reach the boundaries at $z = 1.2$ AU and $z = -0.1$ AU. The area of the source region is taken to be $L_x \times L_y = 5 L_c \times 5 L_c$, which is chosen to be larger than the correlation length in order to obtain the statistical meaningful results. It is also possible that the source regions are small and the transport of SEPs released from those regions are only affected by the field lines connecting to the source regions. The effect is unpredictable and requires a demanding computing resource. Here we only discuss the situation that the source regions are fairly large. The energies for the released protons are $100$, $9.4671$, $3.3057$, $1.6649$, and $1$ MeV, which correspond to the values of $1/v$ (converted to hour/AU): $0.3$, $0.976$, $1.65$, $2.33$, and $3.00$, respectively. The magnetic variances used in the simulations are varied from $\sigma^2 = 0.01$ to $\sigma^2 = 0.6$. The mean-free paths and parallel diffusion coefficients calculated from the quasi-linear theory \citep{Jokipii1966,Giacalone1999} are listed in Table \ref{table_meanfreepath}. In each case, we numerically simulate the intensity-time profiles for test particles arrived at $1$ AU using the magnetic field generated from four different realizations. Each realization is delineated using a new set of random phases, polarizations, and propagation angles, etc. The onset times for different thresholds are recorded when the values of the intensity reach the thresholds $0.1$, $0.01$, and $0.001$ of the peak values, respectively.

\begin{table}
\begin{tabular*}
{0.95\textwidth}{ccccc}
\hline
Energy (MeV) & 1/v (hour/AU) & $\sigma^2 = \delta B^2/B_0^2$ & $\lambda_\parallel$ (AU) & $\kappa_\parallel \: (10^{20} \: cm^2/s)$ \\
\hline
100          & 0.3           & 0.6       &   0.046  & 31.7    \\
100          & 0.3           & 0.3       &   0.092  & 63.4   \\
100          & 0.3           & 0.1       &   0.276  & 190.2   \\
100          & 0.3           & 0.03      &   0.92   & 634  \\
100          & 0.3           & 0.01      &   2.76   & 1902  \\

9.4671       & 0.976         & 0.6       &   0.03   & 6.54  \\
9.4671       & 0.976         & 0.3       &   0.06   & 13.08  \\
9.4671       & 0.976         & 0.1       &   0.18   & 39.24  \\
9.4671       & 0.976         & 0.03       &   0.6   & 130.8  \\
9.4671       & 0.976         & 0.01       &   1.8   & 392.4  \\

3.3057       & 1.65          & 0.6       &   0.026  & 3.24   \\
3.3057       & 1.65          & 0.3       &   0.052  & 6.48   \\
3.3057       & 1.65          & 0.1       &   0.156  & 19.44   \\
3.3057       & 1.65          & 0.03       &   0.52  & 64.8   \\
3.3057       & 1.65          & 0.01       &   1.56  & 194.4   \\

1.6649       & 2.33          & 0.6       &   0.023  & 2.05   \\
1.6649       & 2.33          & 0.3       &   0.046  & 4.1   \\
1.6649       & 2.33          & 0.1       &   0.138  & 12.3   \\
1.6649       & 2.33          & 0.03       &   0.46  & 41   \\
1.6649       & 2.33          & 0.01       &   1.38  & 123   \\

1            & 3.00          & 0.6       &   0.021  & 1.46   \\
1            & 3.00          & 0.3       &   0.042  & 2.92   \\
1            & 3.00          & 0.1       &   0.126  & 9.76   \\
1            & 3.00          & 0.03       &   0.42  & 29.2   \\
1            & 3.00          & 0.01       &   1.26  & 97.6   \\
\hline
\end{tabular*}
\caption[The parallel mean-free paths and diffusion coefficients calculated from quasi-linear theory.]{The parallel mean-free paths and diffusion coefficients calculated from quasi-linear theory \citep{Giacalone1999} for particle energies and magnetic variances used in this study.}
 \label{table_meanfreepath}
\end{table}

\begin{figure}
\begin{center}
\includegraphics[width=1\textwidth]{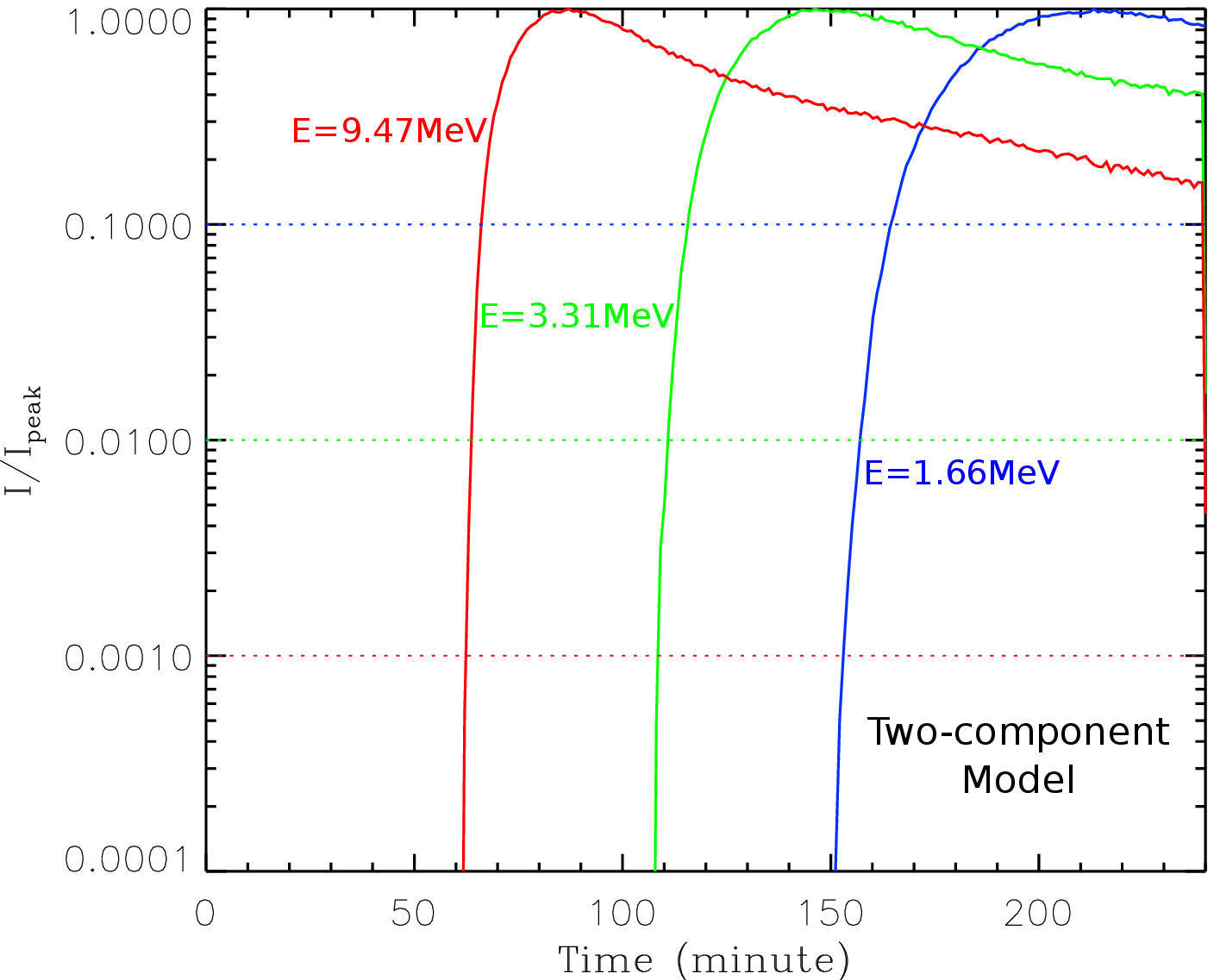}
\caption[The intensity-time profiles of energetic particles at different energies collected $1$ AU normalized using their peak values, in the case of the two-component model.]{The intensity-time profiles of energetic particles collected at $1$ AU normalized by their peak values at energies of $9.4671$ MeV (the red solid line), $3.3057$ MeV (the green solid line) and $1.6649$ MeV (the blue solid line). The magnetic field turbulence is generated from the two-component model and the total variance is $\sigma^2 = 0.3$.}\label{numberprofile1}
 \end{center}
 \end{figure}

\begin{figure}
\begin{center}
\includegraphics[width=1\textwidth]{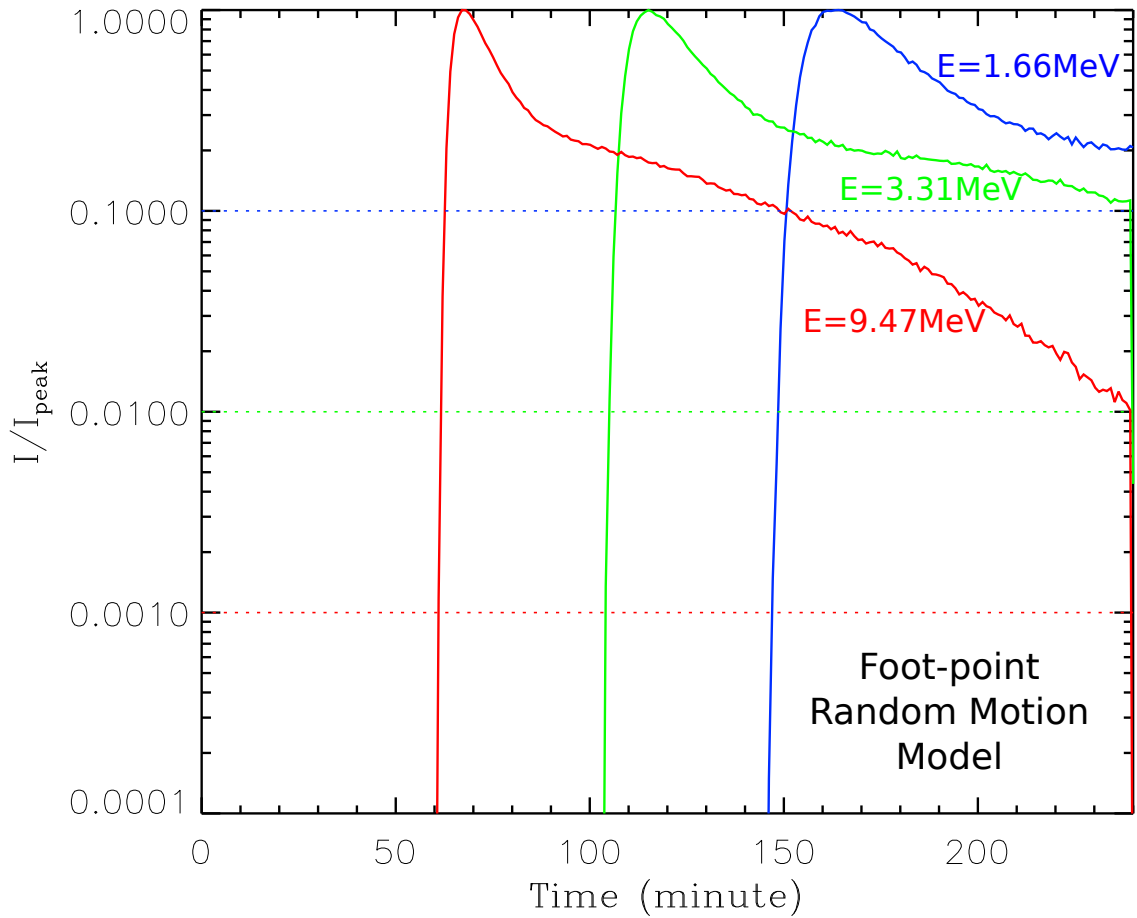}
\caption[The intensity-time profiles of energetic particles at different energies collected at $1$ AU normalized using their peak values, in the case of the foot-point random motion model.]{The intensity-time profiles of energetic particles collected at $1$ AU normalized by their peak values at energies of $9.4671$ MeV (the red solid line), $3.3057$ MeV (the green solid line) and $1.6649$ MeV (the blue solid line). The magnetic field turbulence is generated from the foot-point random motion and the total variance is $\sigma^2 = 0.3$.}\label{numberprofile2}
 \end{center}
 \end{figure}

Figure \ref{numberprofile1} illustrates the intensity-time profiles of energetic particles normalized using the peak values in the case of the two-component model. In this plot the red solid line represents the profile for particles with the energy of $9.4671$ MeV, the green solid line represents the profile for particles with the energy of $3.3057$ MeV, and the blue solid line represents the profile for particles with the energy of $1.6649$ MeV. The thresholds for $0.001$, $0.01$ and $0.1$ of the peak value are labelled using dashed lines. Figure \ref{numberprofile2} displays a similar plot for the case of the foot-point random motion model. The feature of velocity dispersion can be clearly seen from these two plots. The propagation of energetic particles along the $z$-direction depends on their pitch-angles and the scattering they experienced, therefore the energetic particles arrive at $1$ AU at different times. The intensity-time profile usually has a sharp increase when the particles start to reach $1$ AU. For particles with lower energies, the increases are slower compared to the cases for higher energies, presumably because charged particles with lower energies have smaller mean-free paths. However, as we will show below, for the energy range we simulate ($1$ - $100$ MeV), this usually does not introduce a large error in analysing the injection time for energetic particle events under the situations that we study.    

\begin{figure}
\begin{center}
\includegraphics[width=1\textwidth]{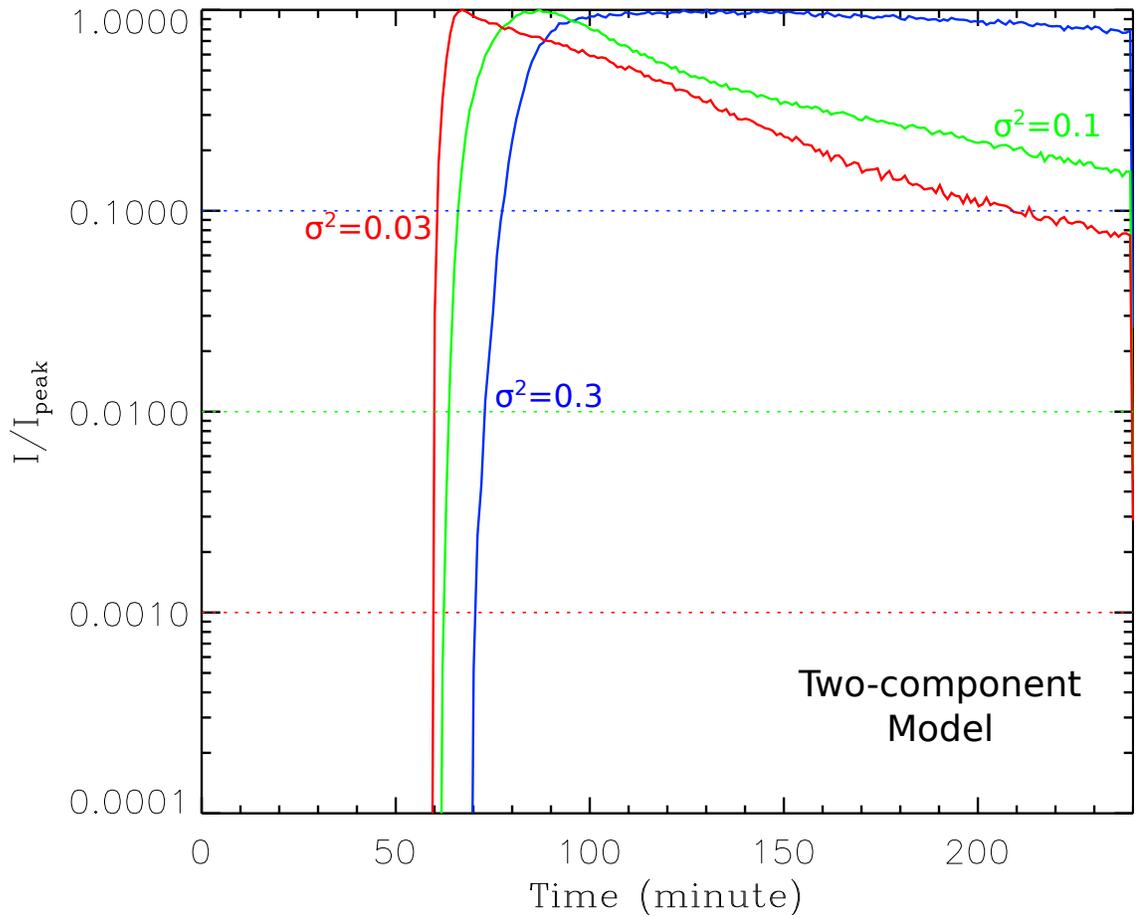}
\caption[The intensity-time profiles of energetic particles collected at $1$ AU normalized using their peak values, in the case of the two-component model for different variances.]{The intensity-time profiles of energetic particles collected at $1$ AU normalized by their peak values at the energy of $9.4671$ MeV. The magnetic field turbulence is generated from the two-component model and the total variance is $\sigma^2 = 0.3$ (the blue solid line), $\sigma^2 = 0.1$ (the green solid line), and $\sigma^2 = 0.03$ (the red solid line). It is shown that the effect of larger magnetic variance results in delayed onsets in the SEP events.}\label{numberprofile3}
 \end{center}
 \end{figure}

\begin{figure}
\begin{center}
\includegraphics[width=1\textwidth]{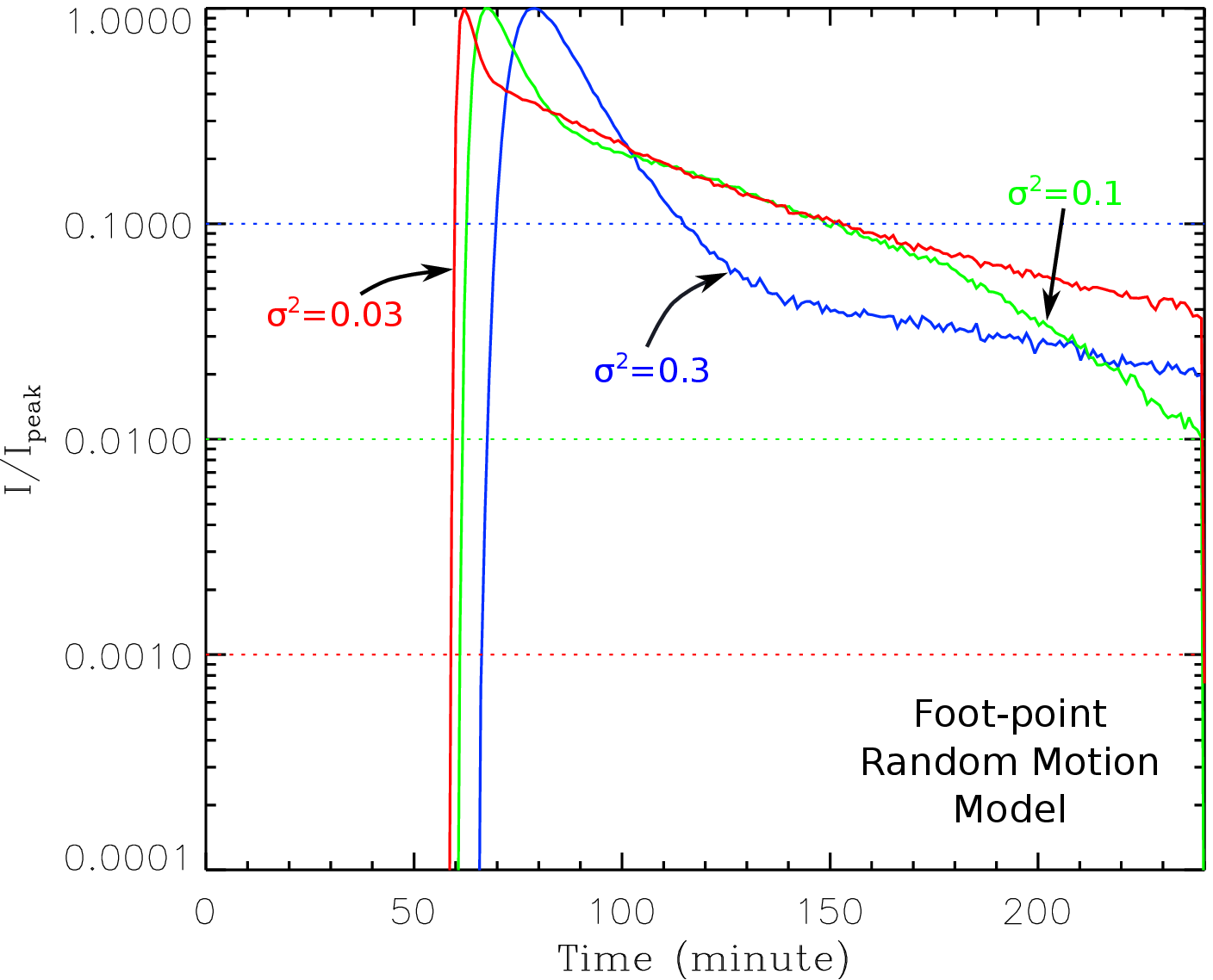}
\caption[The intensity-time profiles of energetic particles collected at $1$ AU normalized using their peak values, in the case of the foot-point random motion model for different variances.]{The intensity-time profiles of energetic particles collected at $1$ AU normalized by their peak values at the energy of $9.4671$ MeV. The magnetic field turbulence is generated from the foot-point random motion model and the total variances are $\sigma^2 = 0.3$ (the blue solid line), $\sigma^2 = 0.1$ (the green solid line), and $\sigma^2 = 0.03$ (the red solid line). It is shown that the effect of larger magnetic variance results in delayed onsets in the SEP events.}\label{numberprofile4}
 \end{center}
 \end{figure}

Figure \ref{numberprofile3} displays the intensity-time profiles of energetic particles at $9.4671$ MeV normalized using their peak values in the case of the two-component model. In this plot the magnetic variances are $\sigma^2 = 0.6$ (the blue solid line), $\sigma^2 = 0.3$ (the green solid line), and $\sigma^2 = 0.1$ (the red solid line). Figure \ref{numberprofile4} displays a similar plot in the case of the foot-point random motion model. It can be seen that in the case of larger magnetic variances, the arriving times for the onsets of energetic particles at $1$ AU are delayed. The delays may be due to two effects: $1$. The first-arriving particles experience more pitch-angle scattering during the propagation and $2$. the lengths of turbulent magnetic field lines are larger in the cases of larger magnetic variances. Since we study the propagation of particles in a Cartesian geometry, the larger magnetic variances only result in longer lengths of magnetic field lines of force. This is different than the field lines of force in a spherical geometry, where the Parker's spiral field lines can be straightened and therefore shortened by the effect of large-scale magnetic turbulence  \citep{Pei2006}. The decays of the intensities of charged particles after the peaks are also different. In the case of larger magnetic variances, the decay is slower because of the enhanced scattering during the propagation.

\begin{table}
\begin{tabular*}
{\textwidth}{ccccc}
\hline
Model         & $\delta B^2/B_0^2$ & Threshold/Peak & Path Length (AU) & $T_{release}$ (min)  \\
\hline
two-component & 0.01               & 0.001       &   1.010            & -0.38                    \\
two-component & 0.01               & 0.01       &   1.012            & -0.3                    \\
two-component & 0.01               & 0.1       &   1.018            & -0.19                    \\

two-component & 0.03               & 0.001       &   1.031            & -0.80                    \\
two-component & 0.03               & 0.01       &   1.042            & -0.69                    \\
two-component & 0.03               & 0.1       &   1.068            & -1.16                    \\

two-component & 0.1               & 0.001       &   1.12           & -2.33                    \\
two-component & 0.1               & 0.01       &   1.15            & -2.73                    \\
two-component & 0.1               & 0.1       &   1.21            &  -3.63                    \\

two-component & 0.3               & 0.001       &   1.29            & -4.23                    \\
two-component & 0.3               & 0.01       &   1.33            & -4.36                    \\
two-component & 0.3               & 0.1       &   1.42            & -4.52                    \\

two-component & 0.6               & 0.001       &   1.50            & 5.67                    \\
two-component & 0.6               & 0.01       &   1.65            & 8.89                    \\
two-component & 0.6               & 0.1       &   2.21            & -24.64                    \\

foot-point  & 0.01               & 0.001       &   1.008            & -0.22   \\
foot-point  & 0.01               & 0.01       &   1.009           & -0.08   \\
foot-point  & 0.01               & 0.1       &   1.011            & -0.07                  \\

foot-point & 0.03               & 0.001       &   1.018            & -0.42 \\
foot-point & 0.03               & 0.01       &   1.022            & -0.35 \\
foot-point & 0.03               & 0.1       &   1.028            & -0.35 \\

foot-point & 0.1               & 0.001       &   1.06            & -0.98                    \\
foot-point & 0.1               & 0.01       &   1.068            & -0.91                    \\
foot-point & 0.1               & 0.1       &   1.08            & -0.91                    \\

foot-point & 0.3               & 0.001       &   1.16            & -1.45                    \\
foot-point & 0.3               & 0.01       &   1.18            & -1.29                    \\
foot-point & 0.3               & 0.1       &   1.21            & 1.29                    \\

foot-point & 0.6               & 0.001       &   1.26            & -1.29                    \\
foot-point & 0.6               & 0.01       &   1.296            & -1.48                    \\
foot-point & 0.6               & 0.1       &   1.34            & -1.09                    \\
\hline
\end{tabular*}
\caption[Results of onset time analyses.]{Results of onset time analyses.}
 \label{table_onset}
\end{table}

\begin{figure}
\begin{center}
\includegraphics[width=1\textwidth]{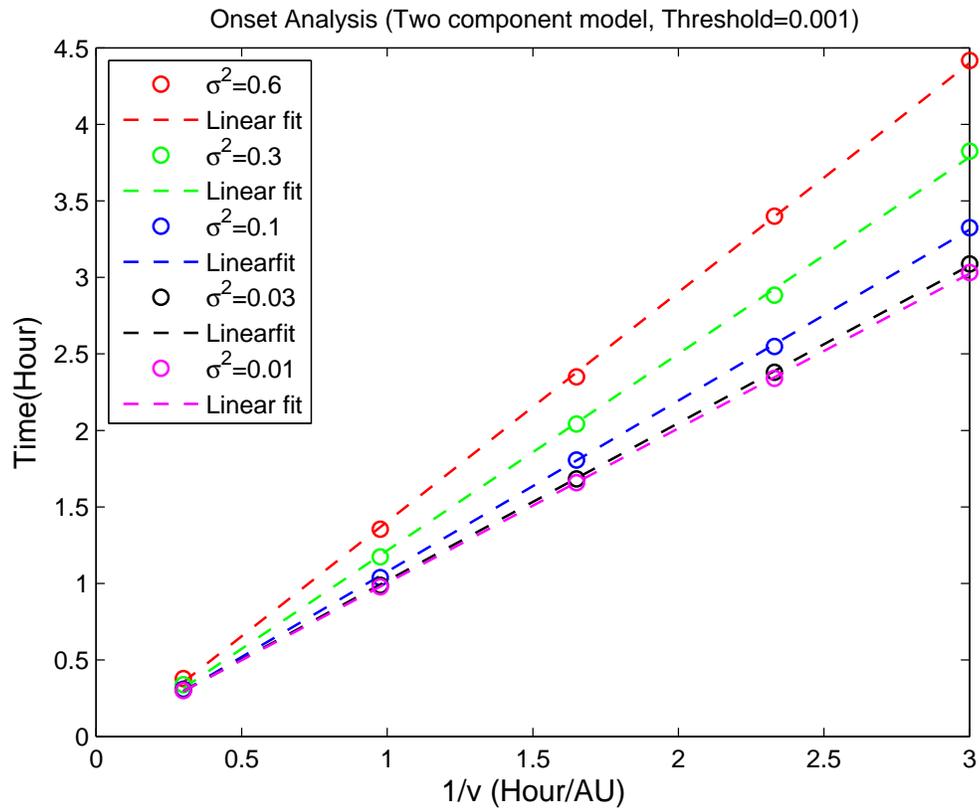}
\caption[Onset analyses for the two-component model and threshold is $0.001$.]{Onset analyses for the two-component model and the threshold is $0.001$. The different markers represent different magnetic variances $\sigma^2$. It is shown that the effect of larger magnetic variances change the slope (increases the path length $L$) of the linear fitting in the onset analyses.}\label{fig-onset1}
 \end{center}
 \end{figure}

\begin{figure}
\begin{center}
\includegraphics[width=1\textwidth]{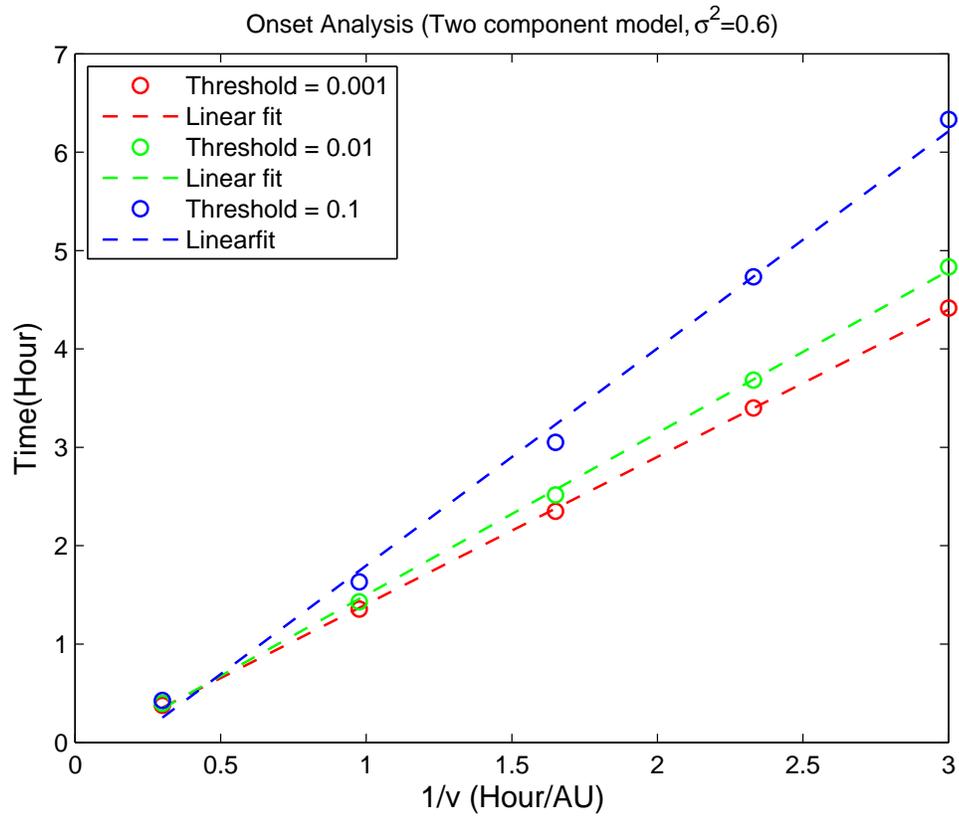}
\caption[Onset analyses for the two-component model and the magnetic variance is $\sigma^2 = 0.6$.]{Onset analyses for the two-component model and the magnetic variance is $\sigma^2 = 0.6$. The different markers represent different threshold. Their corresponding linear fit are labelled using dashed lines. This result shows that effect of the threshold on onset analyses.}\label{fig-onset2}
 \end{center}
 \end{figure}

After obtained the onsets for the intensity-time profiles of energetic particles at all selected energies, we linearly fit the onset time $t$ and $1/v$ based on Equation \ref{equation-onset} and get the values of fitted path length $L$ and release time $t_{release}$. This is similar to the method used to analyse the onsets of SEP events. Since we release particles at $t = 0$, the fitted release times actually represent the errors of the onset analyses. Two examples of these analyses are given in Figure \ref{fig-onset1} and Figure \ref{fig-onset2}. In Figure \ref{fig-onset1} we present the results of the onset analyses for the two-component model, with different magnetic variances varying from $\sigma^2 = 0.01$ to $\sigma^2 = 0.6$ and the thresholds for intensity onsets are taken to be $0.001$ of the peak values. Figure \ref{fig-onset2} shows a similar plot for the onset analyses for the cases that the magnetic variance is $\sigma^2 = 0.6$ and the threshold for intensity onset is varied from $0.001$ of the peak value to $0.1$ of the peak value. One can see that the onset analyses usually have a good linear relation. For a larger magnetic variance and/or a larger threshold value, the onset analyses give relatively large path lengths and more significant errors in the release time $t_{release}$. It has been found that for the case of the foot-point random motion model, the estimated errors for onset analyses are smaller than those of the two-component model. The reason for this result is probably because the parallel diffusion coefficient of particle motion in foot-point random motion model is considerably larger than that in the two-component model. This is illustrated in  Figure \ref{fig-diffusion} and will be further discussed in Section \ref{chapter2-dropout}. The results of the onset analyses for all cases are listed in Table \ref{table_onset}. From the table it is shown that the errors for the released times $t_{release}$ are usually within several minutes unless the variance is large $\sigma^2 = 0.6$ and threshold $= 0.1$. This indicates that although the pitch-angle scattering could play a role, the onset analyses usually have a small error and therefore useful in inferring the release time for energetic particles. However, this method is found to have a relative large error in estimating the path length $L$ for SEP events. The effect of magnetic turbulence on the apparent path lengths is illustrated in Figure \ref{fig-onset1}. When a larger value of the magnetic variance is chosen, the slope of the ``$1/v$ - $t$" line is steepened so the apparent path lengths get larger. This agrees with SEP observations, which usually get a path length that is deviated from the typical path length of the Parker spiral magnetic field \citep{Krucker2000,Tylka2003,Mewaldt2003,Kahler2006,Chollet2008a}. It should be noted that for a realistic heliospheric magnetic field considering the solar rotation, the lengths of the magnetic field lines can occasionally be shorter than that of the nominal Parker spiral because some magnetic field lines are straighten radially by the effect of the random magnetic field \citep{Pei2006}. It is therefore possible that the onset analyses give a path length $L$ smaller than the lengths of the Parker spiral, as they are seen in some observations \citep{Hilchenbach2003,Chollet2007}. Although we use a different numerical model to test the validity of the velocity dispersion analysis, the results are qualitatively consistent with previous studies \citep{Lintunen2004,Saiz2005}. 
 
\section{A Numerical Study of Dropouts in Impulsive SEP Events \label{chapter2-dropout}}

\begin{figure}
\begin{center}
\includegraphics[width=0.8\textwidth]{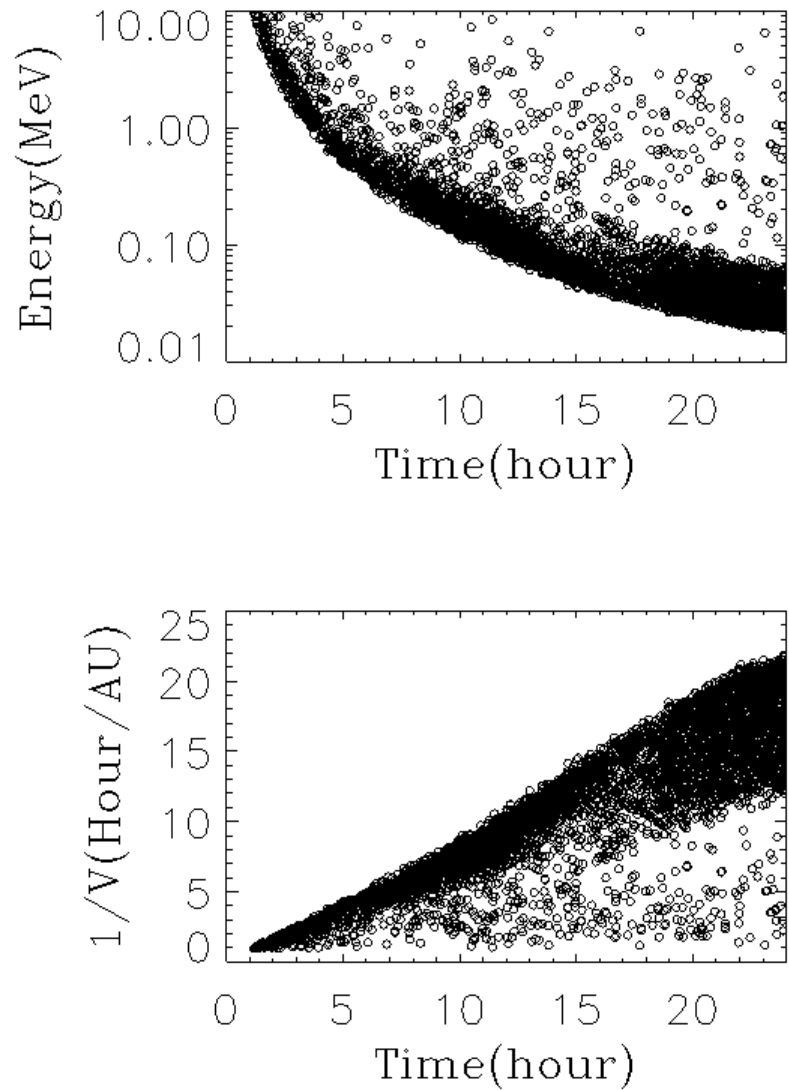}
\caption[An example of SEP event simulated using the foot-point random motion model for the case of the large source region.]{An example of SEP event simulated using the foot-point random motion model for the case of the large source region. \textit{Upper panel}: energy-time plot. \textit{Lower panel}: the inverse velocity $1/v$ versus the time after the release. The simulated event does not show SEP ``dropouts". \label{fig-dropout-large}}
 \end{center}
 \end{figure}

\begin{figure}
\begin{center}
\includegraphics[width=0.8\textwidth]{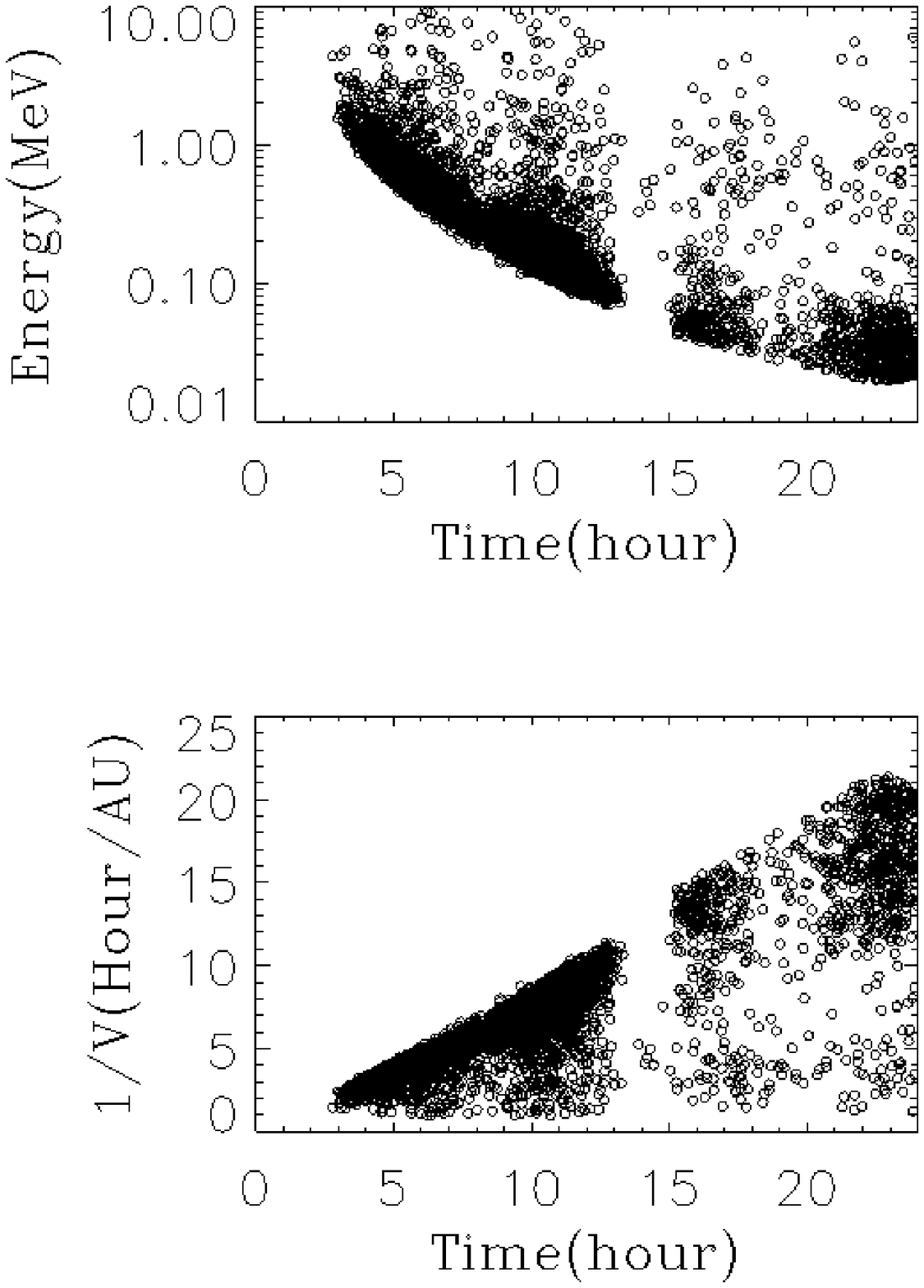}
\caption[An example of SEP dropouts simulated using the foot-point random motion model for the case of the small source region.]{An example of SEP dropouts simulated using the foot-point random motion model for the case of the small source region. \textit{Upper panel}: energy-time plot. \textit{Lower panel}: the inverse velocity $1/v$ versus the time after the release. This example clearly shows dropouts. \label{fig-dropout-small}}
 \end{center}
 \end{figure}

In this section we use turbulent magnetic fields generated from the foot-point random motion model and the two-component model to study the SEP ``dropouts" observed by spacecraft such as ACE and Wind \citep{Mazur2000,Chollet2008b}. In our test-particle simulations the charged particles are released impulsively at $z = 0$ and their trajectories are numerically integrated until they reach the boundaries at $z = 1.6$ AU and $z = -0.1$ AU. The spacecraft observations at $1$ AU are simulated by collecting particles in windows of a size $L_x \times L_y = L_c \times L_c$ when the particles pass the windows at $z = 1$ AU. The record for each window is plotted as a simulated SEP event observed by spacecraft. The source regions are taken to be a circle at the $z = 0$ plane with a radius of \textit{case $1$}: much smaller than the correlation scale ($R_{source} = 0.2 L_c$) and \textit{case $2$}: much larger than the correlation scale ($R_{source} = 5 L_c$). The energy for the released particles ranges from $20$ keV to $10$ MeV. The velocity distribution of released particles is assumed to follow a power law $f = f_0 v^{-4}$ with random pitch angles between $0$ and $1$. The magnetic variance used in the simulations is $\sigma^2 = 0.3$. In Figure \ref{fig-dropout-large} we show a simulated SEP event using the foot-point random motion model for the case of large source region. The upper panel shows the energy-time plot and the lower panel displays the plot of inverse velocity $1/v$ versus the time after the initial release. One can see that in this case the simulated SEP event does not show any dropout. In the small source region case, the dropout can be frequently seen. An example is given in Figure \ref{fig-dropout-small}, which illustrates a simulated SEP event energy-time plot (upper panel) and inverse velocity $1/v$ versus time plot (lower panel). It is shown that two SEP dropouts can be clearly seen at about $t = 13$ - $15$ hour and $t = 17.5$ - $20.5$ hour, respectively. The time intervals of these dropouts are typically several hours, which is similar to that observed in space \citep{Mazur2000}. 

\begin{figure}
\begin{center}
\includegraphics[width=1\textwidth]{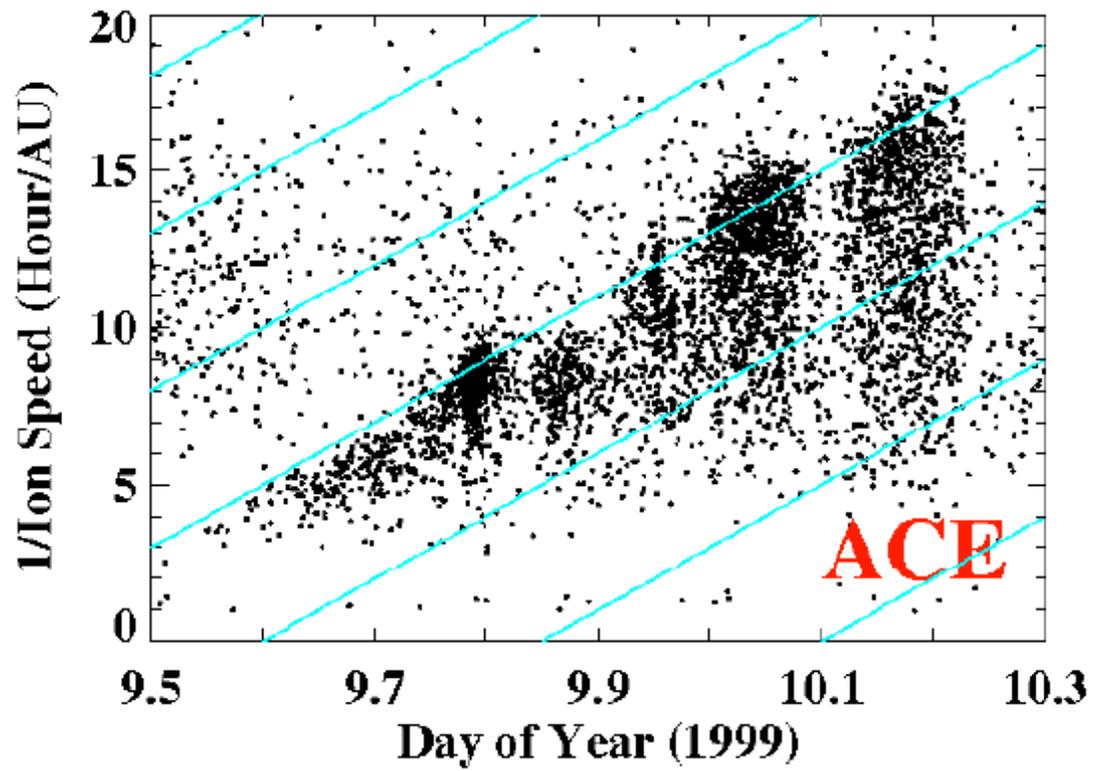}
\caption[An example of the observed SEP dropouts that show different path lengths observed by ACE/ULEIS detector.]{An example of the observed SEP dropouts that show different path lengths observed by ACE/ULEIS detector. Figure provided courtesy Joe Mazur, Aerospace Corporation. }\label{flucedge}.
 \end{center}
 \end{figure}

\begin{figure}
\centering
\begin{tabular}{cc}
\epsfig{file=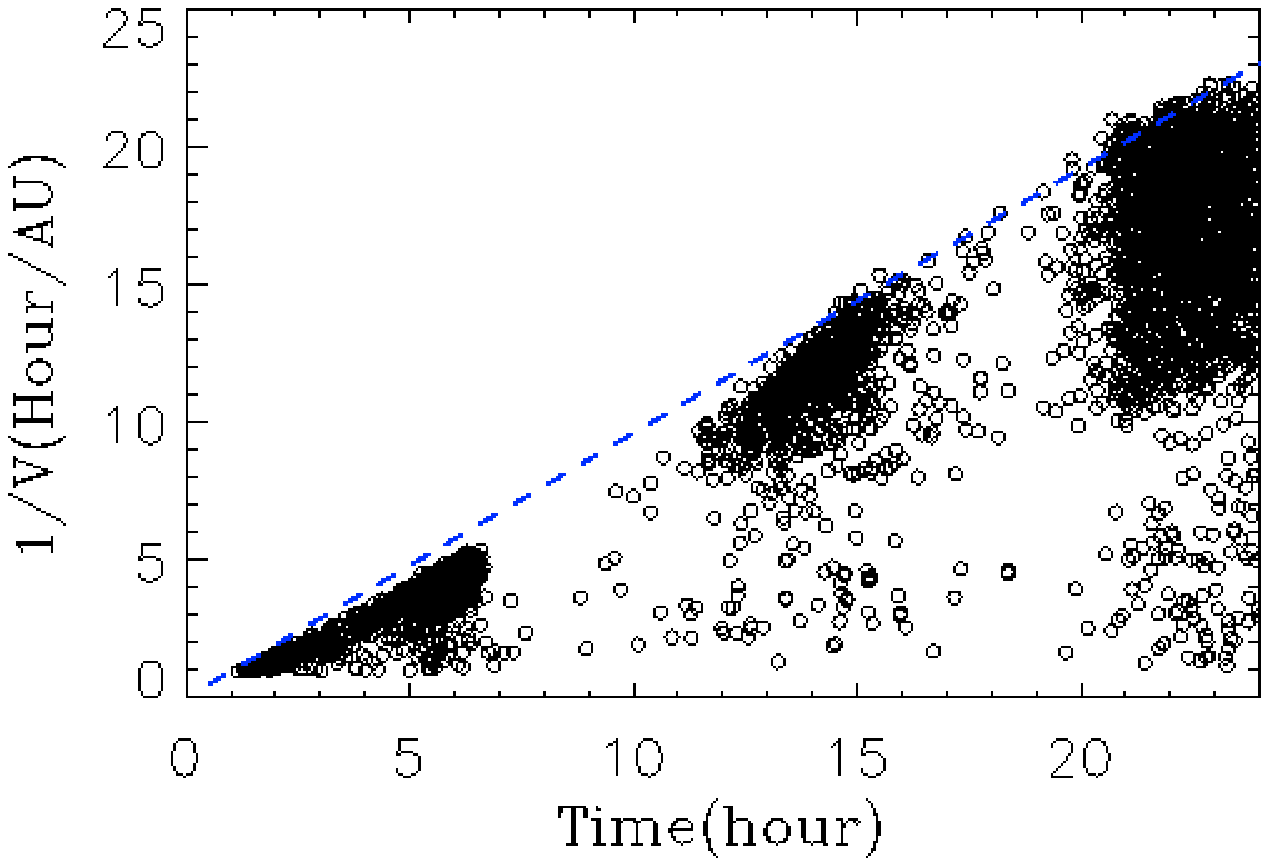,width=\textwidth,clip=} \\
\epsfig{file=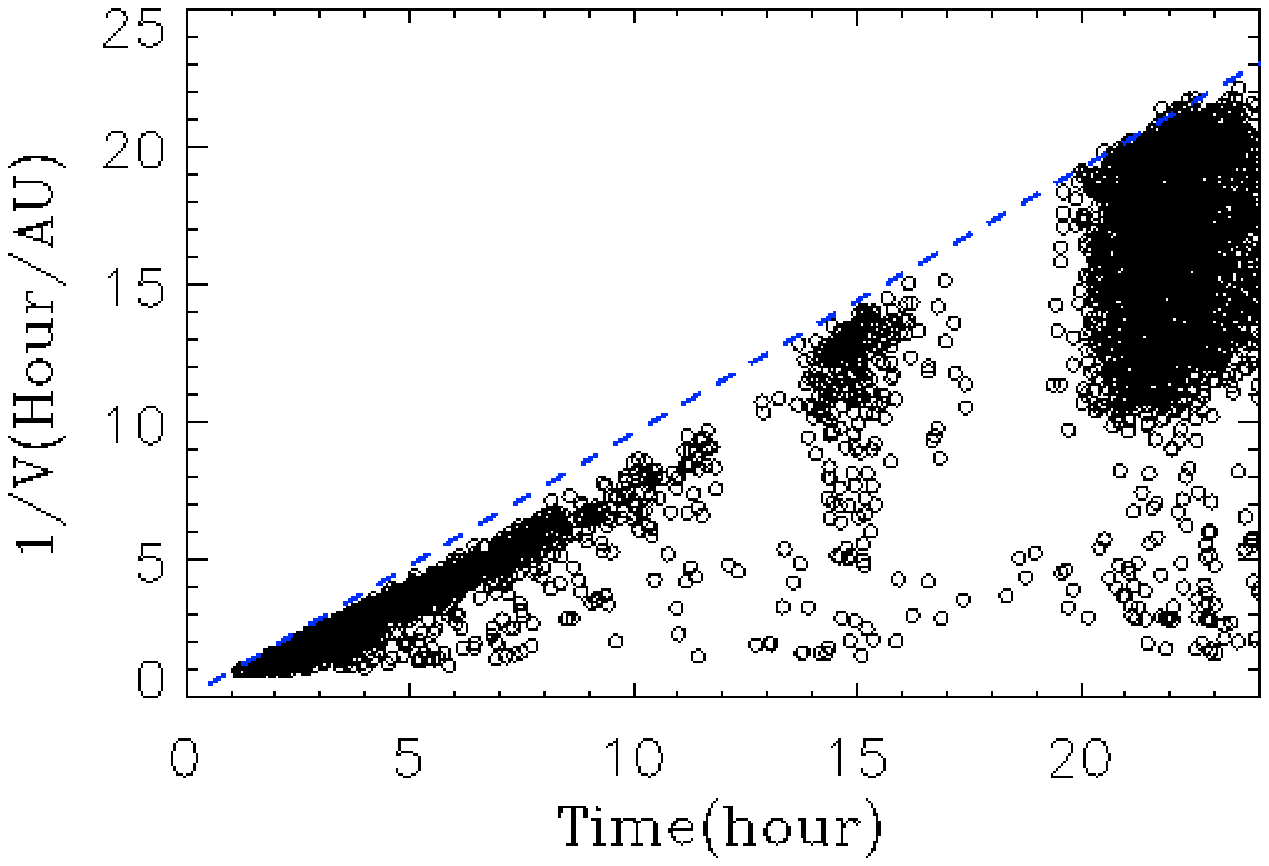,width=\textwidth,clip=}
\end{tabular}
\caption[Examples of SEP dropouts produced from numerical simulations. The result shows a SEP event that observed by two different observers. The two observations have different apparent path lengths.]{Examples of SEP dropouts produced from numerical simulations. The result shows a SEP event that observed by two different observers. The two observations have different apparent path lengths. \label{figure-pathes}}
\end{figure}

The velocity dispersion of particles seen by observers can be varied as the particles travel along different paths and experience different scattering. This can be illustrated by Figure \ref{flucedge}, which shows an impulsive SEP event plotted as $1/v$ versus time. The observation was made by observed by ACE/ULEIS detector in 1999. It displays at least two distinct arrival times at $1$ AU, which indicates that particles follow at least two distinct field-line lengths.
In our simulations, we also find that in some cases the apparent path lengths can be very different. Two examples are presented in Figure \ref{figure-pathes}. In these two plots we use blue lines as a reference, which represent the particle travel along a field line with length $1.04$ AU with pitch angle $\mu = v_\parallel/v = 1$. It can be seen from Figure \ref{figure-pathes} (upper panel) that the edges of the velocity dispersion $t = 15$ hour and $t = 20$ hour indicate particles arriving earlier than that along the blue line. 
In Figure \ref{figure-pathes} (lower panel) the earliest arrival time for particles at about $t = 15$ hour and $t = 20$ are almost the same as indicated by the blue line. In addition, the slopes of the edges of the velocity dispersions are different.

\begin{figure}
\begin{center}
\includegraphics[width=0.8\textwidth]{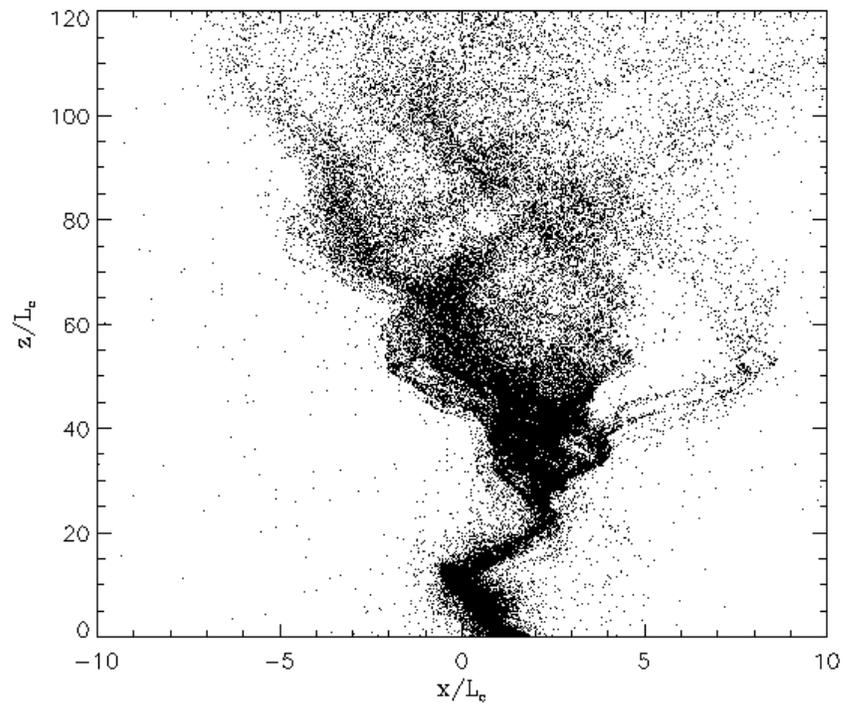}
\caption[The positions of energetic charged particles projected in $x - z$ plane at $t = 12$ hour. The results are from simulations using the foot-point random motion model.]{The positions of energetic charged particles projected in $x - z$ plane at $t = 12$ hour. The results are from the numerical simulations using the foot-point random motion model. It clearly shows that the particles follows the braiding magnetic field lines similar to the Cartoon in Figure \ref{Fluline}. }\label{fig-total-footpoint}
 \end{center}
 \end{figure}

\begin{figure}
\begin{center}
\includegraphics[width=0.8\textwidth]{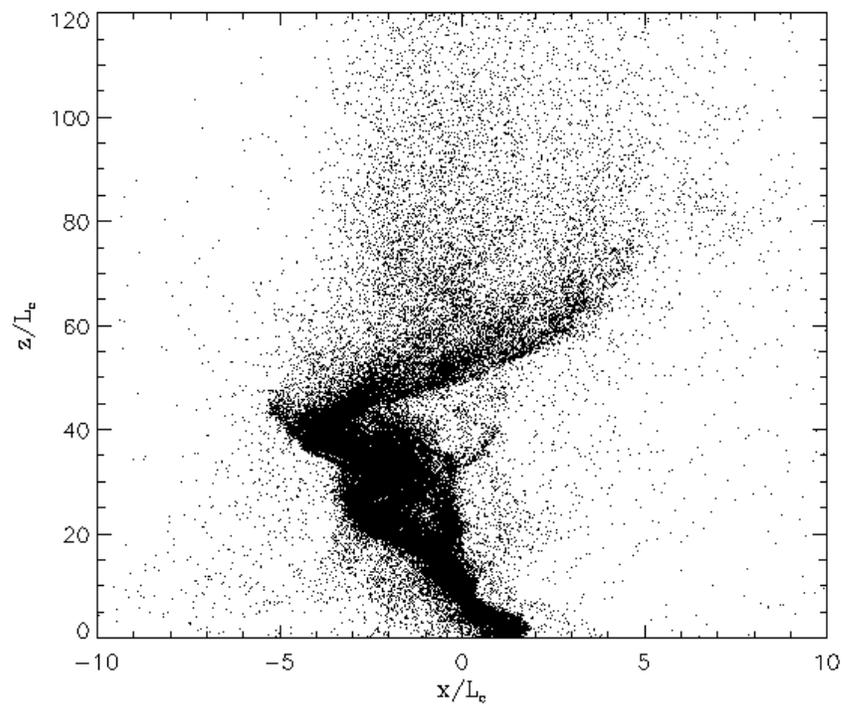}
\caption[The positions of energetic charged particles projected in $x - z$ plane at $t = 12$ hour. The results are from simulations using the two-component model.]{The positions of energetic charged particles projected in $x - z$ plane at $t = 12$ hour. The results are from the numerical simulations using the two-component model (slab and two-dimensional components).}\label{fig-total-twocomponent}
 \end{center}
 \end{figure}

\begin{figure}
\begin{center}
\includegraphics[width=0.7\textwidth]{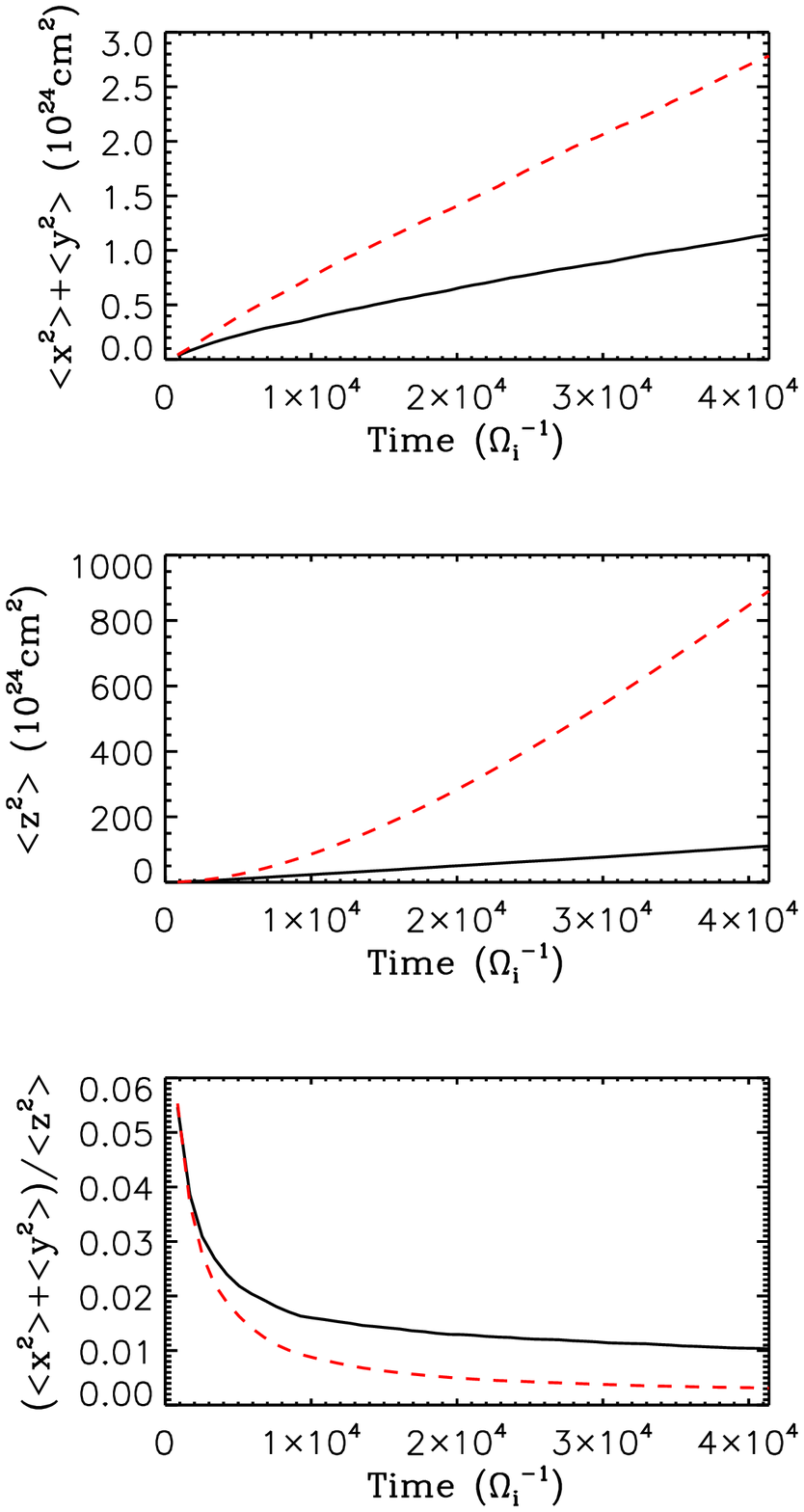}
\caption[Simulated diffusion coefficient of 1 MeV particles parallel and perpendicular to the mean magnetic field.]{Simulated diffusion coefficient of 1 MeV particles parallel (along the z-direction) and perpendicular (along the x- and y-direction) to the mean magnetic field. Black lines represent results from the two-component model and red dashed lines represent results from foot-point random motion model.}\label{fig-diffusion}
 \end{center}
 \end{figure}

We have also attempted to use the two-component model to study the SEP dropouts. However, we did \textit{not} find any clear dropout in our simulations. To further resolve this issue, we have prepared two scatter plots that show the positions for energetic particles  projected on the $x$ - $z$ plane $12$ hours after the initial release. The results are shown in Figure \ref{fig-total-footpoint} for the foot-point random motion model and in Figure \ref{fig-total-twocomponent} for the two-component model. It can be clearly seen in Figure \ref{fig-total-footpoint} that the particles follows the braiding magnetic field lines, and they are therefore separated as the field lines doing random walks. However, this feature is not seen in Figure \ref{fig-total-twocomponent} for the two-component model. A possible reason is that the two-component model contain a slab component that can more efficiently scatter the energetic particles in pitch-angle. To demonstrate this, we measure the diffusion coefficient by implementing the the technique used by \citet{Giacalone1999}. We use the definition of diffusion coefficients $\kappa_{\zeta \zeta} = $ $\langle \zeta^2 \rangle$/$2t$, where $\zeta$ is the spatial displacement in a given time $t$. We calculate the perpendicular and parallel diffusion coefficients for 1-MeV protons in the two turbulence models for the same parameters in the simulation. The results are shown in Figure \ref{fig-diffusion}. For the two-component model, we have $\kappa_\parallel = 1.3 \times 10^{21}$ cm$^2$/s and $\kappa_\perp = 1.1 \times 10^{19}$ cm$^2$/s. For the foot-point random motion model, we have $\kappa_\parallel = 1.5 \times 10^{22}$ cm$^2$/s and $\kappa_\perp = 3.1 \times 10^{19}$ cm$^2$/s. It is shown that the parallel diffusion coefficient for the two-component model is about one order of magnitude smaller than that for the foot-point random motion model, meaning that particles experience more scattering in the two-component model. The calculation also shows a smaller ratio of $\kappa_\perp/\kappa_\parallel$ for the foot-point random motion model ($\kappa_\perp/\kappa_\parallel = 0.002$) compared with that for the two-component model ($\kappa_\perp/\kappa_\parallel = 0.0085$). The results question the popularly used ``two-component" model. If large-scale field line meandering happens to be the explanation for SEP dropouts, pitch-angle scattering due to small-scale scattering should be small so energetic particles can be mostly confined to their field lines of force. When the pitch-angle scattering is large, particles efficiently scatter off their original field lines and observers cannot see the intermittent intensity dropouts. The observational evidence of small scattering has also been shown by \citet{Chollet2011}. They inferred the intensity-fall-off lengths at the edges of the dropouts using ACE/ULEIS
 data and showed that energetic particles rarely scatter off a magnetic field line during the propagation in interplanetary turbulence.

\section{Summary \label{chapter2-summary}}

In this chapter, we presented numerical simulations for the propagation of SEPs in the inner heliosphere. We numerically integrated the trajectories of energetic charged particles in the turbulent magnetic field generated from the commonly used magnetic turbulence models. The observations of SEP events were simulated by collecting charged particles that reach $1$ AU much as a spacecraft detector would. 

Since the initial release of SEPs is highly anisotropic, their propagation cannot be well described by the Parker's transport equation. In Section \ref{chapter2-dispersion} we study the velocity dispersion of SEPs in the turbulent magnetic field and estimate the error involved in the onset analysis commonly used in SEP observations. We find the velocity dispersion can be well produced by this model. For a typical turbulence variation $\delta B^2/B_0^2 \sim 0.1$ observed at $1$ AU and a large source region, we find the  differences between the apparent release time and the actual release time is less than a few minutes, but the apparent path lengths can be significant different than the real path length along the average magnetic field line. For the foot-point random motion model, the error for the inferred release time is smaller than that of the two-component model. This is due to the parallel diffusion coefficient in the foot-point random motion model is considerably larger than that in the two-component model. It should be noted that the energetic particles we study on the onset analysis is fairly energetic ($>1$ MeV) and we assume the energetic particles are released in a large source region. In order to increase the statistics, we collect all the particles that reach 1 AU rather than collect particles using a window of a certain size.

We had also reproduced SEP dropouts in the numerical simulations using the foot-point random motion model, assuming the SEP source region is smaller than the correlation scale. The widths of these dropout are typically several hours, similar to the time scales of dropouts observed in space. The velocity dispersion of the energetic particles appears to have different path lengths, this indicates that the energetic particles travel along different field lines. We have also attempted to use the two-component model to numerically simulate the dropouts of energetic particles. However, we rarely find the evidence of SEP dropouts in our simulation. This is probably because that particle scattering is more efficient in the two-component model compare to that in the foot-point random motion model. This result questions the popular used ``two-component model" in that particles in the turbulence model may scatter off field lines too frequently compared to that constrained by the observed dropouts \citep{Mazur2000}. This explanation is also supported by recent observational analysis \citep{Chollet2011}, which shows that energetic particles are mostly confined to their field lines of forces. 

It has been shown that the slab turbulence model gives a mean-free path of energetic particles much smaller than what observed in SEP events \citep{Palmer1982}. \citet{Matthaeus1990} and \citet{Bieber1996} proposed the two-component model (80\% 2D plus 20\% slab) that can give a mean-free path several times larger than that given by pure slab model. In this study, we provided the evidence that the mean-free path from the two-component model may be need to be even larger to explain the observation of SEP dropouts \citep{Mazur2000}, which is consistent with the observed mean-free path \citep{Palmer1982}.

\chapter{The Effect of Turbulent Magnetic Fields on the Acceleration of Ions at Collisionless Shocks\label{chapter3}}

\section{Particle Acceleration at Collisionless Shocks: An Overview \label{chapter3-overview1}}
Collisionless shocks in space and other astrophysical environments are considered as the main accelerators for energetic charged particles. Diffusive shock acceleration \citep[hereinafter
DSA;][]{Krymsky1977,Axford1977,Bell1978,Blandford1978} is the most accepted
theory for the acceleration of charged particles. The basic conclusions of DSA can be drawn from the Parker's transport equation \citep[Equation \ref{parker_equation},][]{Parker1965} by
considering a fast shock in a 1-D infinite space and steady state system. In the shock frame, we assume that the plasma comes from upstream ($x < 0$) at a flow speed $U_1$, gets compressed and decelerated at the shock ($x = 0$), and flows into downstream ($x > 0$) at a speed of $U_2$. For a source function $Q = \delta(p-p_0)\delta(x)$, which represents the injection of low-energy charged particles with momentum $p_0$ at the shock, the solution of the Parker's equation is

\begin{eqnarray}
f (x, p) &=& (\frac{p}{p_0})^{-\gamma} \exp(-\frac{U_1|x|}{\kappa_{xx,1}}) \qquad \; \; x < 0, p > p_0 \nonumber\\
         &=&  \qquad \quad (\frac{p}{p_0})^{-\gamma}, \qquad \qquad \quad x > 0, p > p_0 \label{Equation-1DDSA}
\end{eqnarray}

\noindent where $U_1$ is the upstream flow speed in the shock frame and $\kappa_{xx,1}$ is the diffusion coefficient of the charged particles normal to the shock surface. The solution naturally predicts a power-law distribution function $f \propto p^{-\gamma}$ with $\gamma = 3U_1/(U_1 - U_2)$. For strong shocks with compression ratios between $2.0$ and $4.0$, the slope index of the power law of the distribution function is between $6.0$ and $4.0$, close to energetic particles observed in many different regions of space. Upstream of the shock front, DSA predicts an exponential drop as a function of the distance from the shock, similar to some spacecraft observations \citep[e.g.,][]{kennel1986}. 

Although DSA has been very successful in explaining the acceleration of charged particles, this theory does have some difficulties. One of the greatest concerns about DSA is how a population of low-energy charged particles gets pre-accelerated at collisionless shocks. Since the Parker's equation does not consider low-energy particles with high anisotropies, how low-energy particles get accelerated at shocks is unclear. This is often referred to as the ``injection problem". There is no current consensus on this issue. We will discuss the injection problem in more detail in Section \ref{chap3-injection}. 

Another important problem for DSA is that the observed energetic particles associated with shock waves are irregular and variable, and they are sometimes very different from what is predicted by the 1-D, steady state solution of DSA (Equation \ref{Equation-1DDSA}). In contrast, DSA is usually considered to be a simple and robust process. It is important to consider how DSA could explain such observations. In recent years, it has been realized that the effects of shock geometries, seed particles, and spatial and temporal variations, etc. can be important and they are considered to be possible solutions for explaining the observed energetic particles. In Section \ref{chap3-variation} we will discuss these observations and list the possible modifications to the $1$-D, steady state solution for DSA to interpret the observations.

There are some other shock-acceleration mechanisms often discussed in the literature. For example, shock drift acceleration \citep[SDA, e.g.,][]{Armstrong1985} and shock surfing acceleration \citep[SSA,][]{Lee1996,Zank1996} at quasi-perpendicular or perpendicular shocks. In SDA, charged particles drift because of the gradient of the magnetic field at the shock front. The direction of the drift motion is in the same direction as motional electric field vector $\textbf{E} = - \textbf{U} \times \textbf{B} /c$, and the particles gain energy during this drift motion. In SSA, it is thought that the cross shock potential electric field could reflect ions upstream and the ions gain energy by the gyro-motion along the motional electric field.   

SDA and DSA are usually considered to be distinct and their relationship requires some clarification.
It has been demonstrated by \citet{Jokipii1982,Jokipii1987} that in a diffusive process, SDA can be unified into DSA by considering the drift term in the Parker's transport equation. In that case, both drift and diffusion play a role and their relative contributions depend on the shock normal angles. \citet{Jokipii1987} showed that the acceleration rate is greatly enhanced at perpendicular shock or highly oblique shocks since the perpendicular diffusion coefficient is usually much smaller than the parallel diffusion coefficient.

In Section \ref{chap3-injection} we discuss the injection problem of ions for DSA. In Section \ref{chap3-paralell} we present a study on the acceleration of thermal ions at parallel shocks using 3-D hybrid simulation (kinetic ions and fluid electron). In Section \ref{chap3-variation} we will discuss a variety of effects that could modify the simple $1$-D diffusive shock acceleration model. In Section \ref{chap3-mag} we present a study on particle acceleration at shocks containing large-scale magnetic variations.

\section{The ``Injection Problem" (The Acceleration of Low-energy Particles) \label{chap3-injection}}
Since the Parker's transport equation \citep[Equation \ref{parker_equation},][]{Parker1965} assumes a quasi-isotropic distribution of energetic particles, it does not include the pre-acceleration process for low-energy particles known as the ``injection problem". It is usually thought that when the charged particles are energetic enough, they can be efficiently scattered by magnetic turbulence close to the shock and get accelerated in DSA. At present there is no consensus on this issue. However, one can work out the condition for the transport equation to be valid when the anisotropy due to diffusive streaming is small. The absolute value of the streaming anisotropy is \citep{Giacalone1999}


\begin{eqnarray}
\xi = \frac{3|S_i|}{vf_0} =  \frac{3 U_1}{v} \left\lbrace 1 + \frac{(\kappa_A/\kappa_\parallel)^2 \sin^2 \theta_{Bn} + (1-\kappa_\perp/\kappa_\parallel)^2 \sin^2\theta_{Bn}\cos^2\theta_{Bn}}{\left[ (\kappa_\perp/\kappa_\parallel) \sin^2 \theta_{Bn} + \cos^2 \theta_{Bn} \right]^2}\right\rbrace^{1/2}.
\end{eqnarray}

\noindent If we define that the particles can be efficiently accelerated by diffusive shock acceleration when $\xi < 1$, this equation gives an injection velocity

\begin{eqnarray}
v_{inj} > 3 U_1 \left\lbrace 1 + \frac{(\kappa_A/\kappa_\parallel)^2 \sin^2 \theta_{Bn} + (1-\kappa_\perp/\kappa_\parallel)^2 \sin^2\theta_{Bn}\cos^2\theta_{Bn}}{\left[ (\kappa_\perp/\kappa_\parallel) \sin^2 \theta_{Bn} + \cos^2 \theta_{Bn} \right]^2} 
\right\rbrace^{1/2}. \label{equation-injection-velocity}
\end{eqnarray}

\noindent For parallel shocks or quasi-parallel shocks, this indicates that $v_{inj} > 3 U_1$ and the injection is relatively easy.  
It is usually thought that Alfven waves excited by the streaming of high-energy, shock accelerated protons can scatter the pitch angle of the particles. When the fluctuations efficiently interact with the particles, these particles are trapped near the shock and gain energy from the plasma compression across the shock. \citet{Ellison1981} first advocated a model for DSA that includes the injection process, where the particles are assumed to be originated from the shock-heated ions and leak freely from downstream to upstream of the shock. The similar models have been developed and extended by a number of authors \citep[e.g.,][]{Ellison1990,Malkov1998,Kang2012}. This is usually referred to as the ``thermal leakage" model.

However, a number of researchers \citep{Quest1988,Scholer1990a,Scholer1990b,Kucharek1991,
Giacalone1992} have found a different scenario for the initial energization at parallel shocks based on the results of self-consistent hybrid simulations. It is found that the accelerated ions originate from the shock layer rather than via leakage from the plasma. The ions can be accelerated from the incident thermal plasma to high energies while they are making gyro-motions in the electric and magnetic fields at the shock layer. Although the average incident magnetic field is parallel to the shock normal, as the enhanced upstream magnetic fluctuations steepen and convect through the shock layer, the angle between the incident magnetic field and the shock-normal right at the shock front can be quite large. A particle can gain the first amount of energy by drifting and being reflected in this ``locally oblique" shock structure \citep{Giacalone1992}. It has been clearly shown by \citet{Kucharek1991} that most of the accelerated particles are reflected and gain the first amount of energy at the shock layer. \citet{Lyu1990} have presented the results of hybrid simulations and they claimed that the leakage protons dominated the accelerated particles. The reason that they obtained a different result from other researchers may be due to the method they used to drive shocks in their simulations. In their simulations, the shocks are initially assumed to be a hyperbolic tangent function with a thickness of several ion inertial lengths $c/\omega_{pi}$, where $c$ and $\omega_{pi}$ are the light speed and proton plasma frequency, respectively. The magnetic fluctuations that are important to reflect ions at shock front are ignored at the beginning of the simulation. It should be noted that previous simulations are restricted to 1-D simulations and occasionally 2-D simulations. In those situations the motions of charged particles are restricted on their original field lines as demonstrated by \citet{Jokipii1993} and \citet{Jones1998}. This restriction motivates us to study the acceleration of ions at parallel shocks using 3-D simulations. In Section \ref{chap3-paralell} we will present a new study on this problem using 3-D hybrid simulations. The results show that energetic particles can move across field lines but the acceleration mechanism is similar to what is found by previous hybrid simulations \citep{Quest1988,Scholer1990a,Scholer1990b,Kucharek1991,Giacalone1992}. Namely, the energetic particles originate in the shock layer and are not due to leakage from downstream.

For perpendicular shocks or quasi-perpendicular shocks, it is thought to be more difficult for particles to be injected into DSA. The required pre-acceleration can be achieved by shock drift acceleration \citep[e.g.,][]{Armstrong1985} or shock surfing acceleration \citep{Lee1996,Zank1996}. Recent progress has been made to distinguish the relative importance for these two processes. It has been found that in order for shock surfing acceleration to be efficient, the thickness of the shock layer has to be very thin (electron scale), which is different from what is observed in space \citep{Bale2003} and in numerical simulations \citep{Leroy1982}. Moreover, it has been discussed that the shock thickness has to be fairly large compared to electron gyroradii to be consistent with the observation of electron heating at shocks \citep{Lembege2004}. More detailed studies using full particle simulations and hybrid simulations have been presented by \citet{Yang2009} and \citet{Wu2009}. 

\citet{Giacalone1999} has demonstrated that the effect of large-scale magnetic turbulence can efficiently lower the injection velocity $v_{inj}$ at perpendicular shocks by increasing the transport of charged particles normal to magnetic field direction. Recent numerical simulations for the acceleration of charged particles in the existence of large-scale magnetic fluctuations show very efficient acceleration, which indicates that there is \textit{no} injection problem \citep{Giacalone2005a,Giacalone2005b}. It should be noted that there is also an injection problem for electrons, which has been considered to be more difficult than that of ions. In Chapter \ref{chapter4} we will discuss the acceleration of electrons. The results suggest that the acceleration of electrons prefers perpendicular shocks.

\section{A 3-D Hybrid Simulation on Particle Acceleration at Parallel Shocks \label{chap3-paralell}}
In order to examine the acceleration of charged particles at parallel shocks, we performed 3-D hybrid simulations to study the initial acceleration process. As pointed out by a few previous works \citep{Jokipii1993,Giacalone1994}, it is important to consider the motions of charged particles in 3-D to avoid the artificial restriction on their gyro-motions.

\subsection{Numerical Methods}
We perform 3-D hybrid simulations for particle acceleration at parallel shocks.
In the hybrid simulation \citep[e.g.,][]{Winske1988}, the ions are treated
fully kinetically and thermal electrons are treated as a massless fluid. It keeps ion-scale kinetic physics but neglect electron-scale kinetic physics. Therefore the relevant spatial scales are ion gyroradius $v_{thi}/\Omega_{ci}$ and ion inertial length $c/\omega_{pi}$, and the relevant time scale is the ion gyroperiod $\Omega_{ci}^{-1}$, where $v_{thi}$ is the ion thermal speed, $\Omega_{ci}$ is the ion gyrofrequency, $c$ is the light speed in \textit{vacuum}, and $\omega_{pi}$ is the ion plasma frequency. This feature is well suited to describe supercritical collisionless shocks, in which the dynamics of ions is important. 

In the simulation, the ions are treated as kinetic particles moving in the simulation domain by solving the equations of motion for each ion $j$:
\begin{center}
\begin{eqnarray}
m_j \frac{d \textbf{v}_j}{d t} &=& q_j (\textbf{E} + \frac{\textbf{v}_j \times \textbf{B}}{c}) - \eta \textbf{J} \label{ion-v}\\
\frac{d \textbf{x}_j}{dt} &=& \textbf{v}_j. \label{ion-x}
\end{eqnarray}
\end{center}

The first two terms in the right-hand side of Equation \ref{ion-v} represent the Lorentz force, and the third term describes the effect of resistive coupling between electrons and ions where $\eta$ is the resistivity and $\textbf{J}$ is the total current. The electric and magnetic fields are defined on grid points, and are interpolated to the locations of particles. After pushing the particles at each time step, the new particle positions $\textbf{x}_j$ and velocities $\textbf{v}_j$ are collected at the grid points to get the quantities such as ion density $n_i$, ion flow velocity $\textbf{V}_i$ and ion current $\textbf{J}_i = q_i n_i \textbf{V}_i$. Since the electrons in the simulation are assumed to be a massless neutralizing fluid. We solve the momentum equation of electron fluid:

\begin{eqnarray}
n_e m_e \frac{d \textbf{V}_e}{dt} = -en_e (\textbf{E} + \frac{\textbf{V}_e \times \textbf{B}}{c}) - \nabla \cdot \overrightarrow{\textbf{P}}_e + \eta \textbf{J} = 0  , \label{equation-electron}
\end{eqnarray}

\noindent where $\textbf{V}_e$ is the electron fluid velocity and $\overrightarrow{\textbf{P}}_e$ is the electron pressure tensor. The electron and ion charge 
density are equal $e n_e = q_i n_i$, where $n_e$ is the electron number density. The pressure tensor $\overrightarrow{\textbf{P}}_e$ in the equation is usually taken 
to be a scalar $\overrightarrow{\textbf{P}}_e = p_e \overrightarrow{\textbf{I}} $. We assume the electron fluid is adiabatic $p_e \propto n_e^{5/3}$.

The electromagnetic fields are described by Ampere's law,

\begin{eqnarray}
\nabla \times \textbf{B} = \frac{4 \pi}{c} \textbf{J} = \frac{4\pi}{c} q_i n_i (\textbf{V}_i - \textbf{V}_e), \label{Ampere}
\end{eqnarray}

\noindent and Faraday's law

\begin{eqnarray}
\frac{\partial \textbf{B}}{\partial t} = -c (\nabla \times \textbf{E}). \label{Faraday}
\end{eqnarray}

At each time step, we get $\textbf{V}_e$ from Equation \ref{Ampere}. The electric field is obtained from Equation \ref{equation-electron} and the magnetic field is updated using Equation \ref{Faraday}.

\begin{figure}
\centering
\begin{tabular}{c}
\epsfig{file=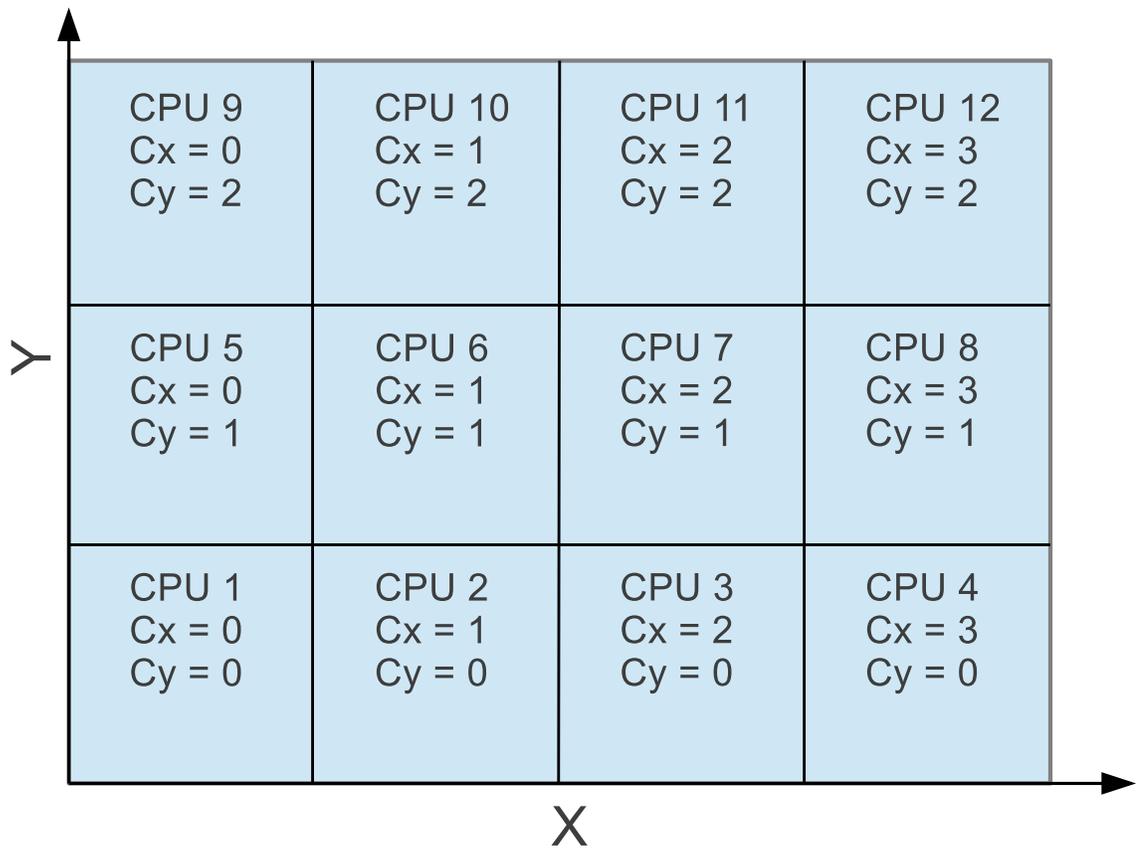,width=\textwidth,clip=}
\end{tabular}
\caption[A 2-D example for the parallization of hybrid simulation.]{A 2-D example for the parallization of hybrid simulation. \label{CPUs}}
\end{figure}

We have improved the parallelization of the $1$-D, $2$-D, and $3$-D hybrid simulation models \citep{Giacalone2000,Giacalone2004,Giacalone2005b}. The new versions of the codes have been implemented and tested on the NASA's Pleiades supercomputer using a few thousand CPU cores. In Figure \ref{CPUs} we use a 2-D example to illustrate the method of parallelization. The 2-D simulation domain is split into several sub-domains along the $x$ and $y$ directions. Each sub-domain is placed on a single CPU core, CPU 1, CPU 2, ... The sub-domain also has two labels that is determined by its location in the global simulation domain. For example, the sub-domain CPU 1 is labelled by $C_x = 0$ and $C_y = 0$, and the sub-domain of CPU 6 is labelled by $C_x = 1$ and $C_y = 1$. The dimension of each sub-domain is $d_x = L_{x}/n_{cx}$ and $d_y = L_{y}/n_{cy}$, where $n_{cx}$ and $n_{cy}$ are numbers of CPUs along the x and y directions, respectively. The information of the kinetic particles in numerical cells $\textbf{x}_j$ and $\textbf{v}_j$ and fields on grid points like $\textbf{B}$, $\textbf{E}$ within each sub-domain is loaded on the corresponding CPU. At each time step, the fields at boundaries are transferred to and from its adjacent sub-domains ($C_x + 1$, $C_y$), ($C_x - 1$, $C_y$), ($C_x$, $C_y + 1$) and ($C_x$, $C_y - 1$). When a particle move across a sub-domain, it will be transfer to the corresponding adjacent CPU. The number of the CPU is determined by its label $C_x$ and $C_y$.

For this work, we consider a three-dimensional Cartesian grid ($x, y, z$). All the physical vectors have components in three directions and also spatially depend on $x, y,$ and $z$. A shock is produced by using the so-called piston method, in which a plasma flow is injected continuously
from one end ($x=0$, in our case) of the simulation box, and reflected elastically at the other end ($x=L_x$). This boundary is also assumed to be a perfectly conducting barrier. The pileup of density and magnetic field creates a shock propagating in the $-x$ direction. In the $y$ and $z$ direction the boundary conditions are periodic for both particles and electromagnetic fields. 

The size of the simulation box for each case is listed in Table \ref{table-hybrid}. The Mach number of the plasma flow in the simulation frame is $M_{A0} = 4.0$. The electron and ion plasma betas are $\beta_e = 1.0$ and $\beta_i = 0.5$, respectively. The grid size is $\Delta x \times \Delta y \times \Delta z = 0.5 c/\omega_{pi} \times 0.5 c/\omega_{pi} \times 0.5 c/\omega_{pi}$, where $c/\omega_{pi}$ is the ion inertial length, and $c$ and $\omega_{pi}$ are the light speed in \textit{vacuum} and proton plasma frequency in the simulation domain at the beginning of the simulation, respectively. The time step is taken to be $\Omega_{ci} \Delta t = 0.01$, where $\Omega_{ci}$ is the gyrofrequency. The ratio between light speed and upstream Alfven speed is $c/v_{A0} = 6000.0$, and the anomalous resistivity is $\eta = 1\times 10^{-6} 4\pi \omega_{pi}^{-1}$. The initial spatially uniform thermal ion distribution is represented using $25$ particles per cell, which is enough to give a reasonable distribution.

Initially the average magnetic field is assumed to be $\textbf{B}_0 = B_0 \hat{x}$, i.e., the averaged shock normal angle is $0^\circ$. We examine four simulation cases with parameters listed in Table \ref{table-hybrid}. The spatial sizes in the $x$ direction are taken to be $300 c/\omega_{pi}$ for all the cases. For the first two cases, the system is assumed to have \textit{no} pre-existing magnetic fluctuation but the size of the simulation box in the $y$ and $z$ direction for Run $2$ ($40 c/\omega_{pi}$) are considerably larger than that for Run $1$ ($10 c/\omega_{pi}$). The fluctuations in the simulation box are self-consistently generated during the simulation. In Run $3$ and Run $4$, the sizes of the simulation boxes are the same as that in Run $2$. Differently, we examine the effect of pre-existing magnetic fluctuations by superposing a random magnetic field on the mean field. The turbulent magnetic field is added at the beginning of the simulation and also injected continuously at the $x=0$ boundary during the simulation. In Run $3$, we assume the pre-existing magnetic fluctuation is one-dimensional and only depends on $x$, but the resulting field can depend on all three dimensions. The amplitude of the fluctuations at wave number $k$ is determined from a Kolmogorov-like power spectrum. The largest and smallest wave lengths are taken to be $40 c/\omega_{pi}$ and $2.5 c/\omega_{pi}$. The total variance is taken to be $\sigma^2 = \delta B^2/B_0^2 = 0.1$. In Run $4$ the magnetic power of the injected fluctuation is assumed to be isotropically distributed in three wave number vectors $k_x$, $k_y$, and $k_z$. The total variance is taken to be $\sigma^2 = 0.1$, the same as that in Run $3$. The difference between Run $3$ and Run $4$ is that in Run $4$ the pre-existing magnetic fluctuation is allowed to be variable in full three dimensions. As we will see below, this has an important effect on the motions of charged particles along the shock front.

\begin{table}
\centering
\begin{tabular*}
{0.86\textwidth}{cccc}
\hline
Run& $L_x (c/\omega_{pi}) \times L_y (c/\omega_{pi}) \times L_z (c/\omega_{pi})$ & $\sigma^2 (\delta B^2/B_0^2)$ & Model \\
\hline
1  & $300\times 10 \times 10$ & 0.0 & - \\
2  & $300\times 40 \times 40$ & 0.0 & - \\
3  & $300\times 40 \times 40$ & 0.1 & 1-D slab\\
4  & $300\times 40 \times 40$ & 0.1 & 3-D isotropic\\
 \hline
\end{tabular*}
 \caption{Some parameters for different cases in the $3$-D hybrid simulations of parallel shocks: the size of the simulation domain in unit $c/\omega_{pi}$, the variance of injected magnetic fluctuation $\sigma^2$, and the model for the injected magnetic fluctuation.}\label{table-hybrid}
\end{table}

\subsection{Results of Numerical Simulations}

\begin{figure}
\centering
\begin{tabular}{c}
\epsfig{file=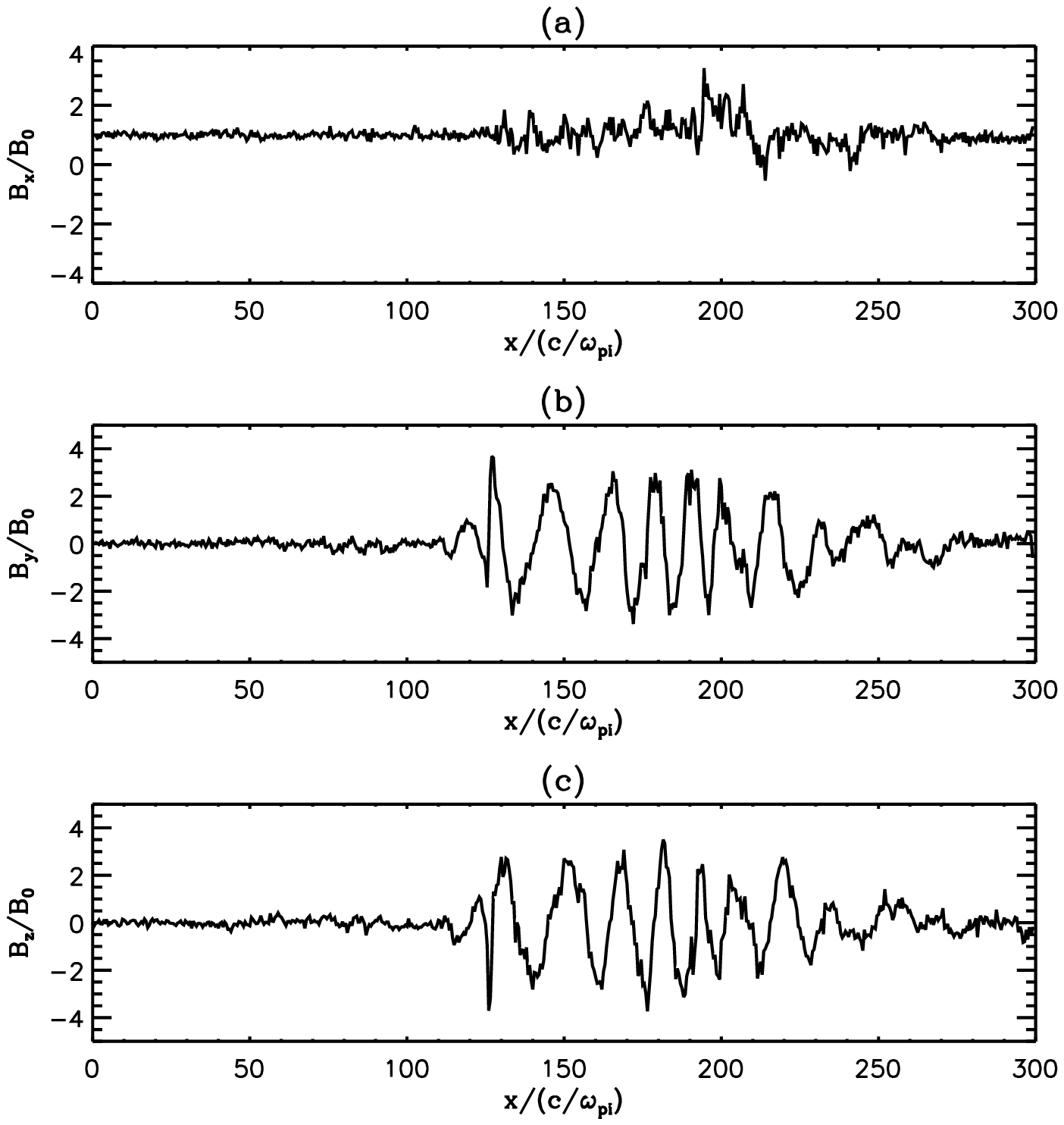,width=0.8\textwidth,clip=}
\end{tabular}
\caption[$1$-D profiles of magnetic field component $B_x$, $B_y$, and $B_z$ for Run $1$ (with a small simulation box in the $y$ and $z$ directions).]{$1$-D profiles of magnetic field components $B_x/B_0$, $B_y/B_0$, and $B_z/B_0$ for Run $1$ (with a small simulation box in the $y$ and $z$ directions), where $B_0$ is the initial upstream magnetic field. \label{profileBsmall}}
\end{figure}

\begin{figure}
\centering
\begin{tabular}{c}
\epsfig{file=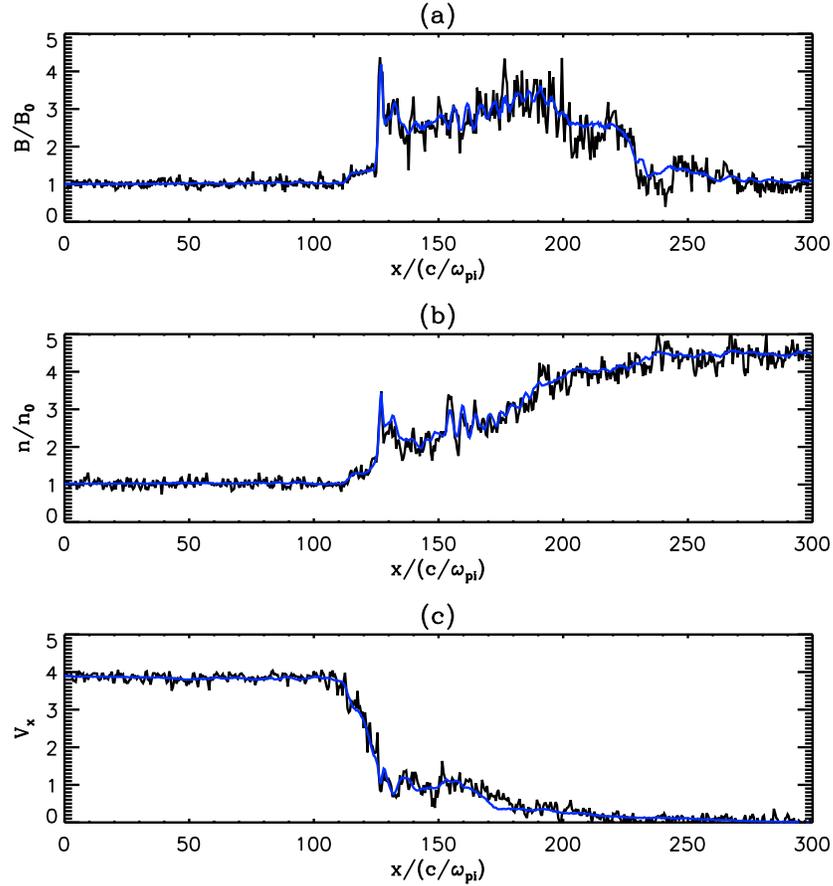,width=0.8\textwidth,clip=}
\end{tabular}
\caption[$1$-D representation of magnetic-field magnitude, plasma density, and velocity for Run $1$ (with a small simulation box in the $y$ and $z$ directions).]{$1$-D representation of magnetic-field magnitude, plasma density, and velocity for Run $1$: (a) the profile of magnitude of magnetic field $B/B_0$ along the same line (black line) and the magnetic field magnitude averaged over the $y$ and $z$ directions (blue line); (b) the profile of plasma number density $n/n_0$ along the same line (black line) and the number density averaged over the $y$ and $z$ directions (blue line), where $n_0$ is the initial upstream plasma number density; (c) the profile of the $x$-component of the plasma flow speed $V_x/V_{A0}$ along the same line (black line) and the plasma flow speed in the $x$ direction averaged over the $y$ and $z$ directions (blue line), where $V_{A0}$ is the initial upstream Alfven speed. \label{profile-small}}
\end{figure}

In Figure \ref{profileBsmall} we present the simulation results for Run $1$ at $\Omega_{ci}t = 120.0$. The figure shows the profiles of the magnetic field components in the $x$ direction $B_x/B_0$ (a), $y$ direction $B_y/B_0$ (b), and $z$ direction $B_z/B_0$ (c) along $y = 5$ $c/\omega_{ci}$ and $z = 5$ $c/\omega_{ci}$, where $B_0$ is the initial upstream magnetic field. It is shown that the magnetic field fluctuations around the shock layer are mostly transverse to the initial magnetic field in the $x$ direction. The fluctuations are circularly right-handed polarized if they propagate along upstream direction, indicating that they are excited by the reflected ions that flow upstream. This result is consistent with previous analytical theories and numerical simulations \citep[e.g., ][]{Quest1988}. Close to the shock layer, the magnetic fluctuations are steepened and enhanced. The fluctuations get compressed and amplified when they are passed by the shock. Different from $1$-D simulations, the fluctuations in the $x$ direction are also observed. In Figure \ref{profile-small} (a) the magnitude of magnetic field $B/B_0$ along $y = 5$ $c/\omega_{ci}$ and $z = 5$ $c/\omega_{ci}$ is plotted using a black line. The magnitude of magnetic field averaged over the $y$ and $z$ directions is overlapped on the plot using a blue line. It can be seen that although there is some small difference between the profile of the magnetic field and the averaged magnetic field, the $1$-D cut is very close to the averaged value of the magnetic field, indicating that the fields are weakly dependent on the $y$ and $z$ directions. The same features can be seen in Figure \ref{profile-small} (b) and Figure \ref{profile-small} (c), which show the profiles and averaged values of plasma number density $n/n_0$ and flow velocity in the $x$ direction $V_x/V_{A0}$, where $n_0$ is the initial upstream plasma number density and $V_{A0}$ is the initial upstream Alfven speed. 

\begin{figure}
\centering
\begin{tabular}{c}
\epsfig{file=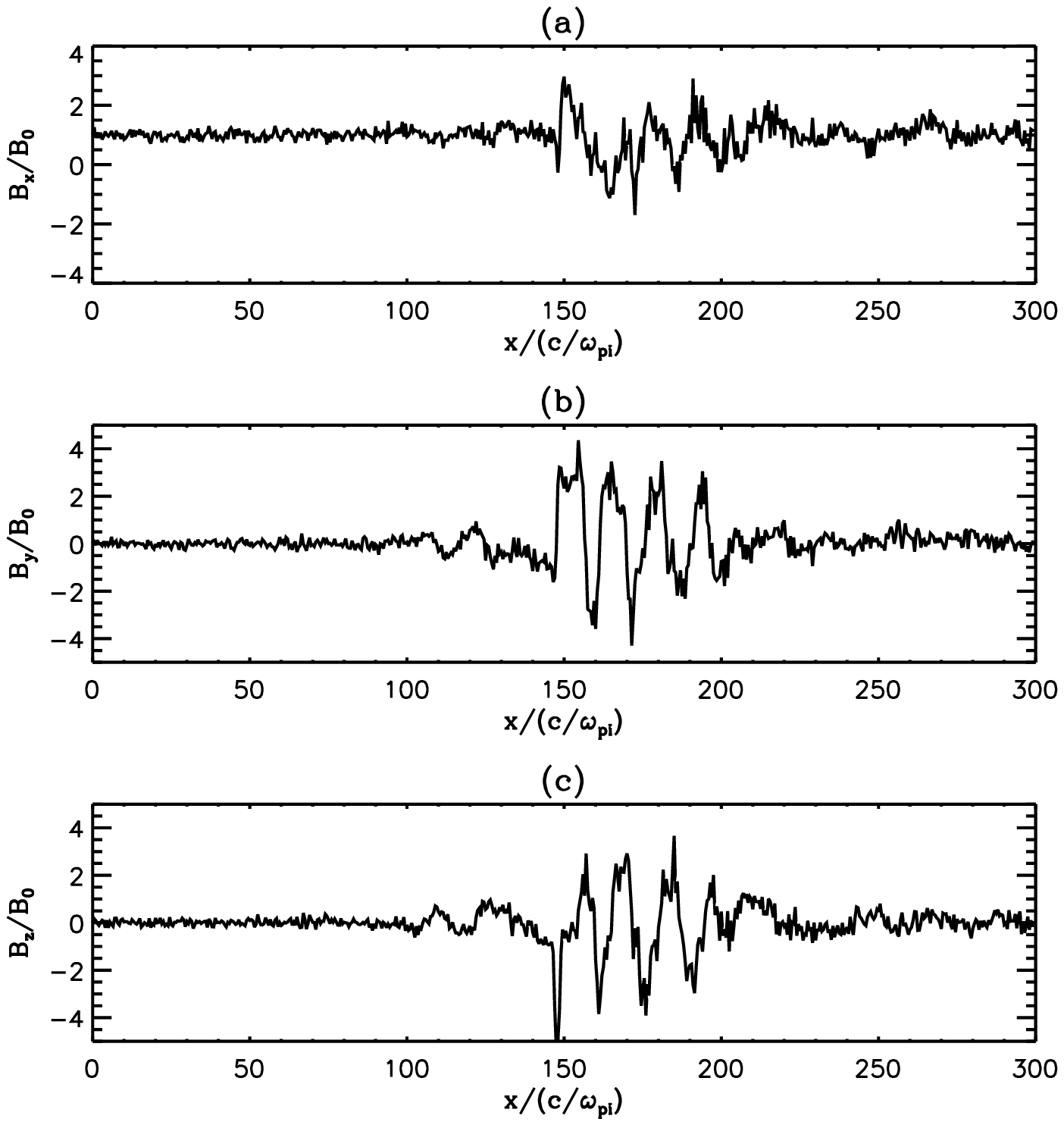,width=0.8\textwidth,clip=}
\end{tabular}
\caption[$1$-D profiles of magnetic field component $B_x$, $B_y$, and $B_z$ for Run $2$ (with a large simulation box in the $y$ and $z$ directions).]{$1$-D profiles of magnetic field components $B_x/B_0$, $B_y/B_0$, and $B_z/B_0$ for Run $2$ (with a large simulation box in the $y$ and $z$ directions), where $B_0$ is the initial upstream magnetic field. \label{profileBlarge}}
\end{figure}

\begin{figure}
\centering
\begin{tabular}{c}
\epsfig{file=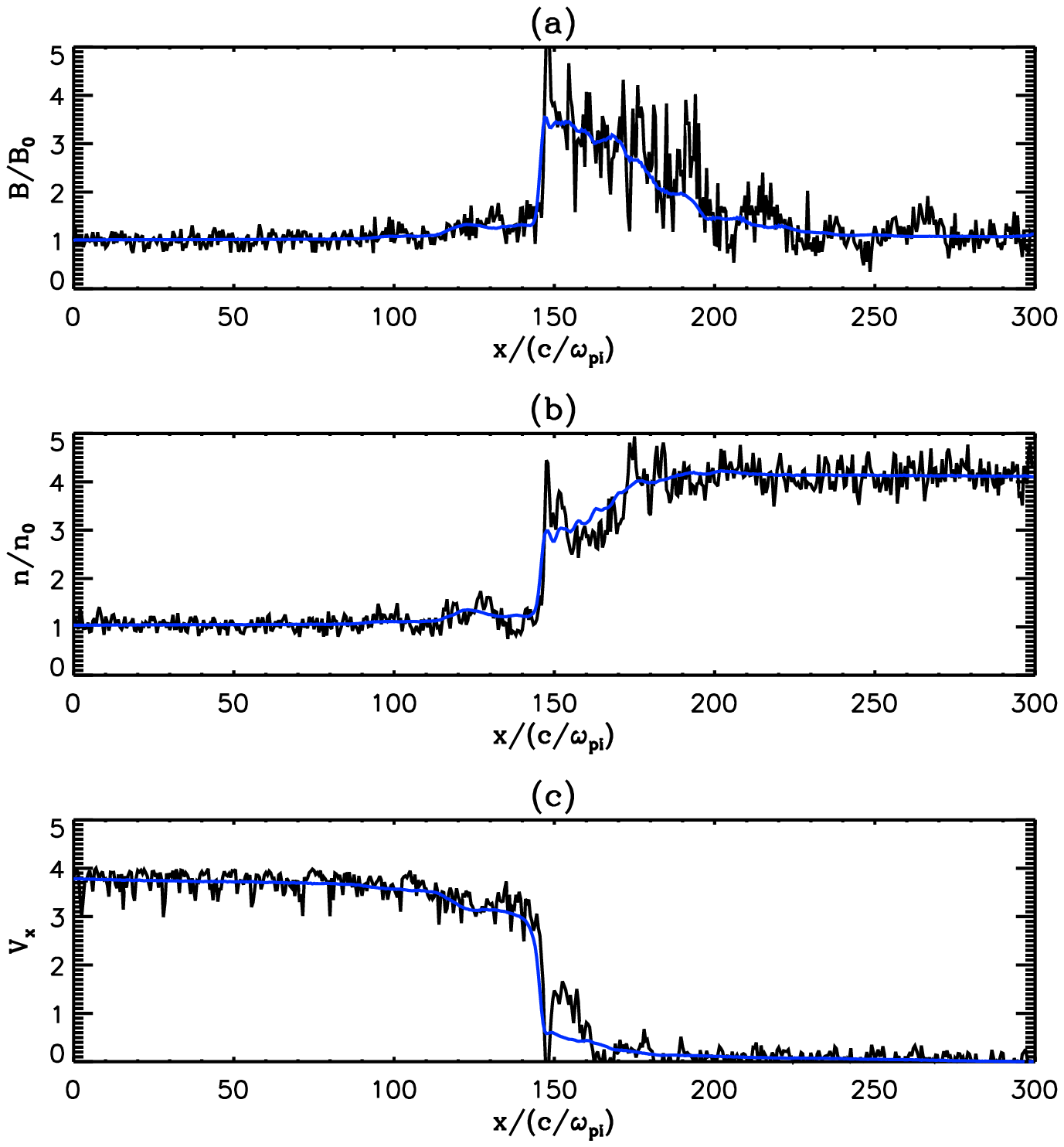,width=0.8\textwidth,clip=}
\end{tabular}
\caption[$1$-D representation of magnetic-field magnitude, plasma density, and velocity for Run $2$ (with a large simulation box in the $y$ and $z$ directions).]{$1$-D representation of magnetic-field magnitude, plasma density, and velocity for Run $2$: (a) the profile of magnitude of magnetic field $B/B_0$ along $y = 35$ $c/\omega_{ci}$ and $z = 35$ $c/\omega_{ci}$ (black line) and the magnetic field magnitude averaged over the $y$ and $z$ directions (blue line); (b) the profile of plasma number density $n/n_0$ along the same line (black line) and the number density averaged over the $y$ and $z$ directions (blue line), where $n_0$ is the initial upstream plasma number density; (c) the profile of the $x$-component of the plasma flow speed $V_x/V_{A0}$ along the same line (black line) and the plasma flow speed in the $x$ direction averaged over the $y$ and $z$ directions (blue line), where $V_{A0}$ is the initial upstream Alfven speed. \label{profile-large}}
\end{figure}

Figure \ref{profileBlarge} shows the simulation results for Run $2$, which has a large simulation domain in the $y$ and $z$ directions than that of Run 1. The physical quantities are plotted in a way similar to Figure \ref{profileBsmall}, except that the profile is selected at $y = 35$ $c/\omega_{ci}$ and $z = 35$ $c/\omega_{ci}$. One can see from Figure \ref{profileBlarge} (a) that the $x$-component of the magnetic field $B_x/B_0$ is more fluctuating compared to that for Run $1$. This is probably due to the fact that a larger number of modes are allowed for the larger simulation domain and more obliquely propagating modes are excited. As shown in Figure \ref{profile-large}, a very different feature for physical quantities in Run $2$ compared to that in Run $1$ is that the $1$-D profiles of physical quantities close to the shock front cannot be well represented by the average values over the $y$ and $z$ directions. This indicates that the physical quantities in Run $2$ is dependent on three spatial dimensions, whereas those quantities in Run $1$ is only weakly dependent on the $y$ and $z$ directions.

\begin{figure}
\centering
\begin{tabular}{c}
\epsfig{file=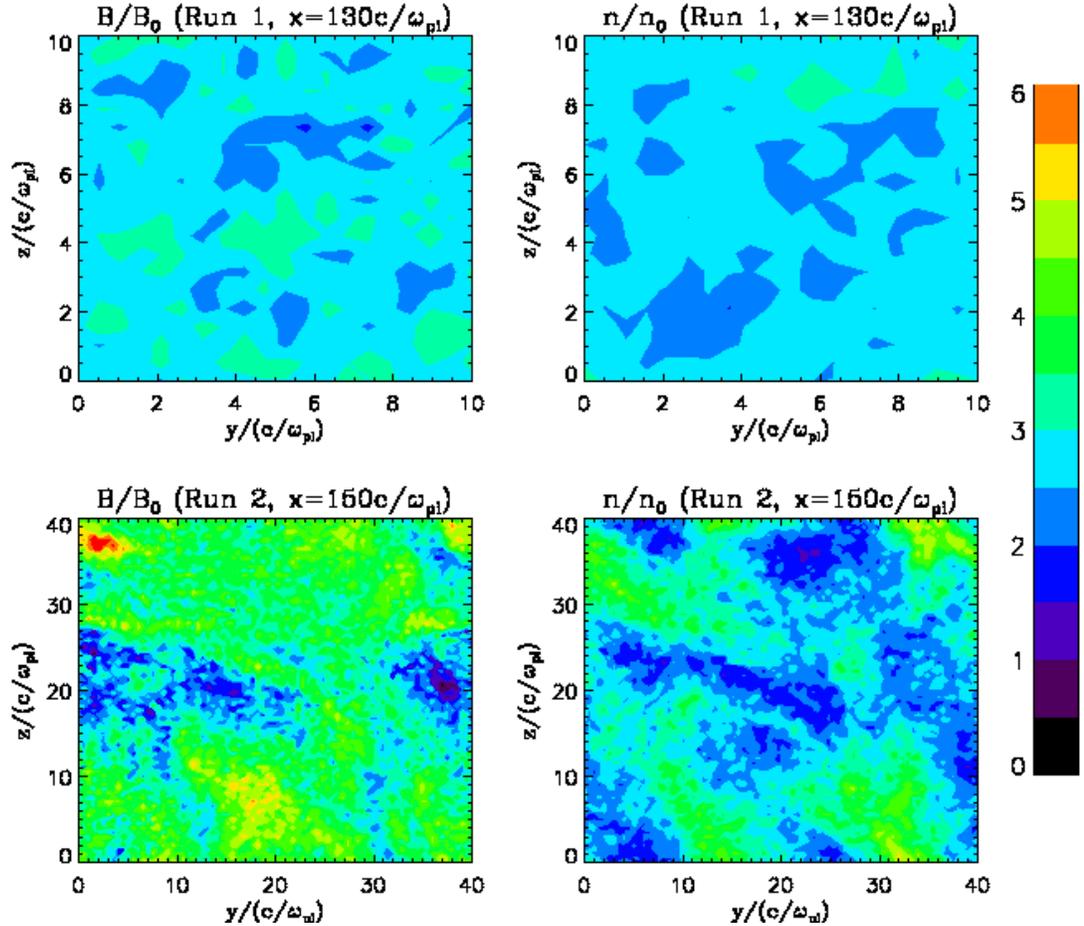,width=\textwidth,clip=}
\end{tabular}
\caption[The cross sections of magnetic-field magnitude and density close to the shock fronts for Run 1 and Run 2.]{The cross sections of magnetic-field magnitude and density close to the shock fronts for Run 1 ($x = 130 c/\omega_{pi}$) and Run 2 ($x = 150 c/\omega_{pi}$) at $\Omega_{ci}t = 120$. \label{bcontour-cross}}
\end{figure}

\begin{figure}
\centering
\begin{tabular}{cc}
\epsfig{file=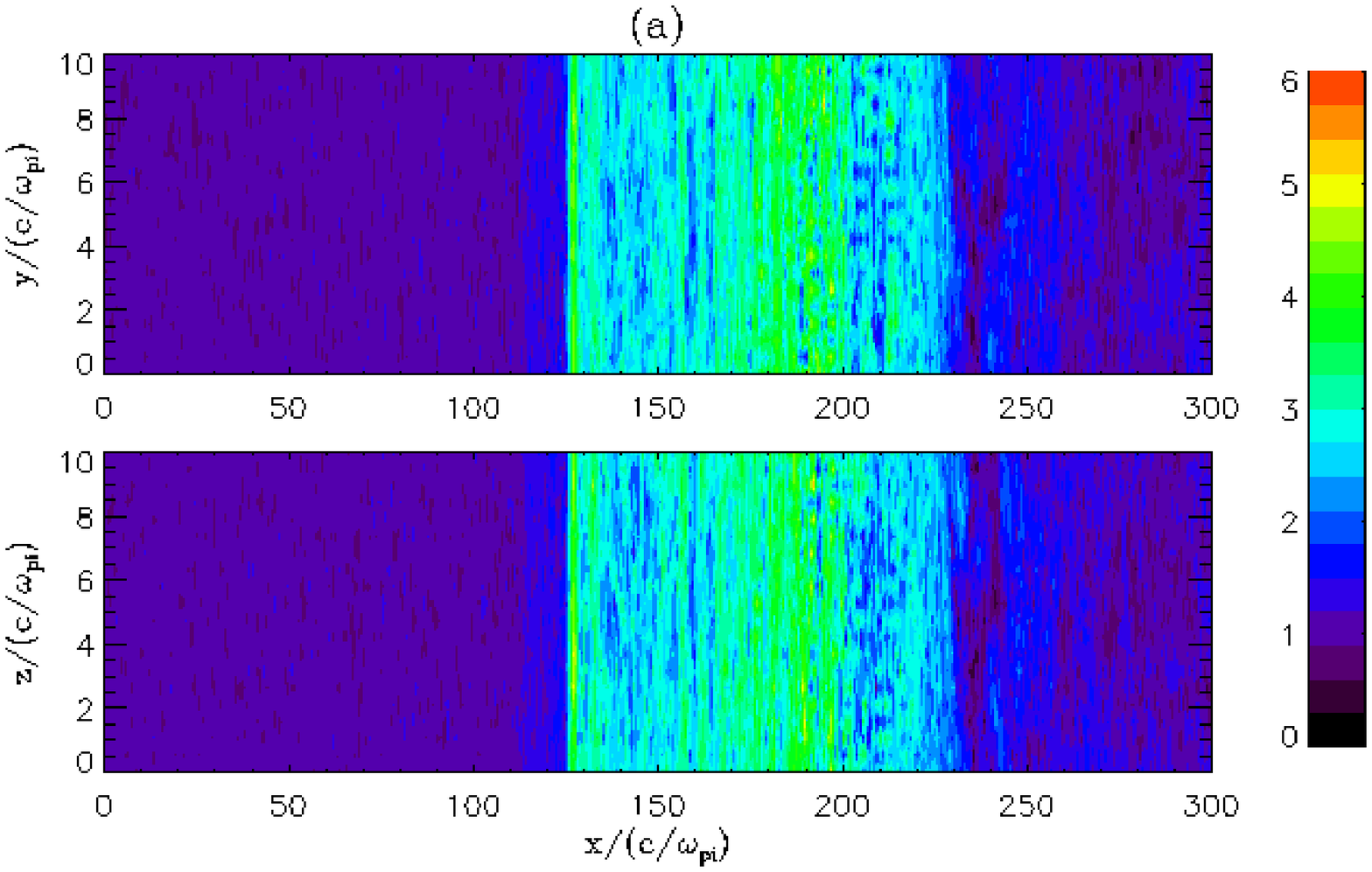,width=\textwidth,clip=} \\
\epsfig{file=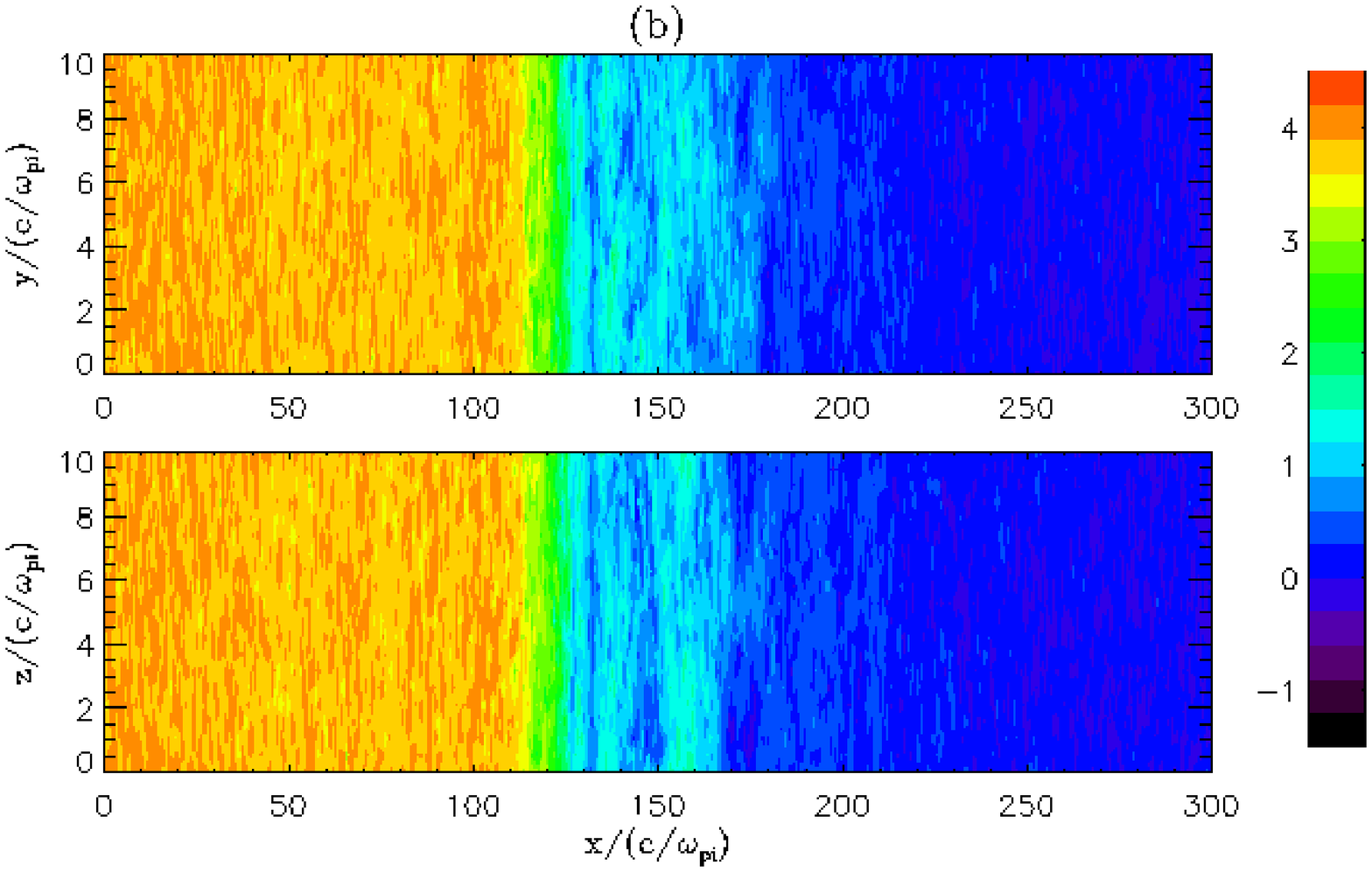,width=\textwidth,clip=}
\end{tabular}
\caption[$2$-D representation of the magnitude of magnetic field $B/B_0$ and the velocity in the $x$-direction $V_x/V_{A0}$ for Run $1$ ($L_y = L_z = 10$ $c/\omega_{pi}$).]{$2$-D representation of (a) the magnitude of magnetic field $B/B_0$ (the $z = 5$ $c/\omega_{pi}$ plane and the $y = 5$ $c/\omega_{pi}$ plane) and (b) the velocity in the $x$-direction $V_x/V_{A0}$ (the $z = 5$ $c/\omega_{pi}$ plane and the $y = 5$ $c/\omega_{pi}$ plane), for the case of Run $1$.  \label{figure-vb-small}}
\end{figure}

\begin{figure}
\centering
\begin{tabular}{cc}
\epsfig{file=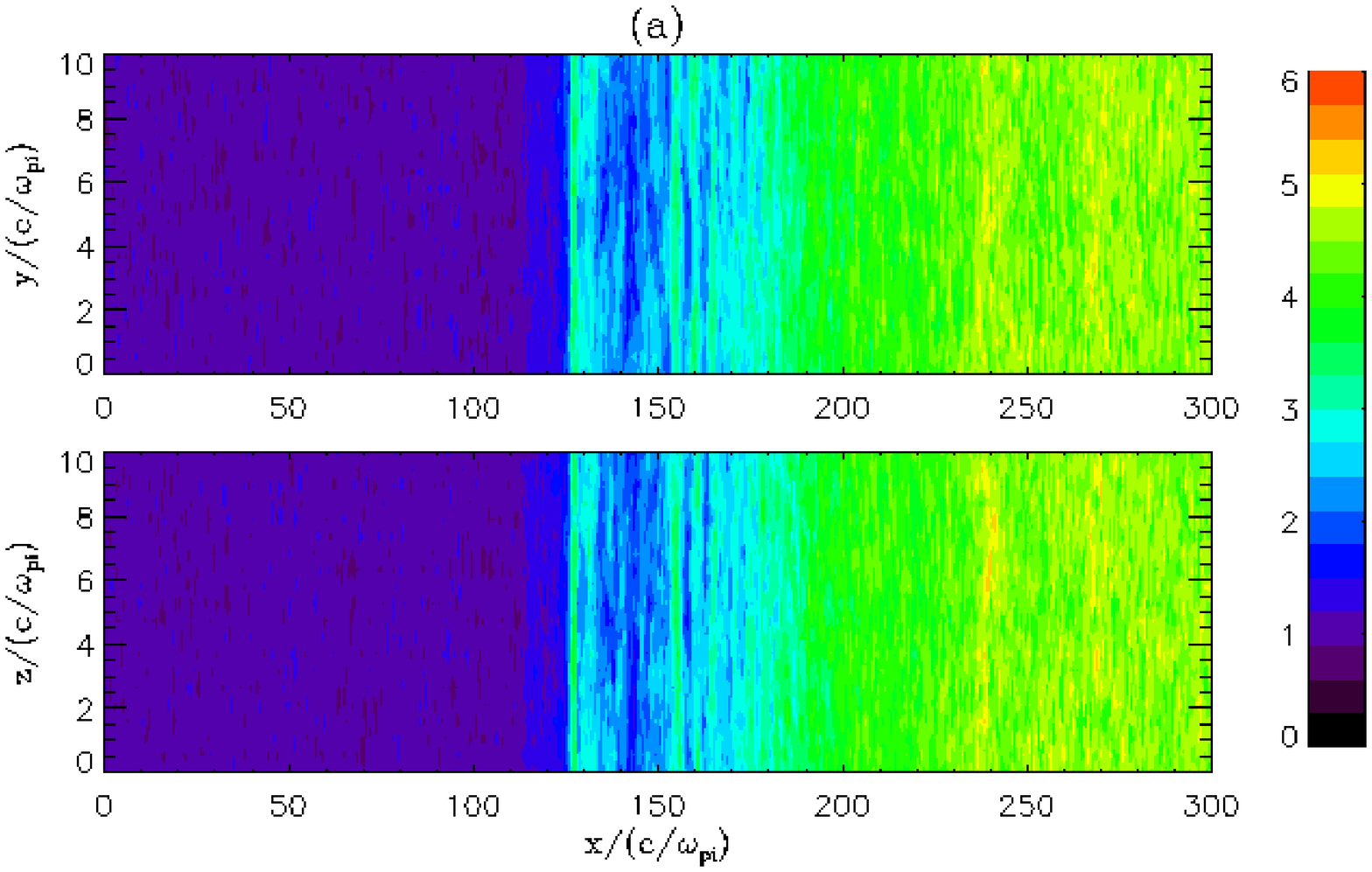,width=\textwidth,clip=} \\
\epsfig{file=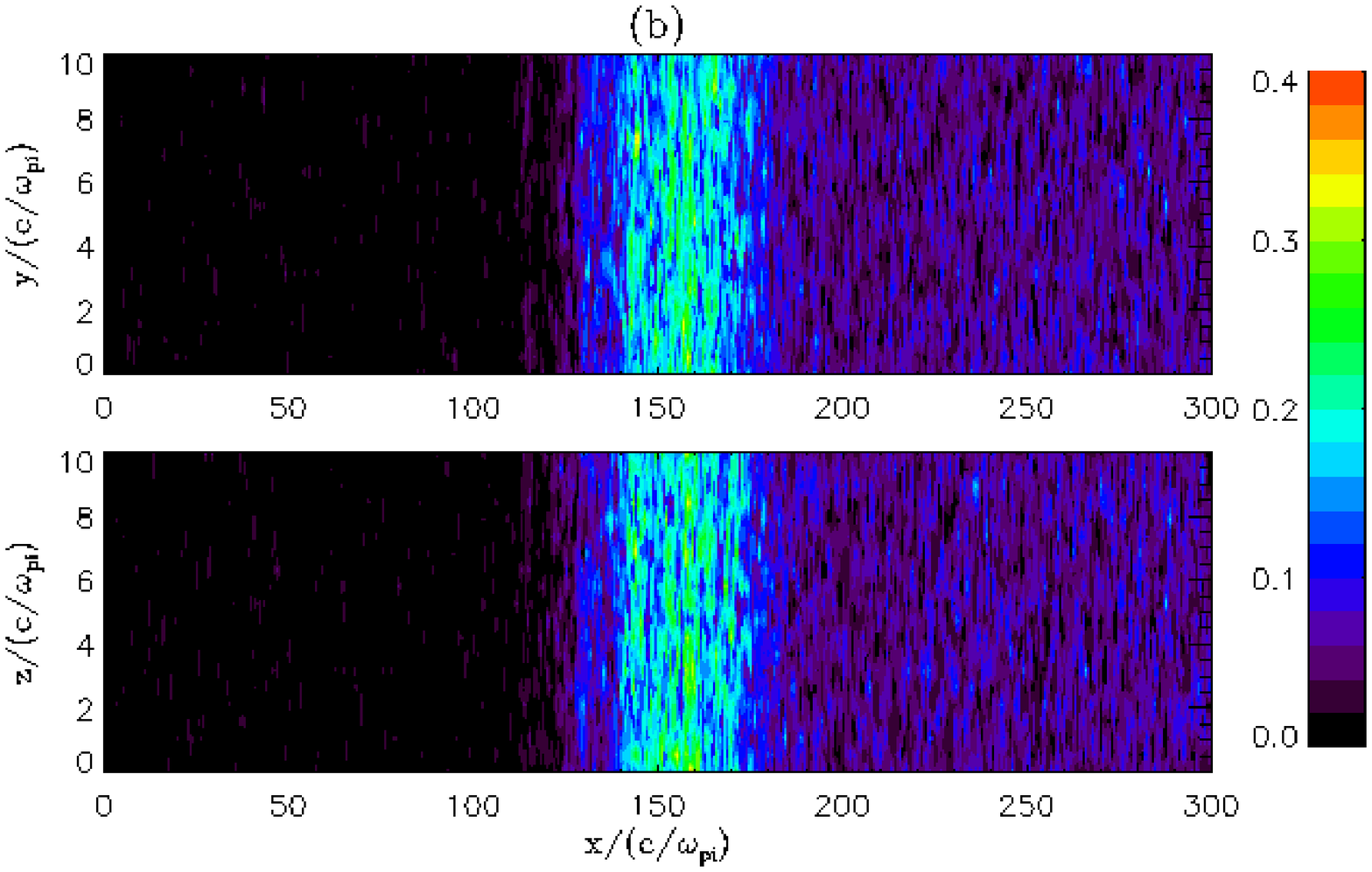,width=\textwidth,clip=}
\end{tabular}
\caption[$2$-D representation of the plasma number density $n/n_0$ and the number density of the accelerated particles for Run $1$ ($L_y = L_z = 10$ $c/\omega_{pi}$).]{Similar to Figure \ref{figure-vb-small}. 2-D representation of (a) the density of the plasma flow and (b) the density of the accelerated particles with energy $3 E_1 < E < 5 E_1$, for the case of Run $1$. \label{figure-den-small}}
\end{figure}

\begin{figure}
\centering
\begin{tabular}{cc}
\epsfig{file=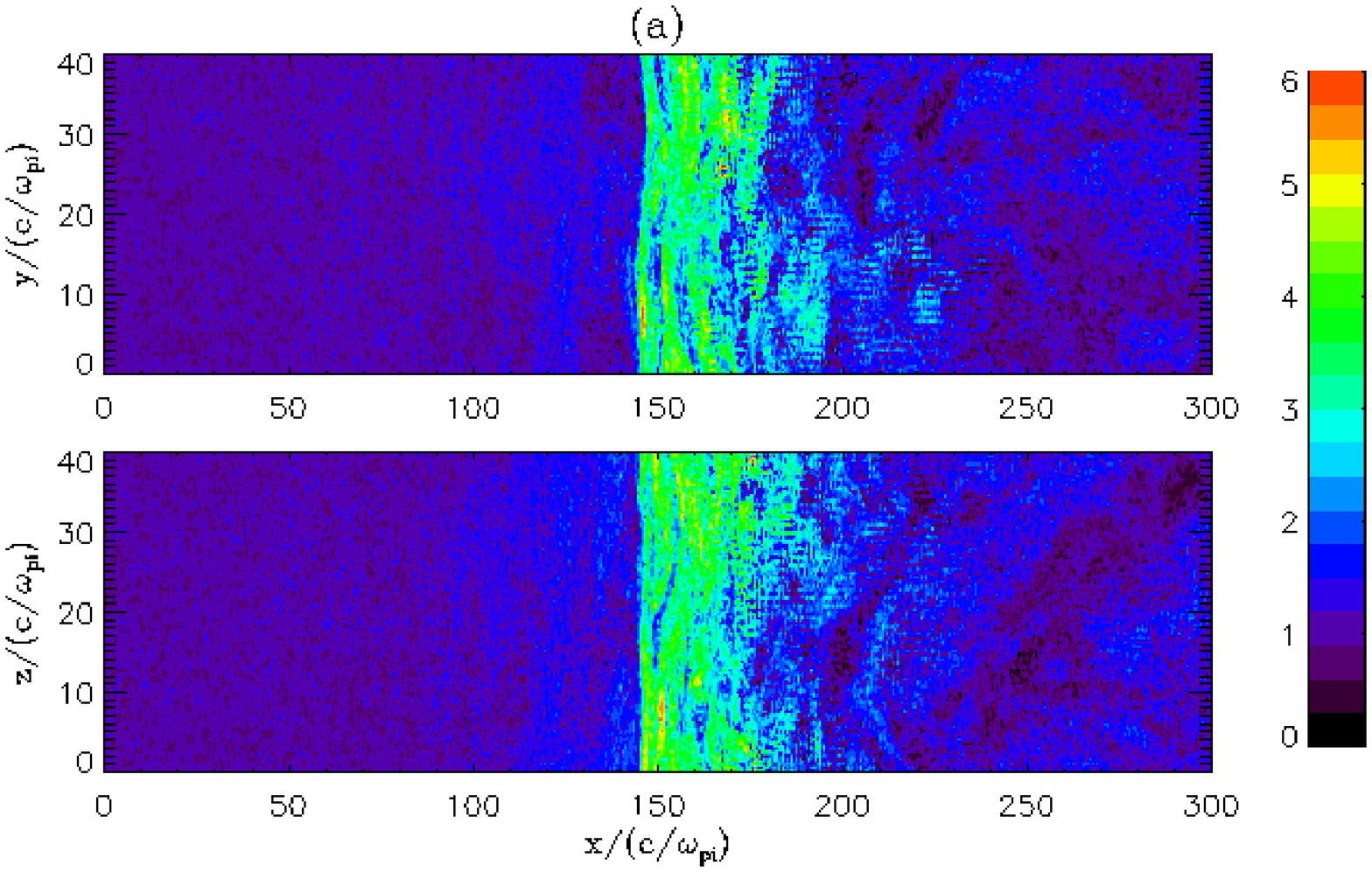,width=\textwidth,clip=} \\
\epsfig{file=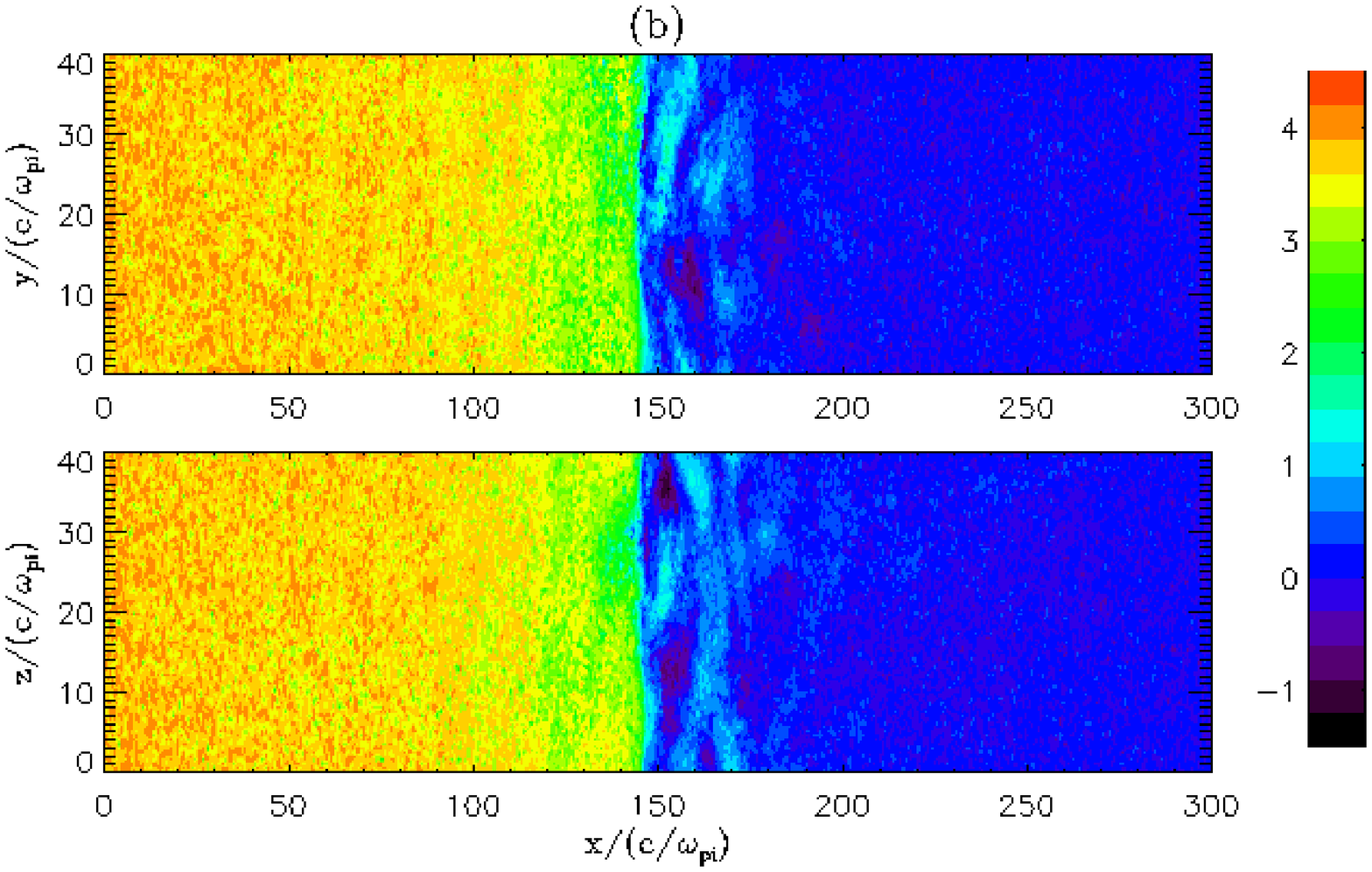,width=\textwidth,clip=}
\end{tabular}
\caption[$2$-D representation of the magnitude of magnetic field $B/B_0$ and the velocity in the $x$-direction $V_x/V_{A0}$ for Run $2$ ($L_y = L_z = 40$ $c/\omega_{pi}$).]{$2$-D representation of (a) the magnitude of magnetic field $B/B_0$ (the $z = 20$ $c/\omega_{pi}$ plane and the $y = 20$ $c/\omega_{pi}$ plane) and (b) the velocity in the $x$-direction $V_x/V_{A0}$ (the $z = 20$ $c/\omega_{pi}$ plane and the $y = 20$ $c/\omega_{pi}$ plane), for the case of Run $2$.  \label{figure-vb}}
\end{figure}

\begin{figure}
\centering
\begin{tabular}{cc}
\epsfig{file=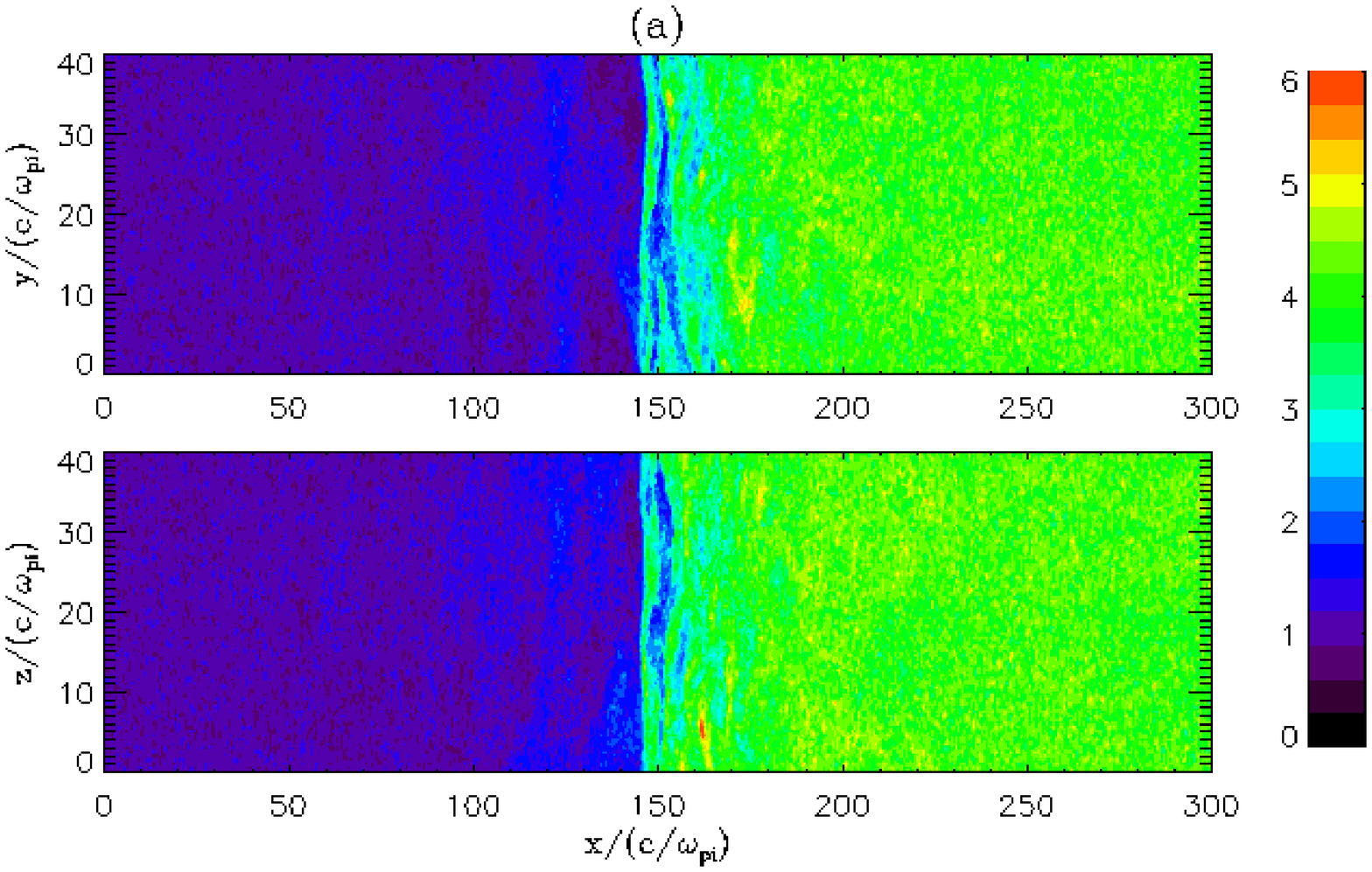,width=\textwidth,clip=} \\
\epsfig{file=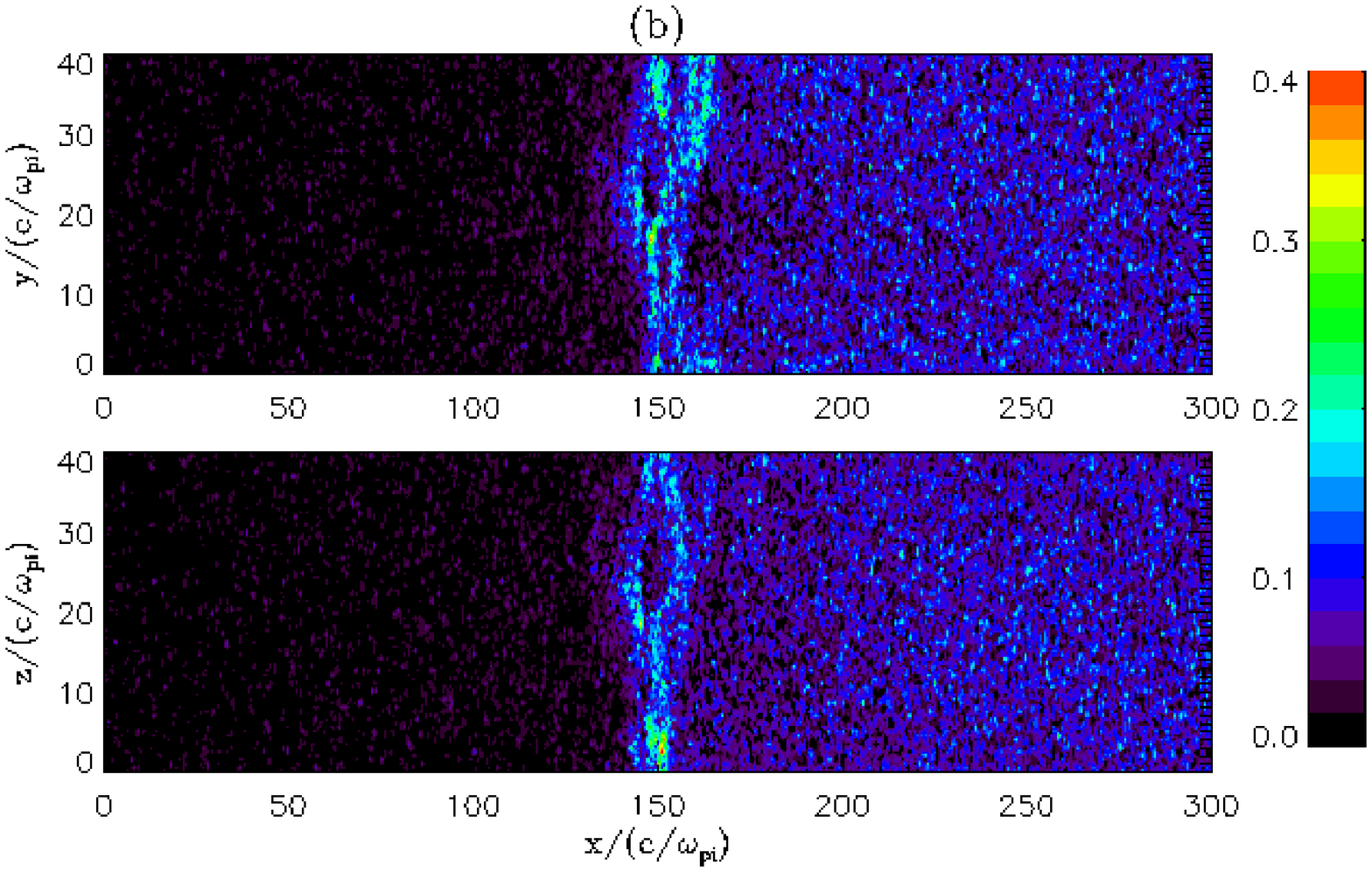,width=\textwidth,clip=}
\end{tabular}
\caption[$2$-D representation of the plasma number density $n/n_0$ and the number density of the accelerated particles for Run $2$ ($L_y = L_z = 40$ $c/\omega_{pi}$).]{Similar to Figure \ref{figure-vb}. 2-D representation of (a) the density of the plasma flow and (b) the density of the accelerated particles with energy $3 E_1 < E < 5 E_1$, for the case of Run $2$. \label{figure-den}}
\end{figure}

\begin{figure}
\centering
\begin{tabular}{c}
\epsfig{file=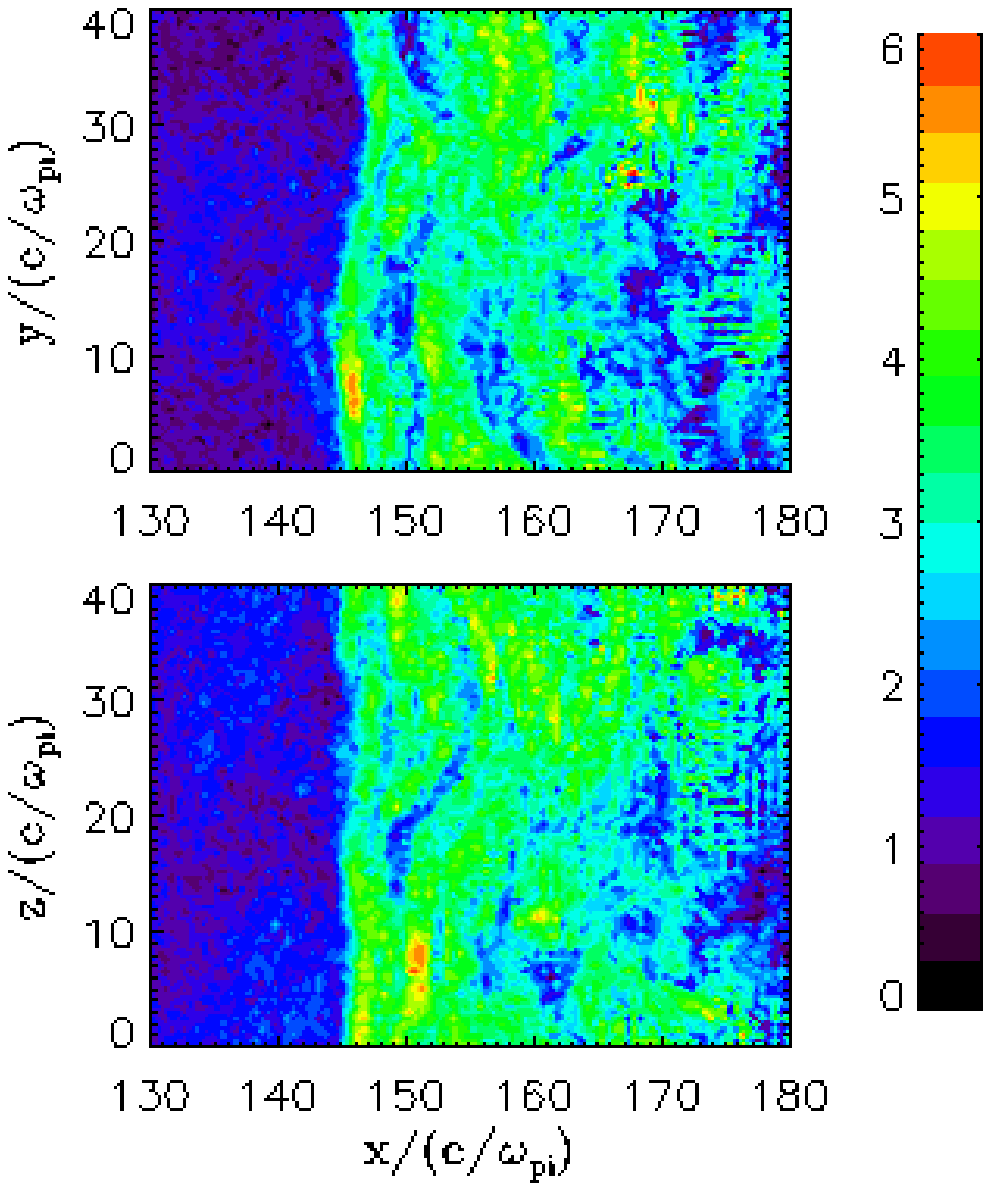,width=0.7\textwidth,clip=}
\end{tabular}
\caption[The magnitude of magnetic field around the shock front for Run 2.]{The magnitude of magnetic field around the shock front for Run 2. \label{bcontour-zoomin}}
\end{figure}

To better illustrate the difference between Run $1$ and Run $2$, in Figure \ref{bcontour-cross} we plot the cross sections of magnetic-field magnitude around the shock front for Run 1 and Run 2. It can be seen that in Run 2 the magnetic-field fluctuation around the shock front has a larger amplitude than that in Run 1. 
The difference between Run $1$ and Run $2$ can also be seen in $2$-D contours. In Figure \ref{figure-vb-small} we show the color-coded contours of the $2$-D cuts of (a) the magnitude of the magnetic field $B/B_0$ (the $z = 5$ $c/\omega_{pi}$ plane and the $y = 5$ $c/\omega_{pi}$ plane), and (b) the flow velocity in the $x$-direction $V_x/V_{A0}$ (the $z = 5$ $c/\omega_{pi}$ plane and the $y = 5$ $c/\omega_{pi}$ plane) for the case of Run $1$.
In Figure \ref{figure-den-small} we show color-coded contours of $2$-D cuts of (a) the density of plasma $n/n_0$ (the $z = 5$ $c/\omega_{pi}$ plane and the $y = 5$ $c/\omega_{pi}$ plane), and (b) the density of accelerated particles with energy $3E_1<E<5E_1$ (the $z = 5$ $c/\omega_{pi}$ plane and the $y = 5$ $c/\omega_{pi}$ plane) for Run $1$, where $E_{1} = \frac{1}{2}m_p V_1^2$ is the upstream plasma ram energy in the shock frame. The simulation time for these figures is $\Omega_{ci} t = 120.0$. From these figures we can see that the shock front is approximately located at about $x = 125$ $c/\omega_{pi}$ where the physical quantities have a jump.
Although the simulation is fully $3$-D, because of the limited size of the simulation box, all the physical quantities appear to be close to $1$-D and only depend on $x$. 
When the size of the simulation box is larger in the $y$ and $z$ directions, the restriction on the motion of the particle can be removed since the wave modes are allowed to grow in $3$-D. This is indicated by Figure \ref{figure-vb} and \ref{figure-den}, which show the similar contours to Figure \ref{figure-vb-small} and \ref{figure-den-small}, but for Run $2$. In this case the lengths of the simulation box in the $y$ direction and the $z$ direction are both $40 c/\omega_{pi}$. In Figure \ref{bcontour-zoomin} the region around the shock front is zoomed in to show the magnetic field structure close to the shock. One can see the $3$-D features around the shock front. 

\begin{figure}
\centering
\begin{tabular}{c}
\epsfig{file=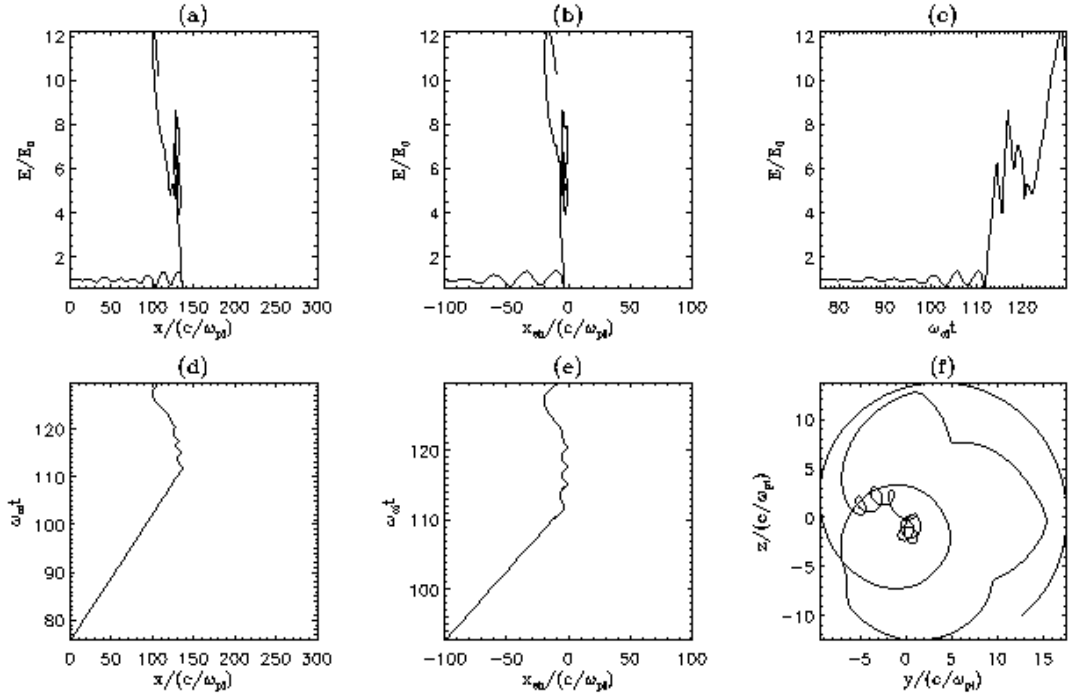,width=\textwidth,clip=}
\end{tabular}
\caption[The trajectory of a representative particle accelerated at the shock layer for the case of Run 1.]{The trajectory of a representative particle accelerated at the shock layer for Run 1. The physical quantities for each panel are: (a) the energy versus the $x$ position in the simulation frame, (b) the energy versus the $x$ position in the shock frame, (c) the energy versus the time in the simulation, (d) the time versus the $x$ position in the simulation frame, (e) the time versus the $x$ position in the shock frame, and (f) the displacement along the $z$ direction versus the displacement along the $y$ direction. \label{figure-traj-small}}
\end{figure}

In Figure \ref{figure-traj-small} we show a trajectory of a representative particle that is accelerated at shock layer for the case of Run $1$. The physical quantities for each panel are: (a) the energy versus the $x$ position in the simulation frame, (b) the energy versus the $x$ position in the shock frame, (c) the energy versus the time in the simulation, (d) the time versus the $x$ position in the simulation frame, (e) the time versus the $x$ position in the shock frame, and (f) the displacement along the $z$ direction versus the displacement along the $y$ direction. One can clearly see that the acceleration happens right at the shock front. The acceleration mechanism is due to the reflection at the shock layer. As shown in this figure, the particle can ride on the shock front and get accelerated for about $15$ gyroperiods and the energy gain can be about $10$ times of the plasma ram energy. This process has been discussed by previous authors \citep{Quest1988,Scholer1990a,Scholer1990b,Kucharek1991,Giacalone1992}. Although we use the 3-D hybrid simulation, in this case the motion of the particle is still restricted by the limited size of the simulation box. This can be seen from Figure \ref{figure-traj-small} (f), which shows that the guiding center of the particle did not move much during the acceleration process. 

\begin{figure}
\centering
\begin{tabular}{c}
\epsfig{file=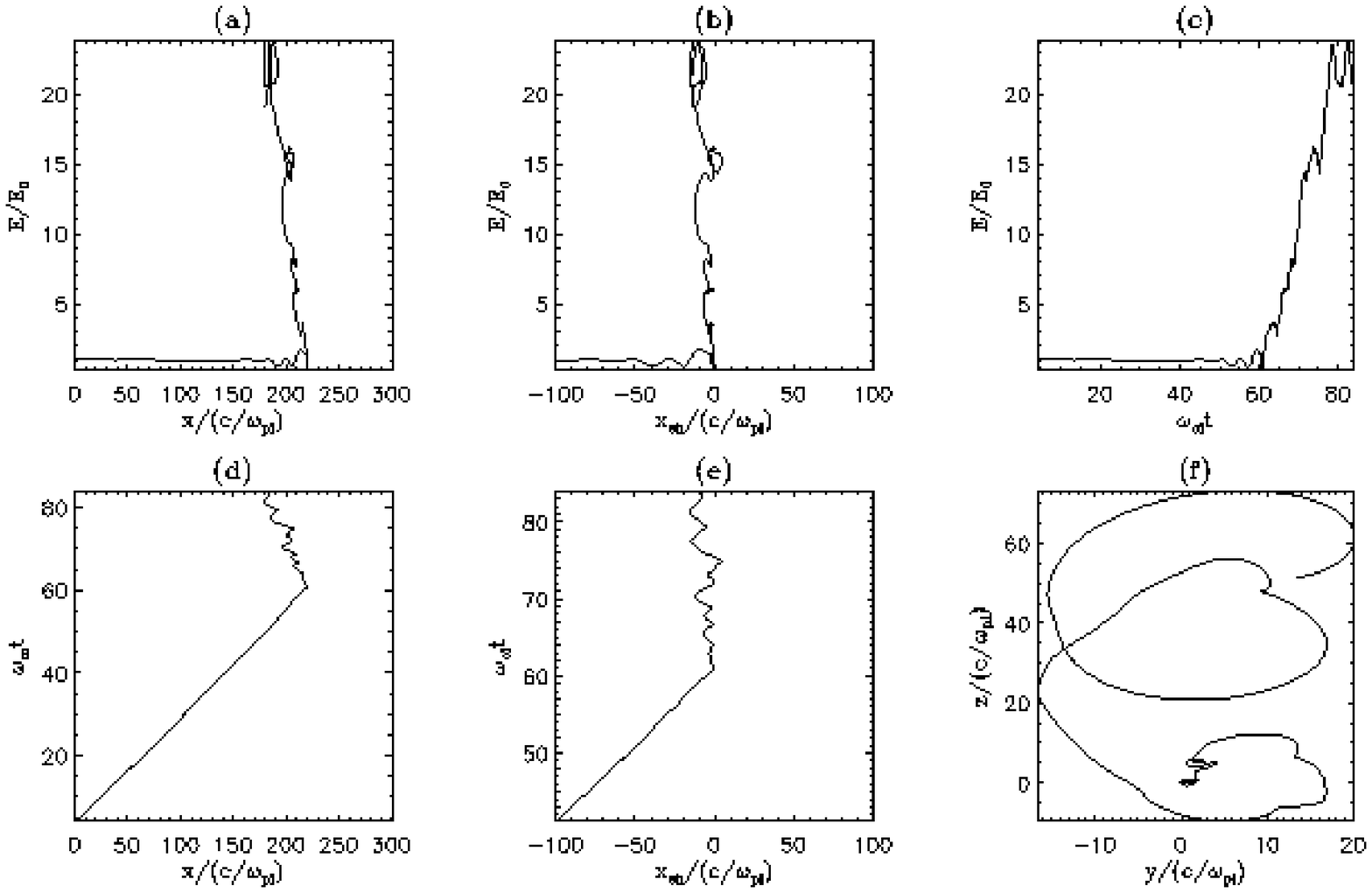,width=\textwidth,clip=}
\end{tabular}
\caption[The trajectory of a representative particle accelerated at the shock layer for the case of Run 2.]{The trajectory of a representative particle accelerated at the shock layer for Run 2. The plotted physical quantities are similar to Figure \ref{figure-traj-small}. \label{figure-traj}}
\end{figure}

\begin{figure}
\centering
\begin{tabular}{cc}
\epsfig{file=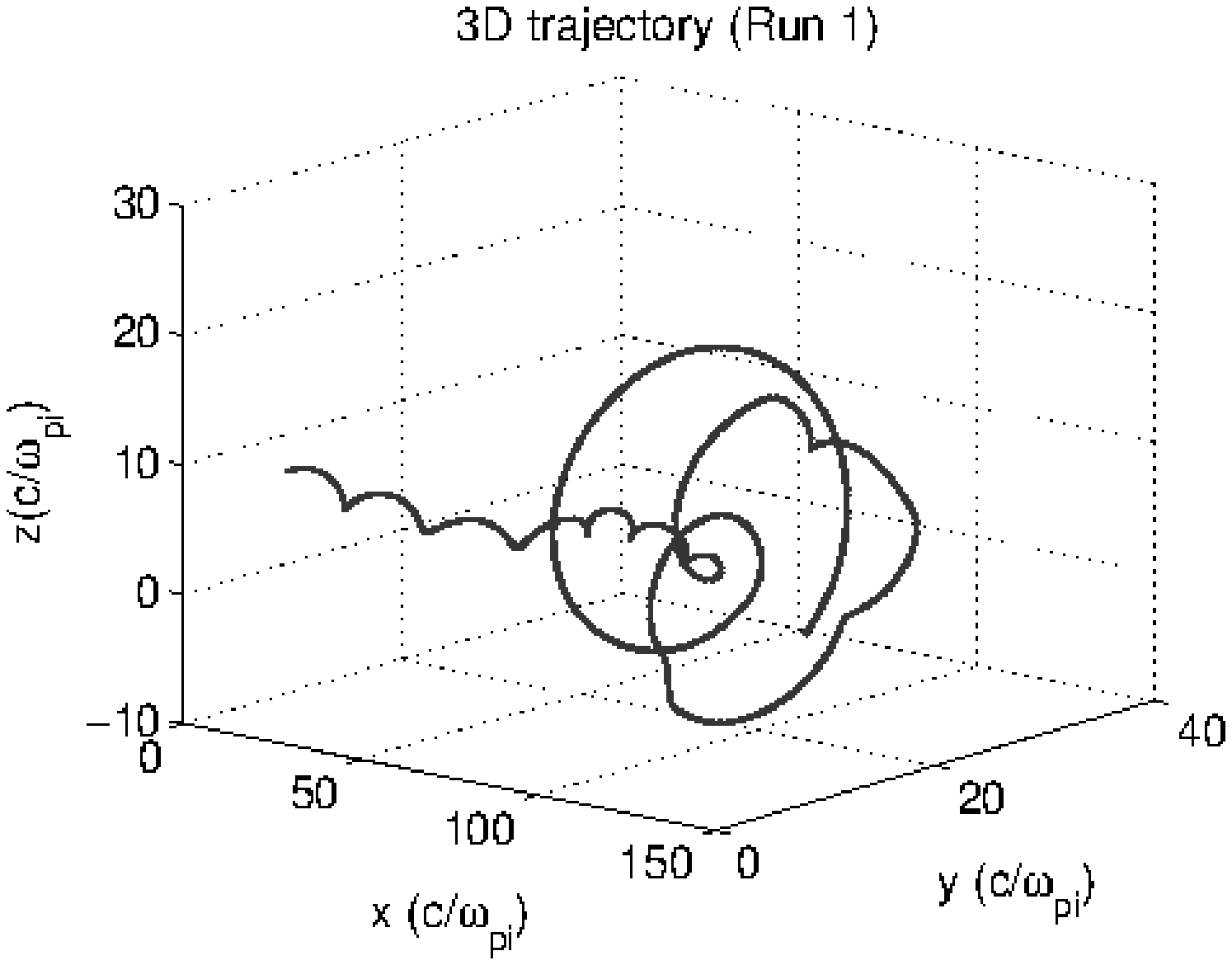,width=0.6\textwidth,clip=}\\
\epsfig{file=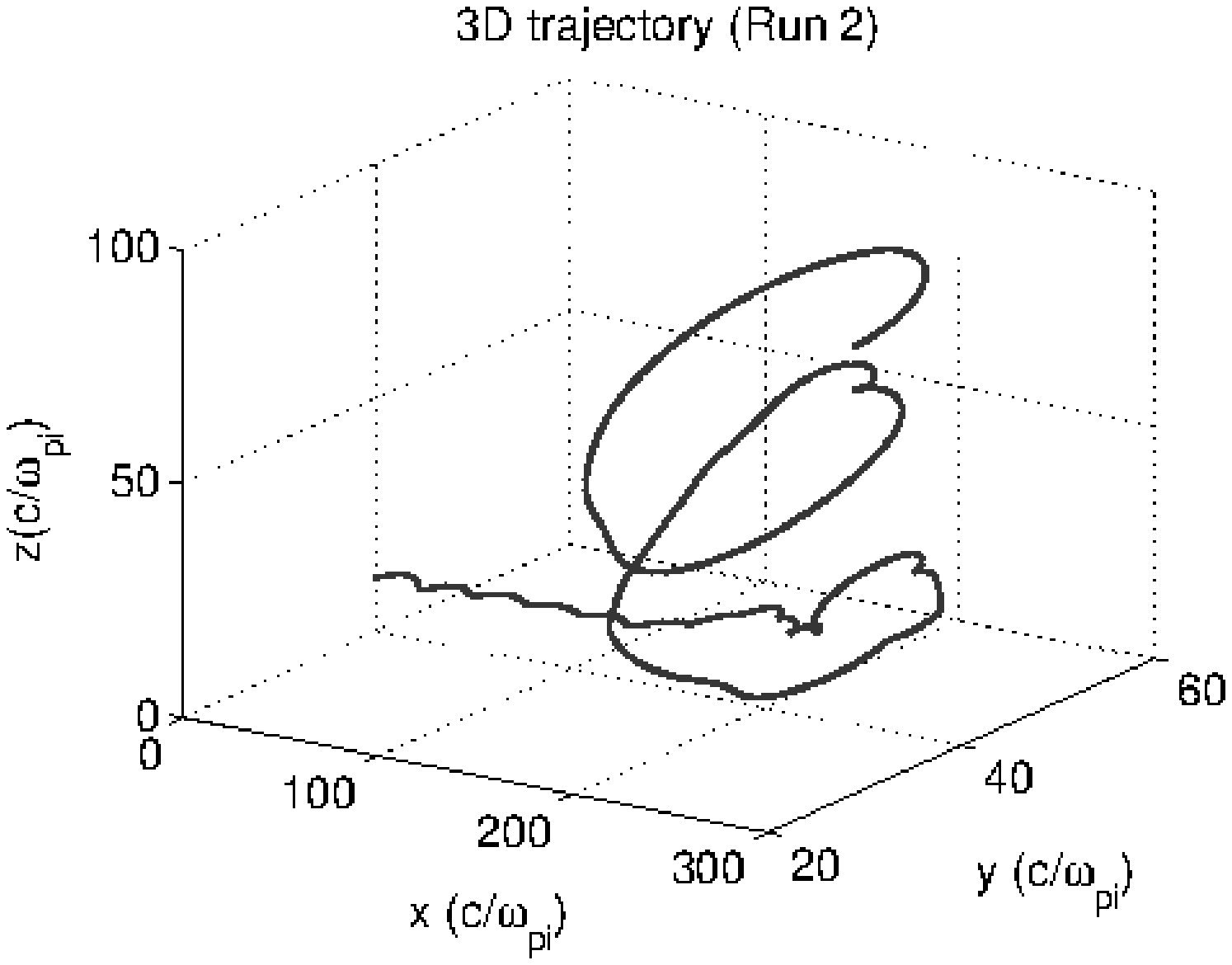,width=0.6\textwidth,clip=}
\end{tabular}
\caption[Three-dimensional plots of representative particle trajectories for Run 1 and Run 2.]{Three-dimensional plots of representative particle trajectories for Run 1 and Run 2. In the y and z directions, the displacements $y = \int v_y dt$ and $z = \int v_z dt$ are used instead of the locations of the charged particles. \label{figure-traj-3D}}
\end{figure}

For the case of Run $2$, it is expected that the trajectories of the accelerated particles could be different since the fluctuations around the shock layer have 3-D structures. Figure \ref{figure-traj} shows an example of the trajectory of the accelerated particles for the case of Run $2$. We find that the acceleration process is quite similar to what is found for the case of Run $1$. The main difference is that the particle is allowed to cross its original field line in this 3-D electric and magnetic fields. This can be seen from Figure \ref{figure-traj} (f), in which the guiding center of the charged particle is drifting in the $y$-$z$ plane during the acceleration. The difference can be seen clearly in Figure \ref{figure-traj-3D}, which shows the three-dimensional trajectories of the two representative particles from Run 1 and Run 2, respectively. It can be seen that the particle from Run 2 drift along the z direction during the acceleration at the shock front, whereas the guiding center of the particle from Run 1 does not move much during the acceleration. 

\begin{figure}
\centering
\begin{tabular}{c}
\epsfig{file=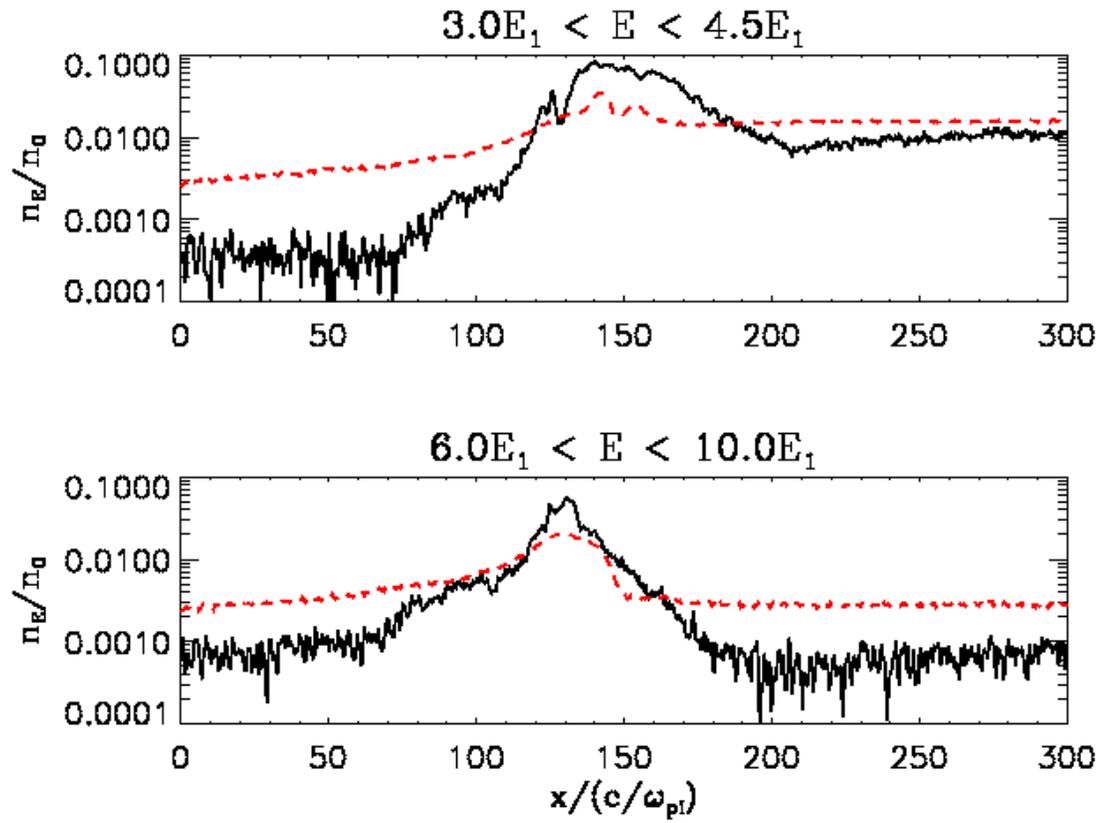,width=\textwidth,clip=}
\end{tabular}
\caption[1-D averages of the density of the accelerated particles in Run 1 and Run 2]{1-D averages of the density of energetic particles in Run 1 and Run 2 at $\Omega_{ci}t = 120.0$ \label{dene1}}
\end{figure}

\begin{figure}
\centering
\begin{tabular}{c}
\epsfig{file=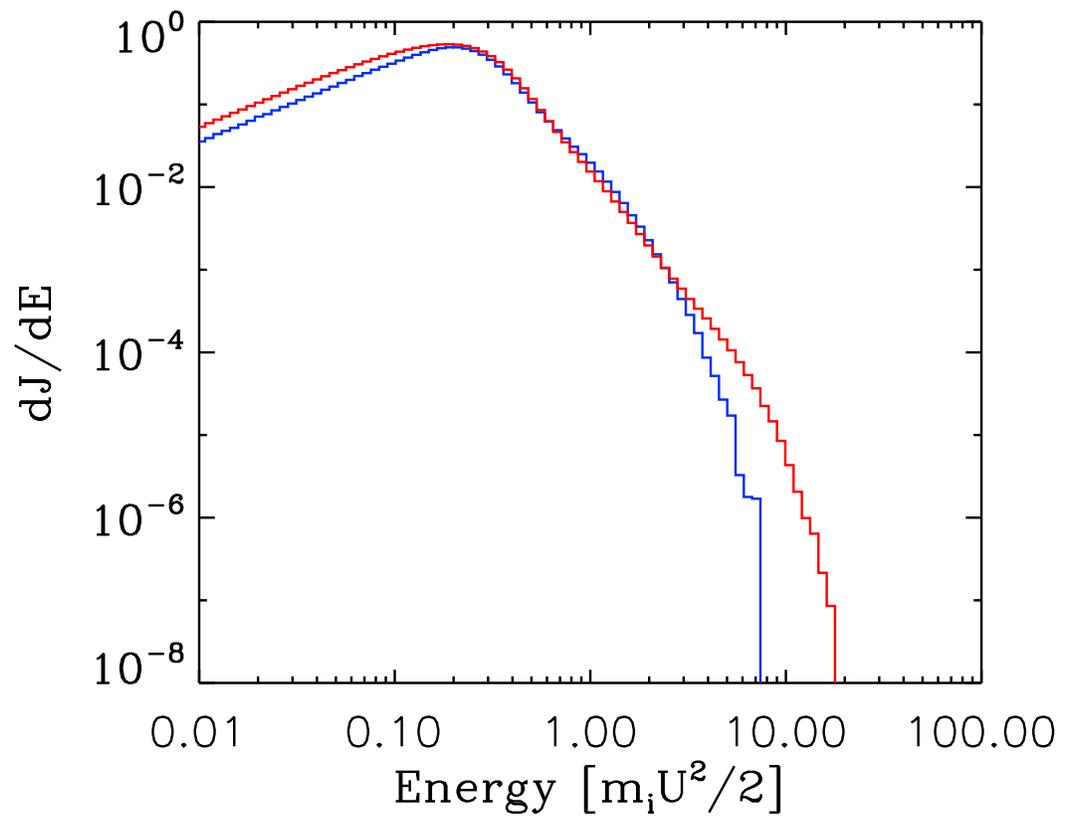,width=\textwidth,clip=}
\end{tabular}
\caption[Downstream energy spectra for ions in Run 1 and Run 2]{Downstream energy spectra for ions in Run 1 (blue) and Run 2 (red) at $\Omega_{ci}t = 120.0$ \label{spectra-ion-compare}}
\end{figure}

The acceleration of thermal protons at the shock front creates a nonthermal population of energetic particles. Figure \ref{dene1} shows the density of accelerated particles (Upper panel: $3.0 E_1 < E < 4.5 E_1$; Lower panel: $6.0 E_1 < E < 10.0 E_1$) averaged over the $y$ and $z$ directions at $\Omega_{ci}t = 120.0$, where $E_1 = 1/2m_iU_1^2$ and $U_1$ is taken to be $5.3 V_{A0}$ for both cases. The black solid lines represent the results from Run $1$ and the red dashed lines represent the results from Run $2$. In Run 1 more accelerated particles are trapped around the shock front than in Run 2. The energetic particles in Run 2 can more efficiently escape upstream and downstream of shock. In Figure \ref{spectra-ion-compare} we show downstream energy spectra in $200 c/\omega_{pi} < x < 300 c/\omega_{pi}$ at $\Omega_{ci}t = 120.0$. The black solid line shows the spectra for Run 1 and the red dashed dot line represents the spectra for Run 2. In both of the two cases, a fraction of thermal protons are accelerated to several times of bulk energy. In the case of Run 2, more high-energy protons are concentrated in downstream region and their energy range is from several times of bulk energy to about $20$ times of bulk energy.

\begin{figure}
\centering
\begin{tabular}{cc}
\epsfig{file=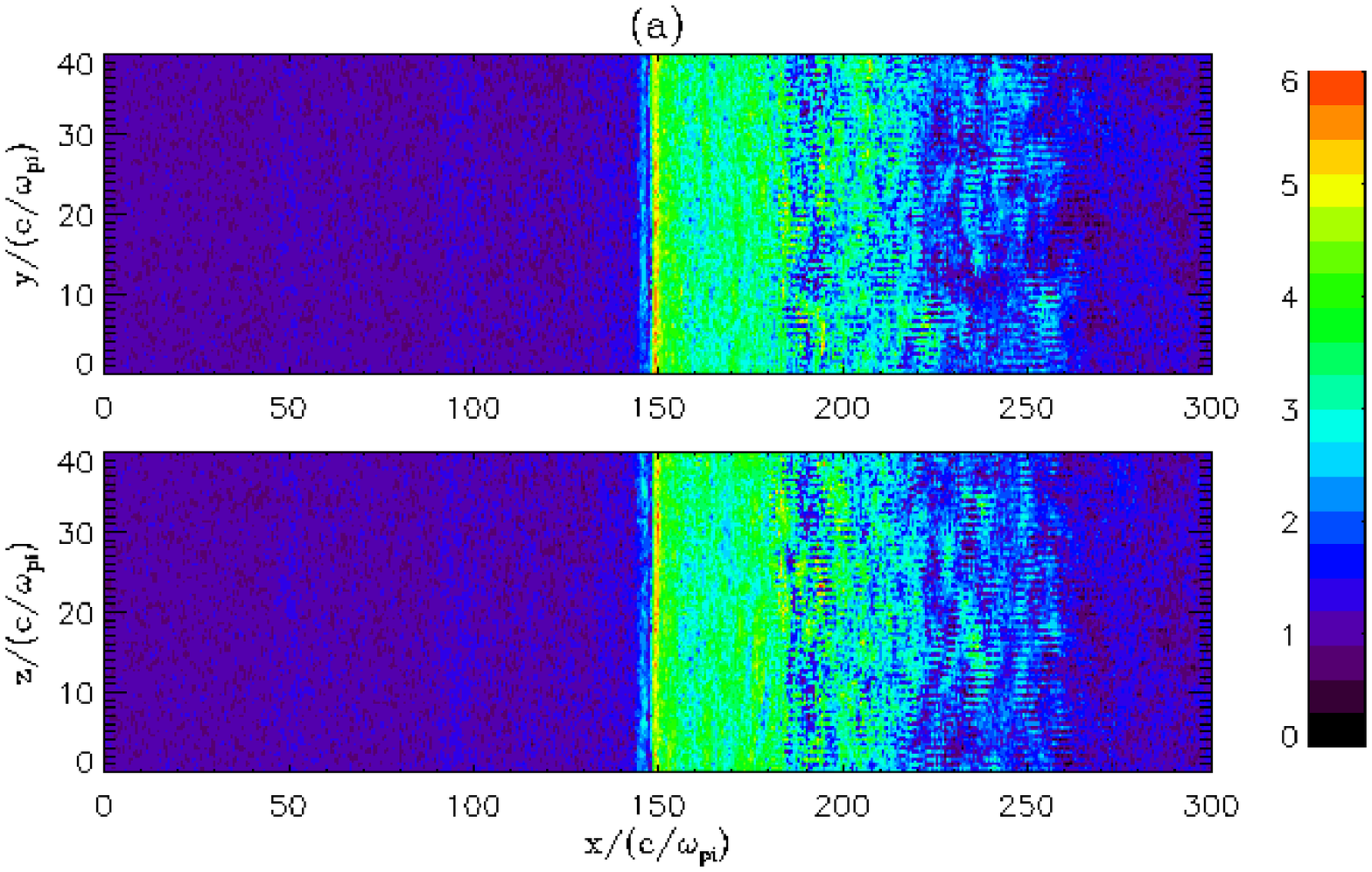,width=\textwidth,clip=} \\
\epsfig{file=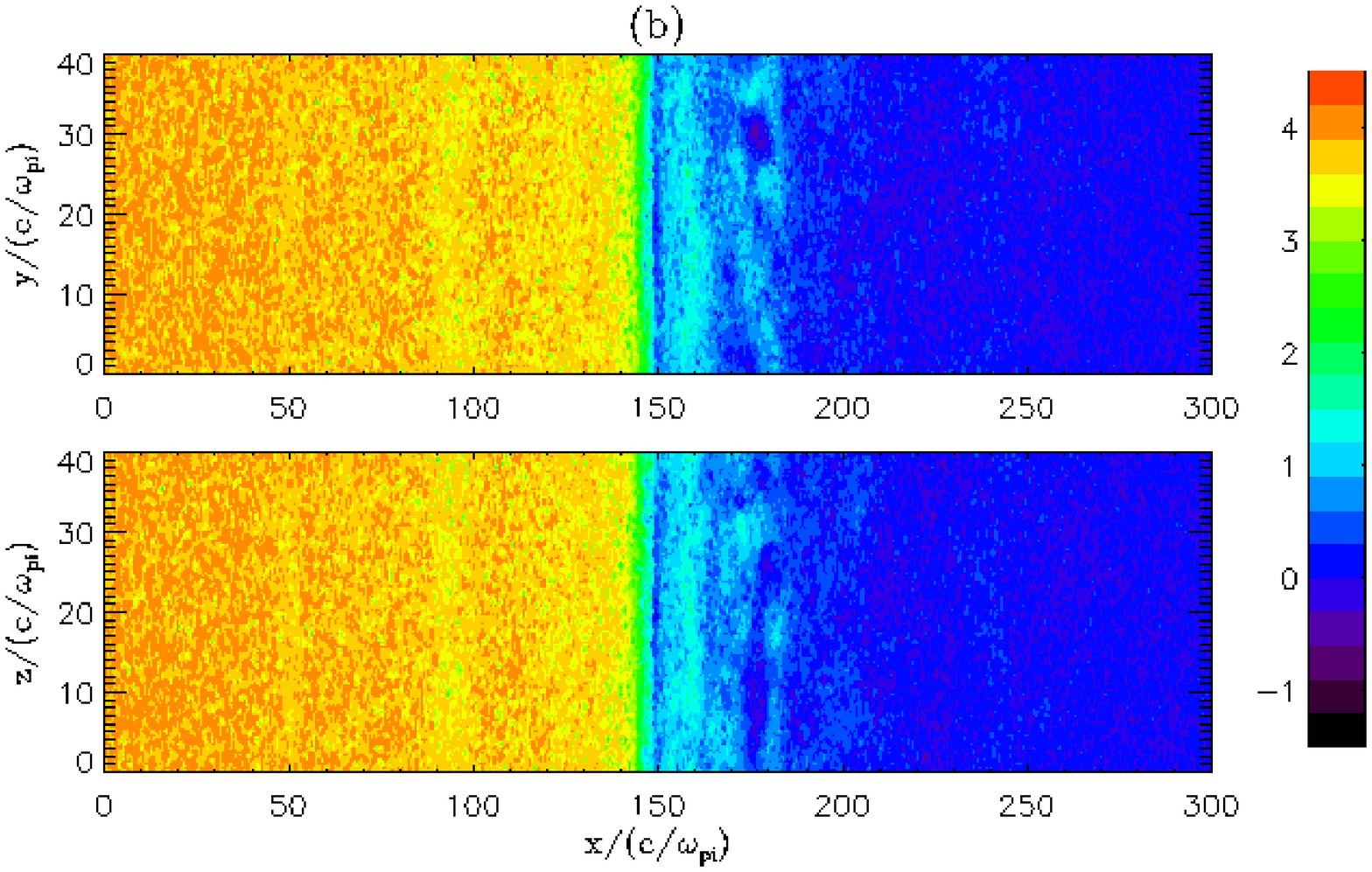,width=\textwidth,clip=}
\end{tabular}
\caption[$2$-D representation of the magnitude of magnetic field $B/B_0$ and the velocity in the $x$-direction $V_x/V_{A0}$ for Run $3$ (1-D pre-existing fluctuations).]{Similar to Figure \ref{figure-vb}, but for the case of Run $3$. \label{figure-vb-slab}}
\end{figure}

\begin{figure}
\centering
\begin{tabular}{cc}
\epsfig{file=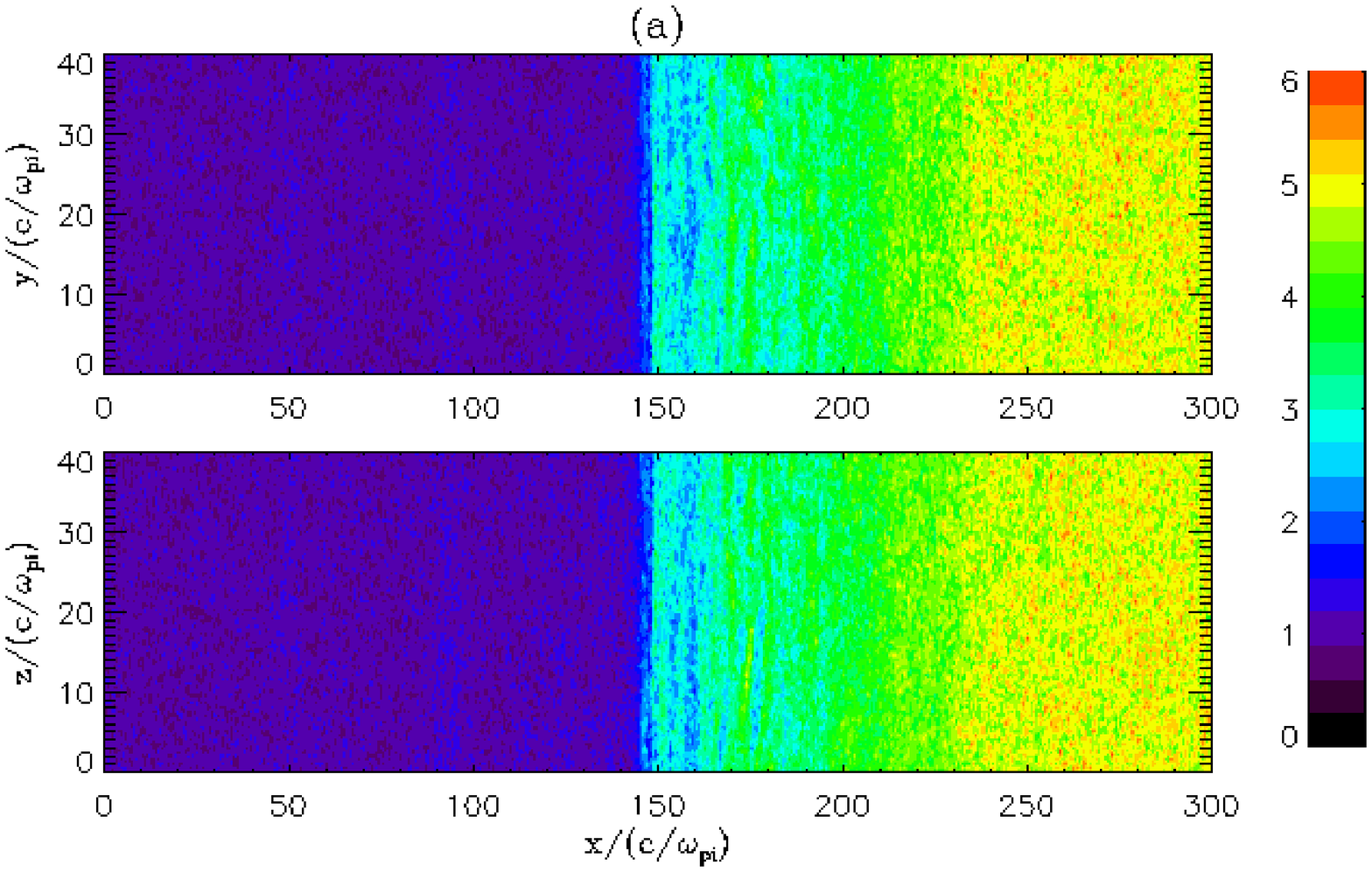,width=\textwidth,clip=} \\
\epsfig{file=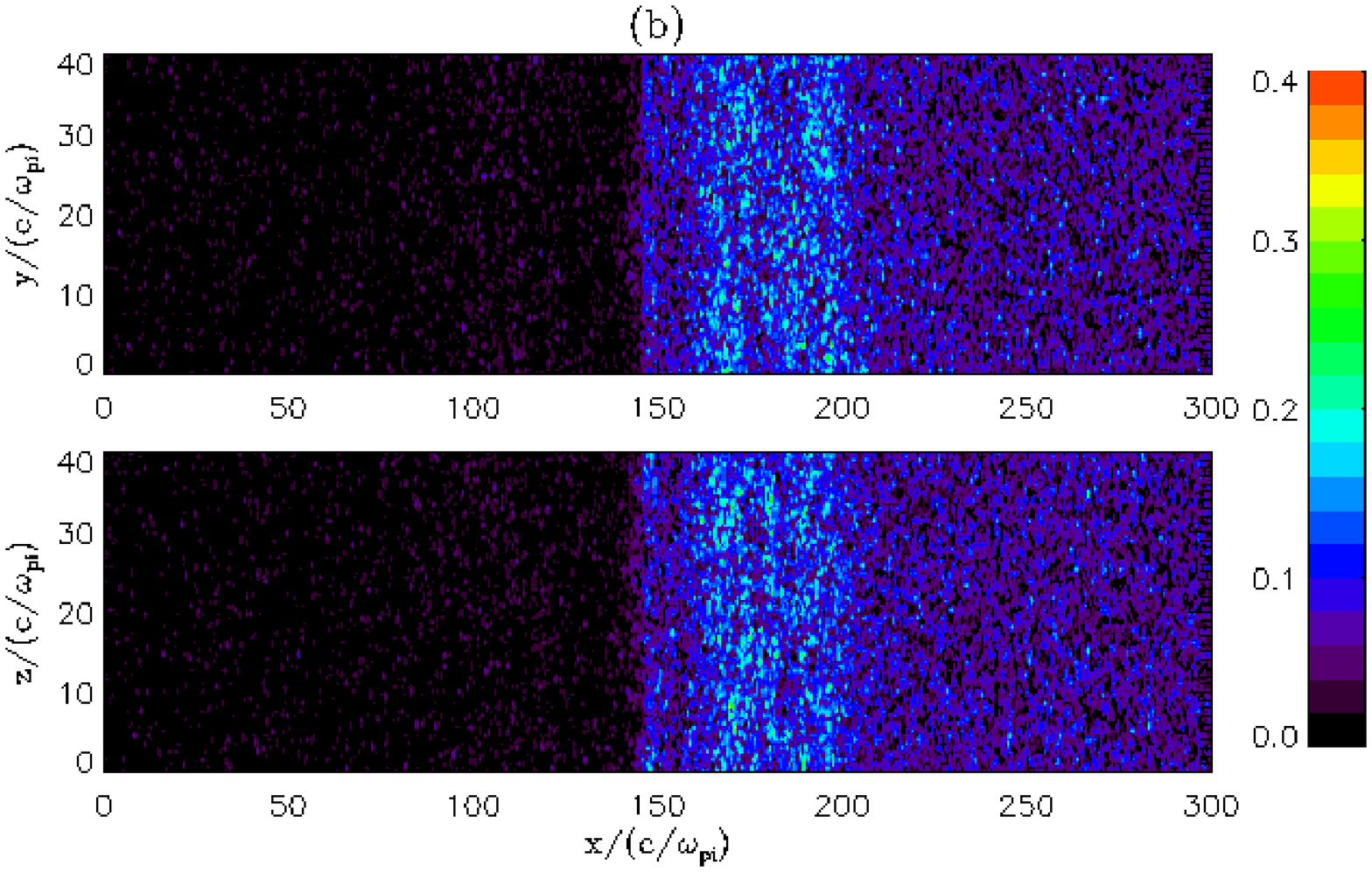,width=\textwidth,clip=}
\end{tabular}
\caption[$2$-D representation of the plasma number density $n/n_0$ and the number density of the accelerated particles for Run $3$ (1-D pre-existing fluctuations).]{Similar to Figure \ref{figure-den}, but for the case of Run $3$. \label{figure-den-slab}}
\end{figure}

\begin{figure}
\centering
\begin{tabular}{c}
\epsfig{file=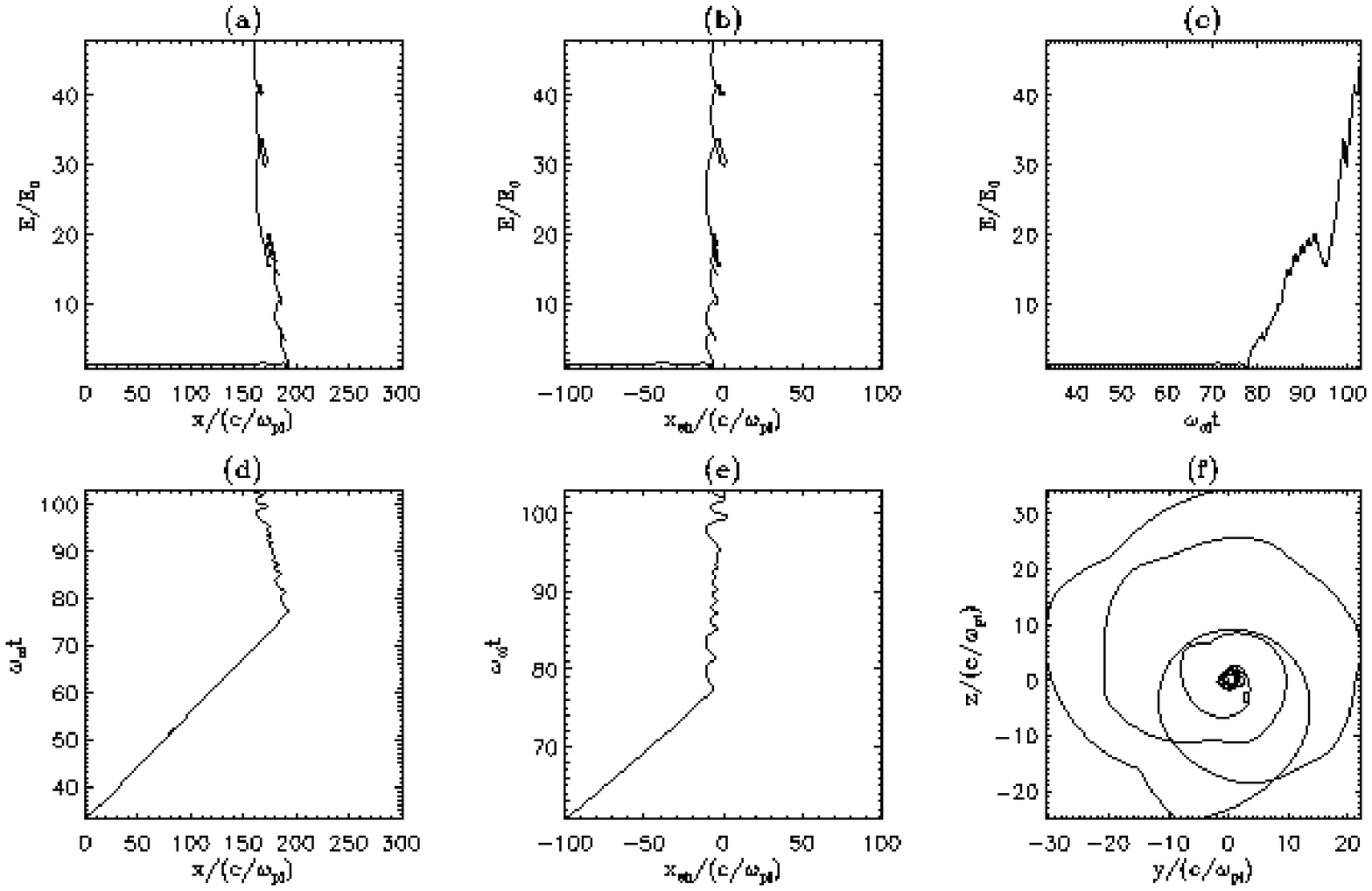,width=\textwidth,clip=}
\end{tabular}
\caption[The trajectory of a representative particle accelerated at the shock layer for Run 3.]{The trajectory of a representative particle accelerated at the shock layer for Run 3. The plotted physical quantities are similar to Figure \ref{figure-traj-small}. \label{figure-traj-slab}}
\end{figure}

We also consider the effect of the pre-existing upstream fluctuations. In Run $3$ we examine the effect of a $1$-D pre-existing magnetic fluctuation that only depends on $x$. Figure \ref{figure-vb-slab}, \ref{figure-den-slab}, and \ref{figure-traj-slab} present analyses the same as the previous cases. We find that in this case all the physical quantities are close to 1-D. The wave excitation processes at the shock front seem to be strongly influenced by the injected 1-D fluctuations. Since the injected fluctuation is only dependent on $x$, it forces the motion of the reflected particle to have the one-dimensional dependence, therefore the fluctuations excited at the shock front are not strongly dependent on $y$ and $z$. The trajectory analysis (Figure \ref{figure-traj-slab}) of an accelerated particle also confirms this idea, which shows that the motion of the particle is close to 1-D and the guiding center is approximately localized at the same position in the $y - z$ plane compared with that at the beginning of the simulation.

\begin{figure}
\centering
\begin{tabular}{cc}
\epsfig{file=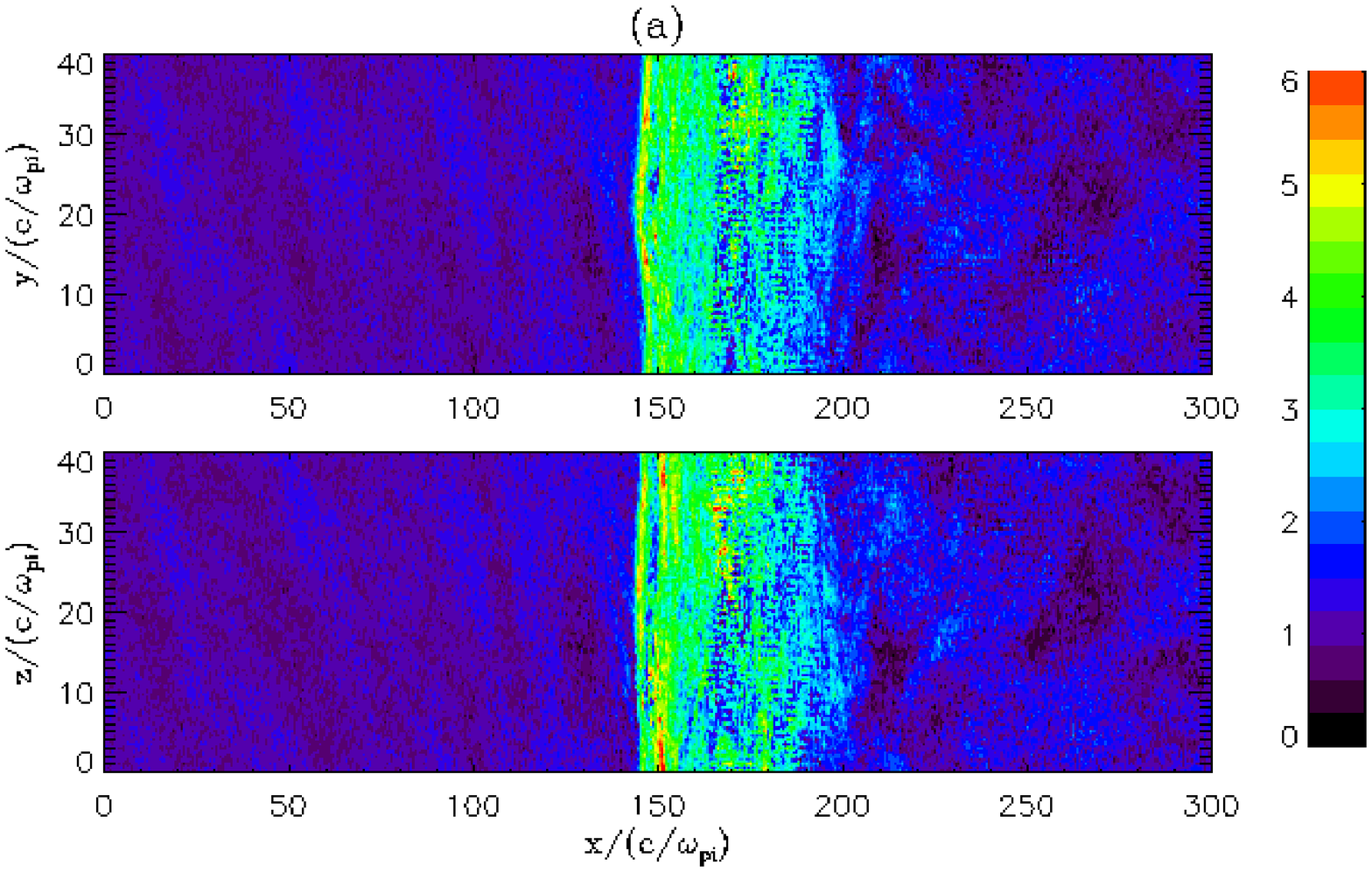,width=\textwidth,clip=} \\
\epsfig{file=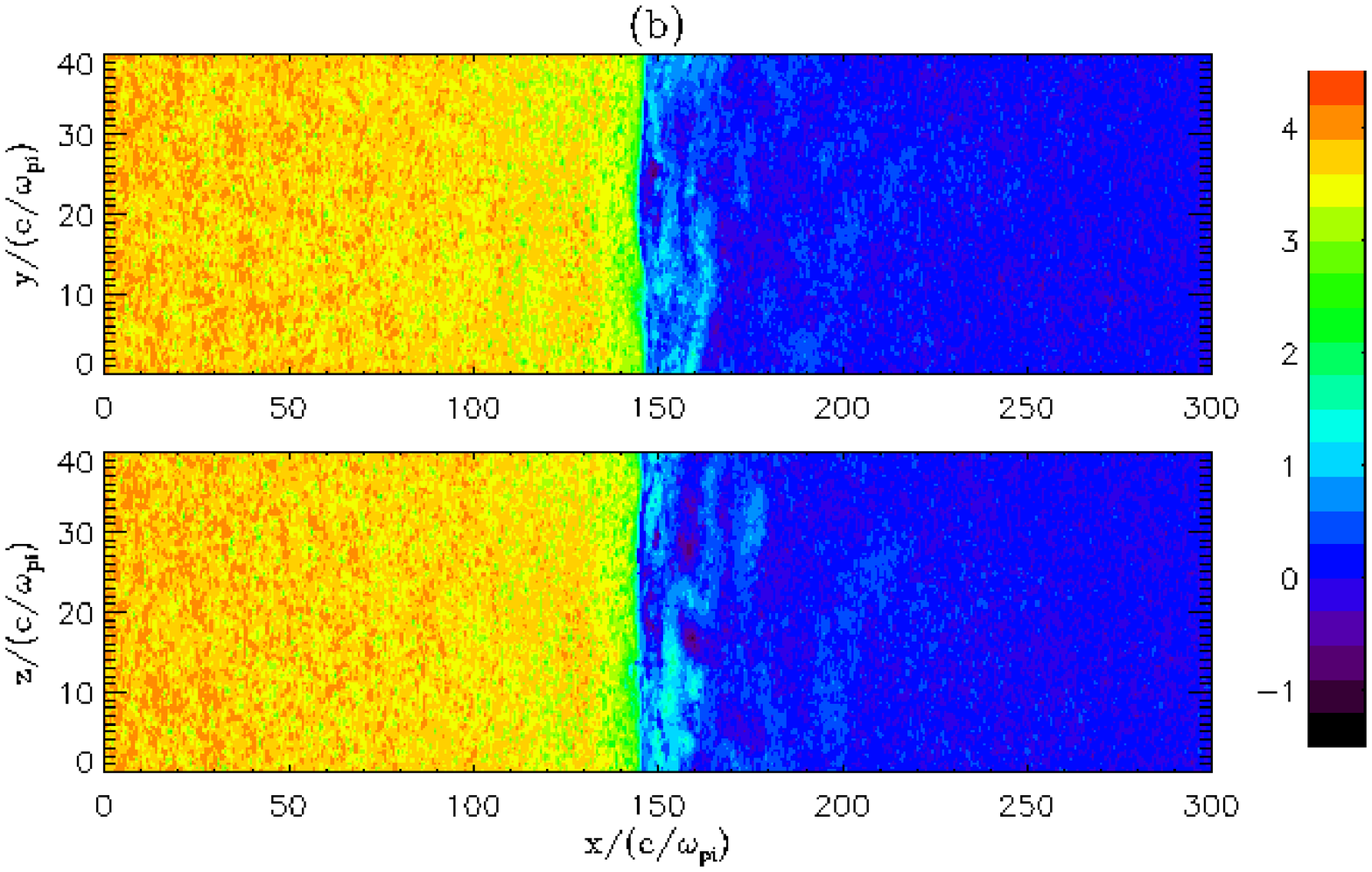,width=\textwidth,clip=}
\end{tabular}
\caption[$2$-D representation of the magnitude of magnetic field $B/B_0$ and the velocity in the $x$-direction $V_x/V_{A0}$ for Run $4$ (3-D isotropic pre-existing fluctuations).]{Similar to Figure \ref{figure-vb}, but for the case of Run $4$. \label{figure-vb-iso}}
\end{figure}

\begin{figure}
\centering
\begin{tabular}{cc}
\epsfig{file=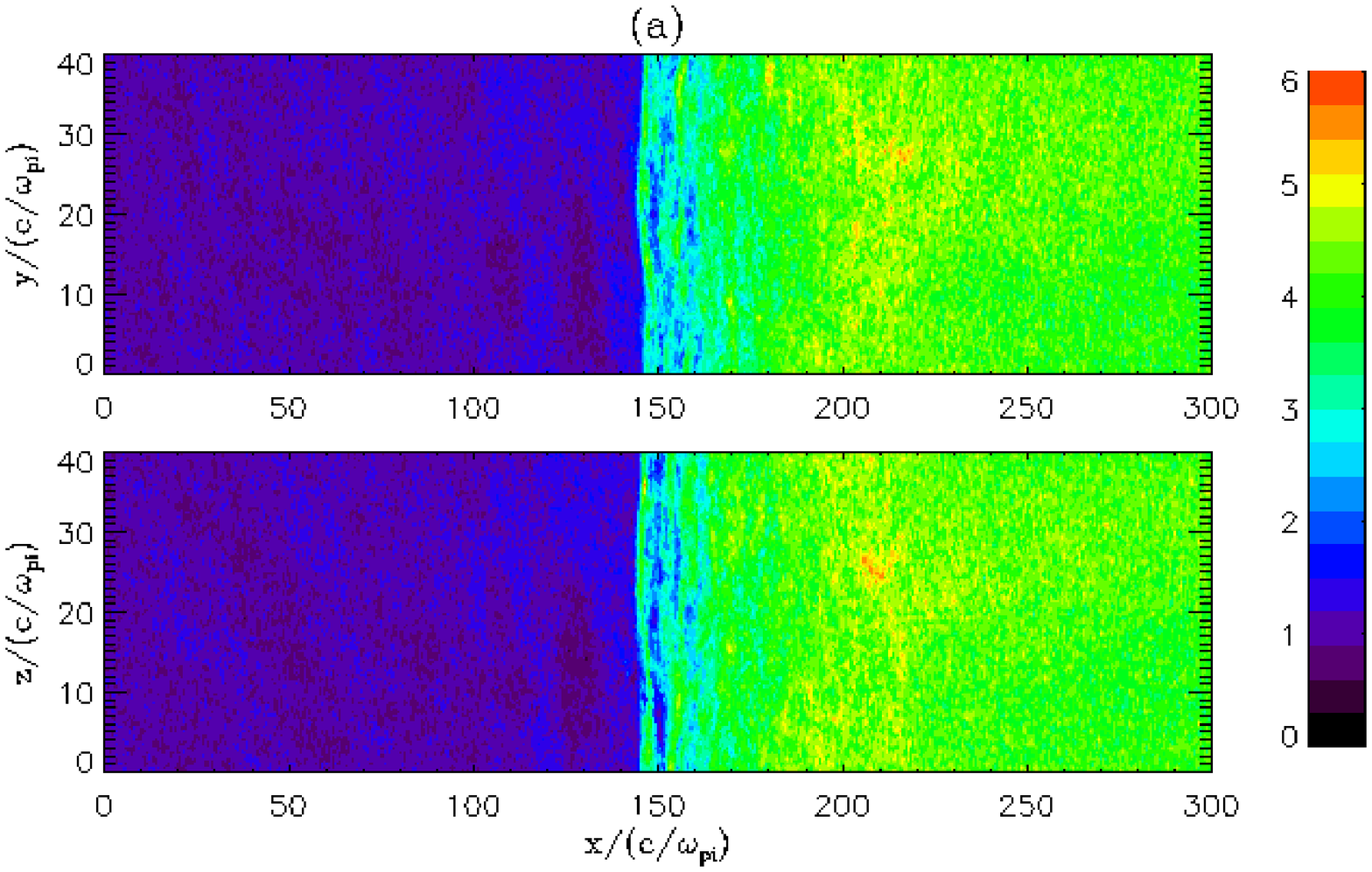,width=\textwidth,clip=} \\
\epsfig{file=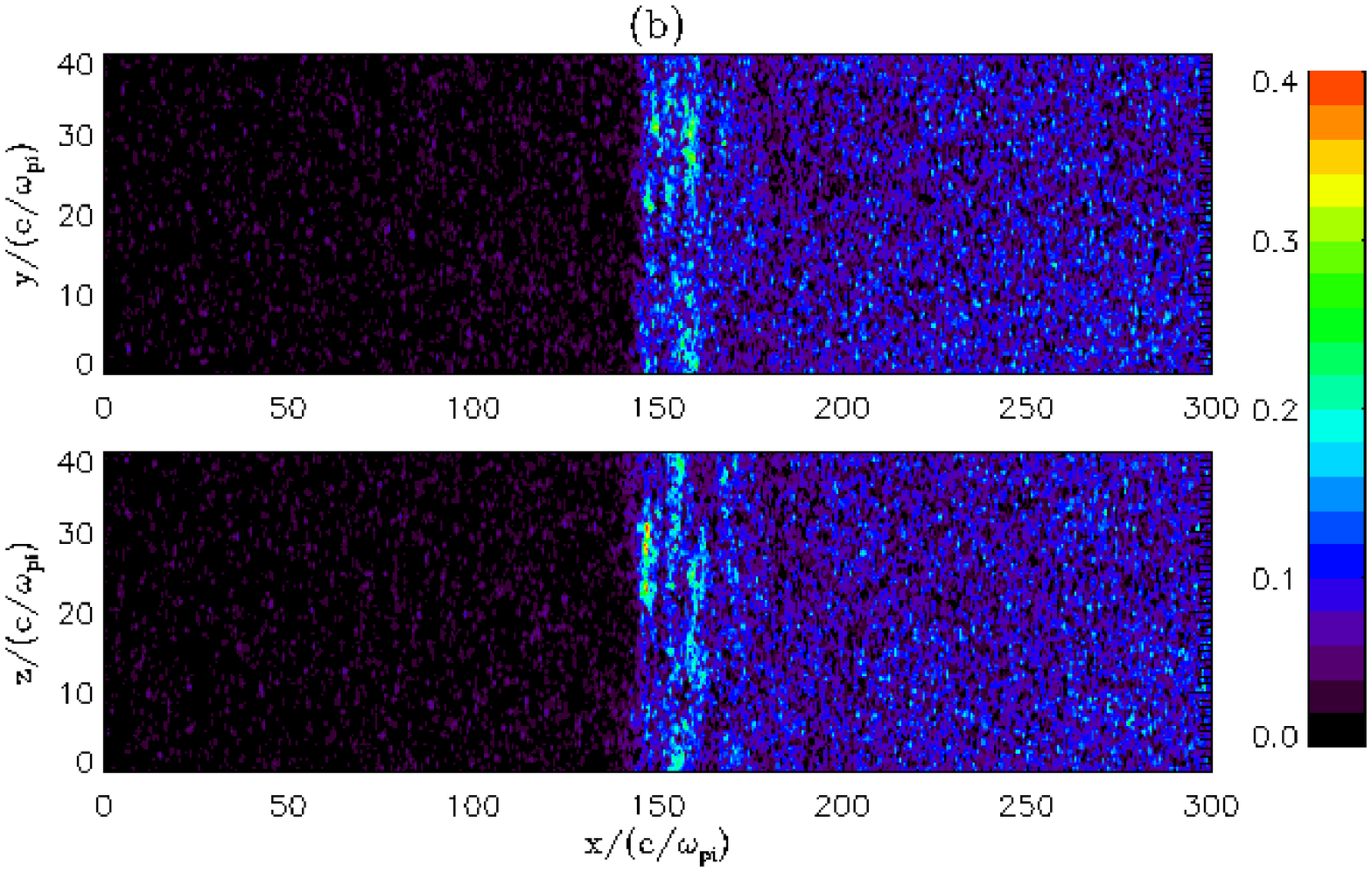,width=\textwidth,clip=}
\end{tabular}
\caption[$2$-D representation of the plasma number density $n/n_0$ and the number density of the accelerated particles for Run $4$ (3-D isotropic pre-existing fluctuations).]{Similar to Figure \ref{figure-den}, but for the case of Run $4$. \label{figure-den-iso}}
\end{figure}

\begin{figure}
\centering
\begin{tabular}{c}
\epsfig{file=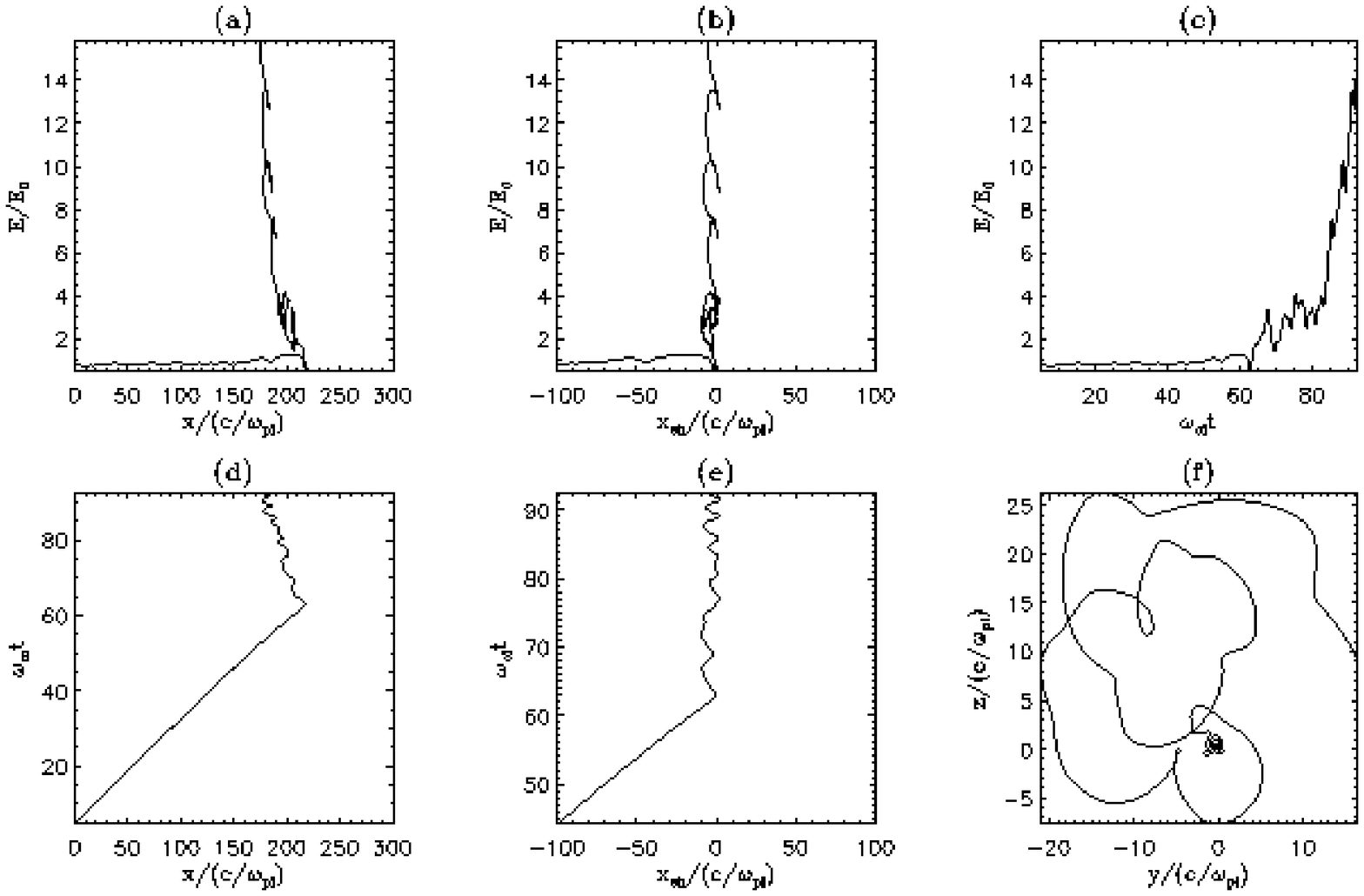,width=\textwidth,clip=}
\end{tabular}
\caption[The trajectory of a representative particle accelerated at the shock layer in the cae of Run 3.]{The trajectory of a representative particle accelerated at the shock layer for Run 4. The plotted physical quantities are similar to Figure \ref{figure-traj-small}. \label{figure-traj-iso}}
\end{figure}

We further examine the effect of pre-existing magnetic fluctuations in Run $4$ by injecting a 3-D isotropic fluctuation. In Figure \ref{figure-vb-iso}, \ref{figure-den-iso}, and \ref{figure-traj-iso} we present similar analyses to Figure \ref{figure-vb-slab}, \ref{figure-den-slab} and \ref{figure-traj-slab}, but for the case that the injected magnetic fluctuation is isotropic in 3-D. The results show that in this case all the physical quantities is dependent on three spatial dimensions. From the analyses of the trajectory of a particle, the guiding center drifts in the $y$-$z$ plane. Note that although the guiding center can freely jump to another field line, the acceleration process is quite similar to the cases we discussed. 

\subsection{Summary and Discussion}

The injection problem is a long standing problem for diffusive shock acceleration. The charged particles are required to be energetic enough and efficiently interact with magnetic fluctuations. In this section we focus on the mechanism for the pre-acceleration, in other words, how a population of thermal or low-energy particles get accelerated at the shock front. Previous studies have proposed two different injection mechanisms for the injection of low-energy ions at parallel shocks, i.e., particle energization by downstream heating \citep{Ellison1981} and by reflection at the shock layer \citep{Quest1988,Scholer1990a,Scholer1990b,Kucharek1991,Giacalone1992}. Although previous self-consistent hybrid simulations have found that the initial energization is due to the ion reflection and acceleration at the shock layer, the results were obtained only for 1-D simulations and occasionally 2-D simulations. As pointed out by \citet{Jokipii1993}, \citet{Giacalone1994}, and \citet{Jones1998}, in a magnetic field that is spatially dependent only on $1$-D or $2$-D, the motions of charged particles are restricted on their original field lines of force. In this sense, the previous simulations are not conclusive.   

In this study, we perform 3-D hybrid simulations to investigate the initial acceleration of thermal protons at parallel shocks. In the case of no pre-existing magnetic fluctuation, we find that the electric and magnetic fields can be generated by the plasma instabilities at the shock front \citep{Quest1988}. When the size of the simulation box normal to the initial magnetic field is large enough ($40$ $c/\omega_{pi}$ in our case), the electric and magnetic fields generated at the shock front show significant 3-D variations. By examining the trajectories of the accelerated particles, we find that the guiding center of the representative particle can move off its original place when the electric and magnetic fields are 3-D, meaning that the particles can move across field lines. We also examine if the effect of pre-existing magnetic fluctuations can modify this results. We find that for the case of the 1-D pre-existing magnetic fluctuations, the electric and magnetic fields close to the shock are weakly dependent on the $y$ and $z$ directions. Since the injected fluctuation is only dependent on $x$, it restricts the motion of the reflected ions and therefore the fluctuations excited at the shock front are not strongly dependent on $y$ and $z$. When we consider a 3-D pre-existing magnetic fluctuation, this restriction is removed. The charged particles can move across field lines, as suggested by analysing the their trajectories. In all the simulation cases, we find that the charged particles can gain energy at the shock layer. The particle can ride on the shock front and gain a large amount of energy (several decades of the plasma ram energy). The results confirm previous hybrid simulations that the initial acceleration of charged particles is right at the shock front \citep{Quest1988,Scholer1990a,Scholer1990b,Kucharek1991,Giacalone1992}, even if in a 3-D electromagnetic field.

\section{Beyond the 1-D Diffusive Shock Acceleration \label{chap3-variation}}

Although the 1-D steady state DSA solution gives a very elegant description for the acceleration of charged particles at the shock front, some other effects could play a role during the acceleration. These effects may help explain the observed variability of energetic particles at shocks. For example, \citet{Ellison1985b} discussed the effects such as adiabatic cooling and limited acceleration time, which can cause turnovers in the power law energy spectra of shock-accelerated charged particles if the diffusion coefficient is proportional to energy. They suggest that the spectra have an exponential rollover $dJ/dE \propto \exp(-E/E_0)$. Recent observations indicate that the spectra are more similar to a double power law spectrum \citep{Mewaldt2006}, i.e., after the spectrum break the spectra still have a power law shape with a steeper slope. \citet{Tylka2005} and \citet{Tylka2006} argue that the variable spectral and compositional characteristics in large SEP events can be produced by considering the effects of shock angles and different species of seed particles. Their model is based on the argument that the injection speed of particles for diffusive shock acceleration is much higher at perpendicular shocks than that at parallel shocks (see Equation \ref{equation-injection-velocity}). As a result, parallel shocks can accelerate charged particles from solar wind particles, whereas perpendicular shocks only accelerate superthermal particles pre-accelerated in solar flares. This model can qualitatively explain the high variable features of spectral characteristics and elemental composition in large gradual SEP events. However, recent numerical simulations show that perpendicular shocks can efficiently accelerate charged particles to high energy in existence of strong pre-existing magnetic fluctuation \citep{Giacalone2005a,Giacalone2005b}. This questions the validity of the Tylka et al. (2005) model. \citet{Li2009} presented a model for particle acceleration at oblique shocks that include both parallel diffusion and perpendicular diffusion. The results can roughly reproduce the observed dependence of break energy on the charge-to-mass ratio. It should be noted that most of these studies consider particle acceleration at a planar shock. In the real situation the particles can sample a range of shock structures with different shock normal angles as they move along shock surface.
 
\begin{figure}
\begin{center}
\includegraphics[width=0.8\textwidth]{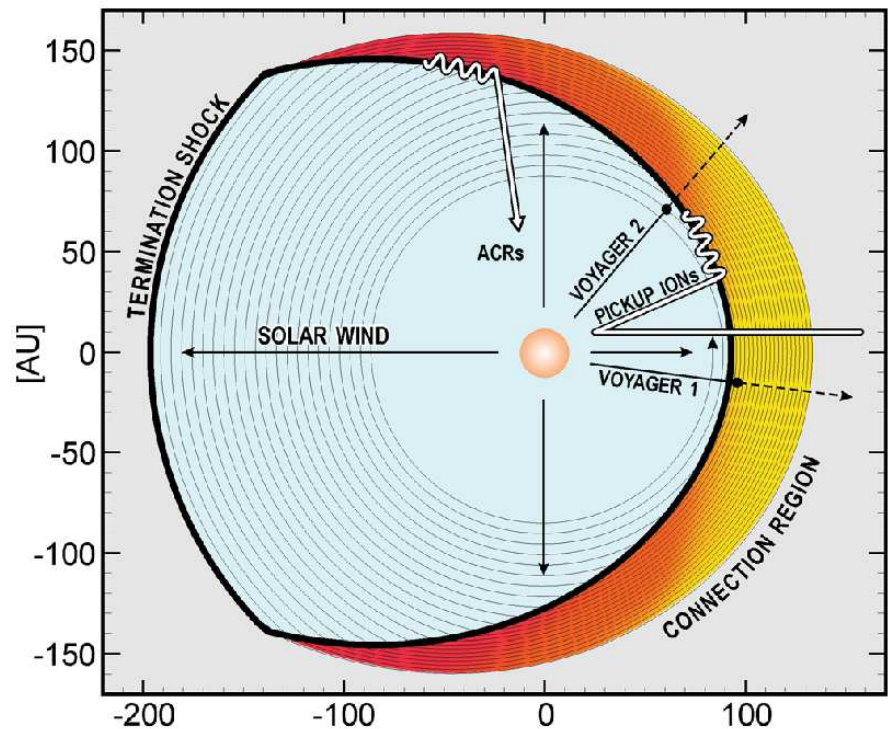}
\caption[\citet{McComas2006}'s idea on the acceleration of anomalous cosmic rays.]
{\citet{McComas2006}'s idea on the acceleration of anomalous cosmic rays: The Cartoon shows an equatorial cut of the termination shock and the heliosphere. The Parker spiral magnetic field get compressed at the termination shock. Pickup ions can get accelerated at the termination shock in all directions but have different acceleration rate. Because charged particles mainly travel along field lines, pickup ions near the nose take a longer time to travel back to the shock than those close to the shock flanks. The figure is adapted from \citep{McComas2006}.
 \label{figure-bluntshock}}
 \end{center}
 \end{figure}

Here we mainly discuss a different mechanism that can modify the 1-D solution of DSA. 
DSA is thought to be the mechanism that accelerates anomalous cosmic rays (ACRs) in the heliospheric termination shock
and also galactic cosmic rays (GCRs) with energy up to at least $10^{15}$ eV in
supernova blast waves. However, recent {\it in situ} observations at the
termination shock and in the heliosheath by \emph{Voyager} $1$ \citep{Stone2005}
found that the intensity of ACRs was not saturated at the place where Voyager 1 crossed 
the termination shock and kept increasing after entering the heliosheath, which
strongly indicates that the simple planar shock model is inadequate to interpret the
acceleration of ACRs. Numerical and analytical studies suggest that the possible
solution can be made by considering the temporary and/or spatial variation
\citep{Florinski2006,McComas2006,Jokipii2008,Kota2008,Schwadron2008}.
In particular, \citet{McComas2006} discussed the importance of the magnetic
geometry of a blunt shock on particle acceleration. The idea can be illustrated by
Figure \ref{figure-bluntshock}. They argued that the
missing ACRs at the nose of the heliospheric termination shock is due to
particle energization occuring primarily back along the flanks of the shock
where magnetic field lines have had a longer connection time to the termination 
shock. \citet{Kota2008} presented a more sophisticated
simulation that gives results similar to that described by
\citet{McComas2006}. \citet{Schwadron2008} also developed a 3-D analytic
model for particle acceleration in a blunt shock, including perpendicular
diffusion and drift motion due to large-scale shock structure.

Large-scale magnetic field line meandering is ubiquitous in the heliosphere and
other astrophysical environments \citep{Jokipii1966,Jokipii1969,Parker1979}.
The acceleration of charged-particles in collisionless shocks has been shown to
be strongly affected by magnetic-field turbulence at different scales
\citep{Giacalone2005a,Giacalone2005b,Giacalone2008}. The large-scale
magnetic field variation will have important effects on the shock acceleration
since the transport of charged particles is different in the direction parallel
and perpendicular to the magnetic field, as shown in early work
\citep{Jokipii1982,Jokipii1987}. Blunt shocks or shocks with
fluctuating fronts \citep{Li2006} that have a similar geometry, are also
relevant to this problem. In this study we analyze the effect of the
large-scale spatial variation of magnetic field on DSA by considering a simple
system that captures the basic physical ideas.

\section{A Study on Particle Acceleration at Shocks Containing Large-scale Magnetic Variations \label{chap3-mag}}

In this section we present a study on the acceleration of particles at shocks containing large-scale magnetic variations. 
The main part of this section has been published in the Astrophysical Journal \citep{Guo2010b}.
Diffusive shock acceleration at collisionless shocks is thought to be the
source of many of the energetic particles observed in space. Large-scale
spatial variations of the magnetic field has been shown to be important in
understanding observations. The effects are complex, so here we consider a
simple, illustrative model. Here, we solve numerically the Parker transport
equation for a shock in the presence of large-scale sinusoidal magnetic-field
variations. We demonstrate that the familiar planar-shock results can be
significantly altered as a consequence of large-scale, meandering magnetic
lines of force. Because the perpendicular diffusion coefficient $\kappa_\perp$ is
generally much smaller than the parallel diffusion coefficient $\kappa_\parallel$,
energetic charged particles are trapped and preferentially accelerated
along the shock front in regions where the connection points of magnetic
field lines intersecting the shock surface converge, and thus create ``hot
spots" of accelerated particles. For regions where the connection
points are separated from each other, the acceleration to high energies will be
suppressed. Furthermore, the particles diffuse away from the ``hot spot" regions
and modify the spectra of downstream particle distribution. These features are
qualitatively similar to the recent Voyager's observation in the heliosheath.
These results are potentially important for particle acceleration at shocks
propagating in turbulent magnetized plasmas as well as those that contain
large-scale nonplanar structures. Examples include anomalous cosmic rays
accelerated by the solar wind termination shock, energetic particles observed
in propagating heliospheric shocks, and galactic cosmic rays accelerated by
supernova blast waves, etc.

\subsection{Basic Considerations and Numerical Model}
The diffusive shock acceleration (DSA) can be studied by solving the Parker
transport equation \citep{Parker1965}. Here we consider a 2-D system, 
the Parker transport equation (\ref{parker_equation}) can
be written in Fokker-Planck form as:

\begin{eqnarray} \frac{\partial f}{\partial t}&=& \frac{\partial^2}{\partial
  x^2}\left(\kappa_{xx} f \right) + \frac{\partial^2}{\partial z^2}\left(\kappa_{zz}
  f\right) + \frac{\partial^2}{\partial x \partial z}\left(2\kappa_{xz}f
  \right) \nonumber \\
   & &-\frac{\partial}{\partial x}\left[\left(U_{x} + \frac{\partial \kappa_{xx}}{\partial x} + \frac{\partial \kappa_{xz}}{\partial z} \right)f\right]
   -\frac{\partial}{\partial z}\left[\left( \frac{\partial \kappa_{zz}}{\partial z} + \frac{\partial \kappa_{xz}}{\partial x} \right)f\right] \nonumber \\
  &&+\frac{\partial}{\partial p^3}\left(\frac{\partial U_{x}}{\partial x} p^3 f \right)
 + Q, \label{Fokker-Planck}
\end{eqnarray}
\noindent where $\kappa_{xx}$ = $\langle \Delta x^2 \rangle/2\Delta t$, $\kappa_{zz}$ = $\langle \Delta
z^2 \rangle/2\Delta t$, and $\kappa_{xz}$ = $\langle \Delta x \Delta z \rangle/2\Delta t$. Following
the usual approach in stochastic integration \citep{Jokipii1977},
the solution can be calculated by successively integrating the trajectories of the pseudo-particles:

\begin{eqnarray}
\Delta x &=& r_1 (2\kappa_\perp \Delta t)^{1/2} + r_3
(2(\kappa_\parallel-\kappa_\perp)\Delta t)^{1/2}\frac{B_x}{B} \nonumber \label{stochastic-x}\\
         & & \;\;\;\;\;\;\;\;\;\;+ U_x\Delta t +
(\frac{\partial \kappa_{xx}}{\partial x} + \frac{\partial \kappa_{xz}}{\partial
z}) \Delta t, \\
 \Delta z &=& r_2 (2\kappa_\perp \Delta t)^{1/2} + r_3
(2(\kappa_\parallel-\kappa_\perp)\Delta t)^{1/2}\frac{B_z}{B} \nonumber \label{stochastic-z}\\
         & & \;\;\;\;\;\;\;\;\;\;+ (\frac{\partial\kappa_{zz}}{\partial z} 
         + \frac{\partial \kappa_{xz}}{\partial x}) \Delta t, \\
\Delta p &=& -\frac{p}{3} \frac{\partial U_x}{\partial x} \Delta t, \label{stochastic-p}
\end{eqnarray}

\noindent where $r_1$, $r_2$, and $r_3$ are different sets of random numbers
that satisfy $\langle r_i r_j \rangle = 0$ and $\langle r_i^2 \rangle = 1$. It can be easily demonstrated that
the ensemble average of stochastic differential equations \ref{stochastic-x}, \ref{stochastic-z}, and
\ref{stochastic-p} is the solution of transport equation \ref{Fokker-Planck}. In order to study diffusive
shock acceleration, we approximate the shock layer as a sharp variation $U_x =
(U_1+U_2)/2 - (U_1-U_2)tanh(x/th)/2$ with a thickness $th$ much smaller
compared with the characteristic length of diffusion acceleration
$\kappa_{xx1}/U_1$. At the same time, we have to make sure that the time step
$\Delta t$ is small enough to resolve the motion in the shock layer.

\citet{Kota2008} and \citet{Kota2010} considered
analytically a model in which the upstream magnetic field was in a plane (say, $x,
y$), with average direction in the $y$ direction. The x-component of the
magnetic field was composed of uniform sections (straight field lines)
alternating in sign, which were periodic in $y$. They find ``hot spots" and
spectral effects that illustrate the effect of an upstream meandering in the
magnetic field.

\begin{figure}
\begin{center}
\includegraphics{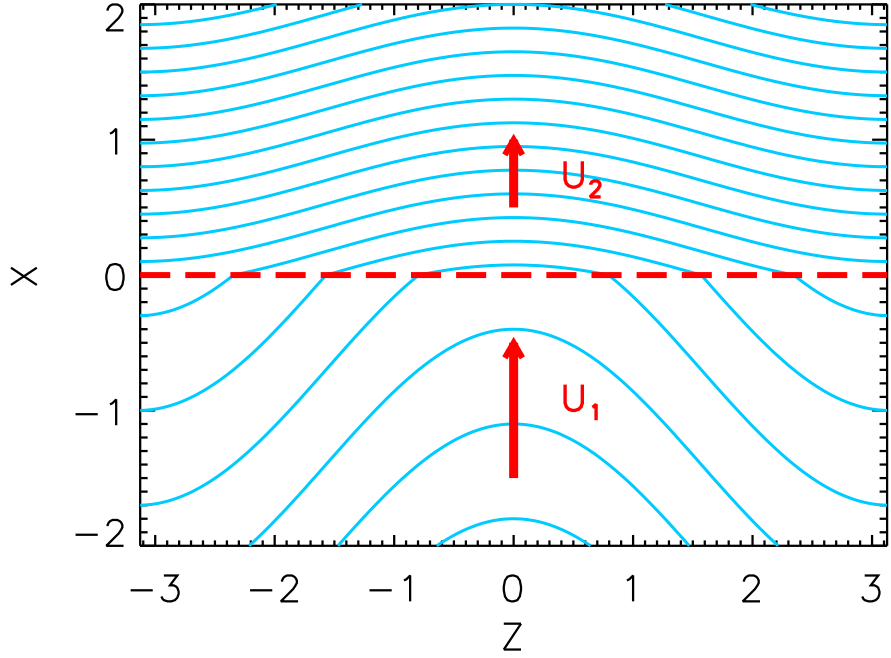}
\caption[The shock and the magnetic field geometry for an upstream average
magnetic field perpendicular to the shock normal.]{The shock and the magnetic field geometry for an upstream average
magnetic field perpendicular to the shock normal. The blue lines
 represent the magnetic field lines and the red dashed line indicates the surface of shock wave.
 The flow velocities are $U_1$ (upstream) and $U_2$ (downstream). \label{figure-shock-fieldline}}
 \end{center}
 \end{figure}

Here, we consider a 2-D $(x,z)$ system with a planar shock at $x=0$, and a
sinusoid magnetic field $\textbf{B} = \textbf{B}_0 + \sin(k z)\delta
\textbf{B}$. For most of the parts in this paper we discuss the case shown in
Figure  \ref{figure-shock-fieldline}. In this figure, the magnetic lines of force are illustrated by blue
lines. The shock is denoted by the red dashed line. Since, in the system of
interest here, the magnetic field is small enough that its effects are very
small (dynamic pressure/magnetic pressure $\sim \rho V_w^2/(B^2/8\pi) \sim 100$
in the solar wind at 1 AU). The system is periodic in the $z$ direction, with
the magnetic field convecting from upstream ($x<0$) to downstream ($x>0$). In
the shock frame, the particles will be subjected to convection and diffusion
due to the flow velocities $U_1$(upstream) and $U_2$(downstream) and diffusion
coefficients parallel and perpendicular to large scale magnetic field
$(\kappa_{1(\parallel, \perp)} $\space and \space$
\kappa_{2(\parallel,\perp)})$, respectively. The gradient and curvature drifts
in this case are only in the direction out of the $x$-$z$ plane and thus
irrelevant to this study. Because of the steady velocity difference between
upstream and downstream, charged particles that travel through the shock layer
will be accelerated. However, since we consider the large-scale magnetic field
variation, transport of energetic particles in the fluctuating magnetic field
become important. The diffusion coefficient in the $x$-$z$ system can be
expressed as:

\begin{equation}
\kappa_{ij} = \kappa_\perp \delta_{ij} -
\frac{(\kappa_\perp-\kappa_\parallel)B_iB_j}{B^2}.
\end{equation}

The normalization units chosen in this study are: the spatial scale $X_0 = 10$
AU, the upstream velocity $U_1 = 500$ km/s, the time scale $T_0 = 3\times 10^6$
sec, and the diffusion coefficients are in unit of $\kappa_0 = 7.5 \times
10^{21}$ cm$^2$/s. The shock compression ratio $r = U_1/U_2$ is taken to be
$4.0$. The shock layer is considered to be a sharp variation $U_x = (U_1+U_2)/2
- (U_1-U_2)tanh(x/th)/2$ with thickness $th = 1 \times 10^{-3}$X$_0$, which is
required to be less than $\kappa_{xx1}/U_1$ everywhere in the upstream
simulation domain, where $\kappa_{xx1}$ is the upstream diffusion coefficient
normal to the shock surface. The simulation domain is taken to be [$-2.0$X$_0<x<$ $2.0$X$_0$,
$-\pi$X$_0$ $<z<$ $\pi$X$_0$ ]. The parallel diffusion coefficients upstream and downstream are
assumed to be the same and taken to be $\kappa_{\parallel1} =
\kappa_{\parallel2} = 0.1$X$_0^2/$T$_0$ at $p = p_0$. The ratio between parallel diffusion
coefficient and perpendicular diffusion coefficient is taken to be
$\kappa_{\perp}/\kappa_{\parallel} = 0.05$, which is consistent with that
determined by integrating the trajectories of test particles in magnetic
turbulence models \citep{Giacalone1999}. The momentum dependence of the
diffusion coefficient is taken to be $\kappa \propto p^{4/3}$, corresponding to
non-relativistic particles in a Kolmogorov turbulence spectrum
\citep{Jokipii1971}. We use the stochastic integration method to obtain the numerical solution of the transport equation. In the Equation \ref{parker_equation}, the source function $Q = \delta (p-p_0)\delta_(x)$ is
represented by injecting pseudo-particles at the shock $x = 0$ with initial
momentum $p = p_0$. The trajectories of pseudo-particles are integrated each
time step to obtain the numerical solution. The pseudo-particles will be
accelerated if they crossing the shock as predicted by DSA. The time step is $1
\times 10^{-7}$T$_0$ and even reduce at the shock front to resolve the variation of $U_x$ at
the finite shock layer. Particles that move past the upstream or downstream
boundaries will be removed from the simulation. The system is periodic in the
$z$ direction, so a pseudo-particle crossing the boundaries in the $z$ direction will
re-appear at the opposite boundary and continue to be followed. A particle
splitting technique similar to \citep{Giacalone2005a} is used in order to
improve the statistics. Although we use very approximate parameters, we note
that the results are insensitive to the precise numbers. Our results are also
qualitatively unchanged after allowing the injection rate to vary as a function
of shock normal angle and different downstream diffusion coefficients. We also
note that our model is simplified to illustrate the physics of the magnetic-field
variation. Other effects such as changes in plasma properties (density, temperature, etc.) is small for
problems of current interest.

\subsection{Results of Numerical Simulations}
\subsubsection{A shock propagating perpendicular to the average magnetic field}
\begin{figure}
\begin{center}

\hfill
\includegraphics[width=0.8\textwidth]{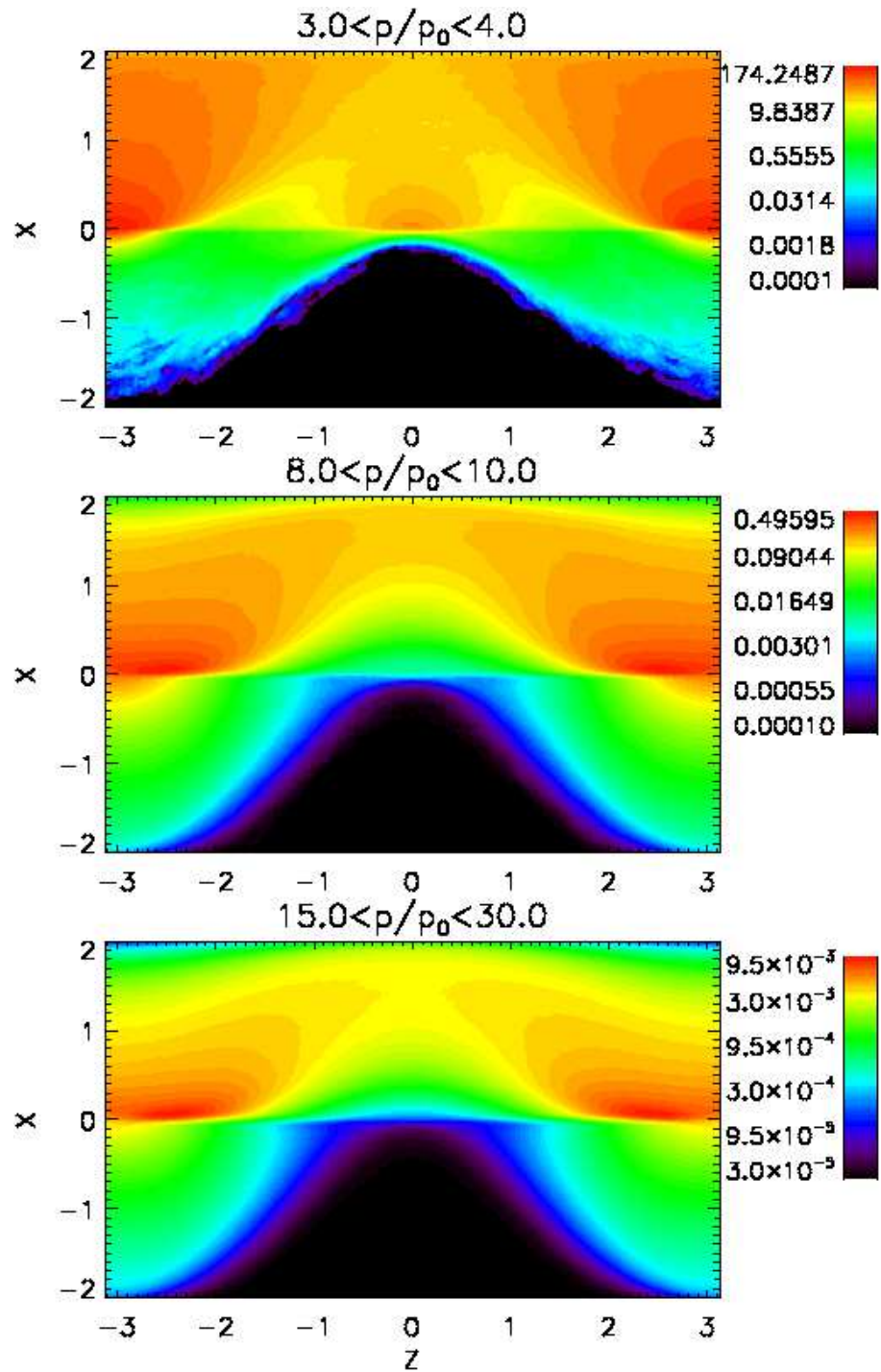}
\caption[The representation of density contour of accelerated particles.]{The representation of density contour of accelerated particles, for
energy range: $3.0<p/p_0<4.0$ (top), $8.0<p/p_0<10.0$ (middle),
$15.0<p/p_0<30.0$ (bottom). It is shown that the hot spots form on the both side
of the system. The acceleration at the center of the shock is suppressed. \label{figure-energy-contour1}}
 \end{center}
 \end{figure}

Consider first the case where the average magnetic field is in the
$z$ direction and the fluctuating magnetic field is $\delta B = B_0$. As shown
in Figure \ref{figure-shock-fieldline}, the magnetic field is convected through the shock front and is
compressed in the $x$ direction, thus $B_{z2} = rB_{z1}$. For the sinusoid
magnetic field considered in this paper, the local angle between upstream
magnetic field and shock normal, $\theta_{Bn}$, will vary along the shock
surface. As a magnetic field line passes through the shock surface, its
connection points (the points where the field lines intersect the surface of
the shock) will be moving apart in the middle of the plane ($z = 0$) and
approaching each other on the both sides of the system ($|z| = \pi$X$_0$). Since,
$\kappa_\parallel \gg \kappa_\perp$, the particles tend to remain on the
magnetic field lines. Because the acceleration only occurs at the shock front,
as the magnetic lines of force convect downstream, the particles will be
trapped and accelerated at places where the connection points converge toward
each other, leading to further acceleration. For the regions where the field
lines separate from each other, the particles are swept away from those
regions. Figure \ref{figure-energy-contour1} displays the spatial distribution contours of accelerated
particles in three energy ranges: $3.0<p/p_0<4.0$ (top), $8.0<p/p_0<10.0$
(middle), and $15.0<p/p_0<30.0$ (bottom). The density is represented by the
number of particles in simulation and its unit is arbitrary. It can be seen
that ``hot spots" form in the regions that connection points approaching each
other at all energy ranges, with lobes extend along the magnetic field lines.
The density of the accelerated particles at the connection-point separating
region (in the middle of the plane) is clearly much smaller, although there is
still a concentration of low-energy accelerated particles there since the
acceleration of low-energy particles is rapid and efficient at perpendicular
shocks. At higher energy ranges (middle and bottom), the lack of accelerated
particles may be interpreted as due to the fact that the acceleration to high
energies takes time.

\begin{figure}
\begin{center}
 \hfill
\includegraphics[width=0.8\textwidth]{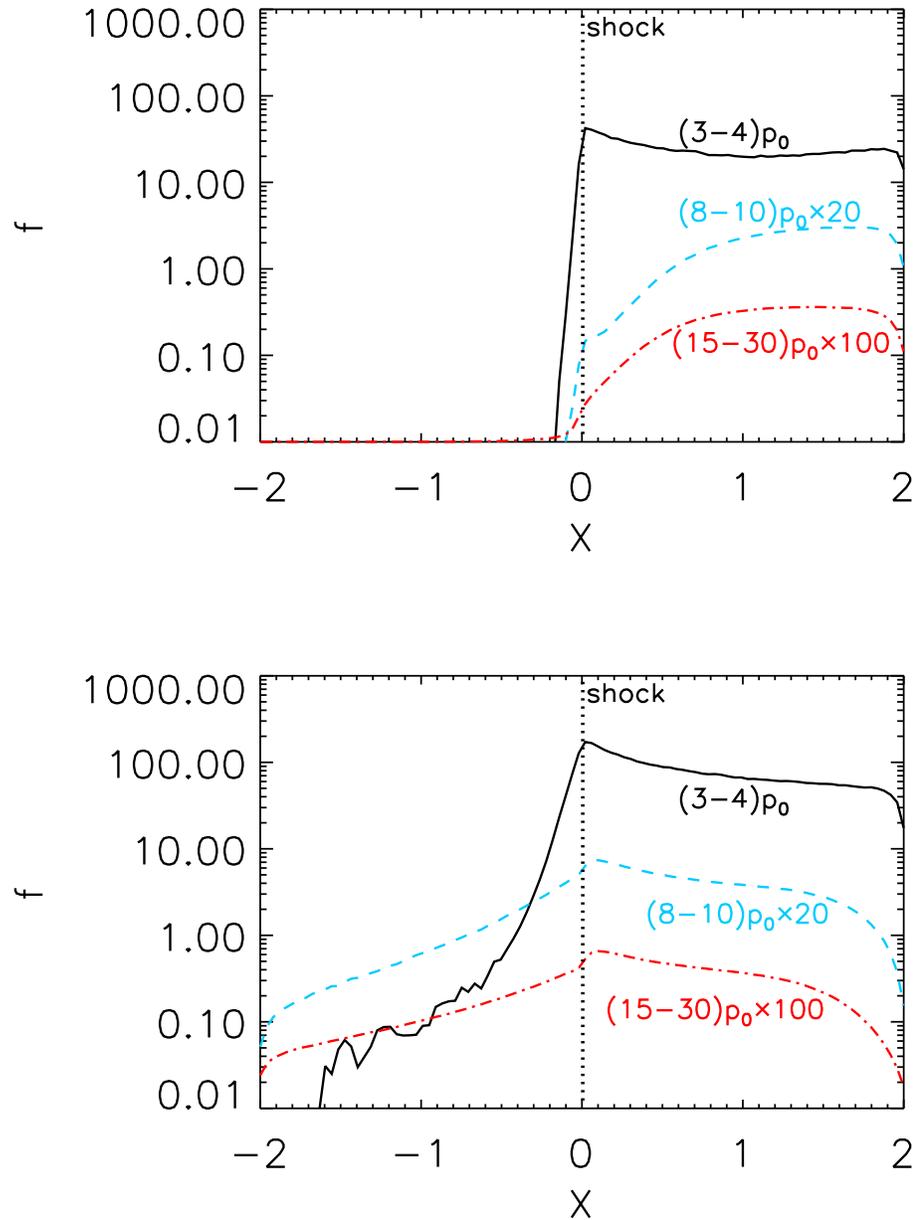}
\caption[The profiles of density of the accelerated particles.]{The profiles of density of the accelerated particles, for energy
ranges: $3.0<p/p_0<4.0$(black lines), $8.0<p/p_0<10.0$(blue dashed lines),
$15.0<p/p_0<30.0$ (red dot dashed lines) at different locations $z = 0.0$ (top)
and $\pi$X$_0$ (bottom), respectively. \label{figure-profile1}}
 \end{center}
 \end{figure}

Figure \ref{figure-profile1} illustrates the profiles of the density of accelerated particles for
different energy ranges at $z = 0 $ (top) and $z=\pi$X$_0$ (bottom). In each panel,
the black solid lines show the density of low-energy particles
($3.0<p/p_0<4.0$), the blue dashed lines show the density of intermediate
energy particles ($8.0<p/p_0<10.0$), and red dot dashed lines show the density
of particles with high energies ($15.0<p/p_0<30.0$). In connection-point
separating regions $z = 0 $ (top), it can be seen that while the downstream
distribution of low-energy particles is roughly a constant, the density of
particles with higher energies increase as a function of distance downstream
from the shock. These particles are not accelerated at the shock layer in the
center of the plane but in the ``hot spots". At $z = \pi$X$_0$ (bottom), the
density of particles of all energies decreases as a function of distance, which
indicates that the accelerated particles diffusive away from the ``hot spots". Since
the high energy particles have larger diffusion coefficients than the particles
with low energy, it is easier for them to transport to the middle of the plane.
The profile at $z = 0$ is similar to \emph{Voyager}'s observation of anomalous
cosmic rays (ACRs) at the termination shock and the heliosheath
\citep{Stone2005,Cummings2008} that shows the intensity of the ACRs is still
increasing and the energy spectrum is unfolding over a large distance after
entering the heliosheath. The same physics has been discussed by
\citet{Jokipii2008}, where ``hot spots" of energetic particles is produced by
the spatial variation of the injection of the source particles. In our current
work, the concentration of energetic particles are a consequence of particle
accelerated in a shock containing large-scale magnetic variation.

\begin{figure}
\begin{center}
\hfill
\includegraphics[width=0.8\textwidth]{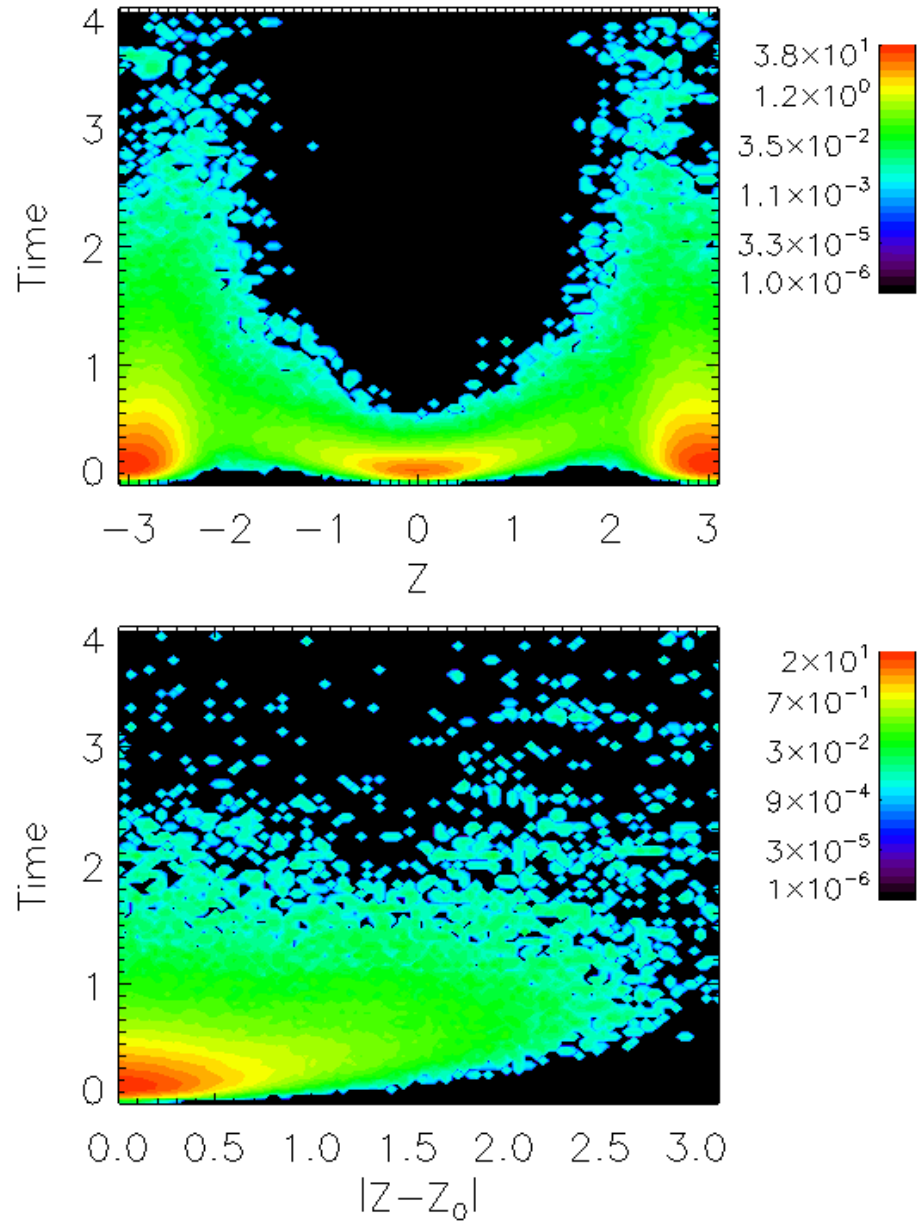}
\caption{$top$: The position in the $z$ direction and time when the particle
momentum reached $p = 3.0p_0$; $bottom$: The travel distance in the $z$ direction
$|z-z_0|$ and time when the particle momentum reached $p = 3.0p_0$. \label{figure-time-acceleration1}}
 \end{center}
 \end{figure}

\begin{figure}
\begin{center}
\hfill
\includegraphics[width=0.8\textwidth]{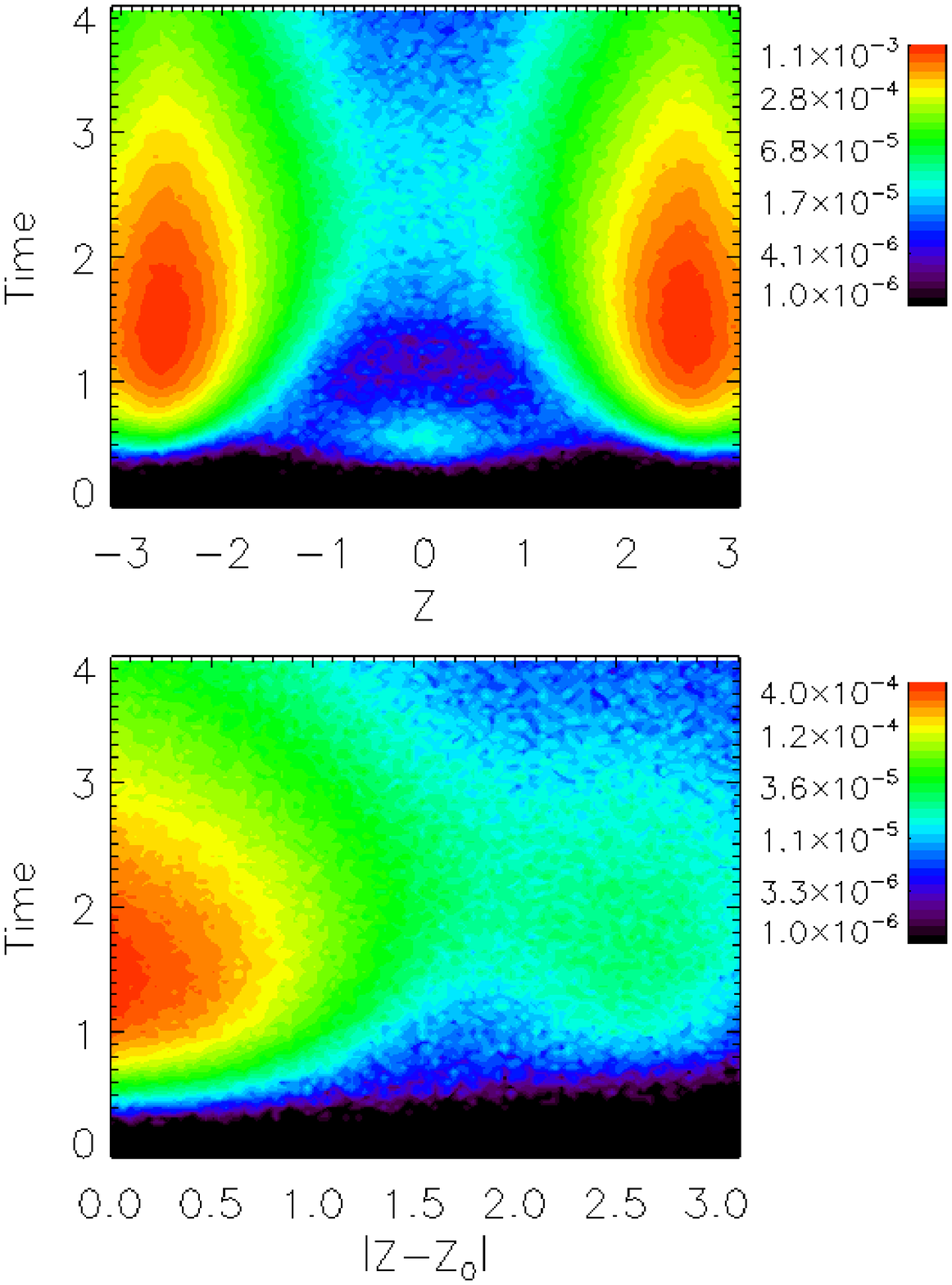}
\caption{$top$: The position in the $z$ direction and time when the particle
momentum reached $p = 10.0p_0$; $bottom$: The travel distance in the $z$ direction
$|z-z_0|$ and time when the particle momentum reached $p = 10.0p_0$. \label{figure-time-acceleration2}}
 \end{center}
 \end{figure}

The top panel in Figure \ref{figure-time-acceleration1} represents the positions in $z$ direction and the
times as soon as the particles reached a certain momentum $p_c = 3p_0$. We show
that particles are accelerated mainly at the connection-point converging
region. There are also particles accelerated at middle of the plane because the
particles can gain energy rapidly at perpendicular or highly oblique shock due
to the smallness of perpendicular diffusion coefficient \citep{Jokipii1987}.
However, the further acceleration is suppressed by the effect that the charged
particles travel away from the connection-point separating region (see also the
top panel in Figure \ref{figure-time-acceleration2}). It is clear that since the particles tend to follow the
magnetic field lines, when the field line connection points separate from each
other as field convects through the shock, the particles travel mainly along
magnetic field and away from the middle of the plane. The characteristic time
for a field line convect from upstream to downstream $\tau_c = D/U_1 \sim $T$_0$,
therefore there is no significant acceleration in the middle of the plane after
$t = \tau_c$. Some of the particles can get more acceleration traveling from
other region to ``hot spots". Figure \ref{figure-time-acceleration1} $bottom$ shows the distance particles
traveled in the $z$ direction from its original places $|z-z_0|$ versus time when
particle get accelerated at a certain energy ($p_c = 3p_0$). It shows that many
particles are accelerated close to their original position, which is related to
the acceleration in the ``hot spot". Nevertheless, there are also a number of
particles travel from the connection point separating region to ``hot spot" and
get further acceleration, which is represented by the particles that travel a
large distance in the z direction.

Figure \ref{figure-time-acceleration2} shows the same plot as Figure \ref{figure-time-acceleration1}, except here the critical momentum is
$p_c = 10.0p_0$. It is shown again in Figure \ref{figure-time-acceleration2} $top$ that most of particles are
accelerated to high energy are in the hot spot. However, as opposite to Figure
\ref{figure-time-acceleration1} $top$, there are very few particles accelerated at the center of plane since
energetic particles more transport away from the middle region and the time
available is not long enough. For a quick estimate, the acceleration time is
approximately,

\begin{eqnarray}
\tau_{acc} &=&
\frac{3}{U_1-U_2}\int^p_{p_0}(\frac{\kappa_{xx1}}{U_1}+\frac{\kappa_{xx2}}{U_2})
d\ln p \nonumber \\
&>& \frac{3}{U_1-U_2}\int^{10p_0}_{p_0}(\frac{\kappa_\perp}{U_1}+\frac{\kappa_\perp}{U_2})
d\ln p \\
&>& \tau_c \nonumber.
\end{eqnarray}

\noindent Therefore for most of particles, they do not have sufficient time to be
accelerated to high energies at the center. A number of particles accelerated
at the center will travel to the ``hot spot" and get more acceleration, as
shown in Figure \ref{figure-time-acceleration2} $bottom$.

\begin{figure}
\begin{center}
\includegraphics[width=25pc]{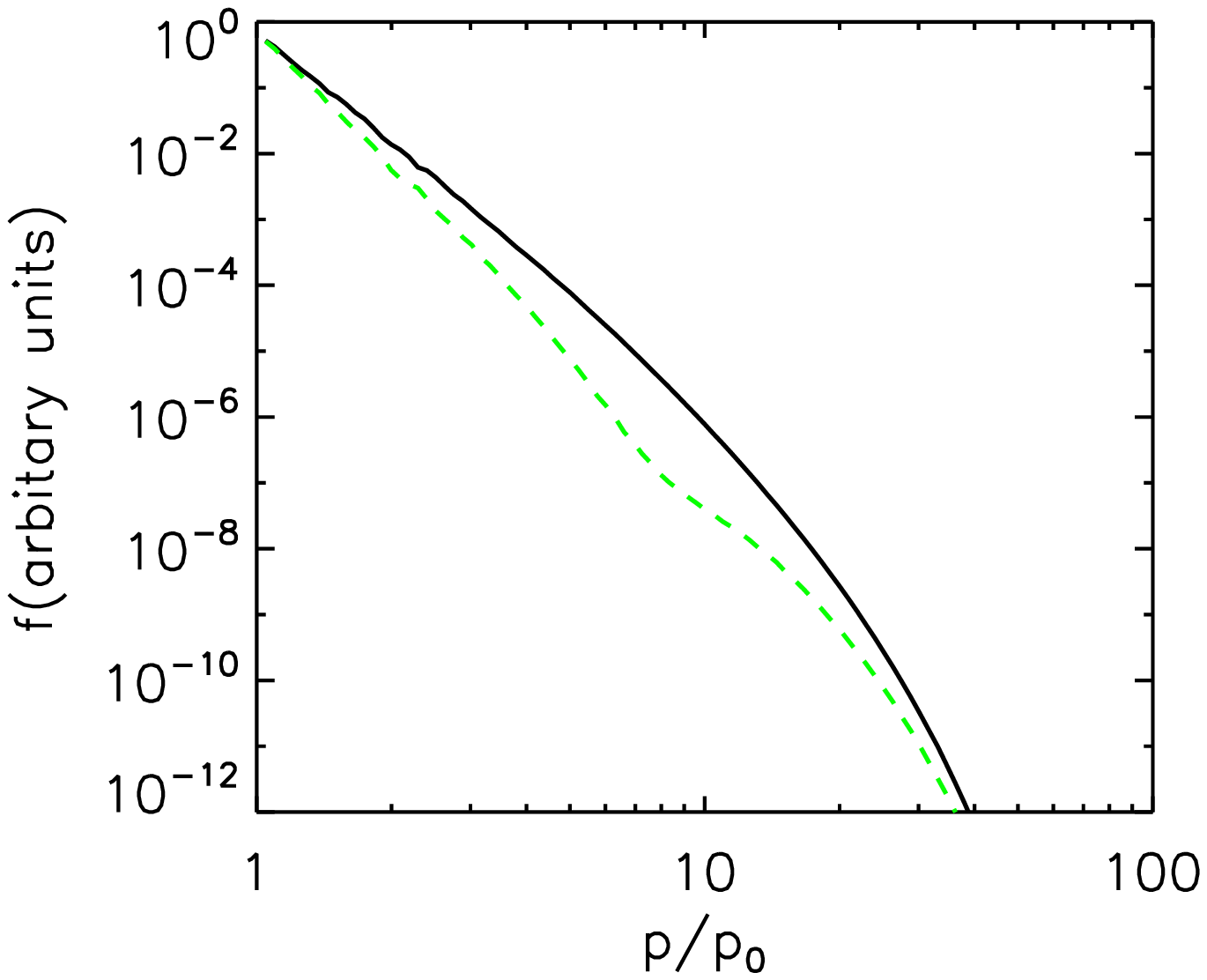}
\hfill
\includegraphics[width=25pc]{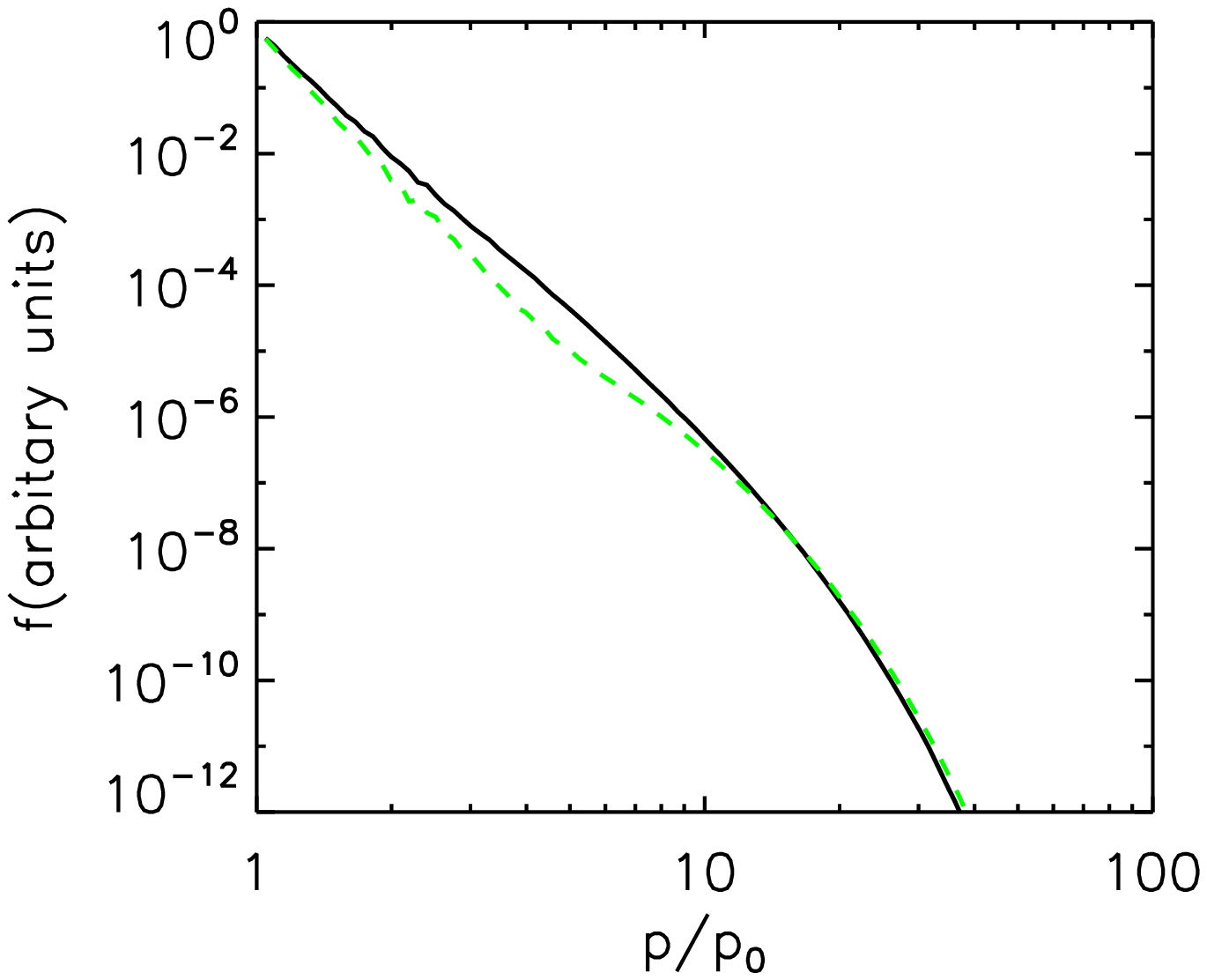}
 \caption[The steady state distribution functions at different regions]{The steady state distribution functions at $top$: [$0.1$X$_0<x<0.3$X$_0$,
$(\pi-0.2)$X$_0<z<\pi$X$_0$] (black solid line), [$0.1$X$_0<x<0.3$X$_0$, $-0.1$X$_0<z<0.1$X$_0$] (green dashed
line) and $bottom$: [$0.8$X$_0<x<1.0$X$_0$, $(\pi-0.2)$X$_0<z<\pi$X$_0$] (black solid line),
[$0.8$X$_0<x<1.0$X$_0$, $-0.1$X$_0<z<0.1$X$_0$] (green dashed line), respectively. \label{figure-spectra1}}
 \end{center}
 \end{figure}

Clearly, the acceleration by the shock containing 2-D spatial magnetic field variations 
shows a different picture than the acceleration by a 1-D planar shock, indicating that the resulting
distribution function is spatially dependent. In Figure \ref{figure-spectra1} we show the steady
state energy spectra obtained in the regions \emph{top}: [$0.1$X$_0<x<0.3$X$_0$,
$(\pi-0.2)$X$_0<z<\pi$X$_0$] (black solid line), and [$0.1$X$_0 <x<0.3$X$_0$, $-0.1$X$_0<z<0.1$X$_0$] (green
dashed line) and \emph{bottom}: [$0.8$X$_0<x<1.0$X$_0$, $(\pi-0.2)$X$_0<z<\pi$X$_0$] (black solid
line), and [$0.8$X$_0<x<1.0$X$_0$, $-0.1$X$_0<z<0.1$X$_0$] (green dashed line). It is shown that
the spatial difference among distribution functions at different locations
caused by large-scale magnetic field variation is considerable. The black lines
in both top plot and bottom plot, which correspond to the ``hot spots", show
power-law like distributions except at high energies. At high energies, the
particles will leave the simulation domain before gain enough energy that
causes the roll over in distribution function, this roll over is mainly caused
by a finite distance to upstream boundary. For other locations, the 2-D effect
we discussed will produce the modification in distribution functions. The most
pronounced effect can be found at the nose of the shock (green lines), in the
top panel the distribution of particles shows a suppression of acceleration at all
the energies. This insufficient acceleration is most prominent in the range of
$6-12 p_0$. At these energies the acceleration time scales are longer than the
time for the field line convection swipe the particles away from the
connection-point separating region, as we discussed above. The bottom plot shows
that deep downstream the spectrum of accelerated particles is similar at high
energies since the mobility of these particles.

\subsubsection{An oblique shock}
The previous discussion has established the effect of a spatially varying
upstream magnetic field on the acceleration of fast charged particles at a
shock propagating normal to the average upstream magnetic field. We next
consider the case where the shock propagation direction is {\it not} normal to
the average magnetic field.

\begin{figure}
\begin{center}
\hfill
\includegraphics[width=0.8\textwidth]{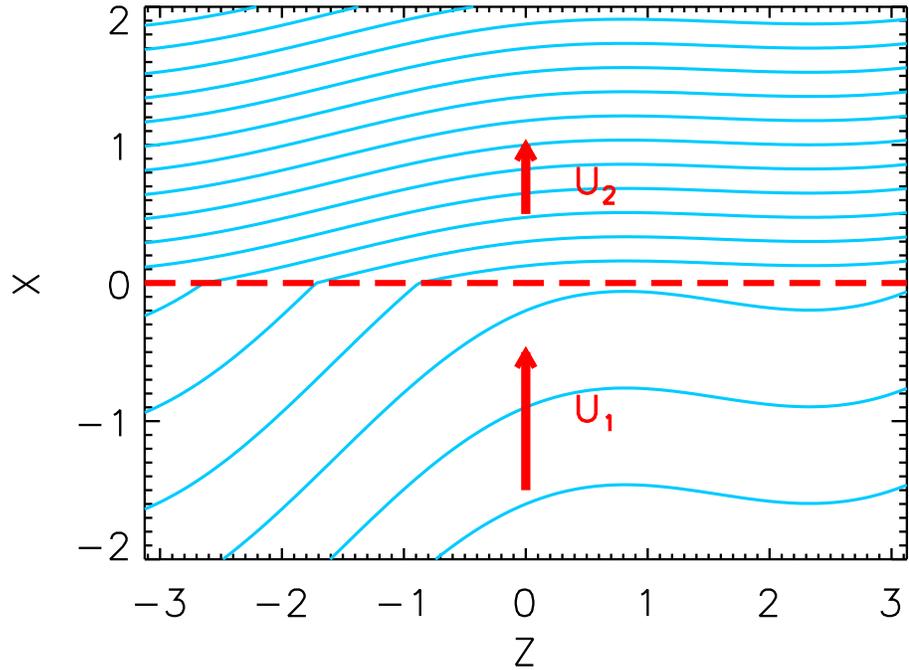}
\caption{The shock and the magnetic field geometry considered for the case of
an average magnetic field is $70^\circ$ of the shock normal. The blue lines
 represent the magnetic field lines and red dashed line indicates the position of shock wave.
 The flow velocities are $U_1$ (upstream) and $U_2$ (downstream). \label{figure-shock-fieldline2}}
 \end{center}
 \end{figure}

\begin{figure}
\begin{center}
\hfill
\includegraphics[width=0.8\textwidth]{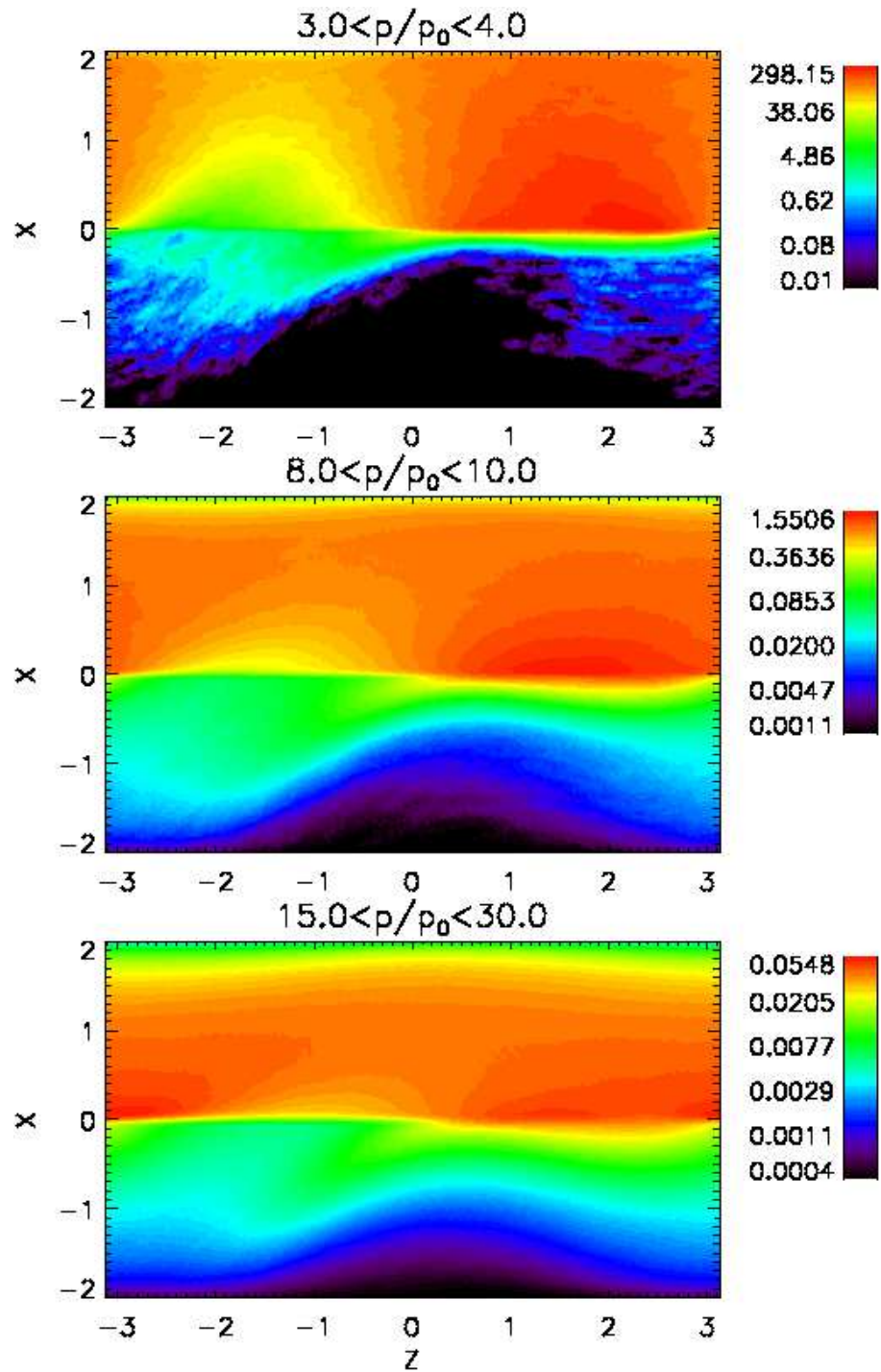}
\caption[The density contour of accelerated particles for the case of oblique
shock and $\delta B/B_0 = 0.5$.]{The density contour of accelerated particles for the case of oblique
shock and $\delta B/B_0 = 0.5$, for energy range: $3.0<p/p_0<4.0 (top)$,
$8.0<p/p_0<10.0 (middle)$, $15.0<p/p_0<30.0 (bottom)$. \label{figure-energy-contour2}}
 \end{center}
 \end{figure}

Clearly, if the varying direction of the upstream magnetic field is such that
at some places the local angle of the magnetic field relative to the average
field direction exceeds the angle of the average magnetic field to the shock
plane, we will have situations similar to that discussed in the previous
sections. There will be places where the connection points of the magnetic
field to the shock move further apart or closer together. Hence we expect that the
same physics can be applicable. An example is given in Figure \ref{figure-shock-fieldline2}. In this case
the ratio of $\delta B/B_0$ is taken to be $0.5$ and the averaged shock normal
angle $\theta_{Bn} = 70^\circ$. It can be seen from this plot that the
connection points can still move toward each other in some regions. Figure \ref{figure-energy-contour2}
shows the density contours of accelerated particles the same as Figure \ref{figure-energy-contour1}, but
for the case of the oblique shock. We find that in this case the process we
discussed in the last section is still persistent, even for an oblique shock and
relative smaller $\delta B/B_0$. The ``hot spot" forms correspond to the
converging magnetic connection points and particle acceleration is suppressed
in the region where connection points separate from each other. We may conclude
that, for a shock that is oblique, if some magnetic field lines can intersect
the shock multiple times, we have ``hot spots" of accelerated particles forms
where the connection points converging together.

\subsection{Summary and Discussion}

\begin{figure}
\centering
\begin{tabular}{cc}
\epsfig{file=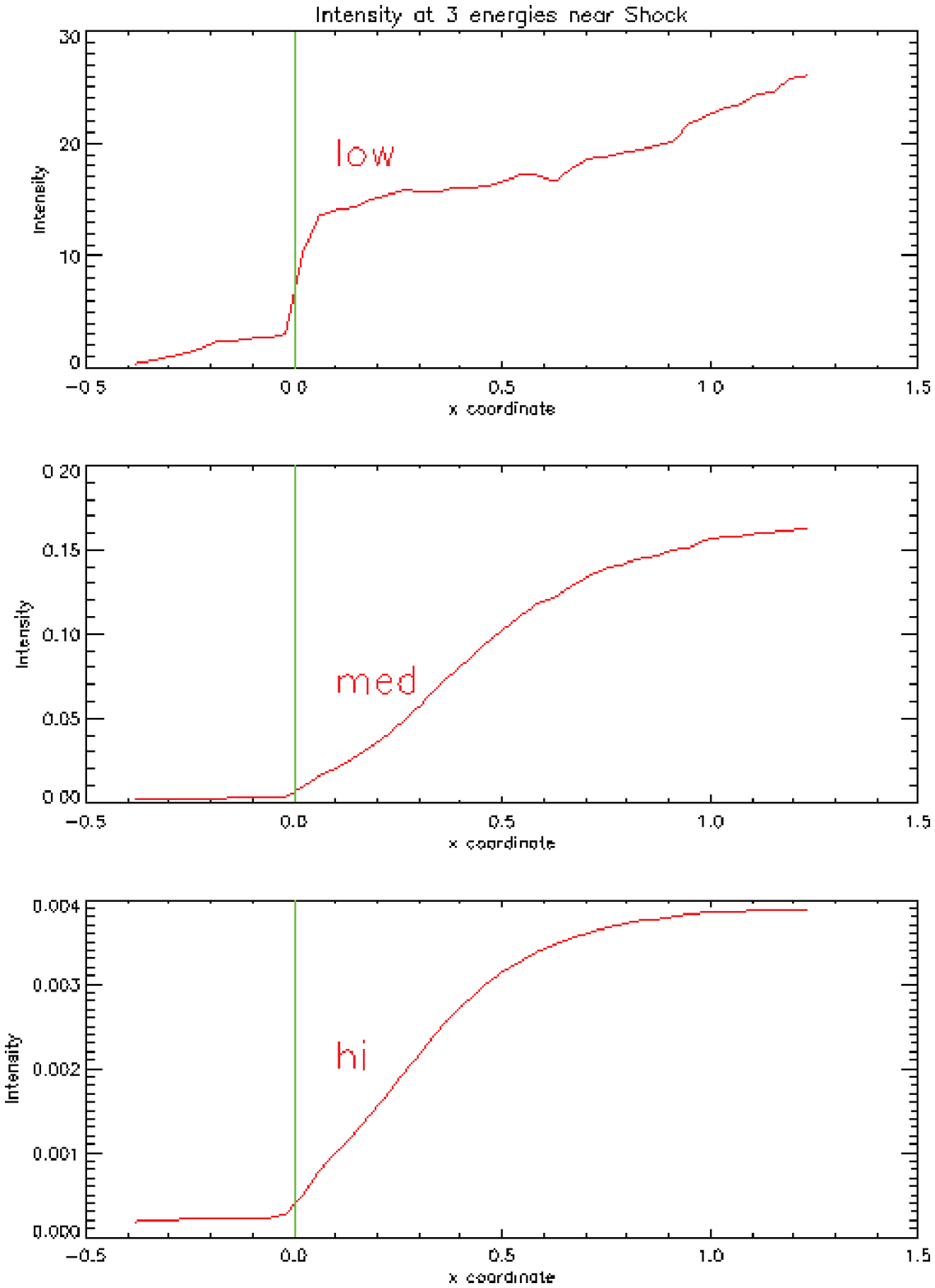,width=0.5\textwidth,clip=} &
\epsfig{file=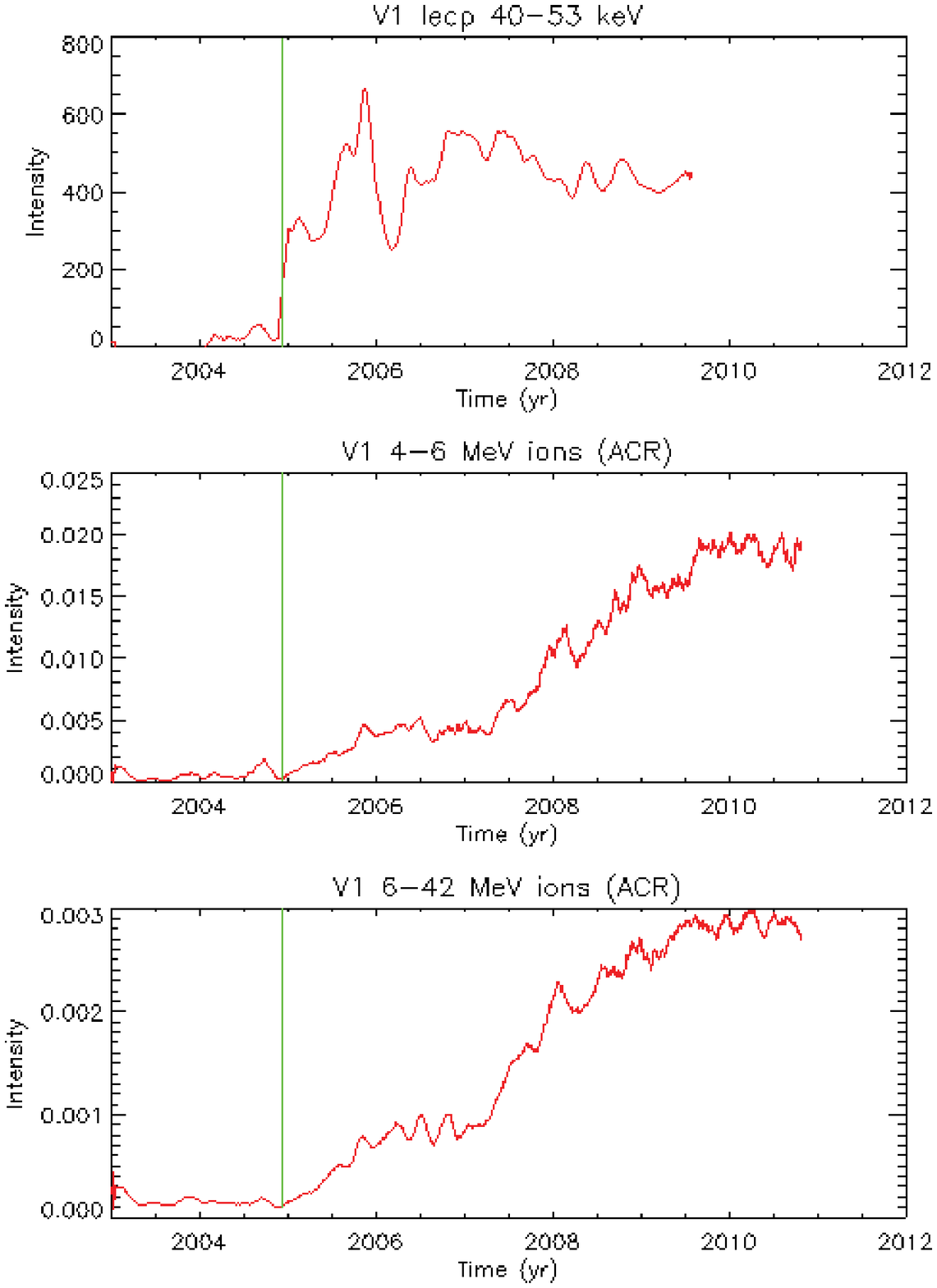,width=0.5\textwidth,clip=}
\end{tabular}
\caption[A comparison between a spatial profile for the density of energetic particles picked from the numerical simulation and observation made by Voyager 1.]{A comparison between a spatial profile for the density of energetic particles picked from the numerical simulation and observation made by \textit{Voyager} 1. \label{figure-compare}}
\end{figure}

The acceleration of charged particles in shock waves is one of the most
important unsolved problems in space physics and astrophysics. Charged
particle transport in turbulent magnetic field and acceleration in shock region
are two inseparable problems. In this section we illustrate the effect of a
large-scale sinusoidal magnetic field variation. This simple model allows a
detailed examination of the physical effects. As the magnetic field lines pass
through the shock, the connection points between field lines on the shock
surface will move accordingly. We find that the region where connection points
approaching each other will trap and preferentially accelerate particles to
high energies and form ``hot spots" along the shock surface, somewhat in
analogy to the ``hot spots" postulated by \citet{Jokipii2008}. The shock
acceleration will be suppressed at places where the connection points move
apart each other. Some of the particles injected in those regions will
transport to the ``hot spots" and get further accelerated. The resulting
distribution function is highly spatially dependent at the energies we studied,
which could give a possible explanation to the \emph{Voyager} observation of
anomalous cosmic rays. Although we have discussed a simplified, illustrative
model, the resulting spectra and radial distributions show qualitative
similarities with the {\it in situ} {\it Voyager }observations. In Figure \ref{figure-compare} 
we compare a spatial profile of energetic particle intensity obtained in our simulation and
the time profile observed by \textit{Voyager} 1 as it passed the termination shock and entered into
the heliosheath. One can see that our simulation qualitatively agree with the observation. This
mechanism gives an interpretation for the observation that the ACR intensity did not saturate at the 
termination shock. 

Thus, the intensities do not in general, peak at the the shock and the energy spectra are
not power laws.  We show that this process is robust even for the case of oblique
shocks with relatively small magnetic field variations. Large scale magnetic
field variation, which could be due to magnetic structures like magnetic
clouds, or the ubiquitous large scale field line random walk, will strongly
modify the simple planar shock solution. This effect could work in a number of
situations for large scale shock acceleration including magnetic variations,
for example, the solar wind termination shock and supernova blast waves.

\chapter{The Acceleration of Electrons at Collisionless Shocks\label{chapter4}}

\section{Overview}

As we discussed in Section \ref{chapter1-particle2} and Section \ref{chapter3-overview1}, collisionless shocks are efficient accelerators for a variety of energetic charged particles observed in the heliosphere. However, the acceleration of electrons at collisionless shocks is generally considered to be more difficult than that of ions. This is primarily due to the fact that the gyroradii of non-relativistic electrons are much smaller compared with that of protons of the same energy 
(by a factor of $\sqrt{m_i/m_e} \sim 43$), therefore low-energy electrons cannot resonantly interact with the large-scale magnetic turbulence or ion-scale waves close to the shock front. Despite the theoretical difficulties, energetic electrons are commonly observed in SEP events. The accelerated electrons are usually observed to be associated with quasi-perpendicular shocks in 
interplanetary shocks and planetary bow shocks \citep[e.g.,][]{Anderson1979}. The acceleration of electrons remains poorly understood.

In this chapter we study numerically the acceleration of low-energy 
electrons at shocks that propagate through turbulent magnetized media, 
using a combination of hybrid simulations (kinetic ions, 
fluid electron) and test-particle electron simulations.
We find that the acceleration of electrons is greatly enhanced due to the effect of large-scale magnetic 
turbulence, which provides a mechanism for accelerating electron to high energy and injecting electrons 
into the diffusive shock acceleration (DSA).

We first review the observations of energetic electrons related to collisionless shocks in Section \ref{electron-observation}. Then we give an introduction for the previous theoretical works on the acceleration of electrons in Section \ref{electron-theory}. In Section \ref{chapter4-method} we discuss the numerical method used in this study. Section \ref{electron-IPS} presents the results on electron acceleration at a shock propagating through a turbulent magnetic field. 
The parameters are similar to observations of electrons associated with interplanetary shocks and also the solar wind termination shock. The results can also be used to understand the acceleration of electrons at CME-driven shocks.
Section \ref{electron-flare} presents the numerical results with parameters similar to solar flare region. We discuss 
and conclude the results in Section \ref{electron-conclusion}.

\section{Observations of Energetic Electrons Associated with Shock Waves 
\label{electron-observation}}

Energetic electrons are often observed to be associated with
collisionless shocks. At the terrestrial bow shock and interplanetary shocks, electrons with energy up to 
 $\sim 100$ keV are often observed close to quasi-perpendicular shocks 
\citep{Fan1964,Anderson1979,Gosling1989a,Oka2006}, i.e.,  
the angle between the incident magnetic field vector and the shock normal $\theta_{Bn}$ is larger than $45^\circ$.
\citet{Anderson1979} showed the \emph{ISEE} spacecraft measurements of upstream electrons ($>16$ keV) 
at the Earth's bow shock that originated from a thin region close to the point of tangency between 
interplanetary magnetic field lines and the shock surface. \citet{Gosling1989a} pointed out that
the accelerated superthermal electrons were often seen close to perpendicular shocks or quasi-perpendicular shocks, 
rather than in the vicinity of parallel shocks. The energy spectra of electrons close to shocks appear to follow
a power law with a slope index between $-3$ to $-4$.
 \citet{Tsurutani1985} reported the observations of energetic electrons associated with interplanetary 
 shocks showing ``spike-like" flux enhancements for energies $> 2$ keV. The spike events
were observed at quasi-perpendicular shocks with $\theta_{Bn} \geq 70^\circ$. Some shock crossings had no
enhancements of energetic electrons that were reported to be associated with low shock speeds and/or
small $\theta_{Bn}$.
\citet{Simnett2005} have presented data showing that energetic electrons ($\sim 50$-$100$ keV)
are accelerated close to interplanetary shocks. They also showed that some
accelerated electrons can escape far upstream of a nearly perpendicular
interplanetary shock. The clear evidence of electron acceleration at interplanetary shocks by 
DSA is rare, but a recent example discussed by \citet{Shimada1999} shows evidence of the importance of whistler waves (a high frequency wave that can resonantly interact with low-energy electrons) close to a quasi-perpendicular shock. The observation by \textit{Voyager} $1$ at the termination 
shock showed a spike-like enhancement of energetic electrons \citep{Decker2005}. 
\textit{Voyager} $2$ observed an exponential increase upstream
of the termination shock and roughly constant downstream in the heliosheath, similar to what is
predicted from DSA \citep{Decker2008}. Both of the \textit{Voyager} spacecraft have observed that 
electrons are accelerated to at least MeV range, indicating that the termination shock can efficiently 
accelerate electrons.

\citet{Bale1999} and \citet{Pulupa2008} observed electron foreshocks and related Langmuir 
waves upstream of interplanetary shocks. These events appeared to be associated with irregular 
shock surfaces with spatial scales $\sim 2 \times 10^5$ km. They 
proposed that the complex upstream electron events result from large-scale irregularities in the shock surface. 
The large-scale ripples have been detected using multiple spacecraft observations \citep{Neugebauer2005,Koval2010} and 
the magnitude of radii of curvature is in the range of $3\times 10^5$ to $10^7$ km with 
an average of $3.5\times 10^6$ km, similar to the correlation length of the interplanetary turbulence \citep[e.g.,][]{Coleman1968}.

In solar energetic particles (SEPs) that originate from CME-driven shocks or solar flares, electrons are 
frequently observed to be accelerated to $10$ keV - $1$ MeV. It is important to point out that in SEP events, 
the electrons and ions are often observed to have tight correlations 
\citep{Posner2007,Cliver2009,Haggerty2009}. 
For example, \citet{Haggerty2009} reported that 175-315 keV electrons are well associated with $1.8$-$4.7$ MeV protons. \citet{Cliver2009} showed that the electrons ($\sim 0.5$ MeV) and the protons ($>10$ MeV) in large SEP events are strongly correlated. The correlation depends weakly on isotope ratio (Fe/O), which indicates that the effect of source regions on the electron-proton correlation is not important. The tight correlation between electrons and ions has also been found in ground base level events (Tylka, private communication), where both electrons and protons are accelerated to relativistic energies. These observations indicate that a common mechanism may exist in the process of particle acceleration during SEP events. The accelerated electrons are believed to produce
type II and type III radio bursts in the solar corona and interplanetary space \citep{Nelson1985}, 
which provides a remote probe for electron acceleration. Herringbone structures featured by fast electron speeds are often observed
in type II radio bursts, which indicates that electrons are accelerated to about $20 - 200$ keV at 
CME-driven shocks \citep{Roberts1959,Cairns1987,Cane1989}. 

In solar flares, the acceleration of electrons can be observed remotely since the accelerated electrons 
are subjected to several radiation processes. Hard X-ray emissions produced by superthermal electrons above 
magnetic loops have been observed \citep{Masuda1994,Krucker2010} and may be associated with flare termination shocks \citep{Forbes1988,Shibata1995}. Using the emissions from neutron-capture 
$\gamma$-ray line and electron bremsstrahlung, \citet{Shih2009} have shown that the emissions of $>30$ MeV 
protons and that of $>0.3$ MeV electrons are proportional to each other. Again, this indicates that the electrons 
and ions are accelerated by closely related mechanisms. 

In astrophysical shocks, electrons are observed to be accelerated to very high energies (may reach $10^{15}$ eV) and produce strong synchrotron emissions in the existence of magnetic fields \citep{Reynolds2011}. At those energies, the electrons are ultra-relativistic and their gyroradii are virtually the same as protons of the same energy. In that situation, both the acceleration of electrons and that of protons may be considered to be DSA. 

As a summary, there is plenty of evidence of energetic electrons associated with fast shock waves. Observations suggest that accelerated electrons are more associated with oblique shocks and rarely seen at parallel shocks. Compared to ions, energetic electrons are more-easily to excite electromagnetic radiation (radio bursts, bremsstrahlung emission, and synchrotron radiations, etc.), which can be used as remote tracers for energetic electrons. In many situations, electron acceleration and ion acceleration are observed to be correlated, which may give a constraint to the acceleration mechanism.

\section{Review of Previous Theoretical Works \label{electron-theory}}

In order to explain the energization of electrons within the shock
layer, \citet{Holman1983,Wu1984} and \citet{Leroy1984} developed analytic
models for electron acceleration from thermal energies by adiabatic
reflection off a quasi-perpendicular shock. This is usually referred to as
shock drift acceleration (SDA) or fast-Fermi acceleration. The theory describes a scatter-free
electron acceleration process in a planar, time-steady shock. It
obtains a qualitative agreement with observations at Earth's bow
shock in terms of the loss-cone pitch-angle distribution and energy
range of accelerated electrons. \citet{Krauss-Varban1989a} used the
combination of electron test-particle simulation and 1-D hybrid
simulation and verified Wu's basic conclusions. The main energy
source of fast Fermi acceleration comes from the $-\textbf{V} \times
\textbf{B}/c$ electric field that is the same as shock drift acceleration
\citep{Armstrong1985}. It can be demonstrated that fast-Fermi
acceleration and SDA are the same process in two different frames of
reference \citep{Krauss-Varban1989b}. Thus one would expect
electrons to drift in the direction perpendicular to the flow and
magnetic field during the acceleration at the shock front.

However it is commonly known that in this process both the fraction and attainable energy of
the accelerated particles are limited \citep[e.g.,][]{Ball2001}. This cannot explain the observed
high-energy electron. For example, observations at Earth's bow
shocks \citep{Gosling1989a} suggested that the accelerated electrons have a power law distribution 
with an exponetial drop, which cannot be produced by this simple mechanism. In a recent work,
\citet{Pulupa2012} presented a comparison between the STEREO measurements of electron acceleration at Earth's
bow shock and the theory of adiabatic fast-Fermi acceleration \citep{Wu1984}. It is clear that the fast-Fermi
acceleration fails to predict the observed energetic electrons in both the accelerated fraction and spectrum shape. Herringbone structures observed in Type II burst \citep{Cairns1987} require a high electron energy, and the energetic electrons cannot be produced by fast Fermi 
acceleration unless electrons are extremely hot.

Some recent progress in shock acceleration is the consideration of nonplanar effects such as shock ripples and magnetic turbulence. The simulations by \citet{Burgess2006} and \citet{Umeda2009} show that small-scale shock ripples can be important in scattering the electrons and facilitating the acceleration. \citet{Savoini2010} discussed the non-adiabatic motion of electrons. Recently, \citet{Jokipii2007} proposed a novel mechanism to solve the injection problem that does not require strong pitch-angle scattering from small-scale fluctuations. The low-energy particles can move along the meandering field lines of force, travel back and forth across a shock front, therefore gain energy from the difference between upstream and downstream flow velocities. In this study (Section \ref{electron-IPS} and Section \ref{electron-flare}), using self-consistent hybrid simulations combined with test-particle simulation for electrons, we have found that efficient electron acceleration can happen after considering large-scale pre-existing upstream magnetic turbulence. The turbulent magnetic field leads to field-line meandering that allows the electrons to get accelerated at a shock front multiple times. The shock surface becomes irregular on a variety of spatial scales from small-scale ripples due to ion-scale plasma instabilities \citep{Lowe2003}, to large-scale structures caused by the interaction between the shock and upstream turbulence \citep{Giacalone2008,Lu2009}. The rippled surface of the shock front also contributes to the acceleration by mirroring electrons between the ripples. The observational evidence of these large-scale ripples has been shown by a number of authors \citep{Neugebauer2005,Bale1999,Pulupa2008,Koval2010}. These results, along with the previous work on acceleration of ions \citep{Giacalone2005a,Giacalone2005b}, suggest that large-scale turbulence has an important effect on the acceleration of both electron and ions at shocks, which is consistent with the correlation between ions and electrons in solar energetic particle events \citep[e.g.,][]{Haggerty2009,Cliver2009}. 

Another scenario that solves this injection problem relies on small-scale waves excited close to the shock front.
\citet{Levinson1992,Levinson1994} has proposed an analytical theory for thermal electron injection due to whistler waves excited by electrons streaming upstream. In this theory efficient electron acceleration requires large Mach numbers $M_A \sim 43/\sqrt{\beta_e}$, where $\beta_e$ is the ratio of thermal electron pressure to magnetic pressure inside the shock. This condition is rarely satisfied for shocks in the heliosphere. \citet{Amano2007} proposed ``electron surfing acceleration'' where electron can reach efficient acceleration by being trapped in electrostatic solitary waves. This mechanism also requires a relative high mach number shock (Alfven mach number $M_A$ is greater than $14$). Recently, \citet{Riquelme2011} used full particle simulations to study electron acceleration at oblique shocks. For the parameters they use, they find that efficient nonthermal electron acceleration when electrons are scattered by oblique whistler waves at a $M_A \sim 7$ quasi-perpendicular shock. However, these full particle simulations generally use unrealistic mass ratio $m_i/m_e$ and the ratio between the speed of light and thermal speed $c/V_{the}$, which may produce some artificial effects. In fact in the simulations they find the electrons are heated at the shock layer to a temperature the same as that of protons \citep{Riquelme2011}. This is not consistent with in situ observations at interplanetary shocks and the relevant theoretical works \citep{Thomsen1987,Schwartz1988,Scudder1995}, in which the the heating of electrons at shocks is significantly less than that of ions \citep{Thomsen1987,Schwartz1988,Scudder1995}. The effect of whistler waves for electron acceleration at shocks still remains to be clarified.


\section{Numerical Method \label{chapter4-method}}
Investigating the motion of charged particles in the vicinity of a collisionless
shock requires a spatial scale large enough for particles to
travel back and forth across the shock, and a spatial resolution
small enough to include the detailed physics for particle scattering
and shock microstructure. We implement a combination of a $2$-D
hybrid simulation to model the fields and plasma flow and a test
particle simulation to follow the orbits of a large number of
energetic electrons. In the first step, we employ a two-dimensional
hybrid simulation \citep{Giacalone2005b}
that includes pre-existing large-scale turbulence. In the hybrid
simulation \citep[e.g.,][]{Winske1988}, the ions are treated
fully kinetically and the thermal (i.e., non-energetic) electrons are
treated as a massless fluid. This approach is well suited to resolve
ion-scale plasma physics that is critical to describe supercritical
collisionless shocks. We consider a two-dimensional
Cartesian grid in the $x-z$ plane. All the physical vector
quantities have components in three directions, but depend spatially
only on these two variables. A shock is produced by using the
so-called piston method \citep[for a discussion,
see][]{Jones1991}, in which the plasma is injected continuously
from one end ($x=0$, in our case) of the simulation box, and
reflected elastically at the other end ($x=L_x$). This boundary is
also assumed to be a perfectly conducting barrier. The pileup of
density and magnetic field creates a shock propagating in the $-x$
direction. To include the effect of large-scale magnetic fluctuations, a
random magnetic field is superposed on a mean field at the beginning
of the simulation and is also injected continuously at the $x=0$
boundary during the simulation. The simplified one-dimensional
fluctuations have the form $\textbf{B}(z, t) = \delta \textbf{B}(z,
t) + \textbf{B}_1$, where $\textbf{B}_1$ is the averaged upstream
magnetic field. The fluctuating component contains an equal mixture
of right- and left-hand circularly polarized, forward and backward
parallel-propagating plane Alfven waves. The amplitude of the
fluctuations at wave number $k$ is determined from a Kolmogorov-like power spectrum:

\begin{eqnarray}
P(k) \propto \frac{1}{1 + (k L_c)^{5/3}},
\end{eqnarray}

\noindent in which $L_c$ is the coherence scale of the fluctuations. For the simulations presented in
this study, we take $L_c = L_z$, which is the size of simulation box in the 
$z$-direction. Note that in addition to magnetic fluctuations,
there are also velocity perturbations with $\delta \textbf{v} = v_{A1}\delta
\textbf{B}/B_1$ (Alfven waves). Different from
previous studies, the consideration of large-scale magnetic fluctuations
enables us to consider the effect of pre-existing magnetic turbulence on electron
acceleration, which has been shown to be important for low-energy ion
acceleration \citep{Giacalone2005a,Giacalone2005b} since particle
transport normal to the mean field is enhanced. However, particle transport
in full $3$-D turbulence cannot be properly treated in a self-consistent way
using available computation. As demonstrated by previous work
\citep{Jokipii1993,Giacalone1994,Jones1998}, in the model with at least
one ignorable coordinate, the center of gyration of particles is confined to
within one gyroradius of the original magnetic field line. The test-electrons
can still move normal to the mean field in our model because of the field-line
random walk.

In the second part of our calculation we integrate the full motion equation of
an ensemble of test-particle electrons in the electric and magnetic fields
obtained in the hybrid simulations (see Figure \ref{electron-mag1}). This part
of the calculation is done separately from the main hybrid simulation as a post
processing phase. As noted by \citet{Krauss-Varban1989a}, high-order interpolation 
of fields is required to ensure numerical accuracy and avoid artificial scattering 
in calculating electron trajectories. In this work we use a second-order spatial
interpolation and a linear temporal interpolation, which ensure
the smooth variations of the electromagnetic fields.  The test-particle
electrons are released uniformly upstream when the
shock has fully formed and is far from the boundaries. The numerical technique
used to integrate electron trajectories is the so-called Bulirsh-Stoer method,
which is described in detail by \citet{Press1986}. It is highly accurate and
conserves energy well. It is fast when fields are smooth compared with the
electron gyroradius. The algorithm uses an adjustable time-step method based on
the evaluation of the local truncation error. The time step is allowed to vary
between $5 \times 10^{-4} $ and $0.1 \Omega_{ce}^{-1}$, where $\Omega_{ce}$ is
the electron gyrofrequency. The ratio $\Omega_{ce}/\Omega_{ci}$ is taken to be
the realistic value $1836$. The total number of electrons in the simulation is
$1.6 \times 10^6$. The electrons that reach the left or right boundary are
assumed to escape from the shock region and are removed from the simulation. The
boundary condition in the $z$ direction is taken to be periodic. The readers
are referred to \citep{Burgess2006} for more details on the numerical methods.

Magnetic field turbulence has already proved to have key effects
on the particle acceleration in collisionless shocks. Unfortunately,
solving the whole problem in three-dimensional space and resolving
magnetic turbulence from coherence scale to electron scale are still
limited by available computation in the near future. This limitation
motivates us to solve these problems approximately. We also note
that in our model the electron test-particle simulation is not
self-consistent since the hybrid simulation does not include the
electron scale physics. The electron scale shock structure
may be important but is neglected here.

\section{The Effect of Large-Scale Magnetic Turbulence on the Acceleration of Electrons: Interplanetary Shocks \label{electron-IPS}}

Much of this section has been published in the Astrophysical Journal \citep{Guo2010a}.
\subsection{Initial Condition and Parameters}
In this section we use the numerical method described in Section \ref{chapter4-method} to study the acceleration 
of electrons at a shock that propagates through a turbulent magnetic field. The parameters are similar to 
interplanetary shocks and the solar wind termination shock. For most part of this study, we consider a turbulence variance $\sigma^2 = \delta B^2/B_1^2 = \delta v^2/v_{A1}^2 = 0.3$,
where $\delta v$ and $v_{A1}$ are the magnitude of velocity perturbation and
upstream Alfven speed, respectively. We also discuss the effect of different
values of turbulence variances.
The size of the simulation box for most of situations is 
$L_x\times L_z = 400 c/\omega_{pi}\times 1024 c/\omega_{pi}$, 
where $c/\omega_{pi}$ is the ion inertial length. We also examine the effect of different size of
simulation box. The Mach
number of the flow in the simulation frame is $M_{A0} = 4.0$ and the averaged Mach
number in the shock frame is about $5.6$. Most of the results presented here
are for the averaged shock normal angle $\langle \theta_{Bn} \rangle = 90^\circ$, but we also
simulate the cases for $\langle \theta_{Bn} \rangle = 60^\circ$ and $75^\circ$ to examine the
dependence of the acceleration efficiency on shock normal angle. The other
important simulation parameters include electron and ion plasma beta $\beta_e =
0.5$ and $\beta_i = 0.5$, respectively, grid sizes $\Delta x = \Delta z = 0.5
c/\omega_{pi}$, time step $\Delta t = 0.01 \Omega_{ci}^{-1}$,
the ratio between light speed and upstream Alfven speed
$c/v_{A1} = 8696.0$, and the anomalous resistivity is taken to be $\eta = 1
\times 10^{-5} 4\pi\omega_{pi}^{-1}$. The initial spatially uniform thermal ion
distribution was generated using $40$ particles per cell. 
Under these parameters, the upstream Alfven speed is about $34.5$ km/s and
the shock speed is about $193$ km/s in upstream frame.

We assume non-relativistic motion that is reasonable because
the highest energy electrons obtained in our study are still non-relativistic.
We release a shell
distribution of electrons with energy of $100$ eV, which
corresponds to an electron velocity $V_e = 30.7 U_1 = 5.7 v_{the}$ in the
upstream frame, where $U_1$ is upstream bulk velocity in the shock frame and
$v_{the}$ is the thermal velocity of fluid electrons considered in the hybrid
simulations. This energy is typical for the halo component of
electron velocity distributions observed in the solar wind.

\subsection{Simulation Results for the Acceleration of Electrons}

\begin{figure}
\begin{center}
\includegraphics[width=1.0\textwidth]{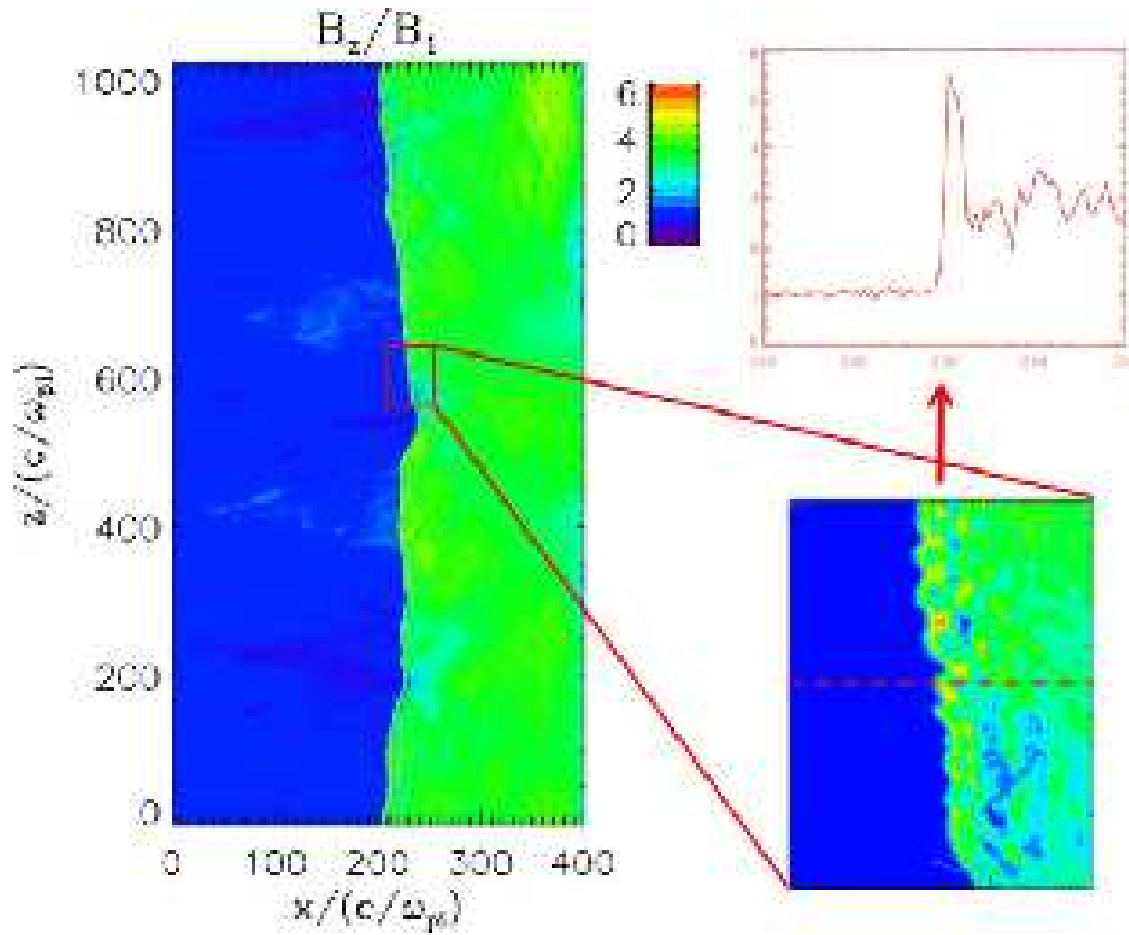}
\caption[A snapshot of the magnetic field in the $z$-direction $B_z/B_1$ represented in color-coded scale]{A snapshot of the magnetic field in the $z$-direction $B_z/B_1$ represented in color-coded scale at $t = 110\Omega_{ci}^{-1}$, where $B_1$ is the averaged upstream magnetic field strength. A region at the shock front is zoomed in on the right bottom and a profile is illustrated on the right upper panel. The shock surface is shown to be rippled and irregular in different scales. \label{electron-mag1}}
\end{center}
\end{figure}

Figure \ref{electron-mag1} shows a snapshot of the $z$ component of the magnetic field,
$B_z/B_1$, at time $110 \Omega^{-1}_{ci}$ in a color-coded contour.  At this
time, the shock is fully developed. In this case, the angle between the average
magnetic field direction and shock normal, $\langle \theta_{Bn} \rangle$ is $90^\circ$. The
position of the shock front is clearly seen from the boundary of the magnetic
field jump. The shock is moving in the $-x$ direction at a
speed dependent on $z$, which is about 1.6 $v_{A1}$ on average. On the right bottom a small 
region of the simulation domain is zoomed in, which shows small-scale irregularities at shock 
front. Because of the effect of large-scale turbulence with the shock, the shock surface becomes
irregular on a variety of spatial scales from small-scale ripples, which could
be due to ion-scale plasma instabilities \citep{Lowe2003}, to large-scale
structure caused by the interaction between the shock and upstream turbulence
\citep{Neugebauer2005,Giacalone2008,Lu2009}. Locally, the structure of the shock, shown in top 
right panel,
is still clearly a quasi-perpendicular shock. The upstream magnetic field
is compressed and distorted as it passes through the shock into the downstream
region. We note that the rippling of the shock and varying upstream magnetic
field leads to a varying local shock normal angle along the shock front. As we
will discuss later, the irregular shock surface and magnetic field geometry
will efficiently accelerate electrons and produce a number of features similar
to observations, such as the electron foreshock and spike-like intensity
increases at the shock front. The meandering of field lines close to the shock
surface helps to trap the electrons at the shock, leading to efficient
acceleration. The shock ripples also contribute to the acceleration by
mirroring electrons between them.

\begin{figure}
\begin{center}
\includegraphics[width=1.0\textwidth]{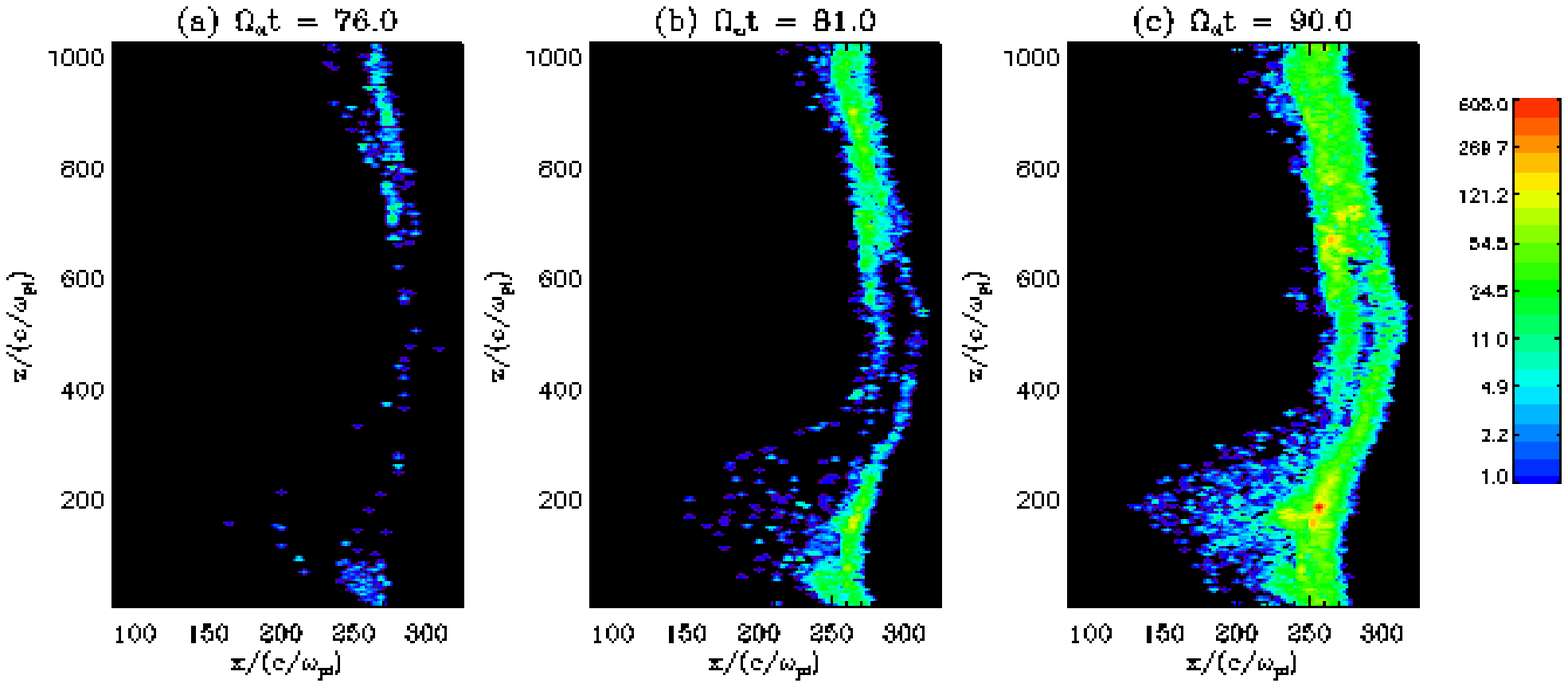}
\caption[The number of energetic electrons with energies $E > 10 E_0$]{The number of energetic electrons with energies $E > 10 E_0$, the initial
release energy $E_0 = 100$ eV, at (a) $\Omega_{ci} t = 76$, (b) $\Omega_{ci} t = 81$,
and (c) $\Omega_{ci} t = 90$, respectively. Initial electrons are released uniformly upstream 
at $\Omega_{ci} = 70$. \label{electron-contour}}
\end{center}
\end{figure}

Figure \ref{electron-contour} shows a color-coded representation of the number of energetic
electrons with energies higher than $10$ times (i.e., $1$ keV) the initial (at
release) energy at three different times (a) $76\Omega_{ci}^{-1}$, (b)
$81\Omega_{ci}^{-1}$, and (c) $90\Omega_{ci}^{-1}$. It is found that
after the initial release, a fraction of the electrons are reflected and
accelerated at the shock front, and then travel upstream along the turbulent
magnetic field lines. These accelerated electrons are then taken back to the
shock by the field line meandering, which provides even further acceleration.
The number of energetic electrons close to the shock surface is highly
irregular because the acceleration efficiency varies along the shock front
depending on the local shock normal angle \citep{Wu1984}. Most of the
electrons concentrate near the shock front since the global magnetic field
is mostly perpendicular to the shock normal. As the field lines convect through
the shock, the electrons eventually are taken downstream. Since the electrons are
tied to individual field lines in 2-D magnetic field, once the electrons are no
longer capable of crossing the shock, there will be no additional significant
acceleration. At this point, once all electrons are downstream, the energy
spectrum no longer changes with time.

\begin{figure}
\begin{center}
\includegraphics[width=1.0\textwidth]{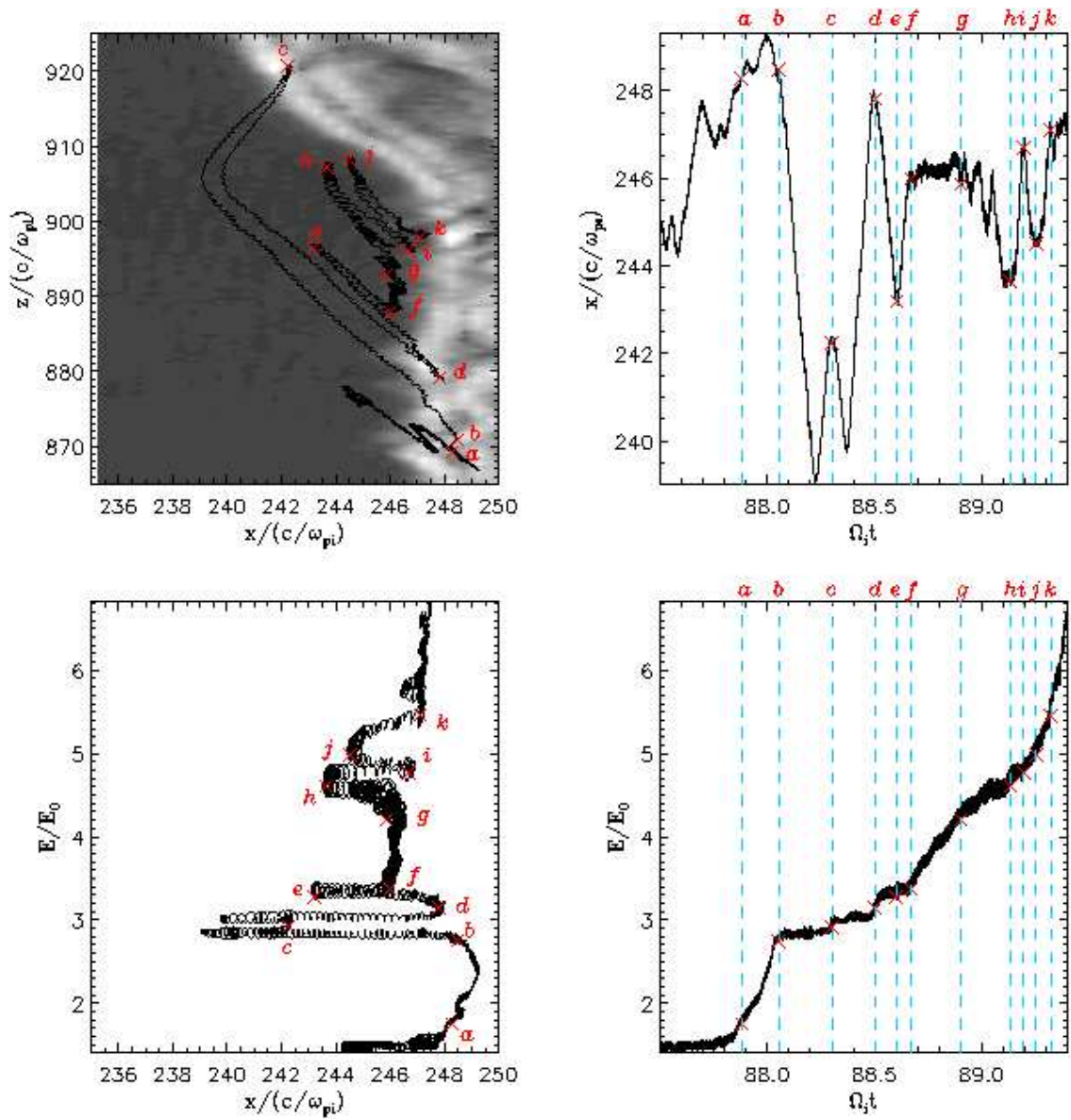}
\caption[Trajectory analysis for electron moving between ripples]{A typical electron trajectory analysis that shows acceleration by multiple mirroring between ripples.
 The top left panel displays the trajectory of the representative electron in $x$-$z$ plane, overlapped with contour
 of $B_z$ magnetic field where the gray-scale is the same as that in Figure 1; The top right panel shows the
 position of the electron in $x$ coordinate as a function of time; The bottom left panel illustrates the energy of
 the representative electron $E/E_0$ as a function of $x$; The bottom right panel shows the dependence of electron energy
$E/E_0$ on time. \label{electron-trajectory}}
\end{center}
\end{figure}

Examination of the trajectories of some electrons shows that the rippling of
the shock front also contributes to the acceleration by mirroring electrons
between the ripples, as illustrated in Figure \ref{electron-trajectory}. 
In this figure, the top left
plot displays the trajectory of a representative electron in the $x$-$z$ plane,
overlapped with the 2-D gray-scale representation of $B_z$ at $\Omega_{ci}t =
89.0$, the gray scale is the same as in Figure \ref{electron-mag1}. The upper right plot shows
the position of this electron (in $x$) as a function of time. The electron
bounces back and forth between the ripples for several times. For example, the
reflections are labeled $a$-$b$, $c$, $d$, $f$-$g$, $h$, and $j$. The energy change
as a function of position, $x$, corresponding to these reflections is shown in
the bottom left panel. We find that there are jumps in energy at each of the
reflections. The panel on the bottom right shows the electron energy as a
function of time that also illustrates the features of multiple accelerations
related to multiple reflections. The trajectory analysis shows that the electron
will be mirrored between the ripples and get
accelerated multiple times. Note that the shock does not move
much during the time scale of this trajectory.

\begin{figure}
\begin{center}
\includegraphics[width=0.8\textwidth]{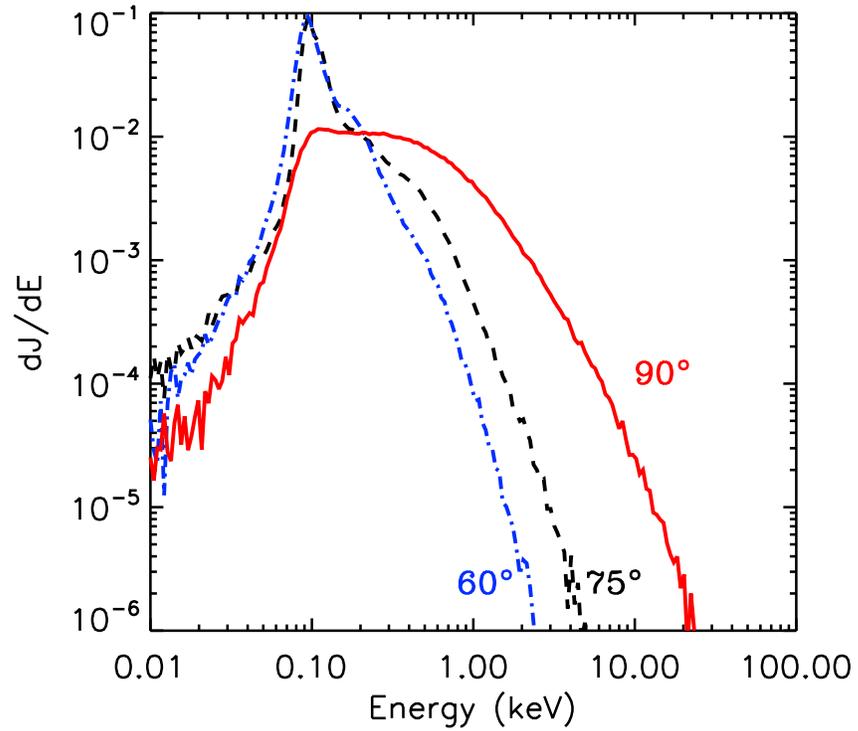}
\caption[Energy spectra of energetic electrons for different shock angles]{The energy flux spectrum of electrons at $\Omega_{ci}t=120$ for different averaged shock normal angles. The red solid line is for the case with shock angle $\langle \theta_{Bn} \rangle=90^\circ$, the blue dot dashed line and the black dashed line are for the cases with $\langle \theta_{Bn} \rangle = 60^\circ $ and $ 75^\circ $, respectively. \label{spectrum-electron-angle}}
\end{center}
\end{figure}

We now consider the effect of varying the angle between the mean magnetic field
and shock-normal. Shown in Figure \ref{spectrum-electron-angle} are the resulting energy spectra for
three different mean shock-normal angles ($\langle \theta_{Bn} \rangle= 60^\circ, 75^\circ$,
and $90^\circ$, respectively) at the end of simulations ($\Omega_{ci}t=120.0$).
It is found that for $\langle \theta_{Bn} \rangle = 90^\circ$, the electrons can be readily
accelerated to up to $200-300$ times the initial energy within $50
\Omega_{ci}^{-1}$. The spectrum is flat between about $0.1$ keV to $0.7$ keV.
This shape is similar to the ``plateau" structure discussed by
\citet{Burgess2006}. Above $1$ keV, the spectrum falls off with energy with a
slope index about $-3$. It can be found that both the number fraction and
highest energy of accelerated particles decrease as $\langle \theta_{Bn} \rangle$ decreases.
We have also tried different values of initial energies (not
shown), and find that the acceleration efficiency decreases for electrons with
higher initial energies, which is similar to the results of
\citep{Burgess2006}.

\begin{figure}
\begin{center}
\includegraphics[width=0.8\textwidth]{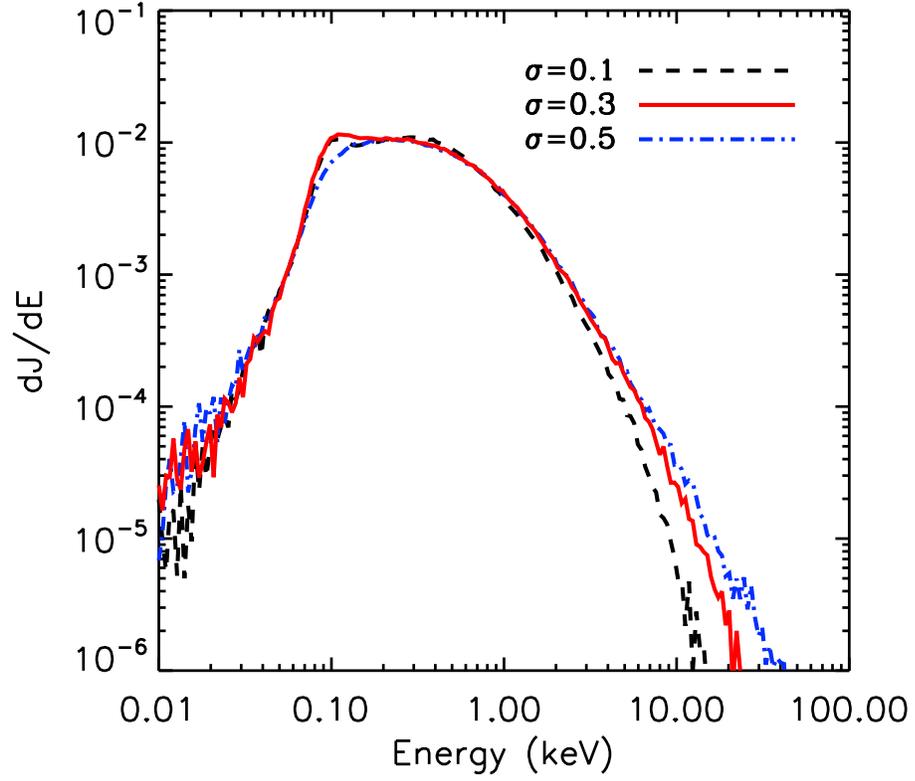}
\caption[Energy spectra of energetic electrons for an averaged perpendicular shock with different turbulence variances]{The energy flux spectra of electrons at $\Omega_{ci}t=120$ for an averaged perpendicular shock with different turbulence variances. The black dashed line, red solid line, and blue dot dashed line are in the cases that $\sigma^2 = 0.1$, $0.3$, and $0.5$, respectively. \label{spectrum-electron-amplitude}}
\end{center}
\end{figure}

The effect of varying level of magnetic turbulence variance is
examined in Figure \ref{spectrum-electron-amplitude}. We compare three cases with different turbulence
variances $\sigma^2 = 0.1$, $0.3$, and $0.5$, respectively. At the end of
simulations, the final energy spectra are similar at low energies, with
significant variations in the spectra only at energies higher than $2$ keV. It
is found that the energy spectrum is hardened at high energies when the
turbulence variance is largest, which indicates that the large-scale turbulence is
more important for accelerating electrons to high energies. We argue that
collisionless shocks that move through magnetic turbulence with significant
power leads to efficient electron acceleration to high energies since the
motion normal to the shock front is enhanced. The reason is that the meandering
of field lines is enhanced, which allows the electrons to have a better chance to
travel though the shock multiple times.

\begin{figure}
\begin{center}
\includegraphics[width=0.8\textwidth]{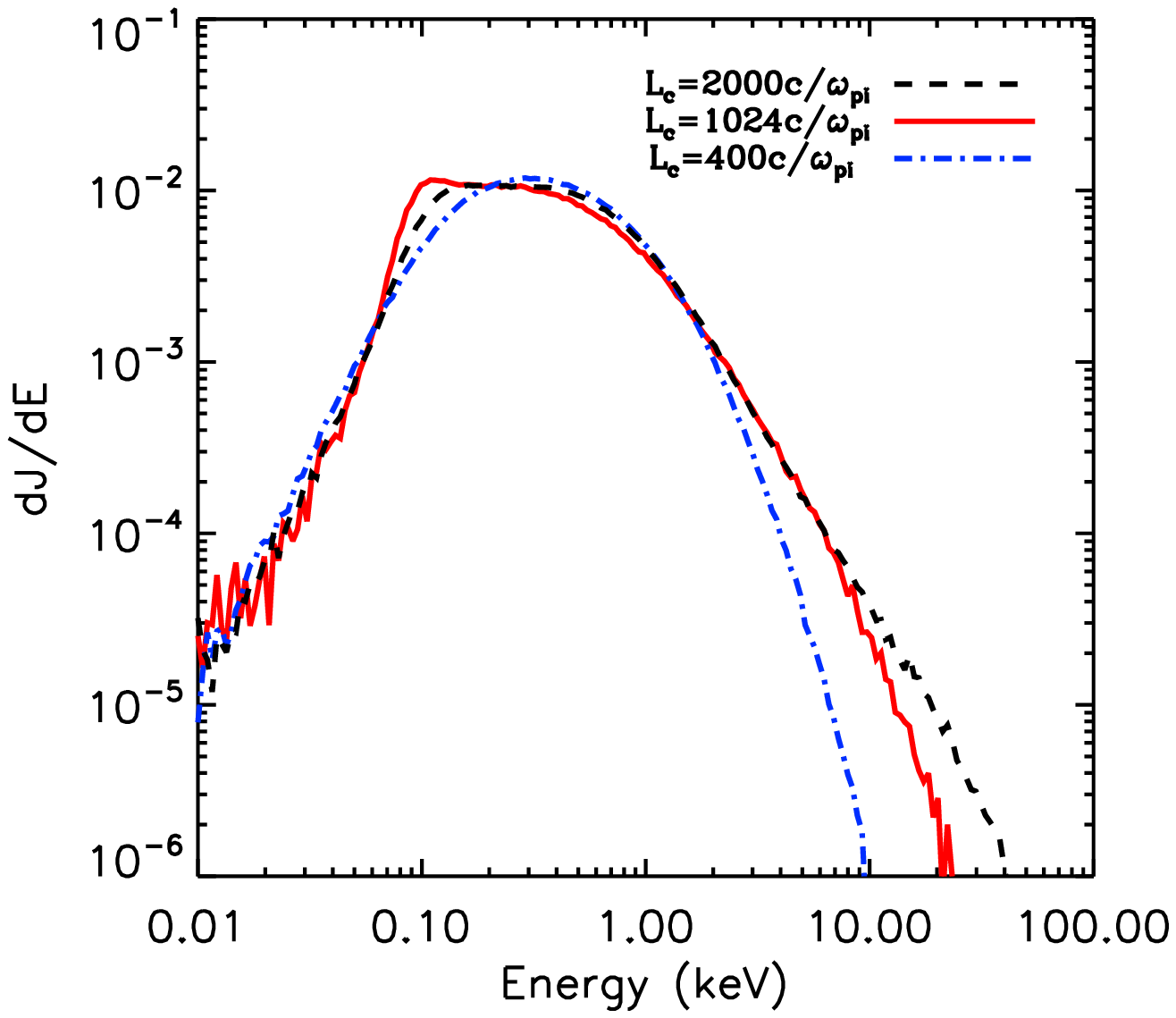}
\caption[Energy spectra of energetic electrons for different coherence lengths]{The energy flux spectrum of electrons at $\Omega_{pi}t=120$ for averaged perpendicular shock with different turbulence coherence lengths. The black dashed line, red solid line, and blue dot dashed line are in the cases that $L_c = 2000c/\omega_{pi}$, $1000 c/\omega_{pi}$, and $400c/\omega_{pi}$, respectively. \label{spectrum-electron-length}}
\end{center}
\end{figure}

We examine the effect of different turbulence correlation lengths in Figure \ref{spectrum-electron-length}. 
We compare three cases with different sizes of simulation box $L_z = 2000$, $1024$, and $400 c/\omega_{pi}$, respectively. In each case, the correlation length is made to be the same as the size of the simulation box in $z$-direction $L_c = L_z$.  It is shown that for the case $L_c = L_z = 2000 c/\omega_{pi}$, more electrons are accelerated to high energy. In the case of $L_c = L_z = 400 c/\omega_{pi}$, less electrons are accelerated and the highest energy is less than the other two cases. 
The more efficient acceleration for the case that $L_c$ is larger can also be understood as the motion of particle normal to the shock front is more enhanced. As derived by Jokipii and Parker (\citeyear{Jokipii1969}), large-scale magnetic fluctuation dominate the random walk of magnetic field lines and can be expressed as 

\begin{eqnarray}
\frac{\langle \Delta x^2 \rangle}{\Delta z} = \frac{P(k=0)}{B_0^2},
\end{eqnarray}

\begin{figure}
\begin{center}
\includegraphics[width=0.8\textwidth]{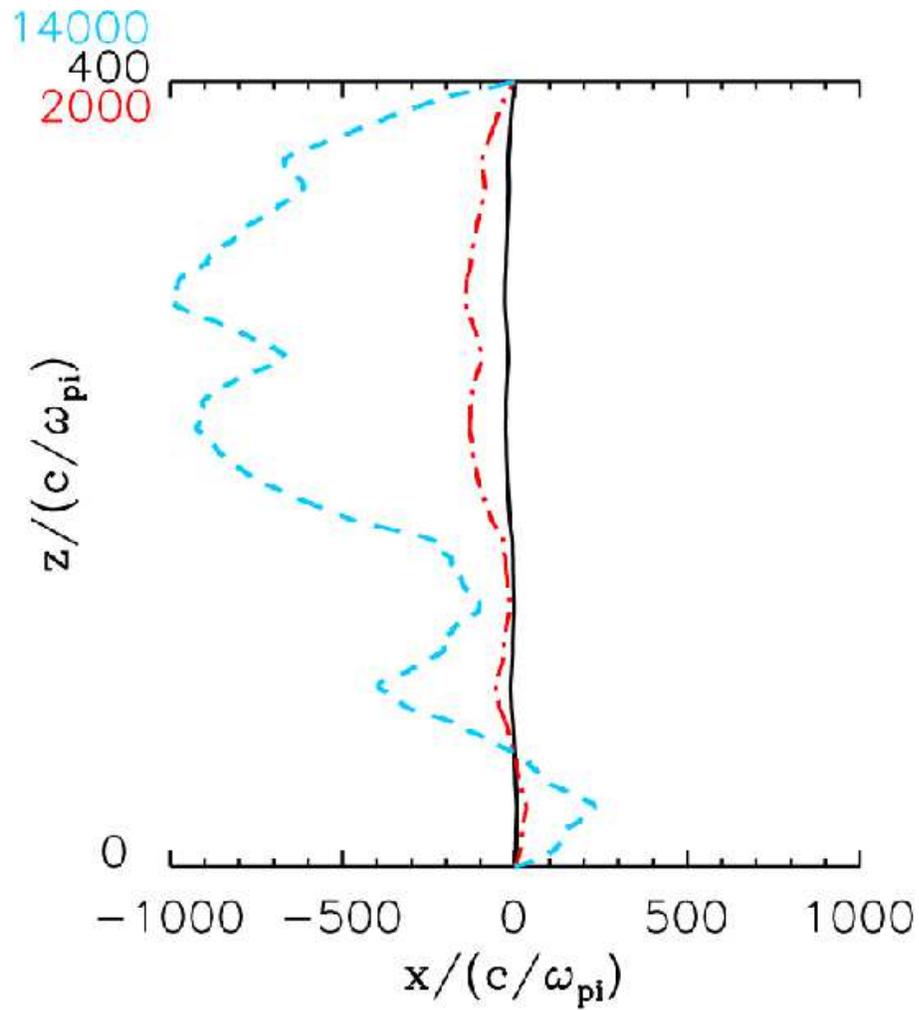}
\caption[Three turbulent magnetic field lines for different correlation lengths.]{Three turbulent magnetic field lines for different correlation lengths. The blue dashed line, red dot dashed line, and black solid line are in the cases that $L_c = 14000c/\omega_{pi}$, $2000c/\omega_{pi}$, and $400c/\omega_{pi}$, respectively. \label{fieldline}}
\end{center}
\end{figure}

\noindent where $\Delta x$ and $\Delta z$ represent the displacement of a field line in the direction transverse and along the mean magnetic field, respectively. $P(k=0)$ represents spectral power at wave number $k = 0$. Although we use a restricted simulation box, our simulation is qualitatively consistent with this picture. In figure \ref{fieldline} we plot three magnetic field lines generated from magnetic turbulence in the cases of three different coherence lengths. In this plot the horizontal axis $x$ represents the spatial distance transverse to mean magnetic field with a plot range from $-1000 c/\omega_{pi}$ to $1000 c/\omega_{pi}$, and the vertical axis $z$ represents the spatial distance along the mean magnetic field. In order to show the difference between these field lines, the spatial range for $z$ axis is changing for different field lines. In the three cases of $L_c = 14000c/\omega_{pi}$ (blue dashed line), $L_c = 2000c/\omega_{pi}$ (red dotted line) and $L_c = 400c/\omega_{pi}$ (black solid line), the spatial ranges in $z$ direction are made to be $0-14000 c/\omega_{pi}$, $0-2000 c/\omega_{pi}$ and $0-400 c/\omega_{pi}$, respectively. One can see that the $x$-direction displacement of the field line in the case of $L_c = 14000c/\omega_{pi}$ is much more enhanced compare to the case of $L_c = 400c/\omega_{pi}$. Because the limited computational resource, the largest simulation is made to be $L_c = L_z = 2000 c/\omega_{pi}$, whereas for typical interplanetary parameters, this value is about $14000 c/\omega_{pi}$ ($0.01$ AU).

\begin{figure}
\begin{center}
\includegraphics[width=0.8\textwidth]{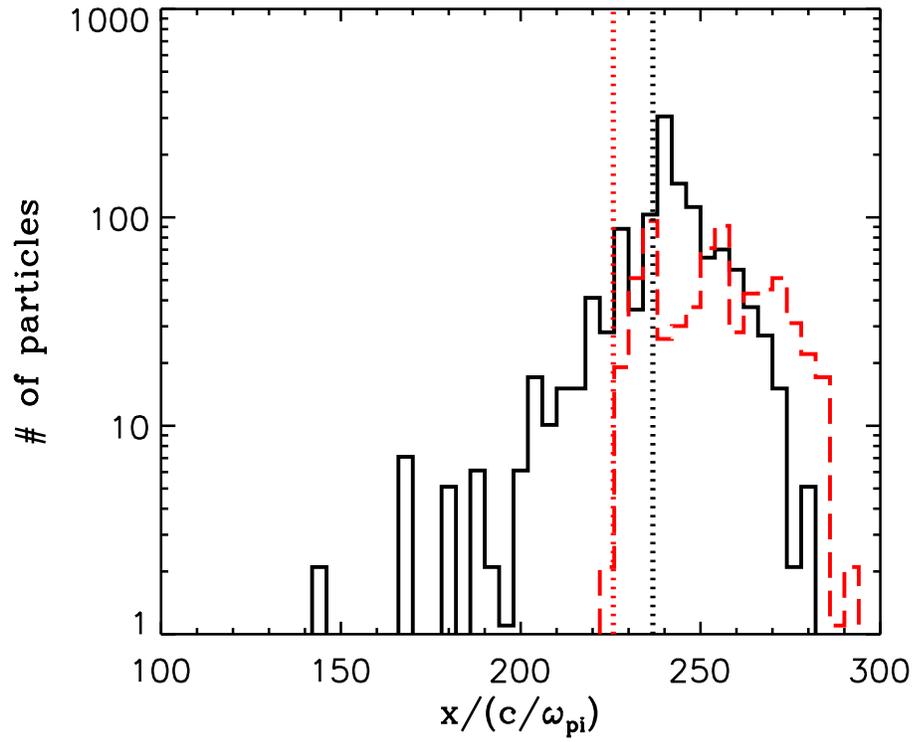}
\caption[The profiles of the number of energetic electrons across the shock]{The profiles of the number of energetic electrons across the shock at $z=200c/\omega_{pi}$ and $z=800c/\omega_{pi}$ and time $\Omega_{ci}t=100.0$, respectively. The red dot line and the black dot line label the correspoding positions of the shock fronts. \label{ch4-electron-profile}}
\end{center}
\end{figure}

The idea of field line random walk (and its related large-scale irregular shock surface) is also useful for interpreting the observation of energetic particles associated with shock waves. An example is shown in Figure \ref{ch4-electron-profile}, which shows the
profiles of the number of energetic electrons at $\Omega_{ci}t=100.0$ as a
function of $x$, for the case of $\langle \theta_{Bn} \rangle=90^\circ$. The black solid line
is the profile at $z = 200 c/\omega_{pi}$, and the red dashed line shows the
profile at $z = 800 c/\omega_{pi}$. The corresponding position of the shock
front at each of these values of $z$ are represented using dot lines. At $ z =
200 c/\omega_{pi}$, it is observed that the energetic electrons travel far
upstream up to about $100 c/\omega_{pi}$ from the shock. However, the profile at $z = 800
c/\omega_{pi}$ shows no significant upstream energetic electron flux. The
upstream energetic electron profiles show irregular features similar to \textit{in-situ}
observations reported by \citet{Simnett2005} (Figure 10). The irregular
features are controlled by the global topology of the large-scale turbulent
magnetic field lines, along which the accelerated electrons could travel far
upstream. Additionally, energetic electron profiles in $x$ direction generally
show ``spike-like" structure close to the shock front, which is usually observed
in interplanetary shocks and Earth's bow shock. This
feature is relatively stable within the simulation time once the upstream
electron structure is developed.

Since the temporal and spatial scales in our hybrid simulation is not large enough compared with the realistic scales of the magnetic turbulence in interplanetary space, we have to estimate how long the upstream electrons can be observed before the shock encounter in realistic parameters. We have carried out an order of magnitude calculation by assuming a planar perpendicular shock whose surface connects to meandering field lines of force at various places. The magnetic turbulence is assumed to be the two-component model \citep[][]{Matthaeus1990}, discussed in Section \ref{chapter2-model}. We take the value of coherence length in solar wind turbulence $L_c = 0.01$ AU and the total variance of turbulence is $\delta B^2/B_0^2 = 0.3$. One can estimate the diffusion of magnetic field lines transverse to the average direction of magnetic field $D_\perp = \Delta x^2/\Delta z$. We take the value from numerical simulation that $D_\perp = 0.14L_c$ \citep{Giacalone1999}. Then we can derive that the field line wandering along the shock normal is $\Delta x \sim 0.4L_c$. Using a shock speed of $800 $ km/s, we estimate that the electron foreshock region for a perpendicular shock propagating into a turbulent upstream is about $10$ minutes. This rough order of magnitude estimate agrees with the \textit{in-situ} observation by \citet{Simnett2005}. The upstream energetic electrons have been also observed by Voyager spacecraft, in a much lager scale \citep{Decker2005,Decker2008}. Voyager II also observed the change of energetic electron intensity, presumbly due to large-scale magnetic field change.

\begin{figure}
\begin{center}
\includegraphics[width=0.8\textwidth]{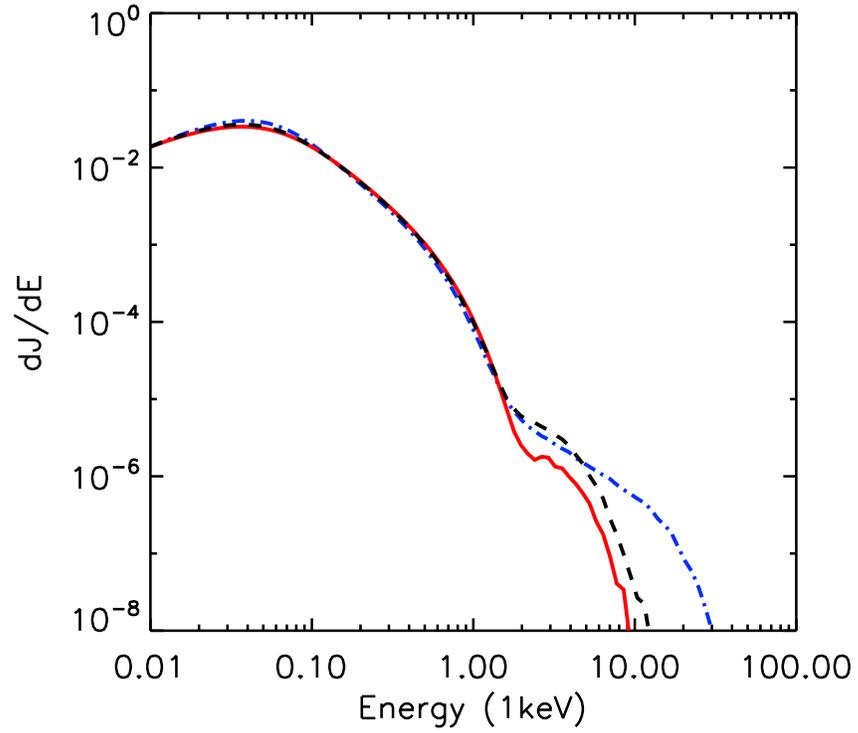}
\caption[Energy spectra of downstream protons for different shock angles]{The energy flux spectra of protons at $\Omega_{ci}t=120$ for different averaged shock normal angle. The red solid line is in the case that the shock angle $\langle \theta_{Bn} \rangle=90^\circ$, the blue dot dashed line and the black dashed line are in the cases that $\langle \theta_{Bn} \rangle = 60^\circ $ and $ 75^\circ $, respectively. \label{spectrum-ion-angle}}
\end{center}
\end{figure}

\begin{figure}
\begin{center}
\includegraphics[width=0.8\textwidth]{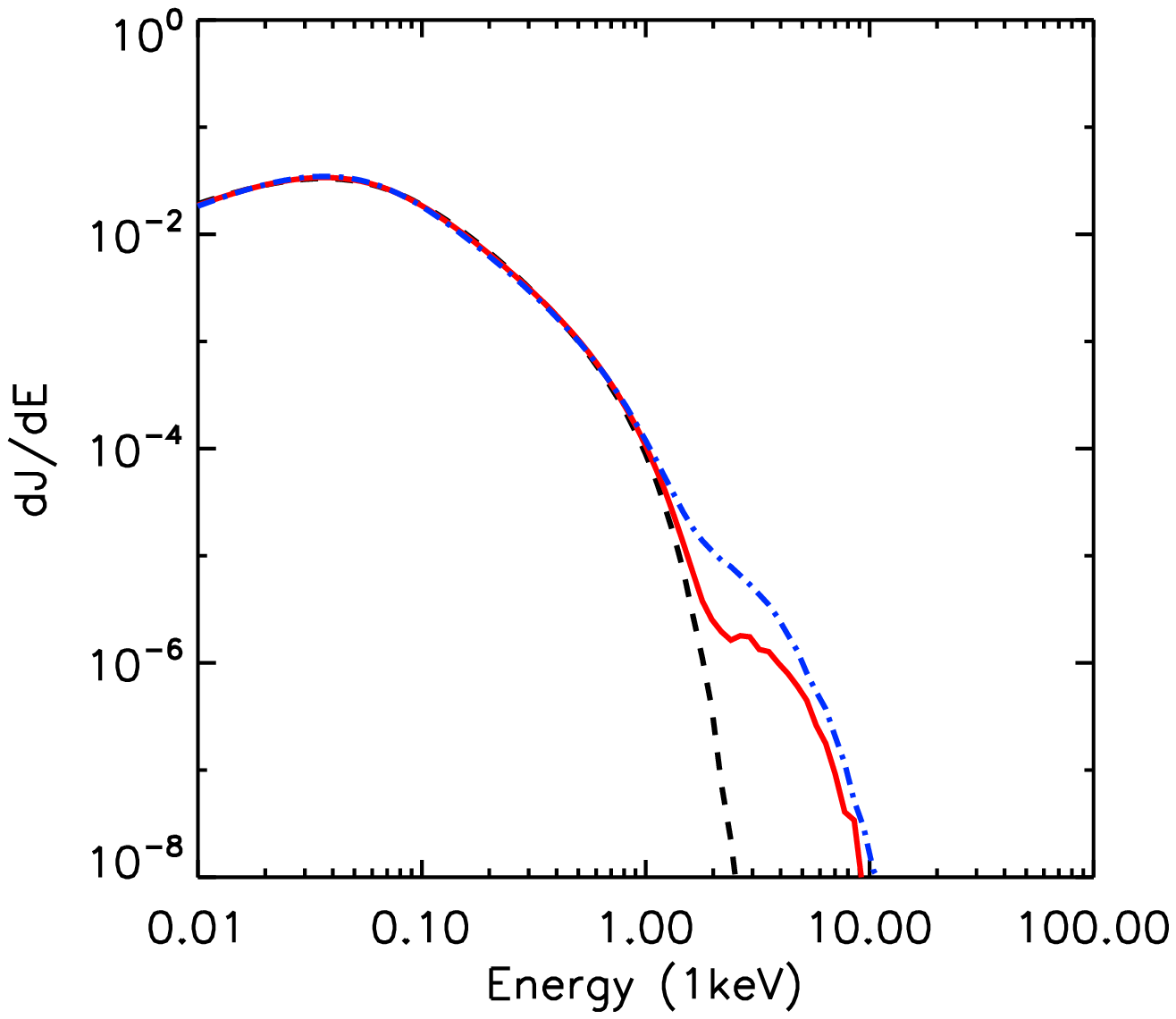}
\caption[Energy spectra of downstream protons for an averaged perpendicular shock with different turbulence variances.]{The energy flux spectra of protons at $\Omega_{ci}t=120$ for an averaged perpendicular shock with different turbulence variances. The black dashed line, red solid line, and blue dot dashed line are in the cases that $\sigma^2 = 0.1$, $0.3$, and $0.5$, respectively. \label{spectrum-ion-var}}
\end{center}
\end{figure}

\begin{figure}
\begin{center}
\includegraphics[width=0.8\textwidth]{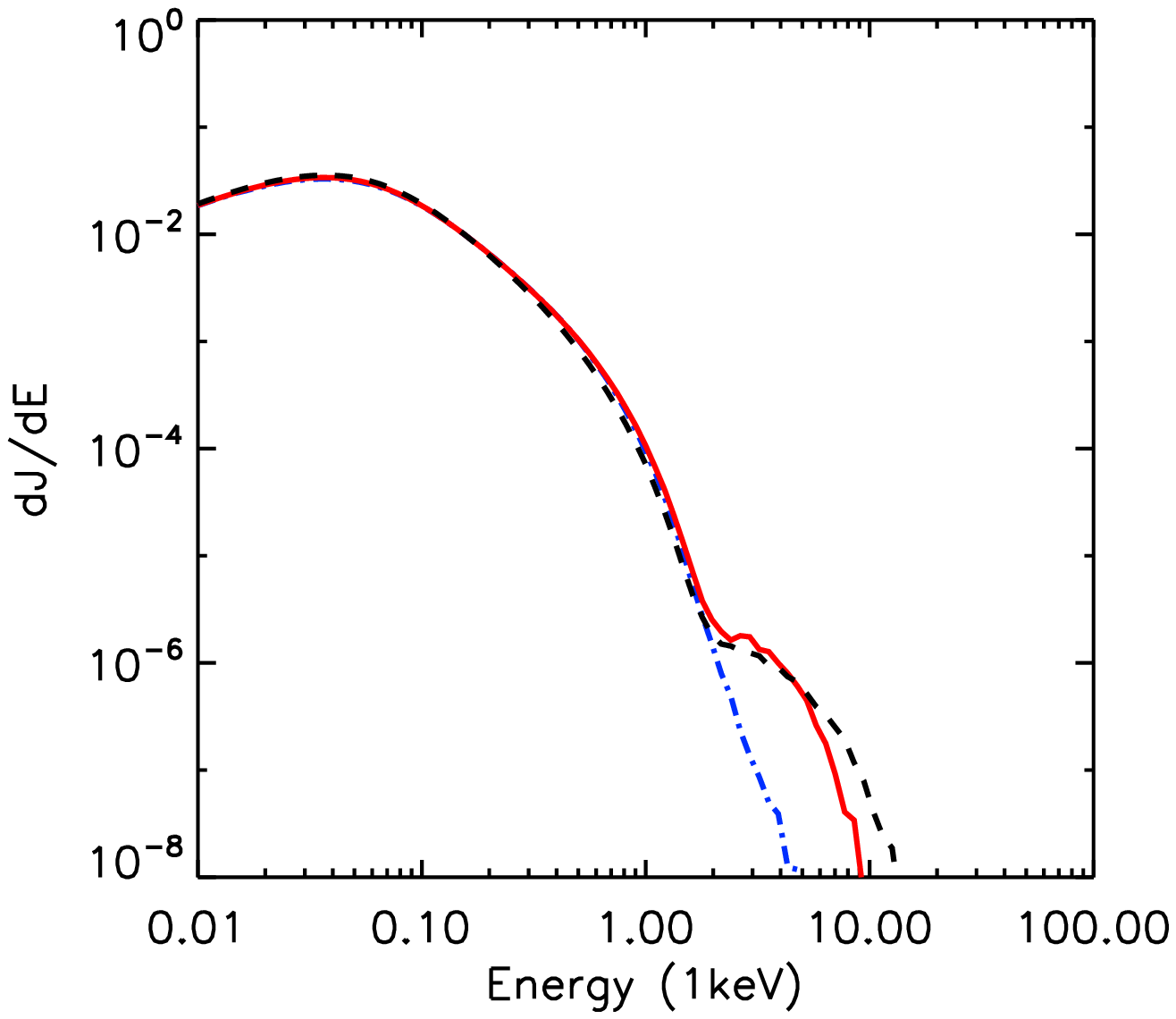}
\caption[Energy spectra of downstream protons for an averaged perpendicular shock with different turbulence correlation lengths]{The energy flux spectrum of protons at $\Omega_{pi}t=120$ for averaged perpendicular shock with different turbulence correlation lengths. The black dashed line, red solid line, and blue dot dashed line are in the cases that $L_c = 2000c/\omega_{pi}$, $1000 c/\omega_{pi}$, and $400c/\omega_{pi}$, respectively. \label{spectrum-ion-length}}
\end{center}
\end{figure}

\subsection{Comparison Between the Acceleration of Electrons and That of Ions and its Implication to SEP Events}

The correlation between electrons and ions in SEP events has been reported and discussed by a number of authors (see Section \ref{electron-observation}). The correlation indicates that electrons and ions are accelerated close to the Sun by similar processes. We have shown that large-scale turbulence has important effects on accelerating electrons to high energy. The result, along with the previous work on the acceleration of ions \citep{Giacalone2005a,Giacalone2005b}, suggests that perpendicular shocks may play an important role in the acceleration of both electron and ions at shocks, which is consistent with the correlation between ions and electrons in solar energetic particle events \citep[e.g.,][]{Cliver2009}. 

Here we explicitly compare the energy spectra of electrons with that of protons. In Figure \ref{spectrum-ion-angle}, \ref{spectrum-ion-var}, and \ref{spectrum-ion-length}, we show the energy spectra of downstream protons from the hybrid simulations, corresponding to the spectra of electrons in Figure \ref{spectrum-electron-angle}, \ref{spectrum-electron-amplitude}, and \ref{spectrum-electron-length}. We find that more efficient acceleration for protons can be obtained in the case of larger values of turbulence variances and correlation lengths. These agree well with the characteristics of the acceleration of electrons. In our simulations, the accelerated protons at the oblique shock with $\langle \theta_{Bn} \rangle= 60^\circ$ are found to reach higher energies, which is different from previous works \citep{Jokipii1982,Jokipii1987,Giacalone2005a,Giacalone2005b}. This is probably due to the limited temporal and spatial scales of our simulations. As shown by \citet{Giacalone2005b} using test-particle simulations, the energy spectra of protons reach the highest energy in perpendicular shock case in a longer time scale $\Omega_{ci}t \sim 50000$ (this corresponds to $5-10$ minutes for typical parameters in solar corona). However the current results from hybrid simulations do show a population of thermal protons can be accelerated to high energies in perpendicular shocks, which supports the idea that both electrons and protons can be efficiently accelerated by shocks with large shock normal angles.

\subsection{Summary}

We have presented the results of the acceleration of electrons (and also protons) at a perpendicular shock that propagates through a turbulent magnetic field. The acceleration of electrons are enhanced due to the effect of large-scale turbulence.
The accompanying results for protons qualitatively show the correlation between accelerated electrons and accelerated ions in oblique shocks with large shock normal angles. 
This indicates that quasi-perpendicular/perpendicular shocks play an important role in SEP events.  

This study will help to explain the correlation between electrons and ions in solar energetic particles from both CME-driven shocks \citep{Cliver2009} and solar flares \citep{Shih2009}. The result also poses a question on the contribution of parallel shocks in SEP events. Since it is difficult for parallel shocks to accelerate electrons, most of SEPs are probably originated from perpendicular shocks. The acceleration of electrons has to be included in current scenario of SEP events \citep[e.g.,][]{Tylka2005}.

\section{The Effect of Large-Scale Magnetic Turbulence on the Acceleration of Electrons: Flare Termination Shocks \label{electron-flare}}

In this section we use the combination of hybrid simulations and test-particle electron simulation (see Section \ref{chapter4-method}) to study the acceleration of electrons at flare termination shocks predicted to exist in the vicinity of solar flares. Most of this section has been published in the Astrophysical Journal \citep{Guo2012b}.

The existence of standing termination shocks has been examined by flare models and numerical simulations \citep[e.g.,][]{Shibata1995,Forbes1986}. Solar flares are observed to be strong sources of energetic charged particles \citep{Aschwanden2002}. The release of magnetic energy by magnetic reconnection is thought to be the driving process \citep{Masuda1994}. While several mechanisms have been proposed to explain the acceleration of charged particles in flares \citep[see review by][and references therein]{Miller1997,Zharkova2011}, there is still no general consensus and this remains an unsolved problem. Recent hard X-ray observations of the non-thermal electron bremsstrahlung emission by \emph{Reuven Ramaty High Energy Solar Spectroscopic Imager} \citep[\emph{RHESSI; }][]{Lin2002} have provided more details of electron acceleration in solar flares. The observations indicate that a large fraction of released energy resides in high-energy electrons during a short amount of time. Hard X-ray sources above the top of magnetic loops have been detected \citep[e.g.,][]{Masuda1994,Krucker2010}, providing important clues to the acceleration process. For example, the loop-top source recently reported by \citet{Krucker2010} shows that a large number of electrons ($> 5\times 10^{35}$) are accelerated to more than 16 keV and the highest energy reaches $\sim$ MeV. Since the observed hard X-ray source requires very efficient acceleration, explaining how such a large number of electrons (probably also ions) are accelerated to high energy poses a challenge to theoretical astrophysics.

\begin{figure}
 \begin{center}
\includegraphics[width=0.8\textwidth]{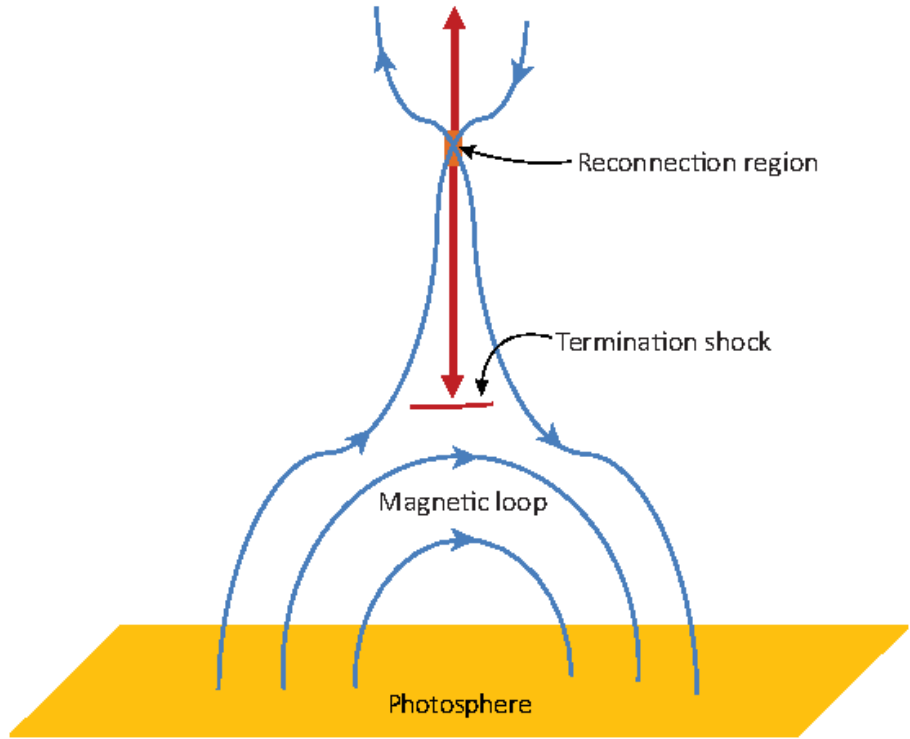}
 \caption{A cartoon illustration of the geometry of flare termination shocks.}
 \label{cartoon}
 \end{center}
 \end{figure}

The existence of fast shocks in the reconnection outflow region has been predicted in flare models \citep{Shibata1995} and numerical simulations \citep[e.g.,][]{Forbes1986,Forbes1988,Shiota2003,Workman2011}. Using MHD numerical simulations, \citet{Forbes1986} studied the formation of a standing termination shock when a high-speed jet driven by reconnection encounters a closed magnetic loop. The geometry of the flare termination shocks can be represented by Figure \ref{cartoon}. The high-speed jet created in the reconnection out-flow region collides with the top of the magnetic loop and produces a fast-mode, standing termination shock. The resulting flare termination shock has a unit normal to its surface that points nearly perpendicular to the magnetic field. This is a perpendicular shock (i.e., the angle between the upstream magnetic field and shock normal vector $\theta_{Bn} = 90^\circ$). \citet{Forbes1986} predicts the existence of this shock with a compression ratio of 2.0 and an upstream Mach number as high as 2.3. A recent study by \citet{Workman2011} shows similar results. The observational evidence of the existence of flare shocks has been presented by \citet{Aurass2002}.

Particle acceleration at flare termination shocks has been considered by a number of authors. It is usually thought that the injection of electrons is a problem at perpendicular shocks. We addressed this in previous sections. There has been some works on facilitating particle acceleration at flare termination shocks. \citet{Tsuneta1998} considered electron heating by a slow-shock pair as a pre-energization process. \citet{Somov1997} considered the role of plasma heating and collapsing magnetic trap at reconnecting magnetic field lines.
In this paper we present results from a combination of hybrid simulations and test-particle electron simulations to study the electron energization at a flare termination shock in the existence of upstream magnetic fluctuations. We consider the nonlinear modification of a planar shock front by upstream Alfvenic fluctuations and its effect on electron acceleration. Although the plasma waves and turbulence in the reconnection outflow region could be considerably different from our simplified model, some intrinsic characteristics of this interaction, such as the braiding of magnetic field lines and shock rippling, should still be preserved. We show that after considering a fluctuating upstream region, the electron acceleration in flare termination shock is a rapid and efficient process. A large fraction of the initial thermal electrons is accelerated to hundreds of keV and even reaches MeV energies in a very short time. This indicates that collisionless shocks may play an important role in particle acceleration in solar flares. In Section \ref{electron-flare-init} we describe the initial conditions and parameters used in this paper. Section \ref{electron-flare-results} discusses the simulation results. 

\subsection{Initial Conditions and Parameters \label{electron-flare-init}}

The size of the simulation domain $L_x \times L_z$ for each case is listed in Table \ref{table}.  The flare termination shock is modeled by injecting plasma continuously from one end $(x=0)$ of the simulation box and colliding the reflecting boundary at the other end $(x=L_x)$. The total variance of magnetic turbulence in each case is listed in Table 1. In all the cases, we take $L_c = L_z$. We expect large-scale magnetic turbulence to exist in reconnection outflow plasma. These fluctuations can be triggered by reconnection, foot-point motion or other processes (see \citeauthor{Krucker2008}, \citeyear{Krucker2008}, for a detailed discussion).  Although this simplified form of magnetic-field fluctuations may not be realistic, any turbulence with large variances should allow strong field-line wandering, which is essential in our particle acceleration model.

In order to produce a low Mach number shock as predicted by other numerical simulations, the inflow Alfven Mach number is taken to be $M_{A0} = 1.0$. After reflection at the right boundary, this produces a shock with averaged Mach number of about $2.0$ in the shock frame, consistent with the flare termination shock predicted by previous MHD simulations \citep{Forbes1986,Workman2011}. The grid sizes are $\Delta x = \Delta z = 0.5
c/\omega_{pi}$ and the time step is taken to be $\Delta t = 0.01 \Omega_{ci}^{-1}$, where $c/\omega_{pi}$ is the ion inertial length and $\Omega_{ci}^{-1}$ is the ion gyroperiod. The plasma beta $\beta_i$ and $\beta_e$ are taken to be $0.03$ and the ratio between light speed and upstream Alfven speed $c/V_{A0} = 410$, which roughly corresponds to an initial situation with temperature $2\times 10^6$ K, number density $8\times 10^9$ cm$^{-3}$ and magnetic field $B_0 = 30$ G, similar to constraints from observations \citep{Krucker2010}. Under these parameters, the average shock speed in the shock frame (also the jet outflow speed) is about $1460$ km/s. This speed is measurable because the outflow plasma from reconnection is moving at the Alfven speed. The estimate from observations \citep{Tsuneta1997} is roughly consistent with this value.

In the second step, we integrate the relativistic equations of motion for an ensemble of test-particle electrons in the two-dimensional time-dependent electric and magnetic fields obtained in the hybrid simulations. These test-particle electrons are treated as a different part from the electron fluid in the hybrid simulations. We use a second-order spatial interpolation and linear temporal interpolation to get the field at the particle position. Initially we release a Maxwellian distribution with $T_e = 2.0 \times 10^6$ $K$ in the upstream frame. The test-particle electrons are released upstream at $\Omega_{ci} t = 30$ after the shock has fully formed and far from the boundaries. The simulation domain in the $x$-direction is large enough so that no test-particle electrons escape from the system. Strictly speaking, this test-particle simulation is only valid when the influence of the accelerated electron to the background fluid is negligible. However, in the end of the simulation, the initial Maxwellian distribution of test-particle elections has been considerably changed due to the energization process at the shock front. This indicates that our approach may not be suitable in studying the long-term evolution of the termination shock. 

\begin{table}
\centering
\begin{tabular*}
{0.7\textwidth}{cccc}
\hline
Run& $L_x (c/\omega_{pi}) \times L_z (c/\omega_{pi})$ & $\delta B^2/B_0^2$ & $\Gamma\%$ (E $\geq$ 15 keV)  \\
\hline
1  & $500\times 400$ & 0.0  & 1.3\\
2  & $500\times 400$ & 0.03 & 4.5\\
3  & $500\times 400$ & 0.1  & 8.0\\
4  & $500\times 400$ & 0.3  & 9.8\\
5  & $500\times 800$ & 0.03 & 4.9\\
6  & $500\times 800$ & 0.1  & 8.9\\
7  & $500\times 800$ & 0.3  & 11.9\\
 \hline
\end{tabular*}
 \caption{Some parameters for different simulation runs. The size of the simulation domain, the variance of injected magnetic fluctuation, and the fraction of electrons whose energy is more than $15$ keV at the end of simulation.}
 \label{table}
\end{table}

\subsection{Simulation Results \label{electron-flare-results}}

In this work we discuss the electron acceleration in flare termination shocks, which have low Mach numbers and high shock speeds. The plasma in solar corona is in a high temperature and low plasma beta (strong magnetic field) regime, which is considerably different from that in interplanetary space. We focus on the modification of shock surface by upstream Alfvenic fluctuations and its effect on acceleration of electrons. Table \ref{table} lists some key parameters for all the simulation runs including the size of the simulation domain, the variance of the injected magnetic fluctuation, and the fraction of electrons whose energy is more than $15$ keV at the end of simulation. For runs 1-4 we consider the effect of different variances of magnetic turbulence. The turbulence variances are from $0.0$ to $0.3$ and the sizes of the simulation domain $L_x \times L_z = 500 c/\omega_{pi}\times 400 c/\omega_{pi}$ ($1.27$ km $\times$ $1.02$ km) for these four cases. For runs 5-7, the magnetic variances are the same as runs 2-4, but the size of the simulation box is changed to $L_x \times L_z = 500 c/\omega_{pi}\times 800 c/\omega_{pi}$ ($1.27$ km $\times$ $2.03$ km) to examine the effect of changing the coherence length. In the flare region, the strong large-scale Alfvenic magnetic fluctuation can be triggered by reconnection, and cascade to small scales. This process is usually assumed to be the source of magnetic turbulence required in many acceleration models \citep[e.g.,][]{Miller1996,Petrosian2004}. Since the size of our largest simulation domain is still much smaller than the observed hard X-ray emission region ($L \sim 10^3$ km), we do not consider the realistic geometry of flare termination shock but approximate it locally as a perpendicular shock that propagates into a plasma containing magnetic fluctuations.

\begin{figure}
\centering
\begin{tabular}{cc}
\epsfig{file=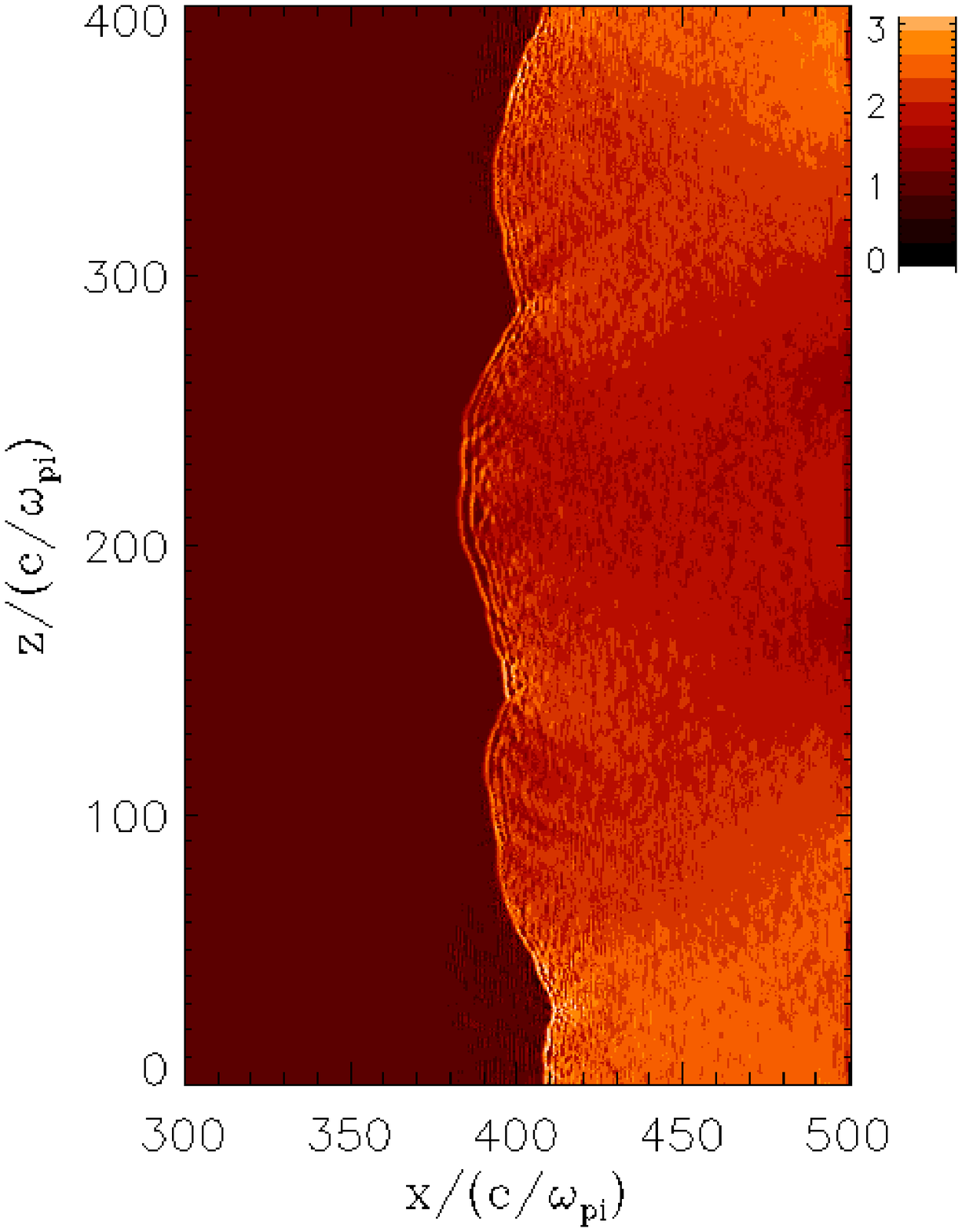,width=16pc,clip=} &
\epsfig{file=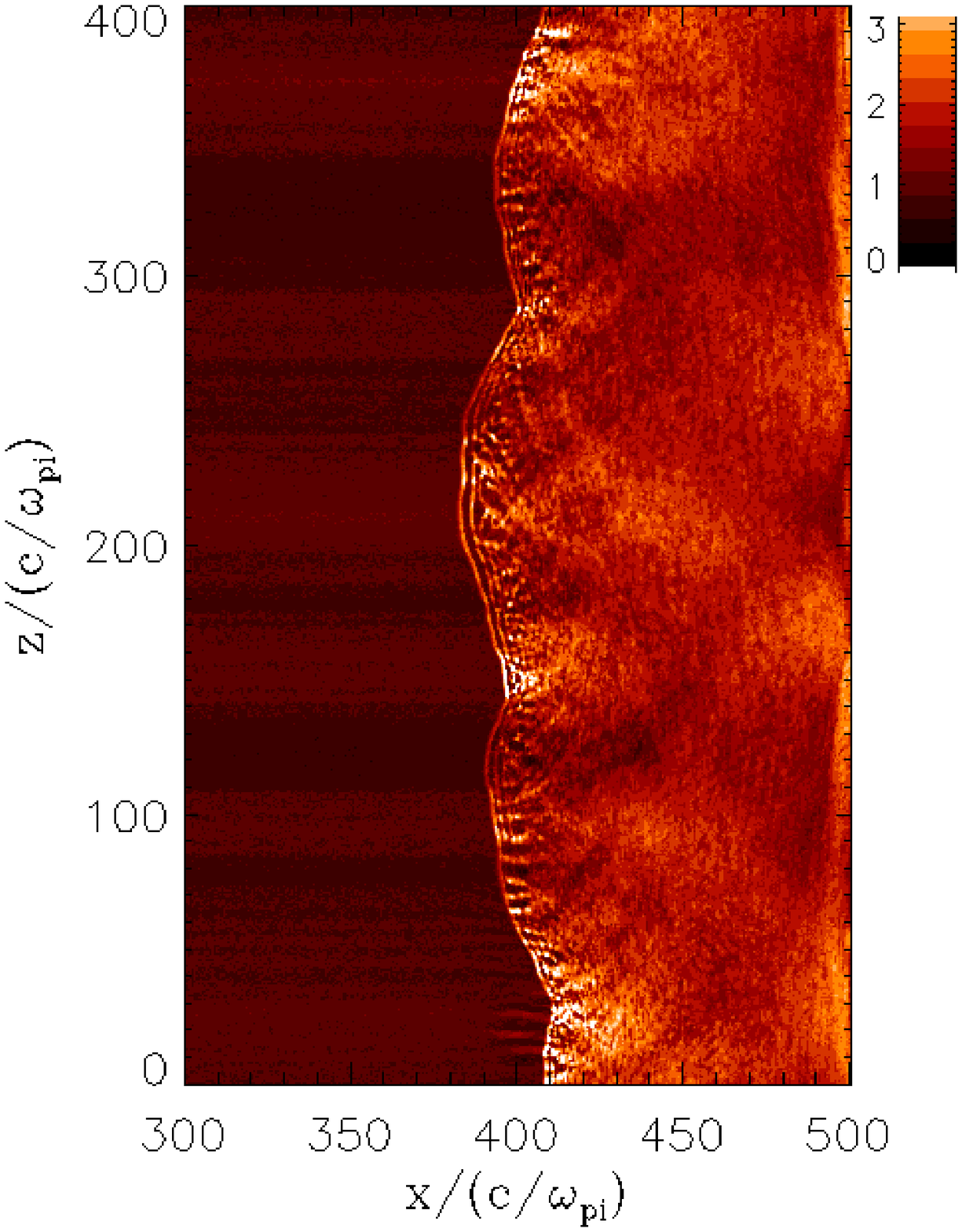,width=16pc,clip=}
\end{tabular}
\caption[The color-coded images of magnetic field $B_z$ and ion density $n_p$.]{The color-coded images of (a) magnetic field $B_z$, (b) ion density $n_p$ for run 3 at $\Omega_{ci} t=110.0$.}
\label{field}
\end{figure}

Figure \ref{field} shows the color-coded contours of (a) magnetic field in the $z$ direction $B_z$ and (b) ion number density $n_p$ from run 3 at $\Omega_{ci} t = 110.0$. The magnetic field and plasma density have been normalized using the average upstream magnetic field $B_{10}$ and in-flow density $n_0$. The averaged Alfven Mach number in the shock frame is about 2.0 and the average compression ratio is about 2.1. As noted in the earlier works \citep{Giacalone2005b,Guo2010a}, the shock surface becomes distorted due to the interaction between the shock front and the upstream turbulence. Meandering magnetic field lines cross the shock front at various locations along the shock, which allows the electrons to cross and/or get reflected at the shock front multiple times. The shock-front rippling has also been shown to contribute to particle acceleration by mirroring electrons between ripples \citep{Guo2010a}.

\begin{figure}
 \begin{center}
\includegraphics[width=0.8\textwidth]{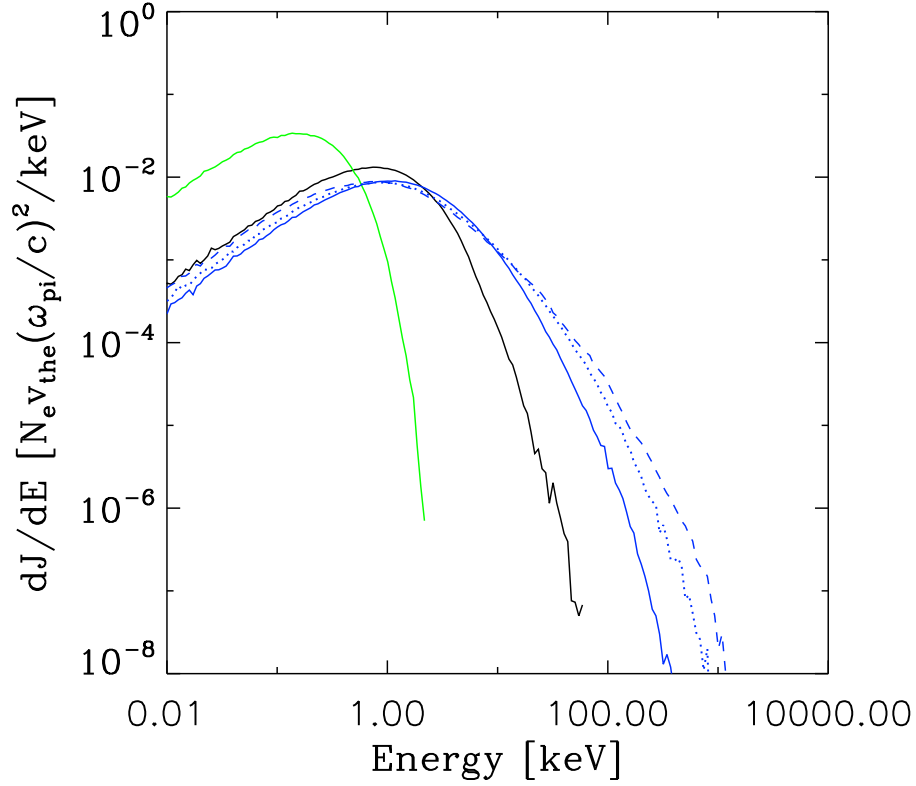}
 \caption[The energy spectra of electrons at the end of the simulation ($\Omega_i t = 130.0$) for runs 1-4.]{The energy spectra of electrons at the end of the simulation ($\Omega_i t = 130.0$). The energy spectra are normalized using $N_e v_{the} (\omega_{pi}/c)^2/keV$, where $N_e$ is the total number of electrons used in the simulations and $v_{the}$ is the initial electron thermal speed. The green solid line shows the initial distribution of thermal electrons in the upstream region. The black solid line displays the energy distribution for all the electrons in downstream region at the end of simulation for run 1.  The blue solid, dot and dashed lines represent results from runs 2, 3, and 4, respectively.}
 \label{spectrum1}
 \end{center}
 \end{figure}

 Figure \ref{spectrum1} presents the energy spectra $dJ/dE$ of electrons. The green solid line shows the initial distribution of thermal electrons in the upstream region. The black solid line displays the energy distribution for all the electrons in downstream region at the end of simulation for run 1. In this case no pre-existing fluctuation is considered and the electron energization is primarily due to heating at shock layer. We have calculated that the effective electron kinetic temperature jump in the downstream region including the superthermal distribution is about 6 times of the upstream temperature. The simulated electron temperature jump is about $40\%$ of the proton temperature jump across the shock layer in our hybrid simulation. This is consistent with the theoretical prediction that the heating of electrons in fast shocks is less than that of ions \citep{Goodrich1984,Scudder1995} and the observational constraints from measurements at planetary bow shocks and interplanetary shocks \citep{Thomsen1987,Schwartz1988}. We note that in our simulation electron heating may not be determined very accurately since the test-particle electrons have no feedback to the electric and magnetic field at the shock layer. In the following we focus on the nonthermal acceleration of electrons at shocks after considering the pre-existing magnetic fluctuations. The blue solid, dot and dashed lines in Figure \ref{spectrum1} represent energy distribution for all the electrons in downstream region at the end of simulation ($\Omega_{ci} t = 130.0$) for runs 2-4, respectively. At this time the energy spectra do not evolve anymore. It can be seen that the electrons are accelerated to high energy after considering the upstream magnetic turbulence. For higher variance of magnetic turbulence, there are more particles accelerated to high energy. For run 4 ($\delta B^2/B_0^2 = 0.3$), $9.8\%$ of electrons are accelerated to more than $15$ keV at the end of the simulation. The efficient electron acceleration can be understood as stronger magnetic turbulence allows stronger field-line meandering, and the electrons can be taken by field lines of force and therefore gain energy at a hock front multiple times.

\begin{figure}
 \begin{center}
\includegraphics[width=0.8\textwidth]{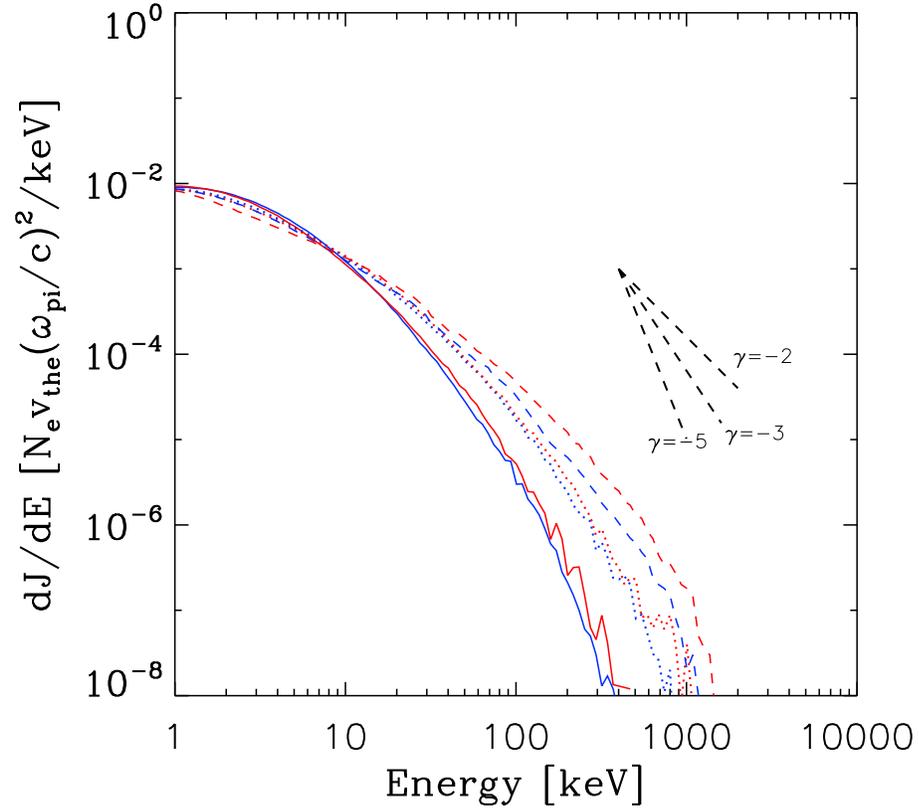}
 \caption[The energy spectra of electrons at the end of the simulation ($\Omega_i t = 130.0$) for runs 2-7.]{The energy spectra of electrons at the end of the simulation ($\Omega_i t = 130.0$). The energy spectra are normalized using $N_e v_{the} (\omega_{pi}/c)^2/keV$, where $N_e$ is the total number of electrons used in the simulations and $v_{the}$ is the initial electron thermal speed. The red solid, dot and dashed lines represent the energy distributions for all the electrons in downstream region at the end of simulation for runs 5, 6, and 7 respectively. The blue solid, dot and dashed lines represent results from runs 2, 3, and 4, respectively.}
 \label{spectrum2}
 \end{center}
 \end{figure}

 In Figure \ref{spectrum2} we examine the effect of changing the coherence length of the magnetic turbulence and focus on the high energy part of the energy spectra. It shows results from runs 5-7 (red lines, $L_z = 800 c/\omega_{pi}$) along with corresponding runs 2-4 (blue lines, $L_z = 400 c/\omega_{pi}$). It is shown that for larger coherence length, the electrons could reach higher energy and the spectral slope tends to be flatter.
The more efficient acceleration in runs 5-7 can be understood as the larger simulation domain in the direction of magnetic field allows more field line wandering normal to the shock ($\Delta x^2 \propto \Delta z$, where $\Delta x$ is the field-line random walk normal to the averaged magnetic field and $\Delta z$ is distance along the field) therefore the electrons move across the shock more easily. This dependence shows that long-wavelength fluctuations are important to accelerate electrons to high energy.

We also analyze the acceleration of protons, which are treated self-consistently in this problem (i.e., they are included in the hybrid simulation). Figure \ref{figure5} shows the differential energy spectra of protons in the shock frame for runs 2-7. Similar to Figure \ref{spectrum2}, the results from runs 5-7 are represented by red lines and the results from runs 2-4 are displayed using blue lines. The accelerated protons show a similar dependence on turbulence variance and coherence length to that of electrons. This dependence has been found previously for the case of higher Mach number shock and larger correlation length \citep{Giacalone2005b}. These results show that both electrons and protons can get efficiently accelerated. However, for the parameters we use the slopes of the energy spectra of protons are considerably steeper than that of the spectra of electrons. This is probably due to the limited temporal and spatial scales of our simulations. As shown by \citet{Giacalone2005b} using test-particle simulations, the energy spectra of protons reach the highest energy in perpendicular shock case in a longer time scale $\Omega_{ci}t \sim 50000$. However the current results from hybrid simulations do show a population of thermal protons can be accelerated to high energies in perpendicular shocks, which supports the idea that both electrons and protons can be efficiently accelerated by shocks with large shock normal angles.

\begin{figure}
\begin{center}
\includegraphics[width=0.8\textwidth]{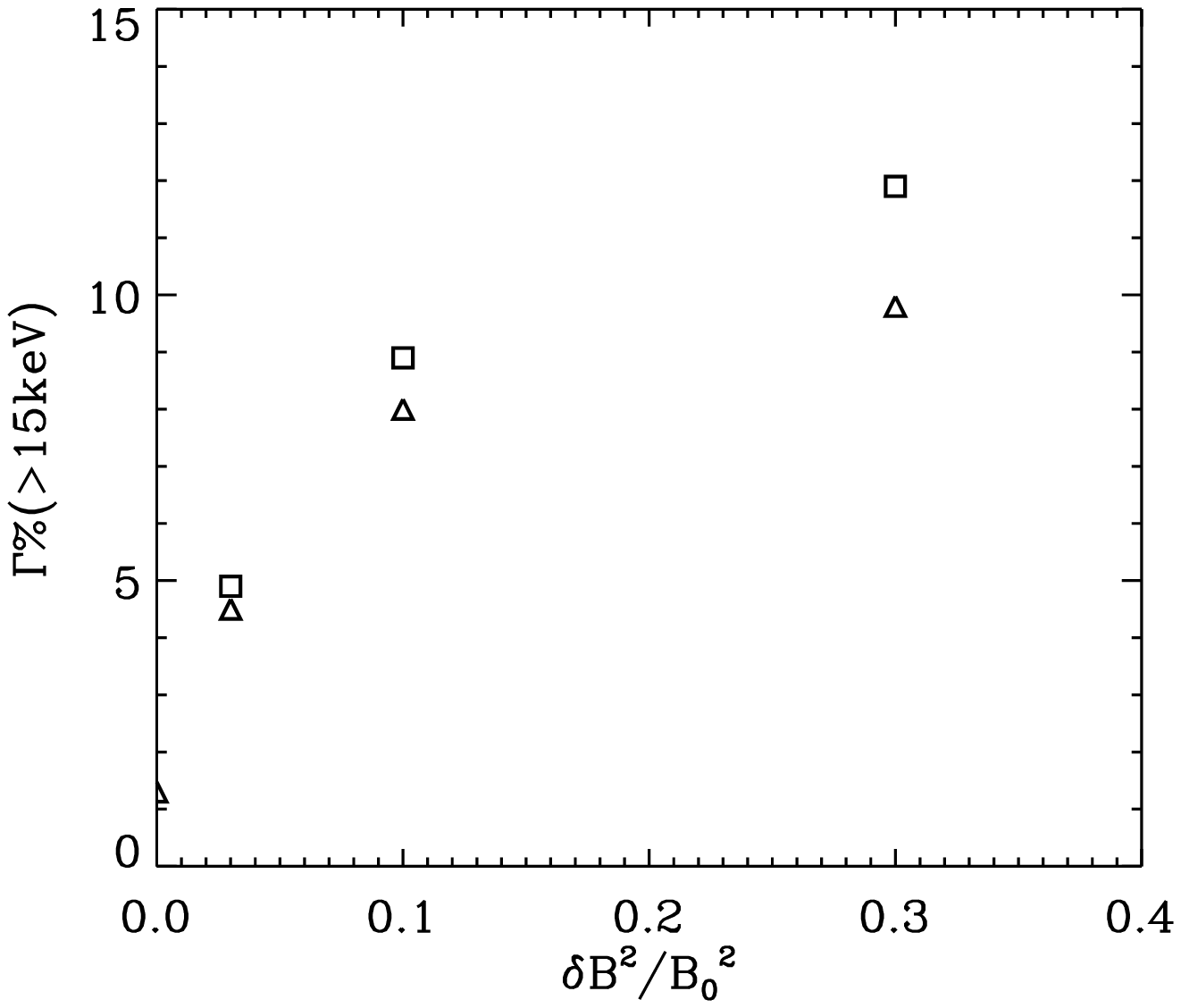}
\caption[The relation between the turbulence amplitude $\delta B^2/B_0^2$ injected in hybrid simulation and the percentage of electrons eventually accelerated to more than 15 keV.]{The relation between the turbulence amplitude $\delta B^2/B_0^2$ injected in hybrid simulation and the percentage of electrons eventually accelerated to more than 15 keV.}
\end{center}
\end{figure}

 \begin{figure}
 \begin{center}
\includegraphics[width=0.8\textwidth]{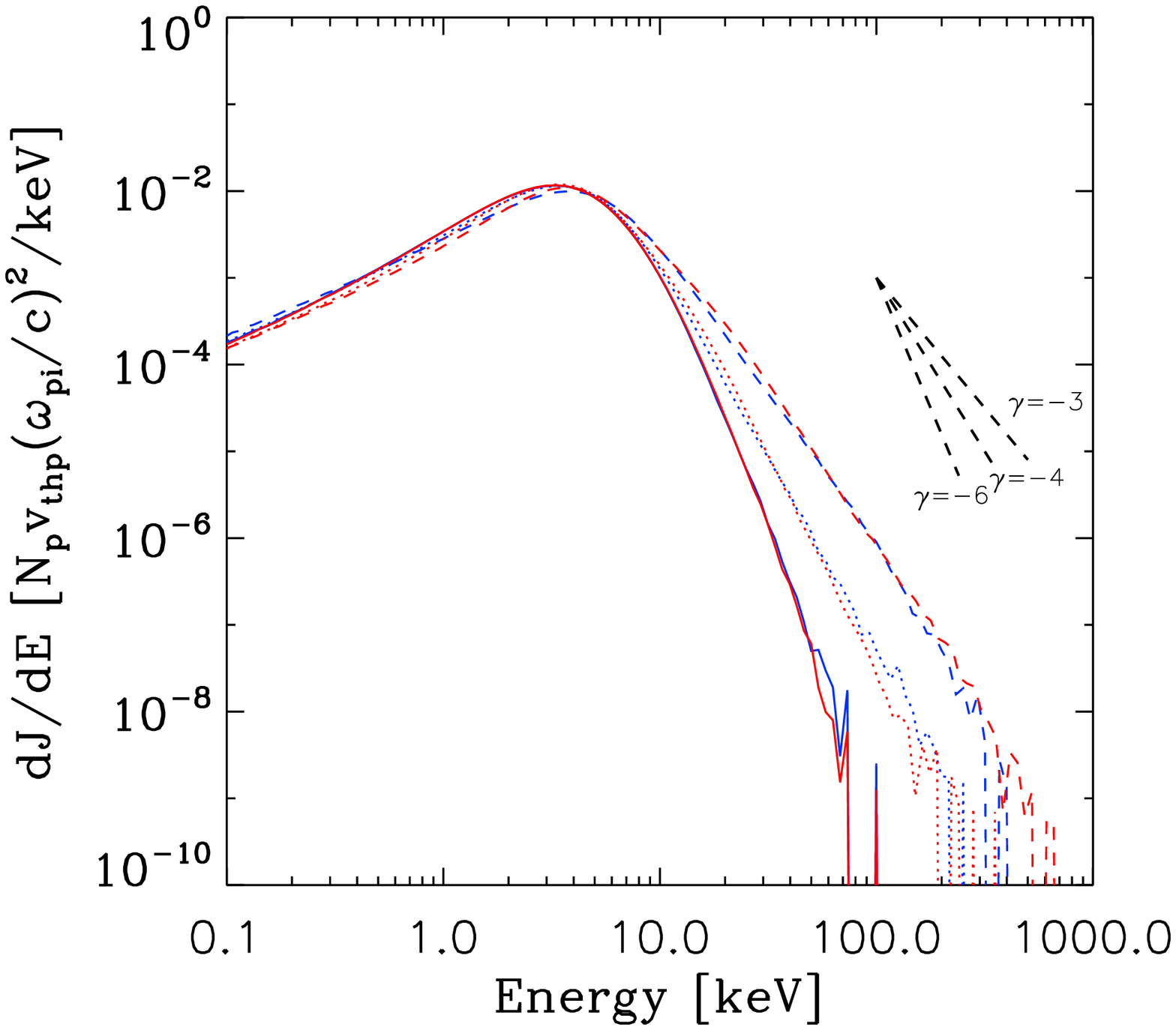}
 \caption[The energy spectra of protons downstream of the shock at the end of the simulation ($\Omega_i t = 130.0$), for runs 2-7]{The energy spectra of protons downstream of the shock at the end of the simulation ($\Omega_i t = 130.0$), normalized using $N_p v_{thp} (\omega_{pi}/c)^2/keV$, where $N_p$ is the total number of protons used to plot the spectra and $v_{thp}$ is the initial proton thermal speed.  The red solid, dot and dashed lines represent the energy spectra for protons in downstream region at the end of simulation for runs 5, 6, and 7 respectively. The blue solid, dot and dashed lines represent results from runs 2, 3, and 4, respectively.}
 \label{figure5}
 \end{center}
 \end{figure}

\section{Discussion and Conclusions \label{electron-conclusion}}

In this chapter we studied the acceleration of electrons at collisionless shocks by
utilizing a combination of a 2-D hybrid simulation to obtain the
shock structure and a test-particle simulation to determine the
motion of electrons. The hybrid simulation provides realistic
electric and magnetic fields within the transition layer of the
shock that effect the motion of test-electrons, which is determined
by solving the equation of motion. The interaction of the shock with
pre-existing upstream fluctuations, and other nonlinear processes
occuring in the hybrid simulation lead to a ``rippling" of shock
surface that also effects the transport of the electrons.
We find that the electrons are efficiently accelerated by a nearly
perpendicular shock.  The turbulent magnetic field leads to field-line
meandering that allows the electrons to cross the shock front many times. The
rippling of the shock front also contributes to the acceleration by mirroring
electrons between the ripples. This acceleration process is more efficient at
perpendicular shocks. As $\langle \theta_{Bn} \rangle$ decreases from $90^\circ$ , both the
number fraction and highest achievable energy of accelerated particles
decreases. Based on our calculations, we conclude that perpendicular shocks are
the most important for the acceleration of electrons. The current study is
helpful in understanding the injection problem for electron acceleration by
collisionless shocks. It is also found that different values of
variances and correlation lengths of the injected magnetic turbulence also strongly affect
the attainable maximum energy and accelerated fraction of electrons.
The cases with larger turbulence variances or larger turbulence correlation lengths have a 
flatter energy spectrum than that with smaller turbulence variance, which suggests the enhanced 
motion of electrons normal to the shock front, due to enhanced field-line random walk, is
of importance for the acceleration of electrons to high energies.

Here we discuss the applications of this process
to interplanetary shocks and SEP events (Section \ref{electron-conclusion-SEP}) and particle
acceleration in solar flares (Section \ref{electron-conclusion-flare}).

\subsection{Application to Interplanetary Shocks and SEP events \label{electron-conclusion-SEP}}

We presented the results of electron acceleration at shocks under parameters similar to interplanetary shocks. 
The results can also be useful for considering acceleration of electrons by shocks in SEP events.
For the case that the averaged shock normal angle $\langle \theta_{Bn}\rangle = 90^\circ$ 
and turbulence variance $\sigma^2 = 0.3$, the electrons can be readily accelerated to up to $200-300$ 
times the initial energy. The resulting spectrum is flat between about $0.1$ keV to $0.7$ keV. At
higher energies, the spectrum falls off with energy following a power law with a
spectral slope of about $-3$. 
We found that the energetic electron density
upstream and downstream of collisionless shocks show filamentary
structures (Figure \ref{electron-contour}). This could help explain electron spike-like
events observed upstream and downstream of terrestrial and
interplanetary shocks
\citep{Anderson1979,Tsurutani1985,Simnett2005}. Observation
by Voyager $1$ at the termination shock and in the heliosheath also
show the evidence of electron spike-like enhancements at the shock
front \citep{Decker2005}. The upstream spatial distribution of
energetic electrons shows irregular features that depend on both
the irregularity in the shock surface and the global topology of
magnetic field lines. At first the electrons are accelerated and
reflected at the shock front, and then they travel upstream along
the magnetic field lines. The electrons could be taken far upstream
by field line random walk. This result can possibly lead to an
interpretation to the complex electron foreshock events recently
observed to be associated with interplanetary shocks
\citep{Bale1999,Pulupa2008}.  \citet{Bale1999} and
\citet{Pulupa2008} proposed that the complex upstream electron events
result from large-scale irregularities in shock surface. In
this paper we have demonstrated that the upstream electron flux may
be controlled by both an irregular shock surface and by large-scale
meandering magnetic field lines.

We have also presented the accompanying results for protons that qualitatively show the correlation between accelerated electrons and accelerated ions in oblique shocks with large shock normal angles. The shocks can efficiently accelerate both electrons and ions. This indicates that quasi-perpendicular/perpendicular shocks play an important role in SEP events.  

\subsection{Application to Particle Acceleration in Solar Flares \label{electron-conclusion-flare}}

Understanding particle acceleration in solar flares is a challenge since only remote observations are available and it is hard to identify the main mechanism. While it is commonly thought that magnetic reconnection drives the energy release, the detailed physical process involved in accelerating the electrons and ions is still not clear. We studied electron acceleration at flare termination shocks that have been predicted by numerical simulations and flare models. We find that electrons are rapidly and efficiently accelerated at a flare termination shock in the presence of pre-existing magnetic fluctuations. Electrons are accelerated to a few MeV in $100$ ion gyroperiod (of the order of a millisecond) and more than $10\%$ of thermal electrons are accelerated to more than $15$ keV given a sufficiently strong magnetic turbulence. We also show that electron acceleration is more efficient for larger turbulence variance $\delta B^2/B_0^2$ and/or a larger turbulence coherence length $L_c$. Both of these indicate that large-scale field-line meandering plays an essential role in accelerating electrons at a shock front. Our simulations show that after considering the magnetic turbulence the flare termination shock could accelerate electrons to much higher energies than usual drift shock acceleration \citep[e.g.,][]{Mann2009}. We note that the similar mechanism has been shown to efficiently accelerate ions and has similar dependence on the turbulence properties. This correlation between ions and electrons is actually commonly observed in solar energetic particle events. For the parameters used in our simulations, the accelerated protons have energy spectra steeper than that of electrons. This is different from the previous results for parameters similar to interplanetary space \citep{Giacalone2005b,Guo2010a}. We note that these results are carried out for energies lower than the injection energy for diffusive shock acceleration. When the pitch-angles of charged particles are scattered sufficiently as to be trapped near the shock, the energy spectra of the accelerated particles are presumably close to that predicted by diffusive shock acceleration.

We also note that for the situation we study, the resulting distribution of electrons is non-Maxwellian. The structure of collisionless shocks may be considerably modified by accelerated particles. While this effect is not considered in our test-particle simulations, it may be important. Due to the limitation of computation, a full particle simulation of collisionless shock with realistic mass ratio $m_i/m_e = 1836$ that includes the influence of turbulent upstream magnetic field is not available so far. The evolution of this flare termination shock remains to be explored.
Also, other plasma effects like emission process in flare region may need to be considered to directly compare with the observations.

\chapter{Conclusions and Future Work \label{chapter5}}

\section{Summary and Conclusions for the Dissertation}

In this dissertation, we studied the effects of magnetic fluctuations on the 
acceleration and transport of charged particles in the heliosphere. We started by reviewing the basics of 
the acceleration and transport of charged particles and relevant physics in the heliosphere. We discussed the Parker's transport equation \citep{Parker1965}, which has been the fundamental equation to study the transport and acceleration of energetic charged particles. Then we focused on the limitation of the transport equation and discussed the possible solutions. Since Parker's transport equation assumes a quasi-isotropic distribution function in momentum space, in the case that the distribution function of energetic particles is highly anisotropic, the evolution of the distribution function cannot be properly described by the transport equation. The examples are the initial release of solar energetic particles and the acceleration of low-energy particles at the shock front. In addition, the observations of impulsive SEP events show fine structures in intensity-time plots on small temporal scales (hours) that cannot be easily described by a large-scale spatial diffusion.

In Chapter $2$ we presented numerical simulations for the propagation of SEPs in the inner heliosphere. We numerically integrated the trajectories of energetic charged particles in the turbulent magnetic field generated from two commonly used magnetic turbulence models (the foot-point random motion model and the two-component model). The observations of SEP events are simulated by collecting charged particles which reached $1$ AU. We study the velocity dispersion of SEPs in the turbulent magnetic field and estimate the error involved in the onset analysis. We find that the the velocity dispersion can be well produced by this model. For a typical turbulence variation $\delta B^2/B_0^2 \sim 0.1$ observed at $1$ AU and a large source region, we find that the difference between the apparent release time inferred from the onset analyses and the actual release time is less than a few minutes, but the apparent path length can be significant different than the real path length along the average magnetic field line. For the foot-point random motion model, the error for the inferred release time is smaller than that of the two-component model. We have also reproduced SEP dropouts in the numerical simulations using the foot point random motion model, assuming the SEP source region is smaller than the correlation scale. The widths of these dropout are typically several hours, similar to the time scales of dropouts observed in space. The velocity dispersion of the energetic particles appears to have different path lengths, which indicates that the energetic particles travel along different field lines. We have also attempted to use the two-component model to numerically simulate the dropouts of energetic particles. However, we rarely find the evidence of SEP dropouts in our simulation for the two-component model. This is because the parallel diffusion coefficient of particles in the two-component model is considerably smaller than that in the foot-point random model. This result questions the popular used two-component model in that it gives more pitch-angle scattering than that constrained by the observation of SEP dropouts. 

In Chapter $3$ we studied two processes for particle acceleration at shock waves. The first problem is associated with the acceleration of low-energy particles at shocks. We presented a numerical study on the acceleration of thermal protons at parallel shocks using 3-D hybrid simulations. The 3-D simulations removed the artificial restriction of the motion of charged particles in previous 1-D and 2-D simulations. The results confirmed the injection mechanism at parallel shocks that the accelerated particles are originated from reflected particles. In the second study we illustrate the effect of a
large-scale sinusoidal magnetic field variation. This simple model allows a detailed examination of the physical effects. As the magnetic field lines pass through the shock, the connection points between field lines on the shock surface will move accordingly. We find that the region where connection points approaching each other will trap and preferentially accelerate particles to high energies and form ``hot spots" along the shock surface. The shock acceleration will be suppressed at places where the connection points move apart each other. Some of the particles injected in those regions will transport to the ``hot spots" and get further accelerated. The resulting distribution function is highly spatial dependent at the energies we studied, which could give a possible explanation to the \emph{Voyager} observation of anomalous cosmic rays (ACRs). This mechanism gives an interpretation for the observation that the ACRs did not saturate at the termination shock. 

In Chapter $4$ we studied the acceleration of electrons at collisionless shocks by utilizing a combination of a 2-D hybrid simulation to obtain the shock structure and a test-particle simulation to determine the motion of electrons. We find that the electrons are efficiently accelerated by a nearly perpendicular shock when the large-scale pre-existing magnetic fluctuations are considered. The turbulent magnetic field leads to field-line meandering that allows the electrons to cross the shock front many times. The rippling of the shock front also contributes to the acceleration by mirroring
electrons between the ripples. This acceleration process is more efficient at perpendicular shocks. As $\langle \theta_{Bn} \rangle$ decreases from $90^\circ$, both the number fraction and highest achievable energy of accelerated particles
decreases. Based on our calculations, we conclude that perpendicular shocks are the most important for the acceleration of electrons. The current study is helpful in understanding the injection problem for electron acceleration by
collisionless shocks. It is also found that different values of variances and correlation lengths of the injected magnetic turbulence also strongly affect the attainable maximum energy and accelerated fraction of electrons. The cases with larger turbulence variances or larger turbulence correlation lengths have flatter energy spectra than the cases with smaller turbulence variances, which suggests that the enhanced motion of electrons normal to the shock front, due to enhanced field-line random walk, is of importance for the acceleration of electrons to high energies. We discussed the applications of this process to interplanetary shocks and SEP events (Section \ref{electron-conclusion-SEP}) and particle acceleration in solar flares (Section \ref{electron-conclusion-flare}). We also discussed the implication of this study to solar energetic particles (SEPs) by comparing the acceleration of electrons with that of protons.  The intensity correlation of electrons and ions in SEP events indicates that perpendicular or quasi-perpendicular shocks play an important role in accelerating charged particles.

\section{Future Work}

\subsection{Effect of Shock Geometry on the Acceleration of Charged Particles in Gradual SEP Events}

The acceleration of solar energetic particles (SEPs) remains to be one of the most important unsolved problems in heliospheric physics. Observations indicate that the acceleration at high energies is highly variable in spectral properties and elemental composition. Many recent works suggest that effects like
seed particles, shock geometries and/or the generation of self-excited waves can play an important role on the observed variable energy spectra \citep[e.g.,][]{Tylka2005}. 

In Section 3.5 we have considered the effect of large-scale shock geometry on the diffusive shock acceleration (DSA) of anomalous cosmic rays at the termination shock. The results show that this effect can significantly modify the well-known 1-D steady state solution of DSA. Similarly, we plan to study numerically the acceleration of particles in CME-driven shocks, including the effect of shock geometry. Since the CME-driven shocks are known to have large-scale non-planar shapes, we expect that the particles can sample shock fronts with various shock normal angles during the acceleration. This effect has not been considered by previous works, which usually assume a planar shock or no cross-field diffusion \citep[e.g.,][]{Lee2005,Tylka2005,Sandroos2007,Li2009}. This work will treat particle acceleration in CME-driven shocks in a more realistic way. 

\subsection{Understanding the Physical Processes in the Acceleration of Low-Energy Particles at Shocks}

Solving the acceleration of energetic particles in the heliosphere requires the knowledge of the acceleration of low-energy particles. Since the DSA is not concerned with low-energy particles that has high anisotropies, most of DSA models do not include the acceleration of low-energy particles or treat it in a \textit{ad hoc} way \citep[e.g.,][]{Ellison1981,Ellison1990}. To fully understand the observed energetic particles, we need to model the acceleration of low-energy protons, electrons, and also heavy ions.

a) Recent observations have indicated that the pre-accelerated particles from solar flares or other source may be important in acceleration of SEPs. In Chapter 3 we have used 3-D hybrid simulations to study the acceleration of low-energy protons at collisionless shocks. We plan to further explore the injection problem using the self-consistent hybrid simulations including the seed particle population pre-accelerated in solar flares or other sources. 

b) We also plan to use full particle simulations (kinetic ions and kinetic electrons) to study the acceleration of electrons at collisionless shocks. In Chapter 4 we have shown that low-energy electrons can be efficiently accelerated when a shock propagates normal to a magnetic field that contains pre-existing large-scale magnetic turbulence. The simulation model include a combination of a hybrid model (kinetic ions and fluid electron) and a test-particle electron model. Since the model does not consider the electron-scale plasma physics and the feedback of accelerated electrons on the shock front, we need to use a full particle simulation model to consider the effects in order to fully solve the problem of acceleration of electrons at collisionless shocks. This work will consider the interplay between large-scale effects (large-scale magnetic fluctuations and shock ripples) and small-scale effects (whistler waves at shock fronts).

\subsection{The Application of the Hybrid Simulations in the Heliospheric Plasma Processes}
During the doctoral study, we have improved the parallelization of the 1-D, 2-D and 3-D hybrid simulation models \citep{Giacalone2000,Giacalone2004,Giacalone2005b}. The new version of the code has good scalability and has been 
tested on NASA's Pleiades supercomputer using a few thousand CPU cores. The high performance of the code on supercomputers allows us to study a variety of problems using multi-dimensional simulations or 1-D simulations with large simulation boxes and long simulation times. While we have presented some results using the new code, plenty of other plasma physics problems in the heliosphere can be studied using this numerical tool.
We have plans to study numerically the ion cyclotron waves excited by fresh pickup ions \citep{Florinski2010} using 2-D hybrid simulations, plasma instabilities and waves upstream and downstream of shocks, and the acceleration of charged particles by multiple shocks \citep{Li2012}. 

\renewcommand{\baselinestretch}{1}		
\small\normalsize						

\bibliographystyle{apj}

\end{document}